%% file: main.tex
\documentclass[twocolumn]{article}
\usepackage{amssymb}
\usepackage{amsmath}
\usepackage{xcolor}
\usepackage{graphicx} 
\usepackage{float}
\usepackage{url}
\usepackage{booktabs}
\usepackage{multirow}
\usepackage{placeins}
\usepackage{siunitx}
\usepackage{hyperref}

\makeatletter
\long\def\@makecaption#1#2{%
  \vskip\abovecaptionskip
  \footnotesize
  \sbox\@tempboxa{#1. #2}%
  \ifdim \wd\@tempboxa >\hsize
    #1. #2\par
  \else
    \global \@minipagefalse
    \hb@xt@\hsize{\hfil\box\@tempboxa\hfil}%
  \fi
  \vskip\belowcaptionskip}
\renewcommand\section{\@startsection{section}{1}{\z@}%
  {-2.2ex \@plus -0.8ex \@minus -.2ex}%
  {0.8ex \@plus .1ex}%
  {\normalfont\normalsize\bfseries\raggedright}}
\renewcommand\subsection{\@startsection{subsection}{2}{\z@}%
  {-1.8ex \@plus -0.7ex \@minus -.2ex}%
  {0.5ex \@plus .1ex}%
  {\normalfont\small\itshape\raggedright}}
\renewcommand\subsubsection{\@startsection{subsubsection}{3}{\z@}%
  {-1.5ex \@plus -0.6ex \@minus -.2ex}%
  {0.4ex \@plus .1ex}%
  {\normalfont\small\itshape\raggedright}}
\def\@listi{\leftmargin\leftmargini
            \parsep 0pt
            \topsep 2pt
            \itemsep 1pt}
\let\@listI\@listi
\def\@listii{\leftmargin\leftmarginii
             \parsep 0pt
             \topsep 1pt
             \itemsep 1pt}
\makeatother
\input{mymacros}

\title{Yinsen: A low power density HTS tokamak fusion reactor for marine and off-grid applications}

\author{}

\newcommand{\paperabstract}{Yinsen is a high-temperature-superconducting (HTS) tokamak reactor concept intended for off-grid applications such as maritime propulsion, remote power, and industrial energy supply. Rather than pursuing economic grid-scale power density, the design is anchored to a materials-limited fusion power density of \mbox{$P_f/S_b=0.7~\mathrm{MW/m^2}$}, obtained from a \mbox{$35~\mathrm{DPA}$} structural limit, a \mbox{$20$-year} plant lifetime, \mbox{$40\%$} utilization, and a first-order geometric damage-peaking correction. The resulting device design has a V-4Cr-4Ti vacuum vessel lifetime of \mbox{$1040~\mathrm{MW\cdot yr}$}. This points to a minimum useful fusion-power of \mbox{$130~\mathrm{MW}$}, sufficient to exceed \mbox{$25~\mathrm{MWe}$} net output for propulsion after realistic recirculating loads are included. Integrated \textsc{FUSE} modeling refines the design into a self-consistent high-field baseline with a shaped \mbox{$9.29~\mathrm{T}$}, \mbox{$9.67~\mathrm{MA}$} plasma, while also showing a broad operating window that extends into higher fusion-power regimes than the minimum \mbox{$130~\mathrm{MW}$} design point and is corroborated by further detailed transport analysis using Astra. The next most limiting attribute associated with next-step fusion devices, power handling in the divertor, is addressed with UEDGE modeling. It is shown that impurity-seeded detached operation is readily attainable with modest neon concentration, reducing peak heat fluxes well below \mbox{$10~\mathrm{MW/m^2}$}. OpenMC neutronics calculations performed with a double-layered WC/W$_2$B$_5$ shield show that the vacuum vessel is the lifetime-limiting solid structure, while all HTS magnets remain comfortably lifetime components: at the \mbox{$130~\mathrm{MW}$} baseline, total TF nuclear heating is only \mbox{$7.4~\mathrm{kW}$} at \mbox{$20~\mathrm{K}$}, and the TF fast-neutron limit corresponds to roughly sixteen vacuum-vessel lifetimes. The same neutronics analysis gives \mbox{$TBR\approx1.1$} with \mbox{$30\%$} \mbox{$^6\mathrm{Li}$} enrichment and no dedicated neutron multiplier. Plant-level studies detail a supercritical CO$_2$ balance of plant and pulsed-power operation for off-grid deployment, using a \SI{34}{kV} medium-voltage backbone with solid-state transformers and local energy storage to manage pulsed magnet transients. Additional plasma physics attributes including pedestal operational regime, ICRH minority scheme, and disruption loads are presented, along with magnet and fuel cycle engineering analysis and a techno-economic study for projected overnight CapEx and fusion-powered shipping economic competitiveness. Taken together, these results suggest that a deliberately low-power-density HTS tokamak, where the vacuum vessel is not designed to be replaced often, offers a near-term path for commercially relevant FOAK fusion reactors where many of the remaining challenges between $Q>1$ and economic grid operation are alleviated.}

\begin{document}

\twocolumn[
\begin{center}
{\LARGE\bfseries Yinsen: A low power density HTS tokamak fusion reactor for marine and off-grid applications\par}

\vspace{1.5em}

{\footnotesize Justin Cohen\textsuperscript{a}, Jason Kaufmann\textsuperscript{a}, Maxim Umansky\textsuperscript{a}, Adriana Ghiozzi\textsuperscript{a}, David Smithe\textsuperscript{a}, Oak Nelson\textsuperscript{b}, Benedikt Zimmermann\textsuperscript{b}\par}

\vspace{1.5em}

{\scriptsize \textsuperscript{a}Maritime Fusion \qquad \textsuperscript{b}Columbia University\par}

\vspace{0.55em}
\end{center}
\vspace{0.1em}
\noindent\rule{\textwidth}{0.4pt}
\vspace{0.1em}
\noindent{\large\bfseries Abstract}\par
\vspace{0.05em}
\noindent\paperabstract\par
\vspace{0.1em}
\par
\vspace{0.1em}
\noindent\rule{\textwidth}{0.4pt}
\vspace{0.15em}
]

\input{intro}
\input{plasma}

\input{engineering}
\input{conclusions}
\clearpage

\input{references}
\clearpage
\appendix
\onecolumn
\input{appendix_pulsed_power.tex}
\input{appendix_facility.tex}
\input{appendix_direct_capex.tex}
\clearpage
\twocolumn
\input{appendix_shipping}
\end{document}

%% file: mymacros.tex
\newcommand{\diff}[2]%
{
  \frac{d#1}{d#2}
}
\newcommand{\diffs}[2]%
{
  \frac{d^2#1}{d#2^2}
}
\newcommand{\pdiff}[2]%
{
  \frac{\partial#1}{\partial#2}
}
\newcommand{\pdiffs}[2]%
{
  \frac{\partial^2#1}{\partial#2^2}
}
\newcommand{\pdiffx}[3]%
{
  \frac{\partial^2#1}{\partial#2 \partial#3}
}

\newcommand{\tdiff}[2]%
{
  {d#1}/{d#2}
}
\newcommand{\tdiffs}[2]%
{
  {d^2#1}/{d#2^2}
}
\newcommand{\tpdiff}[2]%
{
  {\partial#1}/{\partial#2}
}
\newcommand{\tpdiffs}[2]%
{
  {\partial^2#1}/{\partial#2^2}
}
\newcommand{\tpdiffx}[3]%
{
  {\partial^2#1}/{\partial#2 \partial#3}
}

\def\beq{\begin{equation}}
\def\eeq{\end{equation}}
\def\beqar{\begin{eqnarray}}
\def\eeqar{\end{eqnarray}}

\def\gsim{{\lower.3em\hbox{${\>\buildrel > \over \sim\>}$}}}
\def\lsim{{\lower.3em\hbox{${\>\buildrel < \over \sim\>}$}}}

\newcommand{\vect}[1]%
{
  {\bf #1}
}

%% file: intro.tex
\section{Introduction and Design Motivation}\label{sec:introduction}

For most of its history, fusion energy was pursued primarily as a long-horizon scientific mission led by national laboratories and universities. That landscape is now changing: after decades of incremental progress, the field is entering a phase in which commercial programs are beginning to frame fusion not only as a physics challenge, but as an engineering product with a visible path to deployment. Magnetic confinement reached key milestones in the 1990s, when TFTR and JET established deuterium--tritium operation, produced multi-megawatt fusion power, and demonstrated that burning-plasma-relevant conditions were within reach \cite{Hawryluk1998,Keilhacker1999}. In parallel, inertial confinement has recently crossed its own historic threshold, with the National Ignition Facility demonstrating target gain larger than unity in the laboratory \cite{AbuShawareb2024}. With the maturation of high-temperature superconductors, magnetic confinement is now approaching a comparable turning point: compact high-field tokamaks such as SPARC, the first tokamak designed from the outset to fully leverage HTS magnet technology, are explicitly aiming to demonstrate $Q>1$ in a commercially motivated program \cite{Creely2020}. Yet the transition from scientific breakeven to economically competitive, reliable grid-scale electricity may prove to be a larger leap than breakeven itself. Producing a plasma with net energy gain is fundamentally different from operating a plant that can survive high power density, manage neutron damage, sustain repeated pulses or long campaigns, and deliver high availability over decades at an acceptable cost.

That gap is especially acute for the leading compact HTS tokamak approach. The same combination of strong magnetic field and reduced machine size that improves plasma performance and lowers the cost of the magnetic system also intensifies the hardest remaining materials and engineering challenges. In order to recover the capital cost of a complex nuclear plant, economically aggressive compact tokamak concepts generally require high blanket-area-normalized fusion power density, \(P_f/S_b \gtrsim 5~\mathrm{MW/m^2}\). At those loadings, the reactor must simultaneously accommodate extreme divertor power handling, with hundreds of megawatts of power concentrated into the divertor region, high neutron power loading, and rapid cumulative irradiation damage to solid in-vessel structures. While ITER and related facilities will provide critical validation of plasma-facing and structural materials at fluences of order \(0.3~\mathrm{MW\cdot year/m^2}\) \cite{ITER2001}, sustaining multi-\(\mathrm{MW/m^2}\) neutron loading continuously for many years remains well beyond what has been demonstrated. In that regime, even optimistic assumptions on structural survivability imply repeated replacement of major blanket and in-vessel components, making maintenance outages a central determinant of both plant utilization and economic viability.

Yinsen is intentionally framed around a different question: with presently known and viable materials, what power density can a first-of-a-kind (FOAK) fusion power system achieve, rather than what operating point would be required for immediate grid competitiveness? The baseline Yinsen concept is therefore sized around a technologically sustainable fusion power density and a design philosophy in which the major solid blanket structures and primary vacuum vessel are not replaced over the nominal plant lifetime. This sharply reduces dependence on elaborate maintenance architectures involving disassembly of tightly integrated HTS magnet systems, cutting and rewelding vacuum-vessel structures, and routine handling of large masses of highly activated in-vessel hardware. Even when such material is classified as low-level radioactive waste rather than the much longer-lived waste streams associated with fission, its removal, storage, and replacement still impose major cost, schedule, and availability penalties. That choice does handicap conventional grid-centric metrics such as levelized cost of electricity, and Yinsen is not presented as a first-of-a-kind baseload grid plant. Instead, it is aimed at markets where tens of megawatts of carbon-free power, combined with independence from oil-price volatility, geopolitical fuel risk, and weather variability, can command value beyond a simple grid comparison. Candidate applications include marine propulsion, ship-to-shore power, remote industrial energy supply, high-grade heat for chemical processing, hydrogen or ammonia production, and power-hungry facilities such as data centers. In these settings, levelized cost of electricity in isolation is rarely the sole determinant of technology choice; the intrinsic attributes of the energy source; dispatchability, fuel security, islanded operation, and decarbonized energy delivery often play the dominant role in adoption.

\subsection{Lifetime Solid Structure Limited Power Density}

The key design variable in this strategy is the blanket-area-normalized fusion power density \(P_f/S_b\), where \(S_b\) is the blanket-facing surface area of the torus. The central question is not what fusion power density the plasma can momentarily achieve in terms of confinement and stability, but how much fusion power density the solid in-vessel structures can tolerate over the full plant life without forcing major replacement campaigns. Treating structural damage as a power-flux-integrated quantity, and introducing a first-order geometric peaking factor to map blanket-area-averaged loading to the limiting local damage rate, the lifetime damage constraint may be written as
\begin{equation}
0.8 \cdot \left(\frac{P_f}{S_b}\right) \cdot F \cdot f_{\mathrm{pk}} \cdot U \cdot T_{\mathrm{lifetime}}
\leq L_{\mathrm{dpa}},
\end{equation}
where the factor of \(0.8\) accounts for the fraction of D--T fusion power carried by neutrons, \(F\) is the conversion factor between neutron power fluence and accumulated damage in the solid structures \cite{Whyte2024FusionEconomics}, \(f_{\mathrm{pk}}\) is a peak-to-average damage factor, \(U\) is the lifetime-average utilization (average delivered power over plant lifetime divided by nameplate power), \(T_{\mathrm{lifetime}}\) is the plant lifetime, and \(L_{\mathrm{dpa}}\) is the allowable cumulative damage limit, represented by a displacements-per-atom metric. Rearranging gives the corresponding upper bound on sustainable fusion power density, as limited by nuclear damage to solid structures,
\begin{equation}
\frac{P_f}{S_b}
\leq
\frac{L_{\mathrm{dpa}}}
{0.8 \cdot F \cdot f_{\mathrm{pk}} \cdot U \cdot T_{\mathrm{lifetime}}}.
\label{eq:materials_power_density_limit}
\end{equation}

Because Eq.~\eqref{eq:materials_power_density_limit} is written in terms of blanket-area-averaged fusion power density, a first-order correction is needed to estimate the limiting local structural damage. A blanket strip of poloidal arc length \(dl\) has toroidal area \(dA = 2\pi R\,dl\), so if the intercepted neutron power per unit poloidal length is approximately uniform to first order, the local neutron wall loading scales as \(q_n \propto dP_n/dA \propto 1/R\). Relative to the blanket-area average, this gives a peak-loading factor
\[
f_{\mathrm{pk}} \equiv \frac{q_{n,\max}}{\langle q_n \rangle} \approx \frac{R_0}{R_0-a} = \frac{1}{1-\epsilon}
\]
where \(\epsilon = a/R_0\) is the inverse aspect ratio. Because Yinsen is ultimately aimed at a moderately compact advanced-tokamak regime, motivated by improved confinement and higher bootstrap-current fraction, we adopt a representative inverse aspect ratio of \(\epsilon \approx 0.33\) at this scoping stage, which gives \(f_{\mathrm{pk}} \approx 1.5\). This geometric estimate is intended only as a first-order correction; the exact peak damage also depends on source shaping, shielding distribution, and local material stackup, all of which are captured in the detailed neutronics model.

Equation~\eqref{eq:materials_power_density_limit} is therefore the governing first-order design relation for Yinsen. It makes clear that improving economic competitiveness at fixed lifetime requires either reducing the non-recoverable damage per unit neutron energy fluence \(F\), reducing the damage peaking factor \(f_{\mathrm{pk}}\), increasing the allowable damage limit \(L_{\mathrm{dpa}}\), or accepting lower utilization and more frequent maintenance. Here we use DPA as the fundamental lifetime metric for the major solid structures, but it is intended as a proxy for the broader suite of irradiation-driven degradation mechanisms that ultimately bound structural life, including helium production and embrittlement, grain-boundary weakening, weld degradation, swelling, and other unfavorable changes in material properties under neutron exposure. In that sense, the tradeoffs exposed by Eq.~\eqref{eq:materials_power_density_limit} are fundamentally materials and maintenance and operations challenges, not plasma-confinement limitations. This is why high-power-density compact grid-fusion concepts remain challenging: they attempt to maximize revenue by driving \(P_f/S_b\) upward in the same regime where blanket lifetime, maintenance interval, activated waste generation, and availability all deteriorate rapidly in the pursuit of that higher power density.

\begin{figure}[t]
\centering
\includegraphics[width=\columnwidth]{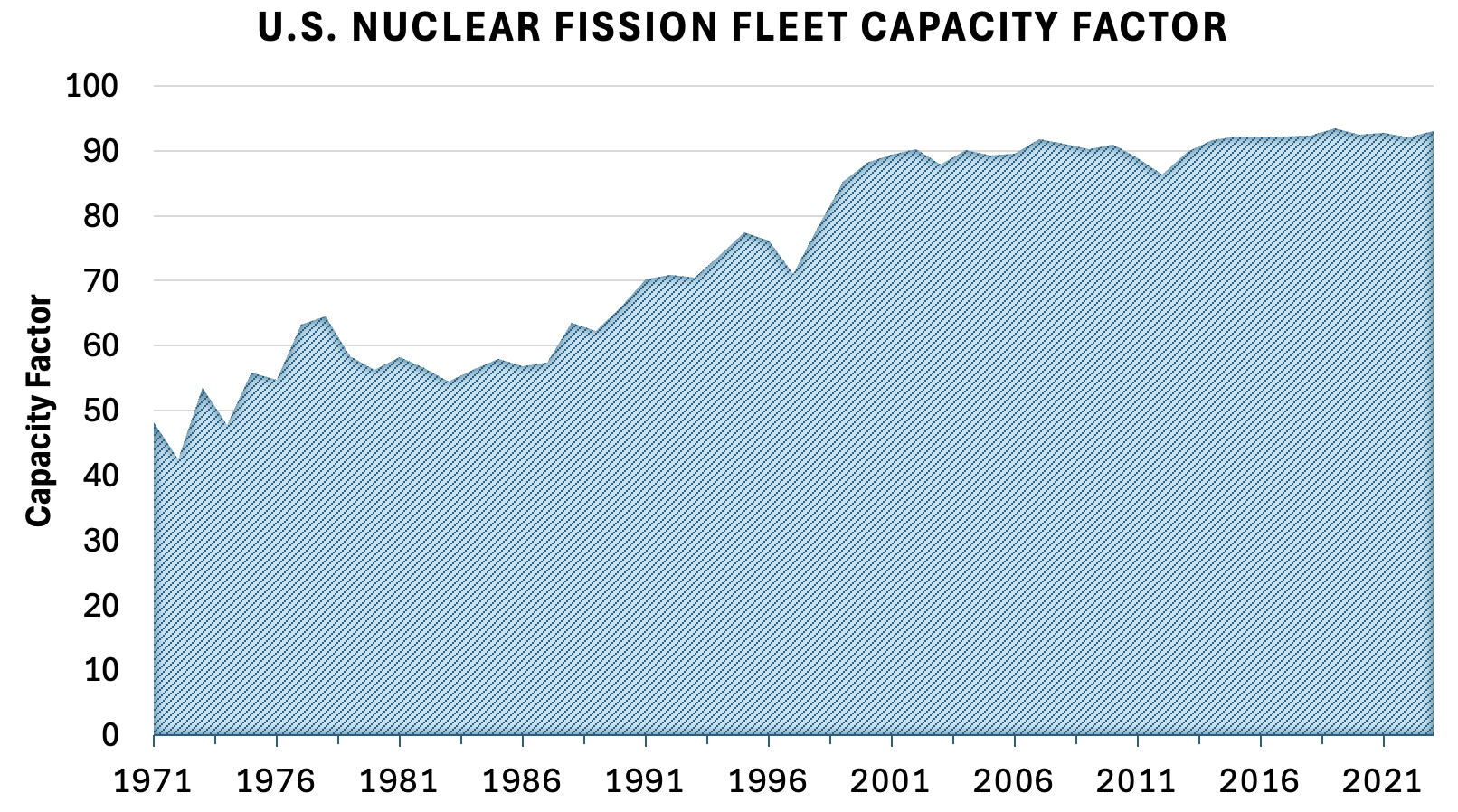}
\caption{Historical U.S.\ nuclear capacity factor \cite{NEIUSNuclearGeneratingStatistics}, used to inform the FOAK utilization floor.}
\label{fig:nppCF}
\end{figure}

For Yinsen, we therefore impose a utilization floor by looking to the early history of its closest technological cousin, nuclear fission, which underwent widespread buildout and adoption in the United States beginning in the 1970s. Figure~\ref{fig:nppCF} shows the historical learning curve of the U.S.\ nuclear fleet from Nuclear Energy Institute statistics \cite{NEIUSNuclearGeneratingStatistics}, whose capacity factor rose from the initial \(40\%\)--\(60\%\) range of the first commercial grid reactors in the 1970s and 1980s to roughly \(90\%\) only after decades of industrial maturation. Civilian grid nuclear power also did not emerge from a blank slate: long before commercial plants were selling electricity to customers, fission power cores had already been developed for defense applications, with the first nuclear submarine, USS \emph{Nautilus}, entering service in 1954, years before Shippingport began civilian grid operation in the late 1950s \cite{USSNautilusHistory,ShippingportDOE}. Even with that head start, fission still required decades to converge toward mature-fleet utilization. It would be unreasonable to expect first-of-a-kind fusion plants with cryogenic superconducting magnets, tighter tolerances on large integrated structures, high-power RF systems, large magnet power supplies, and more demanding high-temperature balance-of-plant systems to climb that learning curve any faster without an Apollo program or Manhattan Project like campaign. Yinsen therefore adopts a minimum viable utilization floor of \(U_{\min}=0.40\) as a lower bound for a commercially relevant FOAK system.

The other critical material parameter informing achievable power density is \(F\), the conversion between neutron power fluence and accumulated solid-structure damage. In the present baseline this value is informed by neutronics calculations (see Section~\ref{sec:neutronics}) and benchmarked against the literature for vanadium-alloy structural materials. For the main vacuum-vessel structure, Yinsen adopts V--4Cr--4Ti, largely motivated by its tolerance for radiation damage and low activation \cite{Pettinari2025}. In this context, a working value of \(F = 5~\mathrm{DPA/(MW\cdot yr / m^2)}\) is chosen.

\begin{figure}[t]
\centering
\includegraphics[width=\columnwidth]{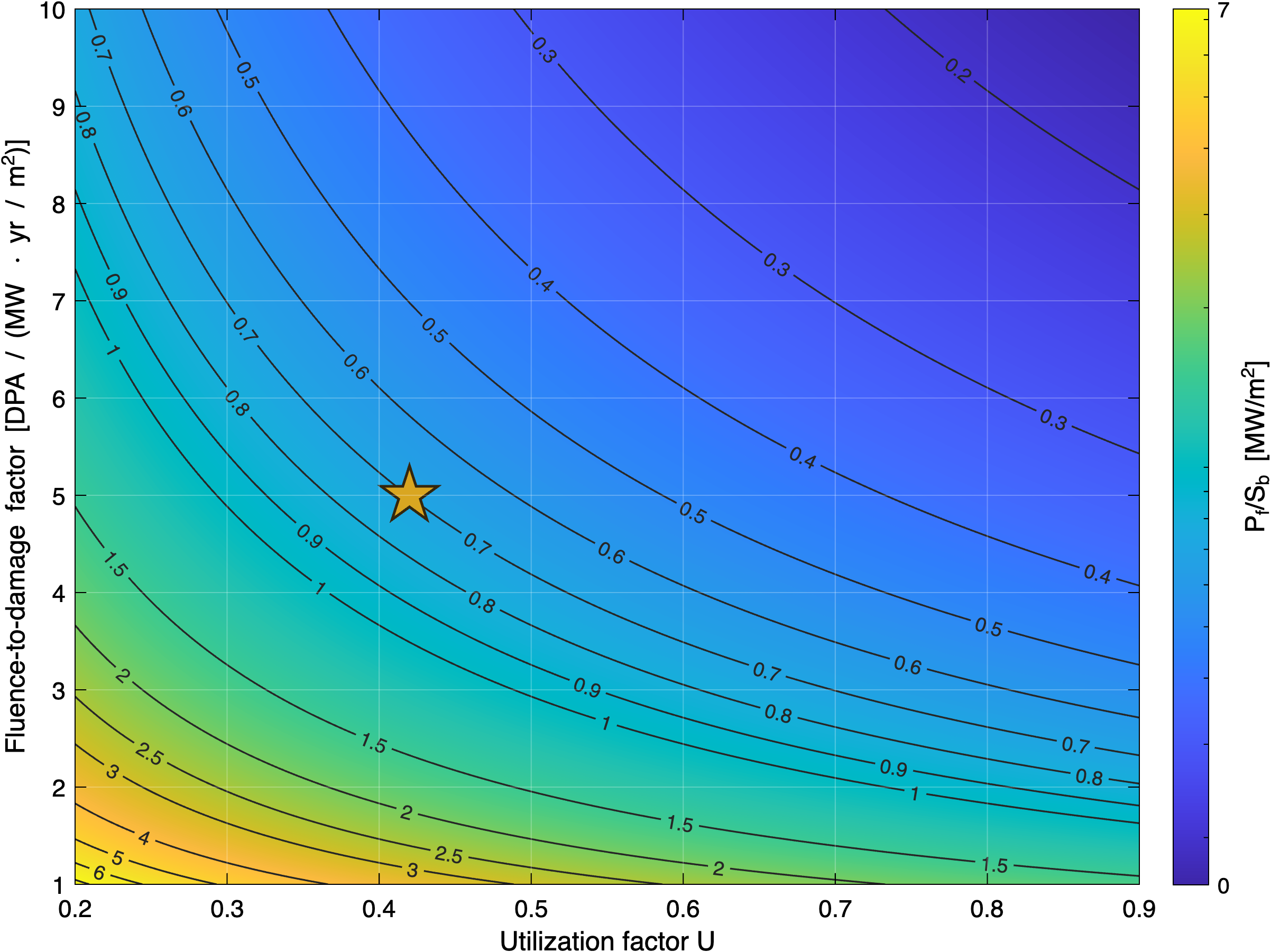}
\caption{Damage-limited upper bound on blanket-area-normalized fusion power density over assumed utilization and fluence-to-damage conversion ranges. The Yinsen baseline adopts \(P_f/S_b = 0.7~\mathrm{MW/m^2}\) as a rounded FOAK ceiling consistent with lifetime structural survivability rather than short-interval blanket replacement.}
\label{fig:tech_power_density}
\end{figure}

Taking a limiting lifetime nuclear damage of \(L_{\mathrm{dpa}} = 35\) before structural performance is degraded beyond acceptable margins for continued safe nuclear operation \cite{ORNLFusionMaterialsProgress25,Chung1995V4Cr4Ti,Chung1995MicrostructuralEvolution,Butt2025VanadiumReview}, further adopting a total facility lifetime of \(T_{\mathrm{lifetime}} = 20~\mathrm{yr}\), and using the representative geometric peaking factor \(f_{\mathrm{pk}} \approx 1.5\), allows determination of a FOAK fusion power density limited by structural component lifetimes. Under those assumptions, Yinsen lands at a fusion power density of \(P_f/S_b = 0.7~\mathrm{MW/m^2}\). Figure~\ref{fig:tech_power_density} shows the sensitivity of the achievable power density to both the energy-fluence-to-damage conversion factor \(F\) and the utilization factor \(U\), assuming a lifetime limit of \(35~\mathrm{DPA}\), a \(20\)-year plant lifetime, and the representative geometric peaking factor described above.

By sizing the plant around a low enough \(P_f/S_b\) to preserve the primary in-vessel solid structures over the plant lifetime, Yinsen avoids making major in-vessel replacement a regular operating requirement. That directly simplifies subsystem design by reducing dependence on elaborate maintenance architectures involving disassembly of tightly integrated HTS magnet systems, cutting and rewelding vacuum-vessel structures, and routine handling of large masses of highly activated in-vessel hardware. Even when such material is classified as low-level radioactive waste rather than the much longer-lived waste streams associated with fission, its removal, storage, and replacement still impose major cost, schedule, and availability penalties over regulatory timescales extending to centuries. Historically, even moderately activated and tritiated devices such as JET and TFTR required substantial remote-handling, tritium-processing, and logistical overhead to maintain the facilities in operable condition \cite{JETRemoteHandling1999,TFTRTritiumMgmt1995}. We therefore view any campaign to swap major blanket and vessel structures as ``open-heart surgery'' for the reactor: a demanding intervention for which there is unlikely to be a true practice device before commercial deployment. This contrasts with fission, where the analogous task of fuel handling and refueling could be learned incrementally on smaller and earlier systems before becoming routine on commercial plants.

\begin{figure*}[t]
\centering
\includegraphics[width=0.88\textwidth]{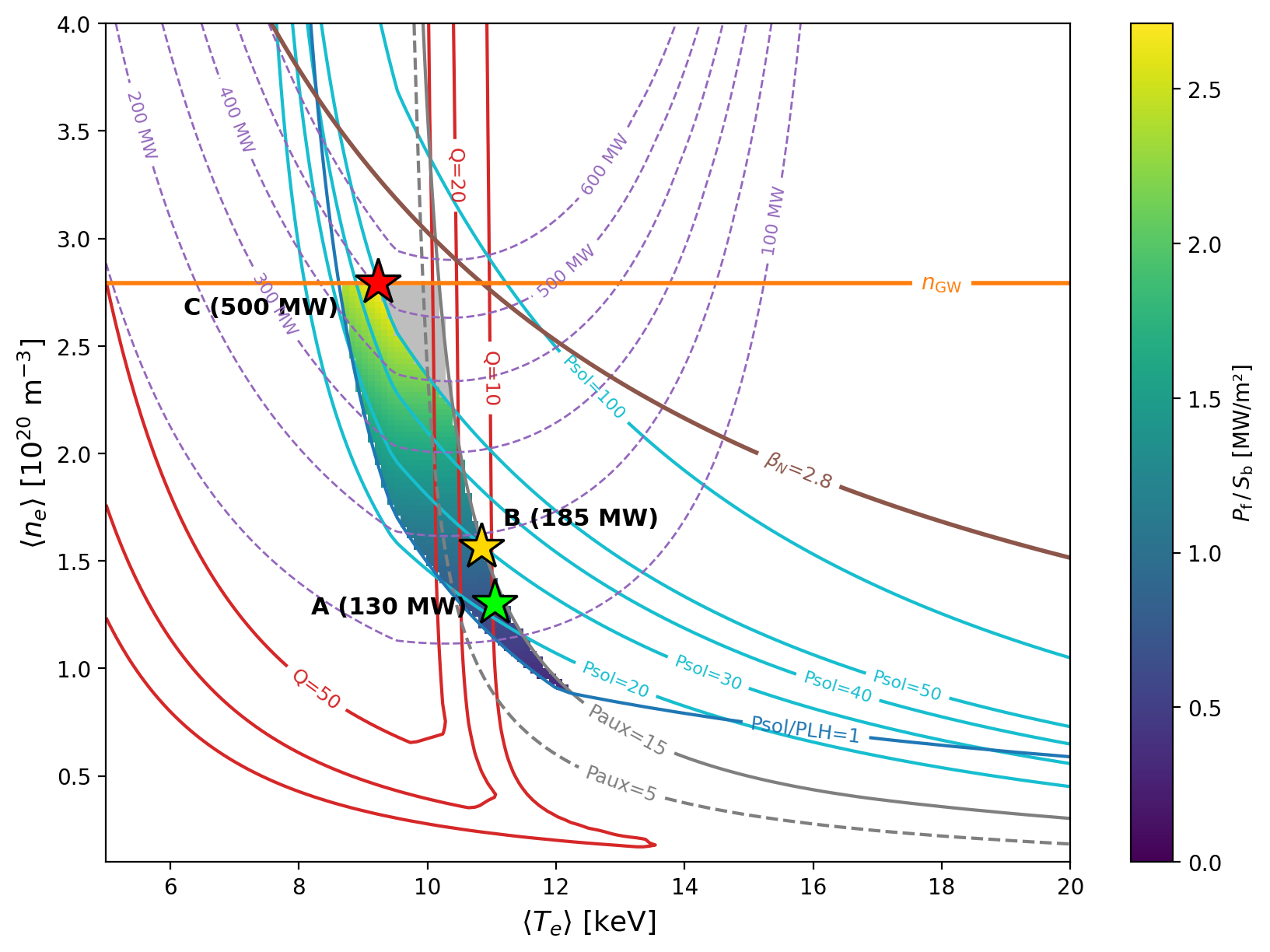}
\caption{POPCON-style operating-space reference for the Yinsen plasma class, with transport informed profiles. The figure shows that the machine can access higher-power-density operating points in principle, but the baseline Yinsen design remains anchored by the cumulative damage limit of the solid blanket and vessel structures rather than by an immediate plasma-operability boundary.}
\label{fig:power_density_popcon}
\end{figure*}

Recognizing that the first Yinsen device is intended to operate for at least \(20\) years, it is also acknowledged that new irradiation-tolerant structural materials may be discovered and industrialized over that operating lifetime, potentially pushing the admissible power density upward by either lowering \(F\) or raising \(L_{\mathrm{dpa}}\). We also note that structural damage is a power-flux-integrated quantity: brief periods of higher power-density operation are manageable with the same machine if one is willing to spend blanket and vessel lifetime more quickly. In that sense, the practical limitation is not immediate operability but cumulative structural burn rate. The operational flexibility of that higher-power-density regime is illustrated by the POPCON map in Figure~\ref{fig:power_density_popcon}. The point of this figure is not that the machine should be operated there continuously, but that the plasma operating space itself does not prevent higher loading; rather, the binding baseline limitation is the long-term survivability of the major in-vessel solid structures. We therefore also analyze the more aspirational ``Case~B'' at \(1~\mathrm{MW/m^2}\) and consider sizing auxiliary systems for higher fusion power. In the case where more advanced in-vessel materials become industrialized during Yinsens operational lifetime, the intent would be to perform such an upgrade at most once over the plant lifetime, such that the baseline design still does not require an easily swappable vacuum vessel. Section~\ref{sec:facility_layout} outlines the corresponding horizontal maintenance concept, in which the vacuum vessel is cut in half, the halves are slid apart, and each half is rotated out from its nine-TF-coil cage while submerged in water to reduce the local radiation field during the operation.

Even operation up to \(500~\mathrm{MW}\) of fusion power appears accessible within Yinsen, with the upper end of that range set primarily by the Greenwald density limit. At that level, the major plant subsystems---balance of plant, cryogenics, fueling, and associated power-conversion hardware---can in principle support the higher throughput, although power exhaust and stable detached divertor operation become more challenging and would likely require substantially higher impurity concentrations. In the present baseline, the auxiliary-heating system is capped at \(15~\mathrm{MW}\) launched, which narrows the accessible operating space, particularly at lower density, but this choice is intentional: it reduces local recirculating power and limits the size and cost of the ICRH plant, while making the reactor more dependent on the strong confinement available at high field to reach high \(Q\). One useful way to view the Yinsen throughput target is that the primary solid structures are effectively designed for a finite lifetime-integrated fusion-energy throughput of about \(1040~\mathrm{MW\cdot yr}\), corresponding to \(130~\mathrm{MW}\times 0.40 \times 20~\mathrm{yr}\). Within that envelope, the power-versus-time trajectory used to extract that total energy can in large part be traded by the customer and plant operator to suit the needs of the application. In a future scenario with more radiation-tolerant structural materials, it is therefore likely that one would not operate at the absolute maximum power-loading limit, but instead trade toward a more modest increase in fusion power together with improved utilization and capacity factor, depending on the target market and the needs of the application.

\subsection{From Fusion Power Density to a Conservative Tokamak Design Point}

Once the allowable blanket-area-normalized fusion power density is fixed, the next question is what amount of fusion power is large enough to matter commercially in the intended deployment environments. For Yinsen we take the minimum relevant net electric output to be \(25~\mathrm{MWe}\), reflecting the scale at which marine propulsion, ship-to-shore supply, remote industrial energy systems, and high-grade process-heat deployments begin to justify the capital complexity of a fusion plant. A simple scoping power balance may then be written as
\begin{equation}
P_{\mathrm{net}} \approx \eta_{\mathrm{BOP}} \cdot P_f - P_{\mathrm{cryo}} - P_{\mathrm{aux}} - P_{\mathrm{pump}} - P_{\mathrm{tritium}}
\label{eq:net_power_floor}
\end{equation}
with \(\eta_{\mathrm{BOP}}\approx 0.40\) for a supercritical CO$_2$ cycle, whose motivation is discussed further in Section~\ref{sec:bop}. In this simple 0D model, the cryogenic load is capped by the design choice to limit total nuclear heating to the \(20~\mathrm{K}\) HTS magnet system. A nominal electric input of \(2~\mathrm{MWe}\), together with an assumed cryogenic coefficient of performance of \(2\%\), corresponds to roughly \(40~\mathrm{kW}\) of available cooling at \(20~\mathrm{K}\), comfortably above the requirement for the full magnet system. The auxiliary power load is taken to scale as
\[
P_{\mathrm{aux}} \approx \frac{P_f}{Q \cdot \eta_{\mathrm{wall\rightarrow plasma}}}
\]
with \(Q>10\) taken as the minimum viable threshold for commercial relevance and \(\eta_{\mathrm{wall\rightarrow plasma}}=0.60\) for ICRH. The tritium-processing and FLiBe pumping loads are also taken to scale linearly with fusion power, using \(7~\mathrm{kW_e/MW_f}\) for tritium processing \cite{Wenninger2017} and \(10~\mathrm{kW_e/MW_f}\) for FLiBe pumping \cite{Kuang2018}. Under these conservative assumptions, a reactor class below about \(130~\mathrm{MW}\) fusion power struggles to retain \(25~\mathrm{MWe}\) net electric output once realistic recirculating loads are included. Yinsen therefore treats \(P_f \gtrsim 130~\mathrm{MW}\) as the minimum relevant fusion-power class. Notably, this remains below the maximum fusion power projected for SPARC, approximately \(140~\mathrm{MW}\) \cite{Creely2020}.

\begin{figure}[t]
\centering
\includegraphics[width=\columnwidth]{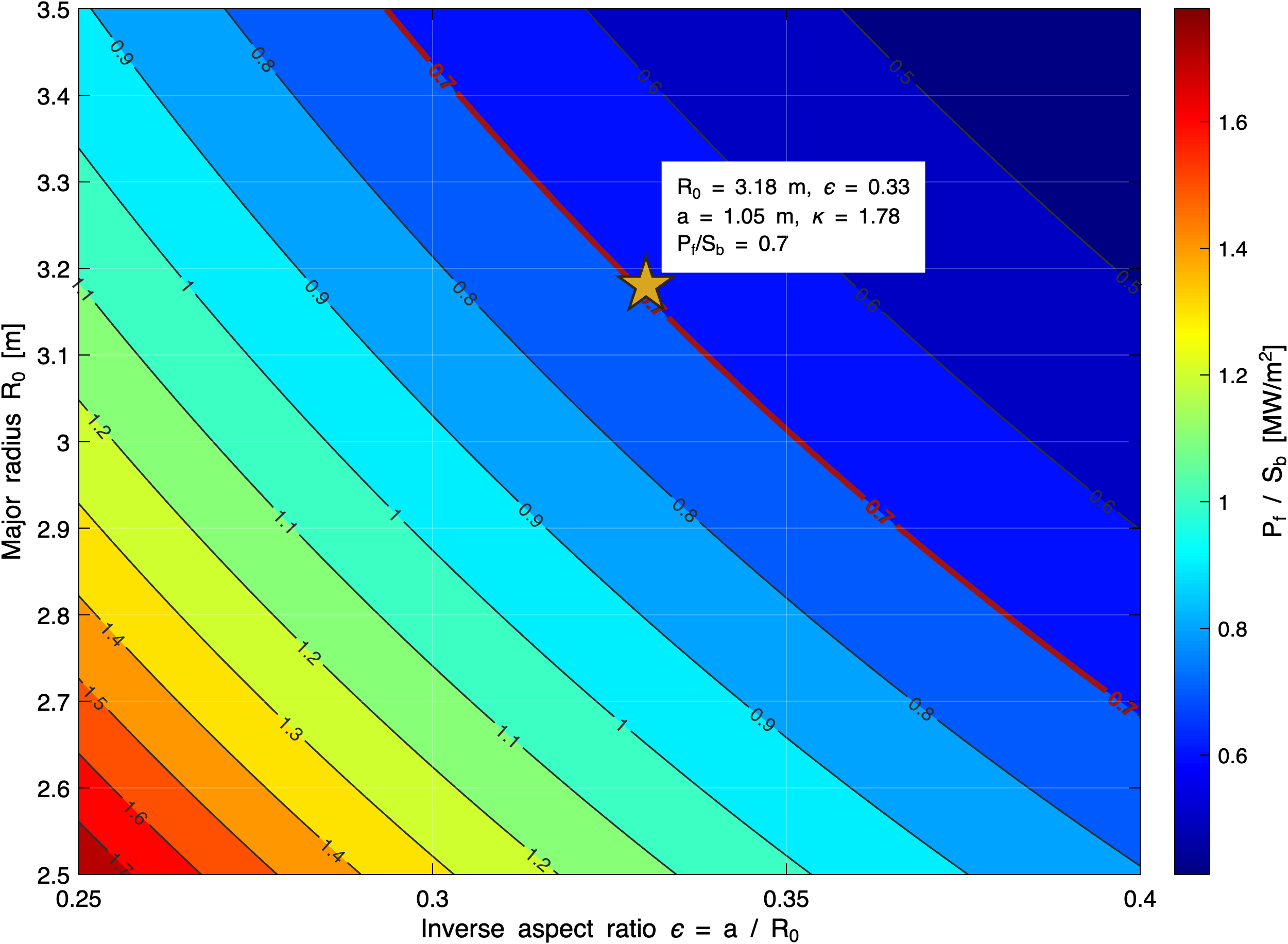}
\caption{Blanket-area-normalized fusion power density for a \(130~\mathrm{MW}\) fusion-power class, plotted against major radius and inverse aspect ratio using the elongation relation \(\kappa \approx 5.4\epsilon\) together with the blanket-area estimate of Eq.~\eqref{eq:blanket_area_estimate}. The red \(0.7~\mathrm{MW/m^2}\) contour marks the intended FOAK power-density ceiling, while the star indicates the representative geometry selected for the baseline Yinsen device.}
\label{fig:power_density_geometry_map}
\end{figure}

For a specified fusion power, the required blanket-area-normalized power density immediately constrains the machine geometry. The blanket surface area is estimated from the tokamak major radius \(R_0\), minor radius \(a\), and elongation \(\kappa\) as
\begin{equation}
\begin{gathered}
S_b \approx 2\pi R_0 P_{\mathrm{pol}} \\
P_{\mathrm{pol}} \approx
\pi a\!\left[
3(1+\kappa)-\sqrt{(3+\kappa)(1+3\kappa)}
\right]
\end{gathered}
\label{eq:blanket_area_estimate}
\end{equation}
and a representative conservative elongation is imposed using the aspect-ratio relation in Eq.~\eqref{eq:kappa_limit} \cite{Stambaugh1992}. Robust vertical stability is especially important for Yinsen given the intended marine deployment environments; this point is discussed further in Section~\ref{sec:vertical_stability}
\begin{equation}
\kappa \approx 5.4\,\epsilon
\label{eq:kappa_limit}
\end{equation}
Figure~\ref{fig:power_density_geometry_map} then applies Eqs.~\eqref{eq:blanket_area_estimate} and \eqref{eq:kappa_limit} to a \(130~\mathrm{MW}\) fusion-power class, depicting the full \((R_0,\epsilon)\) design space with the starred point indicating the representative geometry selected for the baseline Yinsen device.

\begin{figure}[t]
\centering
\includegraphics[width=\columnwidth]{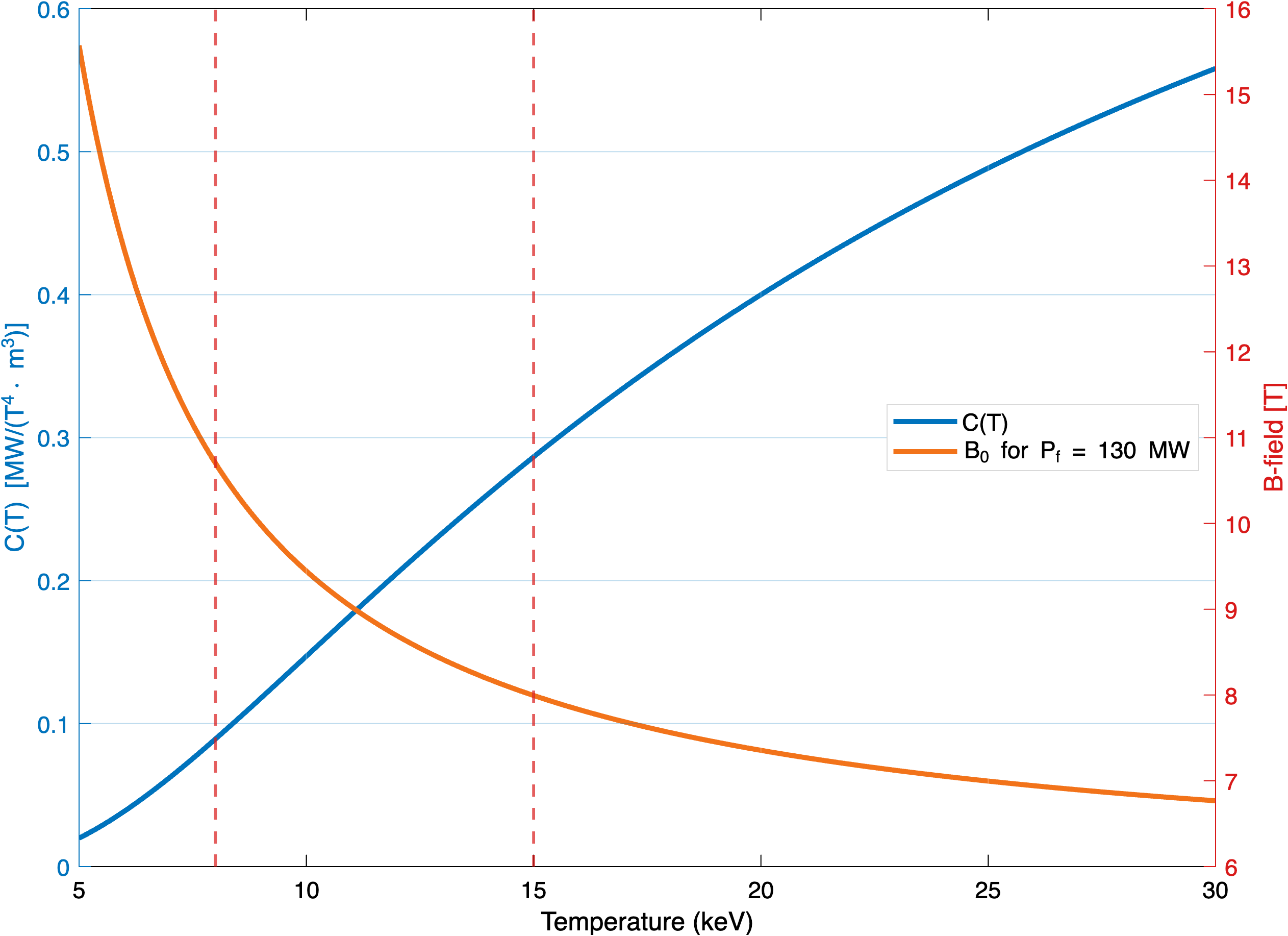}
\caption{Temperature dependence of the beta-limited \(B\)-field requirement for a \(130~\mathrm{MW}\) fusion-power target using \(\beta = 3\%\). Over the \(5\) to \(20~\mathrm{keV}\) reactor regime, the corresponding field spans roughly \(8\) to \(11~\mathrm{T}\).}
\label{fig:beta_temperature_bound}
\end{figure}

With the geometry constrained, the next step is to determine the centerline magnetic-field strength. For scoping purposes we use the conservative beta-limited scaling
\begin{equation}
P_f = C(T)\,\beta^2 B^4 V,
\qquad
C(T) \equiv \frac{E_f\langle \sigma v \rangle}{64\mu_0^2 T^2}
\label{eq:beta_scaling}
\end{equation}
where \(V\) is the plasma volume and \(C(T)\) carries the temperature dependence of the fusion reactivity term. Figure~\ref{fig:beta_temperature_bound} evaluates this relation for a \(130~\mathrm{MW}\) fusion-power target using a representative \(\beta = 3\%\) stability limit. Over the reactor-relevant volume averaged temperature range of roughly \(5\) to \(20~\mathrm{keV}\), the corresponding field requirement spans approximately \(8\) to \(11~\mathrm{T}\). Note that Eq.~\eqref{eq:beta_scaling} is volume averaged, neglects profile effects, and uses a simplified reactivity model, so it is used here only as a conservative field-scaling guide. Within that scoping framework, a baseline field of about \(9~\mathrm{T}\) remains a reasonable target once on-coil \(B\)-field structural limits and overall magnet-engineering margin are taken into account.

\clearpage
\onecolumn
\begin{table}[!t]
\centering
\caption{Yinsen machine parameters}
\label{tab:yinsen_machine_params}
\footnotesize
\renewcommand{\arraystretch}{1.04}
\begin{tabular}{p{6.8cm}cc}
\toprule
\textit{Parameter} & \textit{Symbol} & \textit{Value} \\
\midrule
Major radius & $R_0$ & \SI{3.18}{m} \\
Minor radius & $a$ & \SI{1.05}{m} \\
Inverse aspect ratio & $\epsilon$ & 0.33 \\
Plasma elongation & $\kappa$ & 1.78 \\
Plasma triangularity & $\delta$ & 0.89 \\
Plasma volume & $V_p$ & \SI{99.7}{m^3} \\
Blanket-facing surface area & $S_b$ & \SI{185}{m^2} \\
Toroidal magnetic field & $B_0$ & \SI{9.29}{T} \\
Plasma current & $I_p$ & \SI{9.67}{MA} \\
Vacuum vessel lifetime & $\mathcal{L}_{\mathrm{VV}}$ & \SI{1040}{MW.yr} \\
Flattop duration & $\tau_{\mathrm{flat}}$ & \SI{900}{s} \\
Tritium breeding ratio @30\% Li-6 & $TBR$ & 1.10 \\
\bottomrule
\end{tabular}
\end{table}

\begin{table}[!t]
\centering
\caption{Yinsen operating-point and plasma-state parameters.}
\label{tab:yinsen_operating_params}
\footnotesize
\renewcommand{\arraystretch}{1.04}
\begin{tabular}{p{6.4cm}ccc}
\toprule
\textit{Parameter} & \textit{Symbol} & \textit{130 MW} & \textit{185 MW} \\
\midrule
Fusion power wall loading & $P_f/S_b$ & \SI{0.7}{MW/m^2} & \SI{1.0}{MW/m^2} \\
Neutron power wall loading & $P_n/S_b$ & \SI{0.56}{MW/m^2} & \SI{0.8}{MW/m^2} \\
Utilization (20yr lifetime) & $U$ & 40\% & 28.1\% \\
Thermal power & $P_{\mathrm{th}}$ & \SI{148}{MW} & \SI{206}{MW} \\
Gross electric power & $P_{e,\mathrm{gross}}$ & \SI{62}{MWe} & \SI{87}{MWe} \\
Net electric output & $P_{e,\mathrm{net}}$ & \SI{27}{MWe} & \SI{52}{MWe} \\

Electric gain & $Q_e$ & 1.77 & 2.49 \\
Plasma gain & $Q$ & 9.6 & 13.7 \\
Coupled ICRH power & $P_{\mathrm{ICRH}}$ & \SI{13.5}{MW} & \SI{13.5}{MW} \\
Ohmic heating power & $P_{\mathrm{ohm}}$ & \SI{0.60}{MW} & \SI{0.59}{MW} \\
Normalized beta & $\beta_N$ & 1.23 & 1.49 \\
Safety factor at $\psi_N=0.95$ & $q_{95}$ & 4.05 & 4.13 \\
Bootstrap-current fraction & $f_{\mathrm{BS}}$ & 20.7\% & 24.8\% \\
Internal inductance & $l_i(3)$ & 0.59 & 0.62 \\
Pedestal effective charge & $Z_{\mathrm{eff,ped}}$ & 2.00 & 2.00 \\
Core radiated power fraction & $f_{\mathrm{rad,core}}$ & 32\% & 34\% \\
$P_{\mathrm{SOL}}$ & $P_{\mathrm{SOL}}$ & \SI{26.8}{MW} & \SI{33.4}{MW} \\
$P_{\mathrm{LH}}$ & $P_{\mathrm{LH}}$ & \SI{24.6}{MW} & \SI{28.0}{MW} \\
Energy confinement time & $\tau_E$ & \SI{1.61}{s} & \SI{1.54}{s} \\
$H_{98,y2}$ factor & $H_{98,y2}$ & 1.28 & 1.32 \\
Volume-averaged ion temperature & $\langle T_i \rangle$ & \SI{11.6}{keV} & \SI{11.5}{keV} \\
Volume-averaged electron temperature & $\langle T_e \rangle$ & \SI{11.1}{keV} & \SI{11.2}{keV} \\
Volume-averaged electron density & $\langle n_e \rangle$ & \SI{1.26e20}{m^{-3}} & \SI{1.52e20}{m^{-3}} \\
Greenwald fraction & $f_{\mathrm{GW}}$ & 0.451 & 0.544 \\
On-axis ion temperature & $T_{i,0}$ & \SI{22.7}{keV} & \SI{22.1}{keV} \\
On-axis electron temperature & $T_{e,0}$ & \SI{19.8}{keV} & \SI{20.2}{keV} \\
On-axis electron density & $n_{e,0}$ & \SI{1.6e20}{m^{-3}} & \SI{2e20}{m^{-3}} \\
Pedestal electron temperature & $T_{e,\mathrm{ped}}$ & \SI{10}{keV} & \SI{10}{keV} \\
Pedestal electron density & $n_{e,\mathrm{ped}}$ & \SI{11e19}{m^{-3}} & \SI{15e19}{m^{-3}} \\
Separatrix electron density & $n_{e,\mathrm{sep}}$ & \SI{3e19}{m^{-3}} & \SI{4e19}{m^{-3}} \\
\bottomrule
\end{tabular}
\end{table}
\clearpage
\twocolumn

Finally, the plasma current follows from the edge safety-factor relation
\begin{equation}
q_a = \frac{5\epsilon a B}{I_P}\left(\frac{1+\kappa^2}{2}\right),
\label{eq:q_a_relation}
\end{equation}
so that once \(R_0\), \(a\), \(\kappa\), and \(B\) are selected, the desired safety-factor floor fixes the plasma-current scale. Choosing a conservative edge safety factor of \(q_a \geq 3\) therefore places the Yinsen plasma current naturally in the \(9\) to \(11~\mathrm{MA}\) range. At this point the overall 0D tokamak class is already largely determined from conservative engineering and physics arguments: a low-power-density \(130~\mathrm{MW}\) fusion-power reactor with advanced-tokamak aspect ratio, high-field HTS magnets near \(9\) to \(10~\mathrm{T}\), and plasma current near \(10~\mathrm{MA}\), all selected to remain within a materials-limited blanket-area-normalized power-density envelope.

Within this materials-limited reactor design envelope, the self-consistent baseline design point is then further refined using \textsc{FUSE}. \textsc{FUSE} is an open-source integrated tokamak systems framework developed by General Atomics and built around a centralized data model together with a large library of coupled physics and engineering ``actors'' that solve for self-consistent equilibrium, core profiles resolved by transport and heating/current-drive, radial build, structural constraints, one-dimensional neutronics, and associated plant systems. In the present work it is not used to search for the reactor class from scratch, but to refine and stress-test a class already selected from the power-density, lifetime, geometry, field, and current arguments developed above. Because no validated FOAK fusion costing model yet exists, Yinsen instead uses total HTS tape length requirement as a proxy for magnet-system CAPEX and therefore as a practical optimization target for the reactor.

The constraint that narrows the local design space most strongly is the H-mode access requirement \(P_{\mathrm{SOL}}/P_{\mathrm{LH}} > 1.0\). This is especially restrictive because the same search also pushes toward minimal auxiliary heating power. This makes the scaling law threshold power important for determining design margin: ASDEX Upgrade L--H transition studies show that \(P_{\mathrm{LH}}\) is sensitive to main-ion species, wall and divertor condition/material, magnetic perturbations, and magnetic configuration, with the ion heat channel and edge radial electric field central to the transition physics~\cite{Plank2023LHTransition}. The effects of heavier main-ion species and well-conditioned metallic plasma-facing surfaces should therefore be included when interpreting the \(P_{\mathrm{SOL}}/P_{\mathrm{LH}}\) margin, rather than treating the threshold as a fixed scalar independent of the plasma-facing environment. The objective of minimizing auxiliary heating power is tied directly to marine deployment and packaging requirements, where surrounding plant systems must remain compact enough for integration within large-vessel layouts, and to the need to keep recirculating power low in intermediate-scale fusion plants.

The H-mode-access margin therefore remains an important follow-on issue. Applying the conservative low-density-branch correction to the ITPA threshold scaling gives \(P_{\mathrm{LH}}\) values of order \(30~\mathrm{MW}\) near the minimum-threshold density for Yinsen. Simple L-mode estimates indicate that alpha heating before the transition is only of order \(1\)--\(3~\mathrm{MW}\), although the colder and more resistive L-mode plasma should also provide several megawatts of additional ohmic heating relative to the hot operating point. This motivates further study of access scenarios that use a clean, weakly radiating plasma during the current and density ramp, exploit double-null balance and X-point/divertor geometry to reduce the effective threshold, and then introduce stronger impurity seeding only after H-mode is established. A practical engineering option is to reserve space and interfaces for a second pulsed ICRH plant providing an additional \(10\)--\(15~\mathrm{MW}\) during the access phase, after which alpha heating becomes large enough to support the higher-power H-mode state. The baseline Yinsen point should therefore be understood as a low-power-density, materials-limited reactor whose detailed 0D point and plasma state are refined with \textsc{FUSE} under compactness and recirculating-power constraints, rather than as the product of an unconstrained optimization for maximum performance.

With the machine class fixed in this way, the remainder of the paper begins with the detailed core plasma design in \textsc{FUSE}, Astra and Tokamaker, and then proceeds through ICRH heating, disruption loading, divertor and power handling, HTS magnet engineering, neutronics, pulsed-power operation, balance-of-plant design, fuel-cycle engineering, facility layout, direct overnight capital costing, and shipping-economics analyses that together form the engineering rational for \href{https://marvelcinematicuniverse.fandom.com/wiki/Ho_Yinsen}{Yinsen}

%% file: plasma.tex
\section{Plasma scenario design} 
\begin{figure}[!t]
    \centering
    \includegraphics[width=\columnwidth]{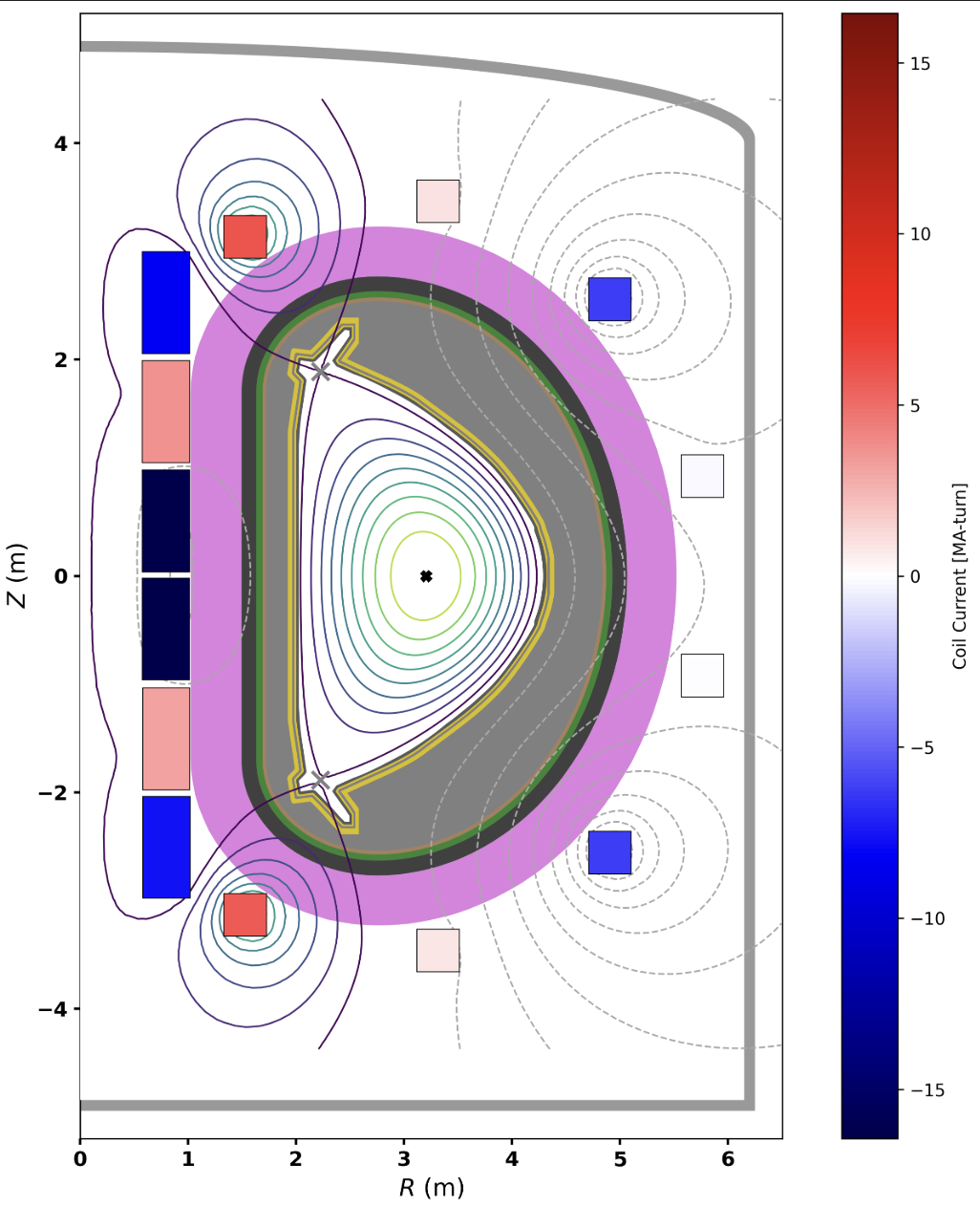}
    \caption{Nominal build and flattop equilibrium for Yinsen. Two layers of neutron shielding (green) are located outside of the main blanket (gray) and inside the TF coil (purple). All coils fit inside the cryostat (gray).}
    \label{fig:build}
\end{figure}

This section introduces the self-consistent baseline plasma state produced with \textsc{FUSE} for the Yinsen reactor class selected in Section~\ref{sec:introduction}. The equilibrium workflow begins with TEQUILA, the fixed-boundary Grad--Shafranov solver used within \textsc{FUSE}. TEQUILA takes the prescribed MXH plasma boundary together with pressure and current-profile inputs and returns the resulting flux surfaces and poloidal-flux distribution. The PF-coil optimizer then determines the currents and placements required to realize that shape. The resulting flattop equilibrium is reconstructed with TokaMaker \cite{hansen_tokamaker_2024} for follow-on control and engineering studies. The integrated build and corresponding flattop equilibrium are shown in Figure~\ref{fig:build}, including the double walled vacuum vessel, blanket, neutron shielding, and TF coil.

\begin{figure}[!t]
    \centering
    \includegraphics[width=\linewidth]{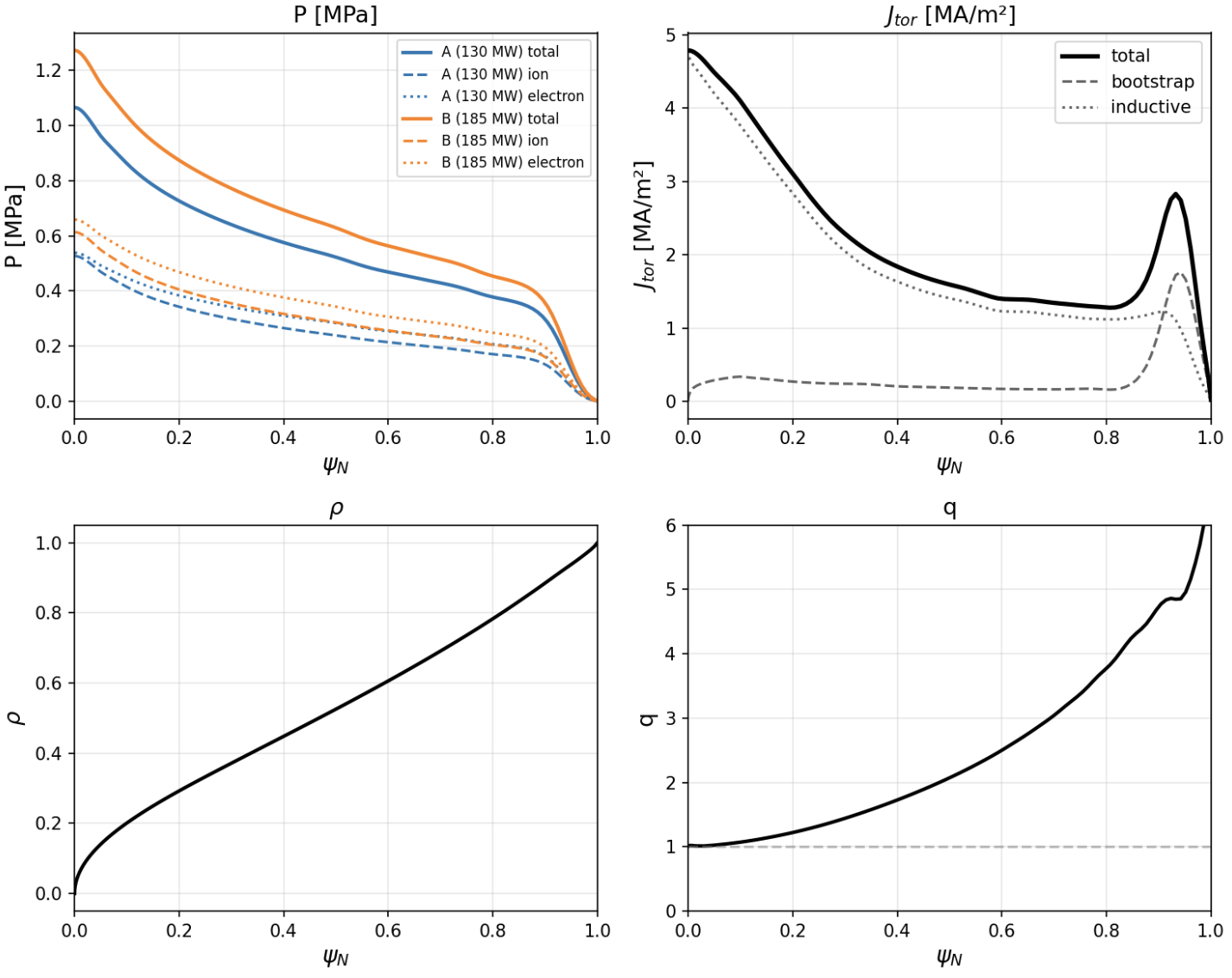}
    \caption{Initial equilibrium summary showing the pressure, toroidal-current-density, normalized radial-coordinate, and safety-factor profiles used in the Yinsen plasma design. Panel (a) compares the Case A (\SI{130}{MW}) and Case B (\SI{185}{MW}) pressure profiles.}
    \label{fig:full_eq}
\end{figure}

The plasma equilibrium attributes are summarized in Figures~\ref{fig:full_eq}, \ref{fig:electron_source}, and \ref{fig:ion_source}. Figure~\ref{fig:full_eq} shows the pressure, toroidal-current-density, normalized-flux, and safety-factor profiles at the selected shaped ($\kappa = 1.78$, $\delta = 0.89$), high-field plasma with a monotonic safety-factor profile and plasma current of \(9.67~\mathrm{MA}\). The pressure profile remains centrally peaked but broad enough to support reactor-relevant fusion power without pushing the edge safety factor toward an aggressive operating regime, while the toroidal-current-density profile shows the expected edge rise associated with the noninductive current contributions near the outer region. This is consistent with the intended Yinsen operating regime, in which bootstrap current provides useful assistance but does not eliminate the need for inductive current drive.

In figures~\ref{fig:electron_source} and \ref{fig:ion_source} the corresponding plasma power balance is detailed. Alpha heating provides the dominant distributed heating source, while ICRH is concentrated more strongly toward the core with nearly negligible ohmic heating. These source profiles provide the input state that the subsequent transport and stability analyses build upon. This initial gaussian deposition profile is refined in the ICRH analysis, particularly regarding the electron absorption tail. At the screening level, \textsc{FUSE} also enforces a basic operating envelope before a case is accepted for follow-on analysis. For the present baseline, those checks include \(q_{95} > 2.0\), \(f_{GW} < 1.0\), an empirical elongation bound, a \(\beta_N\) limit based on ideal-MHD no-wall stability surrogates, and a minimum threshold on vertical growth rate and controllability margin. These internal checks are useful for filtering the design space, but they are not by themselves sufficient to establish reactor viability, which motivates the dedicated vertical-stability analysis that follows in Section~\ref{sec:vertical_stability}.

While a full MHD stability analysis is out of scope here, we note that because the heating system does not include neutral beam injection, the plasma is expected to have limited externally driven \(E\times B\) rotation and to rely primarily on intrinsic torque, plausibly at the order of \SI{1}{N\,m}. \textsc{FUSE} calculates turbulent momentum transport when running TGLF through the gyro-Bohm-normalized momentum flux, \(\Pi = n_e T_e(\rho_s^2/a)\,\Pi_{\mathrm{GB}}\), but the most concerning remaining MHD risks are likely non-axisymmetric tearing and neoclassical tearing modes. Recent ARC MHD studies identify NTM seed-island thresholds and low-torque rotation physics as central uncertainties for RF-heated high-field tokamak scenarios \cite{Leuthold2026ARCMHD}; the corresponding tearing, NTM, error-field, and rotation-braking analysis remains future work for Yinsen.

\addtocounter{figure}{1}
\begin{figure}[!t]
    \centering
    \includegraphics[width=\columnwidth]{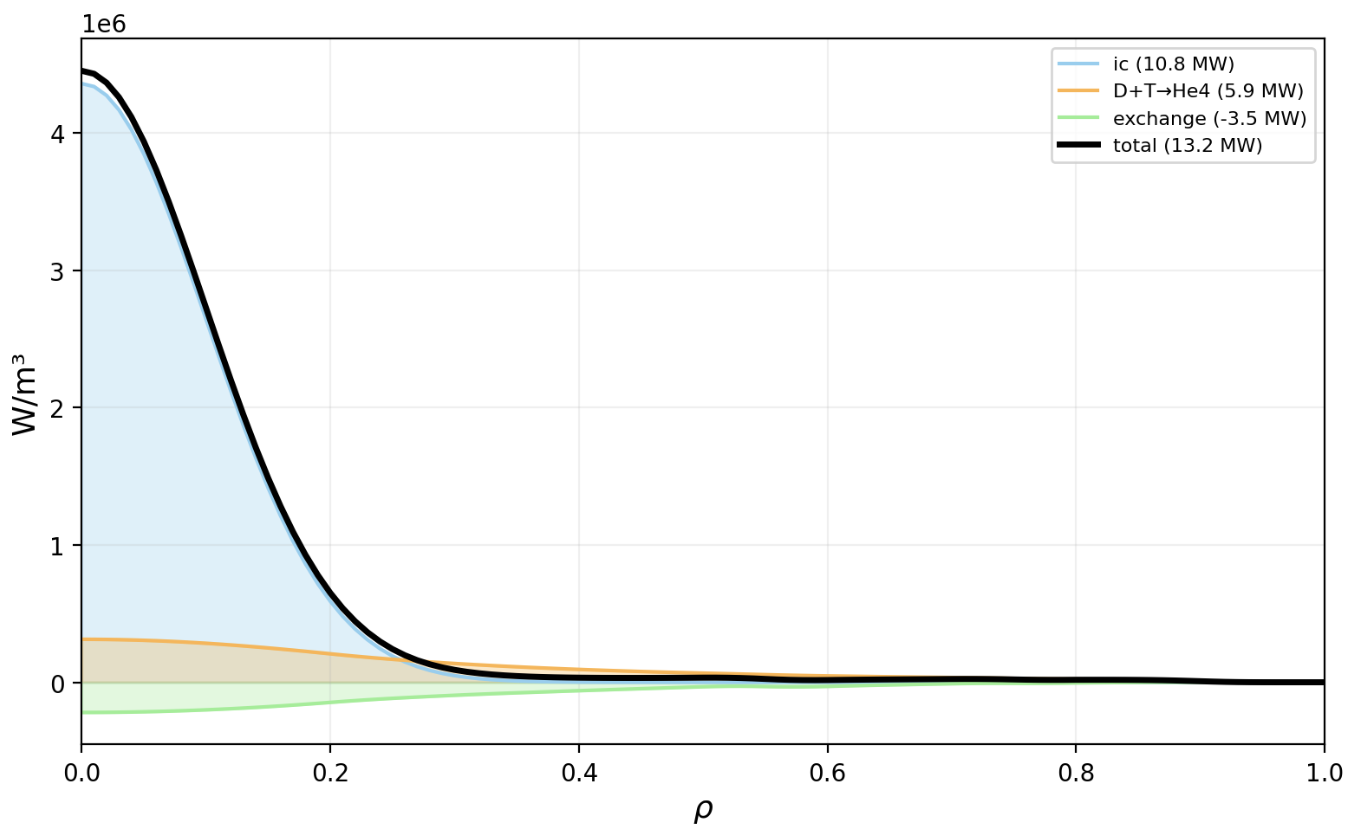}
    \caption{Ion-channel power sources and sinks for the baseline \textsc{FUSE} design point, including ICRH, alpha heating, ohmic heating, exchange, radiative losses, synchrotron loss, and the resulting net total source.}
    \label{fig:electron_source}
\end{figure}

\begin{figure}[!t]
    \centering
    \includegraphics[width=\columnwidth]{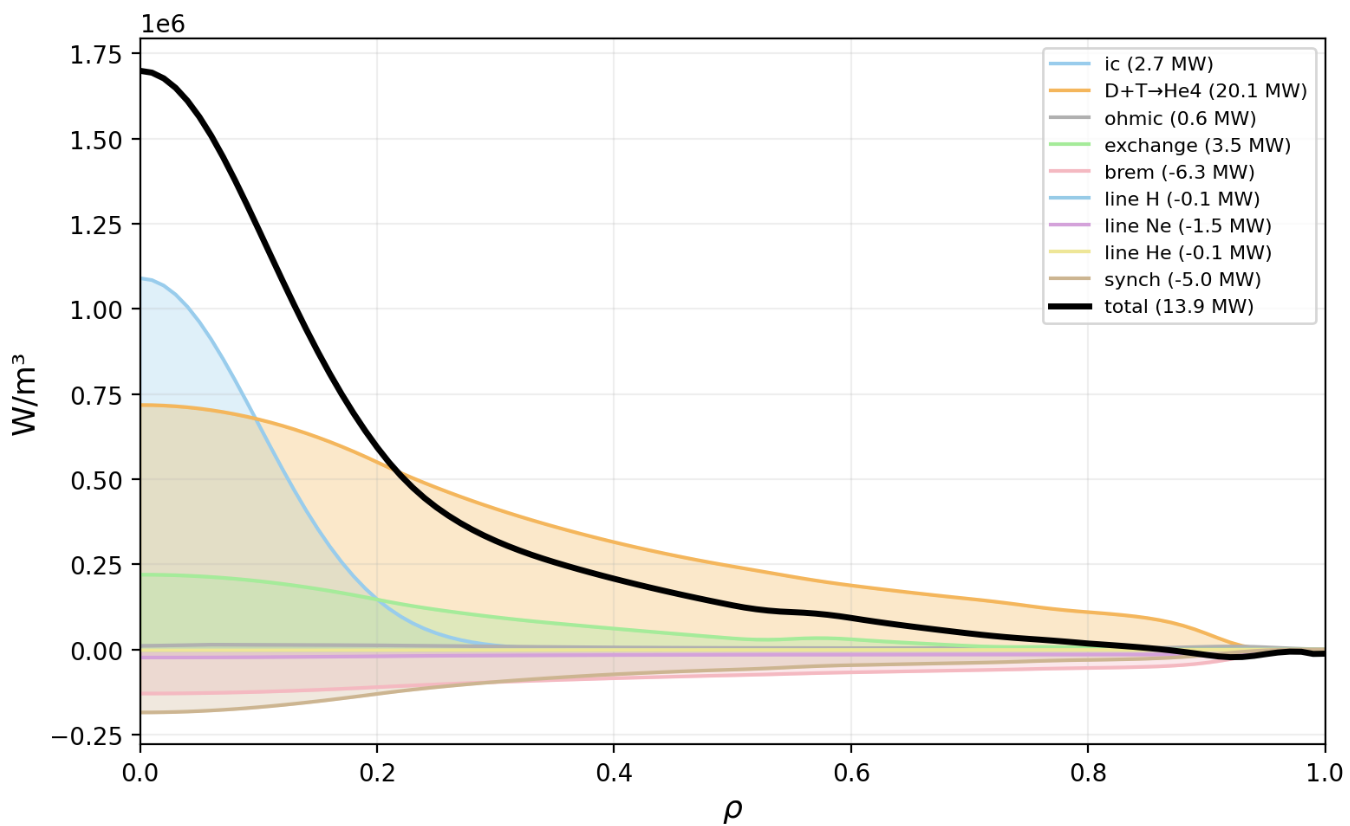}
    \caption{Electron-channel power sources and sinks for the baseline \textsc{FUSE} design point, showing the radial contributions from ICRH, alpha heating, electron-ion exchange, and the net total source.}
    \label{fig:ion_source}
\end{figure}

\subsection{Vertical Stability}\label{sec:vertical_stability} 
While FUSE provides a starting framework with which to asses and optimize the nominal operating conditions for this device, there are several critical elements of FPP design that require additional analysis. For example, vertical stability, which is a combination of the propensity for the baseline plasma to induce a vertical displacement event (VDE) and the capability of the device to control such deviations from nominal operation, requires precise descriptions of the plasma, the conducting structures of the tokamak itself, and the power supplies responsible for vertical stability control. As uncontrolled VDEs will lead to lost of plasma control and disruption, it is essential to ensure the controllability of the nominal operating point prior to device construction.

In this work, we utilize the TokaMaker code set from the open-source Open FUSION Toolkit modeling framework, which has been used extensively for device design \cite{hansen_tokamaker_2024, guizzo_electromagnetic_2025}, to assess the controllability of the baseline plasma. Following previous characterizations of vertical control \cite{humphreys_experimental_2009, nelson_vertical_2023, guizzo_assessment_2024, nelson_implications_2024}, the maximum vertical displacement ($\Delta Z_\mathrm{max}$) is taken as a critical metric for the operational margin of this device. When compared to the minor radius $a$, $\Delta Z_\mathrm{max}$ can be used to characterize ``safe" ($\Delta Z_\mathrm{max}\gtrsim5\%$) and ``robust" ($\Delta Z_\mathrm{max}\gtrsim10\%$) operation. On Yinsen, this corresponds to a requirement that the vertical stability system be able to control vertical instabilities resulting from initial displacements on the order of $\Delta Z_\mathrm{0} = 5-10\,$cm, which may occur from ship motion, ELMs, sawteeth or L-H transitions, for conservative nominal operation.

\begin{figure}
    \centering
    \includegraphics[width=\linewidth]{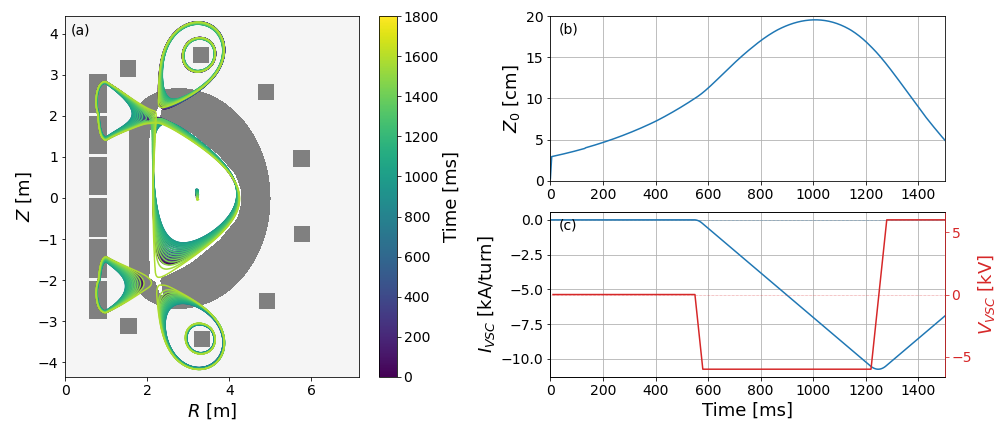}
    \caption{(a) Evolution of the LCFS after a $\Delta Z_\mathrm{0}=10\,$cm initial displacement, subject to control with a $6\,$kV power supply. The evolution of the (b) magnetic axis and (c) current and voltage applied in the vertical stability system ($I_\mathrm{VSC}$ -- blue and $V_\mathrm{VSC}$ -- red, respectively) are also shown.}
    \label{fig:vde}
\end{figure}

To determine $\Delta Z_\mathrm{max}$ for the initial Yinsen design point presented here, all conducting elements of the machine are constructed in TokaMaker and a series of unstable equilibria (acquired by adding fractions of the most unstable eigenmode to the reference equilibrium) are allowed to evolve in time until increasingly large initial displacements are realized, at which point a simple vertical stability feedback circuit is employed to attempt to catch the plasma. An example of this evolution is shown in figure~\ref{fig:vde} for a case with a $\Delta Z_\mathrm{0}=10\,$cm initial displacement, subject to control with a $6\,$kV power supply and a slew rate limit of $2.5 kA$ $(ms)^{-1}$. Due to the thick conducting structures (double-hulled vacuum vessel and blanket walls) present in the Yinsen base design, the passive stabilization in this scenario is sufficient to slow plasma motion to a level at which only the outermost primary PF coils are needed to respond to the plasma motion and catch the equilibrium before it is lost. This results in calculations of $\Delta Z_\mathrm{0,max}>10\,$cm even for smaller ($\sim4\,$kV) power supplies, suggesting that vertical stability may not be a rate-limiting factor for the Yinsen machine.

\begin{figure}
    \centering
    \includegraphics[width=\linewidth]{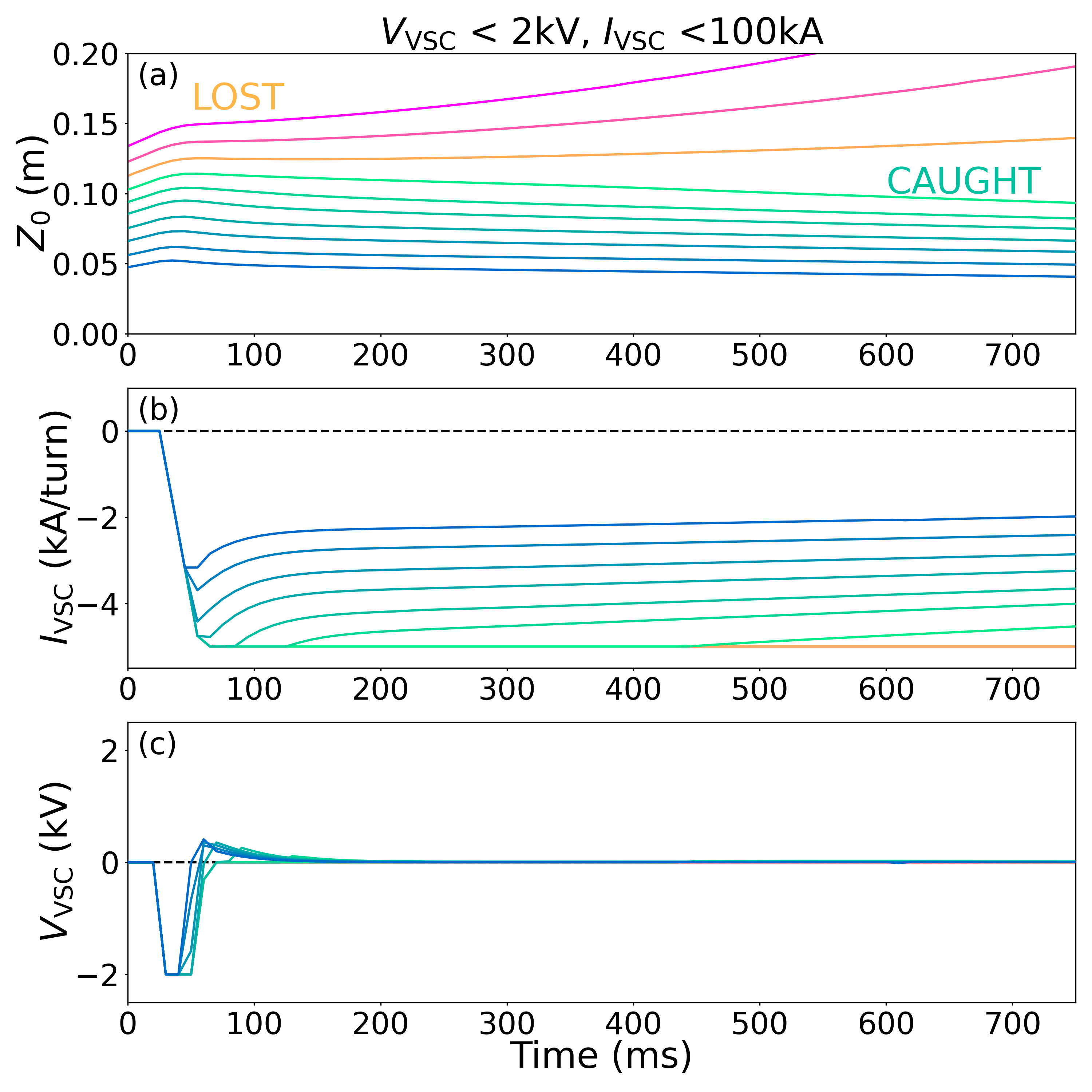}
    \caption{Evolution of (a) the plasma magnetic axis, (b) the VSC coil current and (c) the voltage on the VSC power supply for Yinsen plasmas with varying initial displacements $\Delta Z_\mathrm{0}$. Plasmas with $\Delta Z_\mathrm{0}\sim10\,$cm can be caught with a $2\,$kV voltage and $100\,$kA current limit on the VSC system.}
    \label{fig:vde_invessel}
\end{figure}

However, we note that, as the design basis for Yinsen is refined, vertical stability must remain a core constraint given the application of the device. Removal of all of the conducting structures around the plasma in TokaMaker leads to a machine state in which the PF coils are too far from the plasma to adequately respond to a potential VDE. In these cases, additional passive plates or a dedicated vertical stability coil (VSC) may need to be added to the Yinsen design, depending on the final engineering decisions on power-supply limitations. In all cases, the addition of a dedicated in-blanket VSC drastically improves the vertical controllability of Yinsen. An example of this is shown in figure~\ref{fig:vde_invessel}, where a resistive 100-turn VSC situated closer to the bulk plasma inside the blanket at \(R = 4.0\,\mathrm{m}\) and \(Z = \pm 1.5\,\mathrm{m}\) is utilized to control plasmas up to the robust $\Delta Z_\mathrm{0}=10\,$cm initial displacement with a significantly smaller ($2\,$kV) power supply. A maximum VSC current of $100\,$kA is also imposed during this simulation, showing that there are various flexible options for robust vertical stability systems. This allows for other design factors, including cost, material compatibility and neutronics, to set constraints for this aspect of the design. Placing coils inside the blanket tank remains challenging thermally and for maintenance, as discussed in Section~\ref{sec:invessel_coil_candidates}.

\subsection{Core transport validation}
\begin{figure}
    \centering
    \includegraphics[width=\columnwidth]{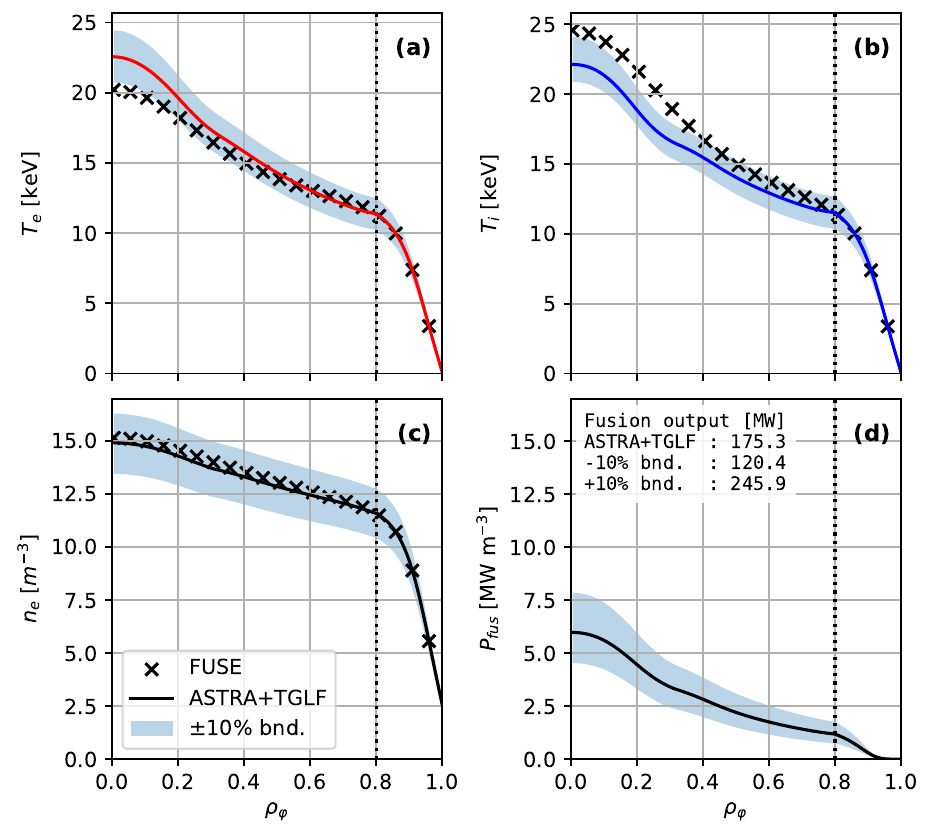}
    \caption{Results of the core transport modeling using ASTRA+TGLF based on the low power scenario. The FUSE results were used as edge boundary conditions up to $\rho_\varphi = 0.8$. The location of the boundary is indicated by the vertical dotted lines. The FUSE results are shown with markers, while the solid lines represent the corresponding ASTRA+TGLF calculations. The uncertainty bands around the modeling are based on a $\pm 10\%$ variation of the FUSE boundary conditions in ASTRA+TGLF modeling. Panel (a) shows the electron temperature, Panel (b) the ion temperature, Panel (c) the electron density, and Panel (d) the fusion power density. The volume-integrated fusion power values are shown in Panel (d).}
    \label{fig:ASTRA+TGLF_modeling_low}
\end{figure}

\begin{figure}
    \centering
    \includegraphics[width=\columnwidth]{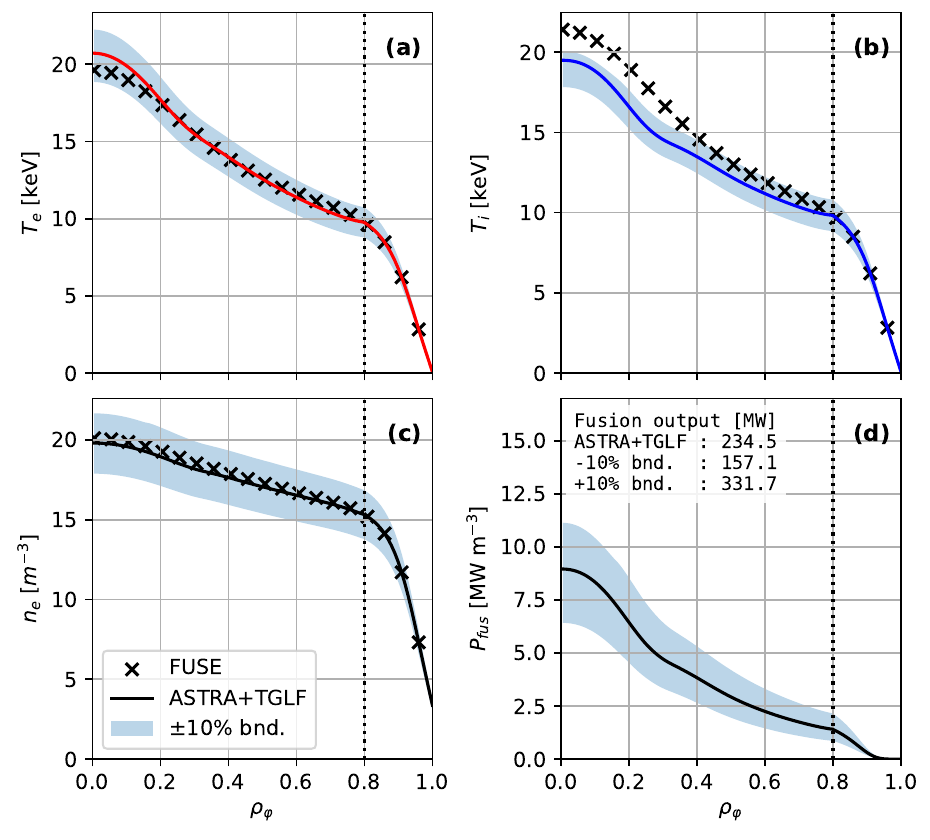}
    \caption{Results of the core transport modeling using ASTRA+TGLF based on the high power scenario. Density in units of $10^{19}$ $m^{-3}$.}
    \label{fig:ASTRA+TGLF_modeling_high}
\end{figure}

The backbone of FUSE core transport studies is the transport solver referred to as \textit{FluxMatcher}. This solver acts as an integrated framework that couples gyrofluid turbulence models and neoclassical transport descriptions in order to reproduce experimentally consistent heat and particle balance. During the initial optimization phase, a neural network surrogate model of the TGLF transport model \cite{Neiser2024TGLF} is utilized to evaluate turbulent transport fluxes at each iteration step. The primary motivation for employing this surrogate is computational efficiency: compared to the full TGLF model, the neural network approximation drastically reduces evaluation time while retaining sufficient accuracy for optimization purposes. This makes it particularly well-suited for large-scale parameter scans, such as those required in multi-objective optimization studies where several thousand FUSE cases must be evaluated. The resulting optimized profiles are presented in Figs.~\ref{fig:ASTRA+TGLF_modeling_low} and \ref{fig:ASTRA+TGLF_modeling_high}, for the low and high power operating points respectively. The FUSE predictions are indicated by discrete markers.

Following the completion of this initial optimization, a higher-fidelity analysis is performed using a stand-alone ASTRA+TGLF workflow (SAT2) \cite{ASTRA,Fable2013,Staebler2007}. In this step, the edge plasma conditions obtained from FUSE are imposed as fixed boundary conditions at the pedestal top, located approximately at $\rho_\varphi \approx 0.8$. Additionally, the heat flux profiles are kept identical to those used in the FluxMatcher optimization to ensure consistency between the two modeling approaches. In addition, gas puff fueling models are used to represent external particle sources, and radiation losses are included through appropriate radiation models for Bremsstrahlung, tungsten (with a prescribed density of $2\cdot10^{-5}\cdot n_e$) and low-$Z$ impurity radiation. The D-T fuel composition is fixed at a 1:1 ratio, and quasi-neutrality is enforced together with the electron density $n_e$. Under these constraints, the ASTRA framework self-consistently solves the coupled heat and particle transport equations across the plasma core.

The resulting ASTRA+TGLF predictions for the low-power case are shown in Fig.~\ref{fig:ASTRA+TGLF_modeling_low} as solid lines. To assess sensitivity to boundary conditions, a $\pm 10\%$ variation is applied to the pedestal-top values of each profile, producing the band structures shown in the figure.

Panel (a) indicates a slight underestimation of the electron temperature $T_e$ in the FUSE results compared to ASTRA. However, the agreement remains good across most of the plasma volume. A similarly strong agreement is observed for the ion temperature $T_i$ in Panel (b) and the density profiles in Panel (c), indicating consistency between the reduced (FluxMatcher+TGLFNN) and higher-fidelity (ASTRA+TGLF) approaches.

Panel (d) presents the resulting fusion power density, which is computed based on the temperature and density profiles in combination with tabulated fusion cross-sections. The inset highlights that the baseline scenario yields an estimated fusion power output of approximately $175~\mathrm{MW}$. Notably, the variation in fusion power due to modest changes in boundary conditions is significant ($120-246~\mathrm{MW}$ for Case A and $157-331~\mathrm{MW}$ for Case B), underscoring the strong sensitivity of fusion performance to edge parameters.

Figure~\ref{fig:ASTRA+TGLF_modeling_high} shows the same set of quantities for the high-power scenario. As expected, this case yields a higher fusion output of $234~\mathrm{MW}$, reflecting the increased boundary conditions and overall pressure. Consistent with the trends observed in the low-power case, the electron temperature $T_e$ is predicted to be slightly higher in ASTRA-TGLF compared to FUSE, while the ion temperature $T_i$ is somewhat lower.

Despite these differences in temperature profiles, the density profile remains in very good agreement between the two approaches, indicating that particle transport is captured consistently. As in the low-power case, the sensitivity to boundary conditions remains significant, with variations in pedestal-top values leading to noticeable changes in the predicted profiles and fusion output. This further emphasizes the critical role of edge conditions in determining core performance, even in higher-power operating regimes.

Interestingly, in both cases the fusion power predicted by ASTRA+TGLF tends to be higher than that obtained with FUSE. This is not readily understood from the profiles, which are matched in density and lower in ion temperature. One possible explanation is the dilution by alpha particles, which are not yet included in the ASTRA calculations. In addition, differences in the treatment of fast-ion pressure may lead to slight variations in the effective plasma volume contributing to fusion, resulting in higher overall fusion output.

However, these simulations clearly indicate that the discrepancy in fusion power between ASTRA+TGLF and FUSE is smaller than the effect of a 10\% variation in the boundary conditions. This highlights that such differences are within the typical uncertainty range of these types of calculations. A similar trend is observed in the confinement enhancement factors, $H_{98}$, which are higher in ASTRA+TGLF, 1.95 for the low-power case and 2.03 for the high-power case. Within these uncertainties, these results demonstrate that the steady-state operating point identified by FUSE is robust when validated against higher-fidelity transport modeling.

To assess additional sources of uncertainty, as well as potential actuators to control the fusion output, the localization and magnitude of the ICRH deposition were varied. First, small shifts in the deposition region, e.g., from the Gaussian heating profile being centered at $\rho_\varphi \approx 0.05$ to $\rho_\varphi \approx 0.25$, did not affect the results discussed here. This indicates that, within this radial range, the core profiles are relatively insensitive to moderate changes in the heating location, likely due to stiff transport effects.

The effect of variations in coupling efficiency was then studied by scaling the deposited power density by $\pm 10\%$ for the low-power scenario (i.e., $\approx 12.3$--$14.7$~MW). This resulted in only minor changes in the fusion output, which varied between $174$ and $176$~MW. The weak sensitivity to input power suggests, again, that the system operates in a stiff transport regime, in which profile gradients are largely constrained by the underlying turbulence and self-consistent $\alpha$-heating. In such a regime, additional heating primarily redistributes energy rather than significantly increasing core temperatures or fusion performance. These findings imply that core heating actuators, such as ICRH localization or moderate power variations, provide limited leverage for controlling the fusion output under these conditions.

\begin{figure}
    \centering
    \includegraphics[width=0.75\linewidth]{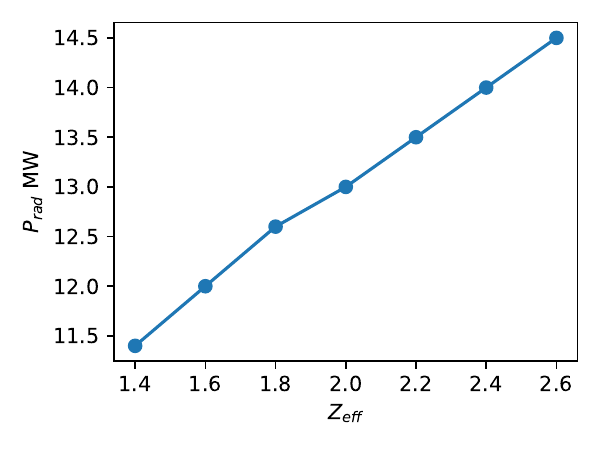}
    \caption{Variation of the radiated power during a scan in the effective charge as a proxy for the core impurity fraction. }
    \label{fig:Zeff_Prad_scan}
\end{figure}

An additional scan, based on the low-power case, was performed by varying the averaged $Z_\text{eff}$ profiles between $1.4$ and $2.6$. In this approach, impurity transport was not modeled self-consistently. Instead, the effective charge was imposed ad hoc as a radially constant quantity. As shown in Fig.~\ref{fig:Zeff_Prad_scan}, this leads to an approximately linear scaling of the radiated power fraction with increasing impurity content, reflecting the proportional increase in impurity concentration, as expected. Nevertheless, the total radiated power remains comparatively small, as low-$Z$ impurities are largely fully ionized in the hot core and therefore radiate inefficiently, with only edge impurities contributing significantly to radiation losses.

The assumption of a constant $Z_\text{eff}$ constitutes a strong simplification. In realistic scenarios, impurity sources at the edge typically lead to radially localized accumulation, with higher concentrations in the peripheral region where temperatures are lower and partial ionization enables efficient radiation. This spatial structure is expected to enhance edge radiation and modify both the power exhaust and pedestal conditions. Consequently, future modeling efforts should incorporate self-consistent impurity transport across the full radius to capture these effects and provide more reliable quantitative predictions.

Despite the variation in impurity content, the fusion output remains nearly constant across the scan, at around $175$~MW. This weak sensitivity indicates that the core plasma conditions are largely insensitive to changes in impurity fraction within the considered range. This behavior is consistent with the limited sensitivity observed in the heating power scan. As the high-power case exhibits a similar transport regime, these findings are expected to remain valid for both the heating deposition and $Z_\text{eff}$ scans.

Overall, the results indicate that the pedestal-top boundary condition has a significant impact on the predicted fusion output, and variations of $\pm 10\%$ in the pedestal-top parameters are adopted as a representative uncertainty interval. Differences between the FUSE and ASTRA workflows are smaller than these uncertainties, yet remain non-negligible and are not surprising given the differences in the model implementations. Rather, these differences can be interpreted as an additional benchmark, ultimately supporting the validity of the discussed concepts.

Neither moderate variations in heating power (deposition) nor substantial changes in core impurity content provide effective actuators for controlling the fusion output. Instead, boundary conditions play a more dominant role. In particular, scans of the pedestal-top density demonstrate a strong influence on overall performance, highlighting edge fueling as an effective control knob. By modifying the pedestal-top density, edge gas fueling affects the core profiles through transport coupling, enabling more efficient control over fusion performance. Another viable actuator is the adjustment of the D-T ratio, which allows for fine-tuning of the fusion output without significantly altering the global confinement properties or transport regime. Thus, edge fueling and adjusting the D-T ratio are proposed as the main actuators for tuning the fusion output power of Yinsen.

Future work will extend this validation to higher-fidelity approaches, such as fully nonlinear gyrokinetic simulations, as well as integrated modeling frameworks that self-consistently couple core transport, edge physics, and impurity dynamics, to further increase confidence in the predicted operational space.

\subsection{Pedestal predictions} 

\begin{figure}[!t]
    \centering
    \includegraphics[width=\columnwidth]{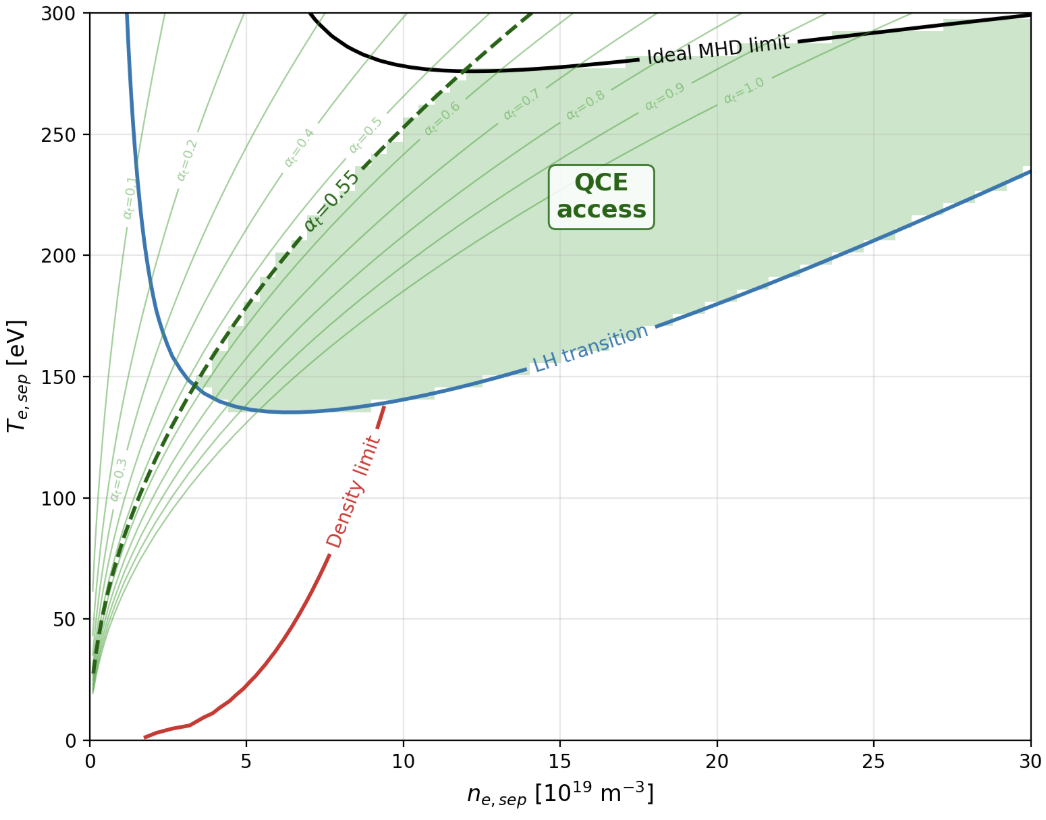}
    \caption{Separatrix operating-space (SepOS) scan for Yinsen, shown in terms of separatrix density and temperature. The plot highlights the H-mode access boundary, the density limit, and the $\alpha_t$ contour relevant to quasi-continuous exhaust (QCE) access.}
    \label{fig:yinsen_sepos}
\end{figure}

The separatrix operating-space view in Figure~\ref{fig:yinsen_sepos}
adds an important edge-physics interpretation to the baseline core
solution. The dimensionless parameter $\alpha_t$, which characterizes
the normalized pressure gradient at the separatrix, determines whether
the edge plasma can sustain quasi-continuous exhaust (QCE) rather than
discrete ELMs. For a strongly shaped plasma such as Yinsen, the
principal requirement for QCE access is $\alpha_t \gtrsim 0.55$
\cite{Eich2025SepOS,Eich2021SepOS,Faitsch2023QCE} within the H-mode
existence region, bounded by the L--H transition on the low-density
side and the H--L back-transition at high density. In practice, this
means the natural route from the present minimum-$T_{e,\mathrm{sep}}$
operating point is not toward lower edge temperature but toward somewhat
higher separatrix density, consistent with the SepOS interpretation
developed for SPARC, where a candidate QCE point was identified at high
$n_{e,\mathrm{sep}}$, moderate $T_{e,\mathrm{sep}}$, and
$\alpha_t \sim 0.7$ \cite{Eich2025SepOS}.

For Yinsen QCE is attractive not only
as a pedestal regime, but also to extend divertor-lifetime. A modest bias toward higher
$n_{e,\mathrm{sep}}$ appears necessary to enter the QCE-accessible
region. Achieving that would replace large Type-I ELM excursions with a
more continuous filamentary exhaust channel, which is much more
compatible with the detached, radiative divertor operation targeted for
Yinsen than an impulsive high-energy ELM regime
\cite{Faitsch2023QCE,Harrer2022QCE}.

\raggedbottom
\subsubsection{Edge boundary conditions}

The pedestal boundary conditions in FUSE are set using EPEDNN but we can also use the empirical \texttt{TE\_MAST} pedestal actor \cite{Meneghini2024FUSE} to provide additional approximate bounding ranges for the edge parameters. Starting with

  \begin{equation}
      \beta_{p,\mathrm{ped}} = \frac{C_1^{1.6}\,
  C_2^{1.2}}{i_{p,\mathrm{norm}}^{0.533}},
      \qquad
      w_{\mathrm{ped,full}} = \frac{C_1^{0.8}\,
  C_2^{1.6}}{i_{p,\mathrm{norm}}^{0.267}},
  \end{equation}

  \noindent where $i_{p,\mathrm{norm}} = I_p\,[\mathrm{MA}] / (a\,B_T)$
  is the normalized plasma current. The coefficients $C_1 = 4.0$ and
  $C_2 = 0.11$ are the MAST-data calibration coefficients adopted for
  this analytic scaling, and the pedestal half-width used in the profile
  construction is $w_{\mathrm{ped}}=w_{\mathrm{ped,full}}/2$ in
  normalized poloidal flux. The pedestal-top pressure then follows from

  \begin{equation}
      p_{\mathrm{ped}} =
  \beta_{p,\mathrm{ped}}\,\frac{B_p^2}{2\mu_0},
  \end{equation}

  \noindent with $B_p = \mu_0 I_p / L_{\mathrm{pol}}$ the
  poloidal-perimeter-averaged poloidal field. The pedestal density is prescribed independently of the pressure
  through
  the Greenwald fraction at the pedestal top,

  \begin{equation}
      n_{e,\mathrm{ped}} \;=\;
  f_{\mathrm{GW,ped}}\;\frac{I_p\,[\mathrm{MA}]}{\pi a^2}
      \;\times\;10^{20}\;\mathrm{m}^{-3},
  \end{equation}

  \noindent Because the pedestal pressure is fixed by the
  scaling law and the density is set by the Greenwald fraction, the
  pedestal electron temperature is fully determined:
  \begin{equation}
      T_{e,\mathrm{ped}} =
  \frac{p_{\mathrm{ped}}}{n_{e,\mathrm{ped}}\,(1 + T_i/T_e)}.
  \end{equation}
  For the present device parameters ($I_p = 9.67$\,MA, $B_T = 9.29$\,T,
  $a = 1.05$\,m, $L_{\mathrm{pol}} \simeq 9.5$\,m, and $T_i/T_e = 1$),
  the analytic pedestal scaling gives $i_{p,\mathrm{norm}} \simeq 0.99$,
  $\beta_{p,\mathrm{ped}} \simeq 0.65$,
  $p_{\mathrm{ped}} \simeq 0.43$\,MPa, and
  $w_{\mathrm{ped}} \simeq 0.044$ in $\psi_N$. To remain consistent with
  the FUSE/ASTRA transport boundary conditions near $\rho_\psi \simeq
  0.8$, we consider two pedestal-top ranges. Case A corresponds to a
  lower-density, hotter pedestal top with
  $n_{e,\mathrm{ped}} = (10.5\text{--}12.5)\times10^{19}\,\mathrm{m^{-3}}$,
  $f_{\mathrm{GW,ped}} \simeq 0.38$--0.45, and
  $T_{e,\mathrm{ped}} \simeq 10.6$--12.6\,keV. Case B corresponds to a
  higher-density pedestal with
  $n_{e,\mathrm{ped}} = (14\text{--}16)\times10^{19}\,\mathrm{m^{-3}}$,
  $f_{\mathrm{GW,ped}} \simeq 0.50$--0.57, and
  $T_{e,\mathrm{ped}} \simeq 8.3$--9.5\,keV.

  The separatrix values are then treated as the edge endpoint of the
  same hyperbolic-tangent profile construction. We take
  $T_{e,\mathrm{sep}} \simeq 100$-200\,eV, spanning conservative
  attached-divertor conditions and hotter-edge operation relevant to
  possible QCE access. The corresponding separatrix density ranges are
  $n_{e,\mathrm{sep}} \simeq (3\text{--}4)\times10^{19}\,\mathrm{m^{-3}}$
  for case A and $(4\text{--}5)\times10^{19}\,\mathrm{m^{-3}}$ for case B.

\section{ICRH Heating Deposition and Minority Scheme} 

Yinsen will rely on ion cyclotron resonance heating (ICRH) for auxiliary heating and plasma-temperature control. Only one resonance-heating mode is presently anticipated, namely minority(fundamental)-hydrogen / second-harmonic deuterium, with the two resonances overlapping. The resonance zone is placed at the magnetic axis, which fixes the operating frequency at \SI{140}{MHz}. The primary requirement for the ICRH system is complete first-pass absorption of the launched RF power as the wave propagates from the antenna to the resonance zone, so that deposition remains strongly peaked in the central core with little or no transmitted power available to damp off-axis.

\begin{figure}
    \centering
    \includegraphics[width=\columnwidth]{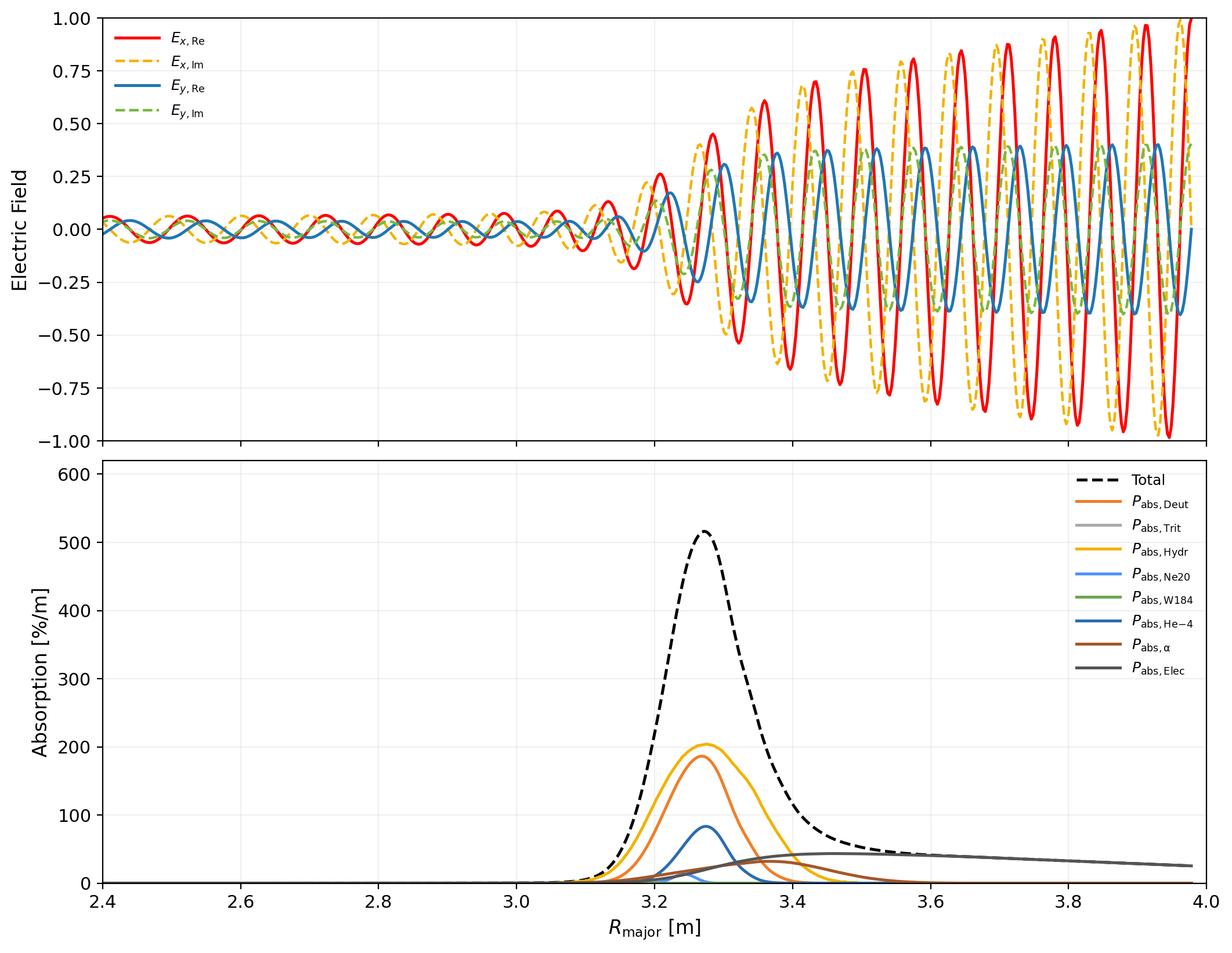}
    \caption{ICRH wave propagation and absorption on plasma species from a 1D slab CARDS analysis. The upper panel shows the wave electric field, with the wave launched from the right (low-field side). The lower panel shows the absorption profile versus major radius for electron Landau damping together with hydrogen fundamental minority resonance and deuterium, neon, and helium second-harmonic absorption channels.}
    \label{fig:yinsen_cards_run_nominal}
\end{figure}

To predict those absorption profiles, we use the 1D full-wave modeling code CARDS \cite{Smithe1987CARDS}. For the present application, the 1D midplane treatment is an accurate representation of the dominant propagation path because the scenario is intentionally designed for strong first-pass absorption. CARDS was originally used at PPPL for ICRH studies on TFTR and has been modernized for the present study, with benchmarking used to verify that the rewrite did not change the underlying results. The model includes a first-order finite-Larmor-radius treatment for warm-plasma absorption, together with mode-conversion physics and electron Landau damping, the latter being significant for a reactor-relevant high-temperature plasma.

Figure~\ref{fig:yinsen_cards_run_nominal} shows a representative CARDS run using the high-power electron-density and electron-temperature profiles from Figure~\ref{fig:ASTRA+TGLF_modeling_high}. In that case, the peak electron density is \(2.02\times10^{20}\,\mathrm{m^{-3}}\) and the peak electron temperature is 19.9~keV, with the ion temperature taken to be comparable to the electron temperature. The species mix is \(50{-}50\) deuterium--tritium, with \(n_D/n_e = n_T/n_e = 0.345\), together with significant neon, \(n_{\mathrm{Ne}}/n_e = 0.0087\), and helium-4 ash, \(n_{\mathrm{He}}/n_e = 0.1065\). Approximately 92\% of the helium-4 population is treated as thermal, with the remaining 8\% treated as a fast-ion population having a core temperature of \SI{300}{keV} that falls to \SI{9}{keV} at the edge. For this case, the minority-hydrogen concentration is \(n_H/n_e = 0.01\). The figure reports the electric field and absorption profiles versus major radius. The first-pass absorption in this case is 99.4\%.

One important feature of these profiles is the broad electron deposition caused by electron Landau damping. This acts along the full wave-propagation path, in contrast to the ion absorption, which remains localized near the resonance zone. In practice, this electron damping is largely unavoidable at fusion-scale temperatures and depends only mildly on the launched toroidal \(k_{\parallel}\) spectrum.

In Figure~\ref{fig:yinsen_cards_run_nominal} it is evident that the deuterium second-harmonic damping profile is narrower than the hydrogen-minority damping profile. Thus, minority concentration can be used to adjust the width of the absorption profile. At the same time, second-harmonic damping is strongly dependent on ion temperature, and if that is too cold, then complete first-pass absorption may not occur.

\begin{figure}
    \centering
    \includegraphics[width=\columnwidth]{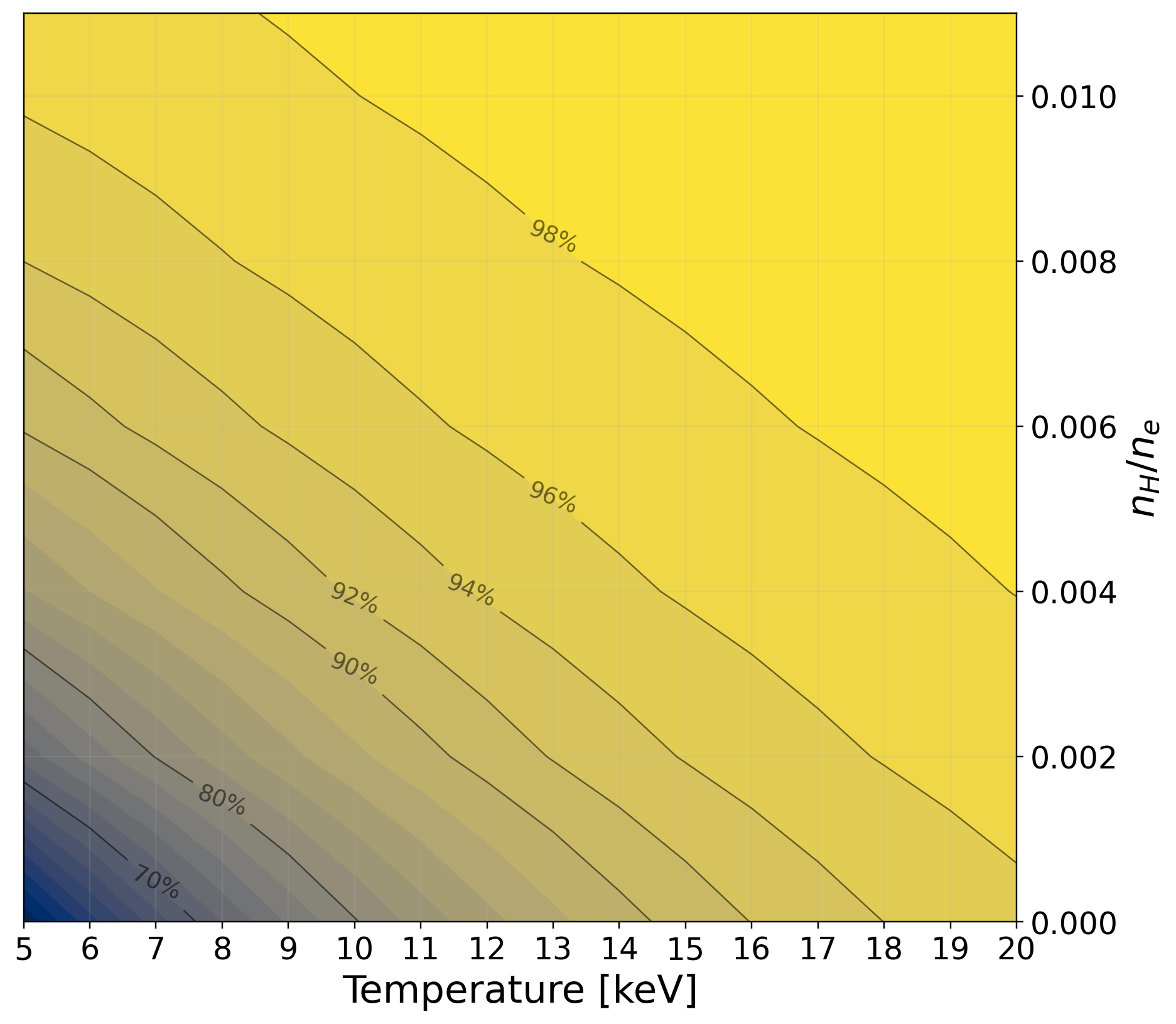}
    \caption{Variation of first-pass absorption with temperature and minority concentration. For the nominal Yinsen point at \(n_H/n_e = 0.01\), complete first-pass absorption is maintained across the scanned temperature range; by contrast, zero minority concentration at lower temperatures does not achieve complete first-pass absorption.}
    \label{fig:eta_temp_study}
\end{figure}

Figure~\ref{fig:eta_temp_study} shows the variation in first-pass absorption percentage for temperatures from 5~keV to 20~keV and minority concentration from 0 to 0.02. At the low end, 5~keV and zero minority, the first-pass absorption is only 50\%. However, for nominal parameters, with \(n_H/n_e = 0.01\), first-pass absorption is assured for the entire temperature range. At the steady operating temperature of 20~keV, first-pass absorption occurs even with little to no hydrogen minority. In one possible scenario, a higher minority fraction is used during startup, with that concentration tapering down during flattop as the ion temperature rises and second-harmonic absorption becomes more effective.

Neon and helium-4 also experience second-harmonic absorption at the same location as deuterium. Of particular concern is the fast helium-4 ash produced by D--T fusion reactions, because its much higher effective temperature can create an unwanted parasitic absorption channel. For that reason, the hot helium-4 species is included explicitly in the CARDS model. In the present runs, fast-ion absorption is small but still visible, at approximately 7.7\%. Additional runs including very small concentrations of fast helium-3 and tritium ash from the lower-probability D--D branch showed negligible effect.

It is also worth noting that the tritium third-harmonic resonance is coincident with the minority-hydrogen / second-harmonic-deuterium resonance. CARDS does not presently include that resonance, because doing so would require either higher-order Larmor truncation or a more difficult all-orders treatment. Since tritium is a thermal species in this scenario, its third-harmonic absorption is nevertheless expected to be small. The minority-hydrogen / second-harmonic-deuterium scenario also places the second and fourth harmonics of tritium just outside the inboard and outboard midplane scrape-off layers, respectively. Since the SOL temperatures are much lower, these higher-order edge resonances are not presently expected to be important.

An alternate resonance-heating mode for high-field tokamaks is helium-3 minority / second-harmonic tritium \cite{Migliore2026SPARCICRF}. That approach would reduce the RF frequency to about \SI{93}{MHz} in the Yinsen case. We do not presently envision using that scenario for Yinsen, primarily because Helium-3 availability is prohibitive for mass market adoption. A secondary issue is that it places the hydrogen fundamental and deuterium fundamental resonances just outside the inboard and outboard midplane SOLs, respectively. The deuterium fundamental resonance near the antenna launchers is especially concerning because the associated Bernstein-wave physics could create a channel for parametric decay and parasitic edge loss.

Investigation of launcher design, edge coupling, RF sheaths, and impurity production is outside the scope of the present study, but clearly belongs in future work. There is already a substantial body of prior work on multi-strap, Faraday-shielded antennas that should carry over directly to the Yinsen problem. In addition, the narrow-SOL, high-density configuration of a high-field tokamak may permit alternate launcher concepts \cite{Smithe2025NovelCoupler}. Finally, the \SI{140}{MHz} operating frequency lies above the band of a widely used high-power tetrode for tokamak ICRH, which tops out near \SI{130}{MHz} \cite{Moriyama1992X2242X2274}. In the absence of further development of that vacuum-electronics source, Yinsen will therefore rely on power combination from kilowatt-scale solid-state RF sources, which are not restricted at \SI{140}{MHz}.

\section{Disruption forces modeling with TokaMaker} 

Disruptions are a central engineering risk for compact high-field tokamaks because large magnetic and thermal stored energy can be released on short timescales. Disruption severity also scales strongly with plasma current through current-quench rate and halo-current loading, making mechanical survivability of in-vessel structures a first-order design constraint \cite{maris_impact_2022}. To quantify this for Yinsen, we use the TokaMaker workflow established for vertical controllability to calculate the forces induced in the vacuum vessel during representative unmitigated disruption scenarios. While steps will be taken to avoid disruptivity limits during routine operation \cite{wang_active_2025}, it is still essential that Yinsen, as a first-of-its-kind device, be able to withstand the associated disruption loads.

\begin{figure}[!t]
    \centering
    \includegraphics[width=\columnwidth]{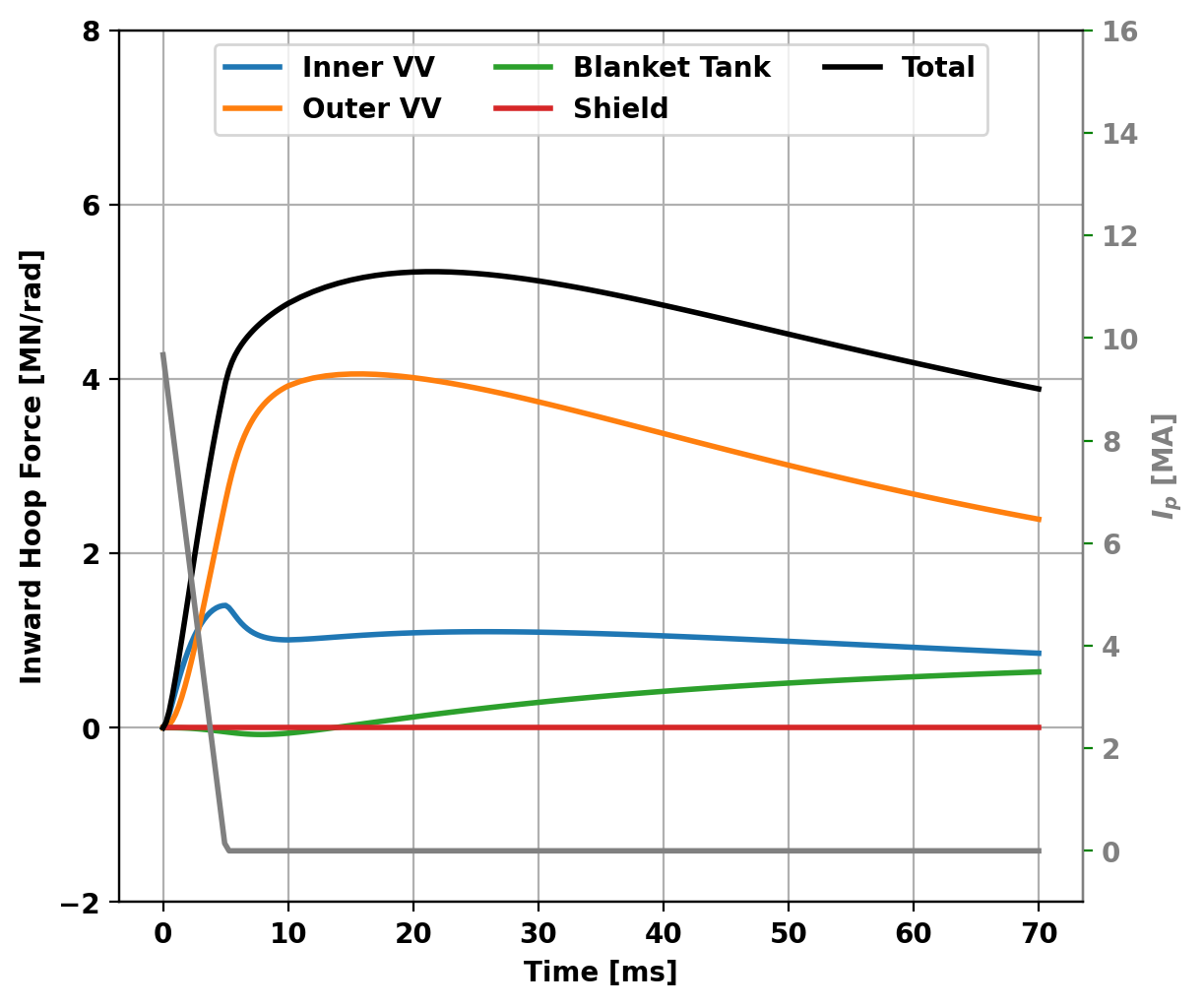}
    \caption{Time evolution of the integrated inward hoop-force contributions from the inner and outer vacuum vessels, blanket tank, and shield, together with the imposed plasma-current quench on the secondary axis. The total inward hoop force peaks at approximately \SI{5.23}{MN/rad}, with the outer vacuum vessel carrying the dominant share.}
    \label{fig:disruption_hoop_force}
\end{figure}

To simulate the effects of a disruption with TokaMaker, a model current quench (CQ), which would result from VDE-induced and symmetric disruptions, is applied during a time-dependent simulation of plasma evolution. Here we set a linear current quench time of 10ms based on the ITPA disruption database \cite{eidietis2015itpa}. Figure~\ref{fig:disruption_hoop_force} summarizes the resulting electromagnetic loading on the conducting structures, including the inner and outer vacuum-vessel shells, blanket vessel, and surrounding shield structures. The total integrated inward hoop force rises rapidly during the first few milliseconds of the CQ and peaks at approximately \SI{5.23}{MN/rad}. That peak is dominated by the outer vacuum vessel, which reaches about \SI{4.06}{MN/rad}, while the inner vessel peaks near \SI{1.40}{MN/rad}. The shield contribution remains negligible on this scale, which is also consistent with the fact that the shield is not toroidally continuous by design and therefore does not provide a strong continuous current path around the machine. The blanket tank contribution grows more gradually to about \SI{0.64}{MN/rad} as induced currents diffuse outward. This is physically consistent with the present geometry, since the thicker outer vessel provides the primary global load-bearing shell during the induced-current response, while the inner vessel participates more as a plasma-facing boundary than as the main structural member.

\begin{figure}[!t]
    \centering
    \includegraphics[width=0.92\columnwidth]{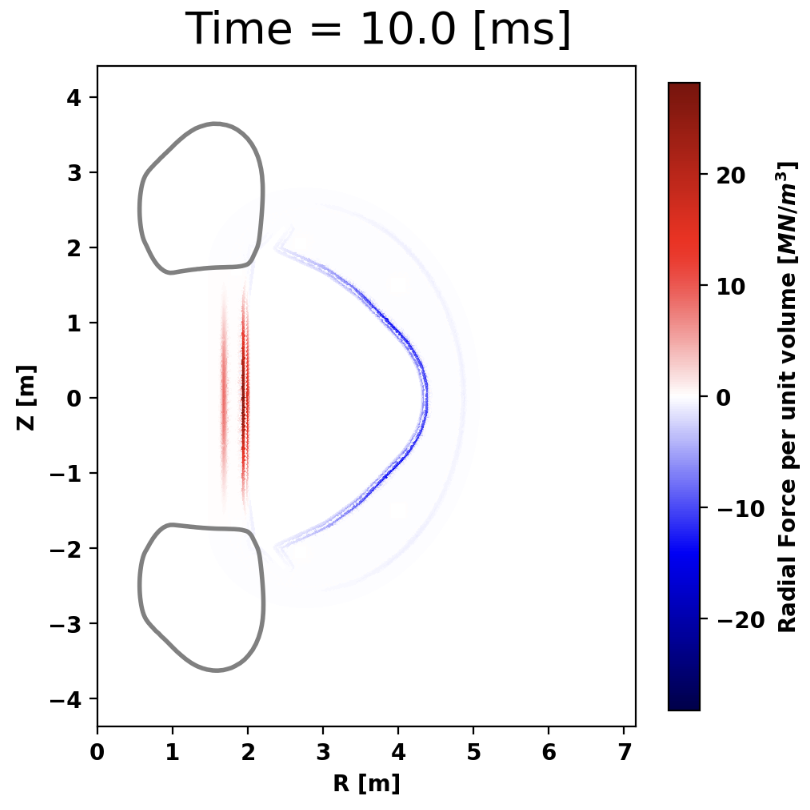}
    \caption{Radial force density at \SI{10}{ms} during the imposed current quench, showing that the strongest loading is concentrated on the inboard straight leg and the outboard curved nose of the vessel.}
    \label{fig:disruption_radial_force}
\end{figure}

The force history also shows that the total structural load remains elevated after the plasma current itself has collapsed, which reflects the finite decay time of eddy currents in the surrounding conducting shells. The radial force-density snapshot in Figure~\ref{fig:disruption_radial_force} shows the same story spatially: the strongest loading is concentrated in the inboard straight section and the outboard curved nose of the vessel, matching the expected halo-current and eddy-current pattern for a compact high-field tokamak. From a design perspective, the important result is not that disruption forces are small, but that the dominant global load path runs through the outer vacuum vessel, which is also the thicker (\SI{3.5}{cm}) compared to the inner vacuum vessel (\SI{1.5}{cm}).

While a full three-dimensional FEA of resulting deformation, local stress concentration, weld loading, and plastic strain is beyond the scope of the present work, the TokaMaker results indicate that the disruption challenge for Yinsen is of the same general character as that reported for devices such as ARC and ITER, rather than representing a qualitatively new regime. A useful first-pass screening can nevertheless be obtained by treating each vessel shell as a thin cylinder. If $F'_i$ is the integrated inward hoop force per radian from TokaMaker for shell $i$, and $L_i$, $R_i$, and $t_i$ are the effective loaded height, mean radius, and wall thickness of that shell, then the equivalent radial pressure is $p_{\mathrm{eq},i}\approx F'_i/(L_iR_i)$ and the corresponding membrane hoop stress is
\beq
\sigma_{\theta,i}\approx \frac{p_{\mathrm{eq},i}R_i}{t_i}=\frac{F'_i}{L_it_i}.
\eeq
Using the peak values from Fig.~\ref{fig:disruption_hoop_force}, together with the disruption-model shell dimensions, gives \(\sigma_{\theta,\mathrm{inner}}\approx \SI{20}{MPa}\) for the inner vacuum vessel and \(\sigma_{\theta,\mathrm{outer}}\approx \SI{25}{MPa}\) for the outer vacuum vessel. These are membrane-stress estimates only, so they do not capture local bending, weld stress concentration, buckling, or the larger asymmetric halo-current loads that would require a dedicated structural analysis. In that sense, the present results should be interpreted as showing that disruption loading remains a first-order structural design constraint for Yinsen, but one that can be addressed through conservative vessel sizing, deliberate load sharing into the outer shell and supports, and future disruption-mitigation provisions. The application of additional disruption mitigation techniques, such as the inclusion of a dedicated runaway electron mitigation coil, may be the subject of future investigation.

\FloatBarrier
\input{edge_divertor}

%% file: edge_divertor.tex
\section{Divertor and power exhaust}\label{sec:divertor_power_exhaust}

\subsection{Projections for SOL width}

Empirical scalings for the SOL power width $\lambda_q$ are used to
estimate the expected SOL width in Yinsen. Using representative
parameters, $T_{sep}$=100 eV and $B_{pol}$=1 T, with the scaling
proposed in Ref. \cite{Eich2020},

\beq
\lambda_q [mm] \approx 1.6 \times \frac{a}{R} \rho_{s,pol},
\eeq

where

\beq
\rho_{s,pol} = \frac{\sqrt{m_D T_{sep}}}{e B_{pol}}
\eeq

one finds $\lambda_q \approx$ 0.5 mm.

Alternatively, with the scaling proposed in Ref. \cite{Brunner2018},

\beq
\lambda_q [mm] = (0.63 \pm 0.08) \times B_p^{-1.19 \pm 0.08}
\eeq

one finds $\lambda_q \approx$ 0.6 mm.

These numbers provide the ballpark range of the expected SOL width
which guides detailed edge and divertor modeling described further.

\begin{table*}[!t]
  \centering
  \caption{Comparison of divertor metrics between Yinsen,
    ARC~V1~\cite{Sorbom2020ARCRecent}, ITER~\cite{Pitts2019ITERDivertor},
    EU-DEMO~\cite{Reimerdes2020DEMODivertor},
    Japan-DEMO~\cite{Asakura2017JapanDEMODivertor}, and
    CFETR~\cite{Li2020CFETRConfig}. $^*$ARC $n_{sep}$ value is
    estimated by assuming $n_{sep} \sim 0.35\langle n \rangle$ for
    typical H-mode, where $\langle n \rangle$ is the core
    volume-average density, as done in
    \cite{Miller2024MantaPowerHandling}.}
  \label{tab:yinsen_divertor_metric_comparison}
  \scriptsize
  \setlength{\tabcolsep}{4pt}
  \renewcommand{\arraystretch}{1.05}
  \resizebox{\textwidth}{!}{%
  \begin{tabular}{lccccccc}
    \toprule
    Parameter & Yinsen A & Yinsen B & ARC V1 & ITER & CFETR & EU-DEMO & Japan-DEMO \\
    \midrule
    $P_{\mathrm{fus}}$ (MW) & 130 & 185 & 500 & 500 & 558 & 2000 & 1694 \\
    $R$ (m) & 3.18 & 3.18 & 3.65 & 6.2 & 7.2 & 8.8 & 8.5 \\
    $B_T$ (T) & 9.29 & 9.29 & 11.6 & 5.3 & 6.5 & 5.8 & 5.94 \\
    $P_{\mathrm{SOL}}$ (MW) & 26.8 & 33.4 & 83 & 100 & 91 & 150 & 258 \\
    $n_{sep}$ ($10^{20}\,\mathrm{m^{-3}}$) & 0.273 & 0.328 & 0.61 & 0.45 & 0.25 & 0.25 & 0.2 \\
    $P_{\mathrm{SOL}} B_T / R$ & 78.3 & 97.6 & 263 & 85.5 & 82.2 & 98.9 & 369 \\
    $\left(P_{\mathrm{SOL}} B_T / R\right) / n_{sep}^2$ & 1051 & 907 & 707 & 422 & 1310 & 1580 & 9230 \\
    \bottomrule
  \end{tabular}%
  }
\end{table*}

The two simple screening metrics used below separate the parallel
heat-flux challenge from the dissipation challenge. In the first
metric, $P_{\mathrm{SOL}} B_T / R$, Yinsen is helped by the fact that
$P_{\mathrm{SOL}}$ is intentionally kept relatively low: the machine is
designed near the minimum useful fusion-power class, with only modest
fusion-alpha power and minimal auxiliary heating, so the lower bound
on SOL power is set mainly by the requirement to remain above
$P_{LH}$ for H-mode access rather than by a large exhaust burden from
high core power. The second metric,
$\left(P_{\mathrm{SOL}} B_T / R\right) / n_{sep}^2$, which captures the challenge of heat dissipation, remains more
challenging because H-mode operation still implies
relatively low edge and separatrix density. In that sense, Yinsen is
not an especially severe parallel heat-flux problem for a high-field
device, but it is still a demanding detached-divertor problem because
the available volumetric dissipation margin is limited by the low edge
density.

\subsection{UEDGE modeling}

Detailed edge and divertor simulations are carried out for Yinsen with
edge plasma transport code \cite{Rognlien1992}. UEDGE solves the
Braginskii fluid equations for magnetized plasmas in
the tokamak SOL to determine the steady-state plasma density,
temperature, power and particle flows. It also utilises a fluid model
to include computation of neutral dynamics and particle sources over
the simulation domain. UEDGE has been previously applied to study
divertors in high-field reactor-class devices and is here used to
refine the point-design and quantitatively assess divertor and power
exhaust performance.

For the desribed modelings, the basic UEDGE physics model is used, as
described in \cite{Rognlien1992}; not including plasma drifts and
electric currents at this stage. For impurity radiation, impurity ions
are included in the fixed-fraction model \cite{Hulse1983,Post1995}.
Furthermore, to simplify the model, a diffusive neutrals model is used
for hydrogen gas instead of the more detailed Navier-Stokes model.

The UEDGE baseline case is configured for major radius
$R_{maj}$=3.18 m, the exhaust power into the SOL in the range of $P_{sol}$=30 MW,
which corresponds to the characteristic P/R ratio of 9.4. The
exhaust power from the core $P_{sol}$ is applied at the core
interface, distributed uniformly in the poloidal direction, and evenly
split between electrons and ions. Plasma density is set at the core
interface to match the density at the separatrix in the expected edge density range of 2-5 $\times$ 10$^{19}$ m$^{-3}$. The magnetic geometry for Yinsen is close to a symmetric double null,
and therefore for the present modeling, UEDGE is set up for the
lower-half domain, assuming that the midplane is the up-down symmetry
plane. While the model does not include full details of the first wall,
the target-plate locations and angles represent the current divertor design. Note that in previous tokamak experiments, the tilt of divertor target plates affects the threshold of divertor detachment\cite{Wang2022}. The design used here is a first pass at a geometry that is not yet fully optimized with respect to target-plate tilt and neutral baffles, which can further increase divertor performance and detachment access in future iterations.

Since the SOL width is much smaller than the
distance to the outer wall, the details of the outer-wall geometry
likely do not affect the plasma equations being solved in
UEDGE. However, the shape of the wall may still have an effect on
the recycling neutrals as the wall serves as neutral
baffles. Including a more detailed description of the neutral baffles
represents a potential avenue for refinement of the model in future
iterations.

In the model, the recycling coefficient on all material surfaces is
set to 0.99, to account for some neutral gas pumping at the wall. The
anomalous thermal transport model is set using globally uniform
thermal transport coefficients $\chi_{\perp,e,i} = 0.01$ m$^2$/s to
match the expected ~0.5 mm power width of the SOL. The anomalous
density diffusion coefficient is taken $D$=0.2 m$^2$/s.

Fig. (\ref{fig:ue_base_case}) describes the UEDGE base case with
electron density on the separatrix $n_{sep}$ = 3 $\times$ 10$^{19}$
m$^{-3}$ and $P_{sol}$ = 27 MW. In the base case, there is no impurity radiation in the
model, and the divertor heat flux arrives essentially unmitigated on
the target plates. Without impurity radiation, the peak heat flux on
the outer target plate is extremely high, close to 100 MW/m$^2$ on the
outer plate, see Fig. (\ref{fig:ue_base_case}).
{
  \begin{figure}[!t]
    \includegraphics[scale=0.40]{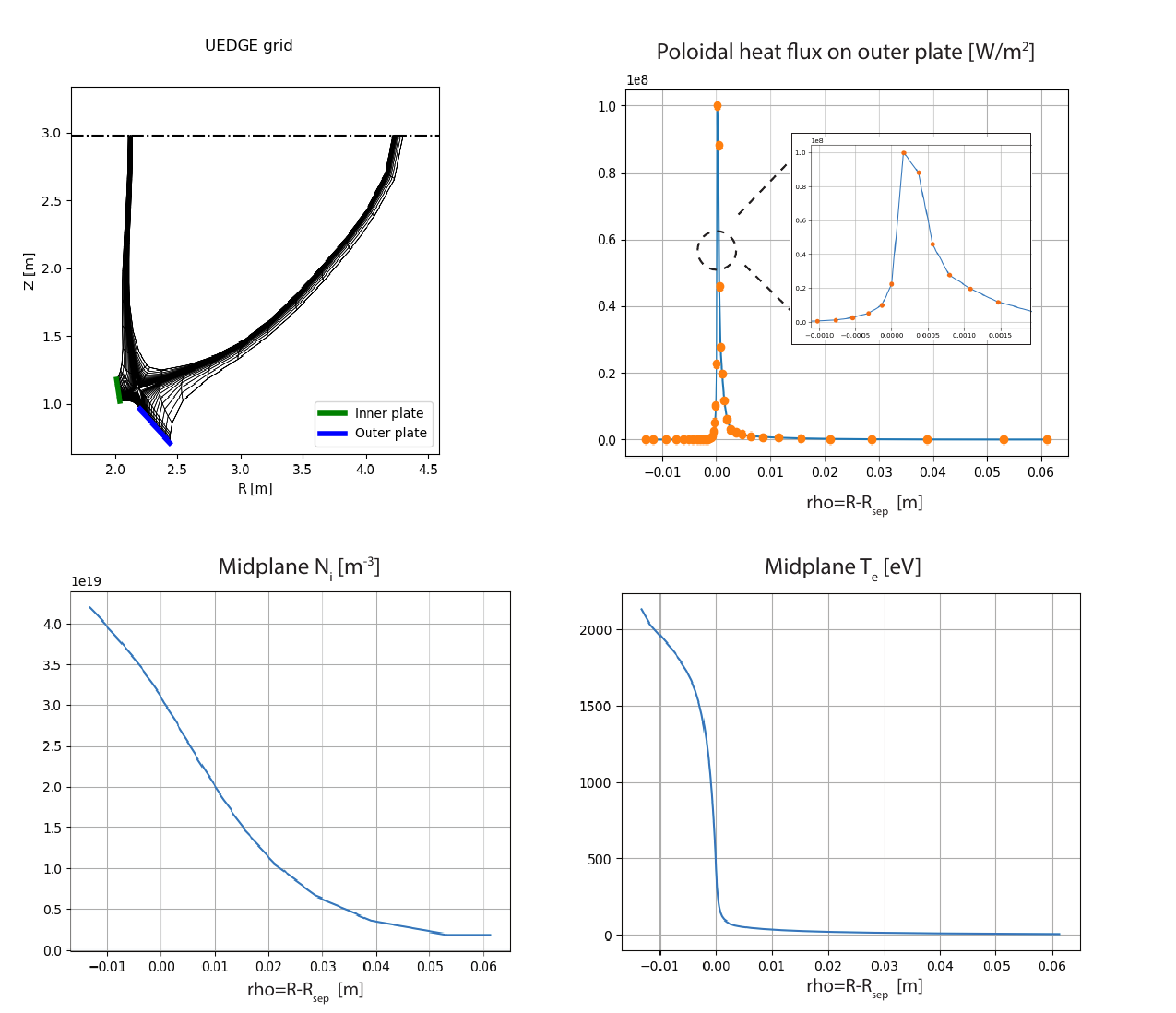}
    \caption{ UEDGE base case with $n_{sep}$=3$\times$10$^{19}$
      m$^{-3}$, $P_{sol}$ 27 MW, and no impurity
      radiation. The top left plot shows the model geometry and the top right plot shows the
      normal heat flux profile on the outer target plate. The heat profile
      width (mapped to the midplane) is close to 0.5 mm, consistent
      with the Eich scaling projection. \hfill }
    \label{fig:ue_base_case}
  \end{figure}
}

Next, neon impurity is introduced in the ``fixed-fraction'' model
\cite{Hulse1983,Post1995}. Keeping $n_{sep}$ and $P_{sol}$ fixed, the
impurity fraction is increased, and one can observe that the peak
power on the target plates and the peak plasma temperature on the
plates are monotonically reduced as the neon impurity fraction is
ramped up, as shown in Fig. (\ref{fig:ue_zimp_sweep}).
  
{
  \begin{figure}
    \includegraphics[scale=0.40]{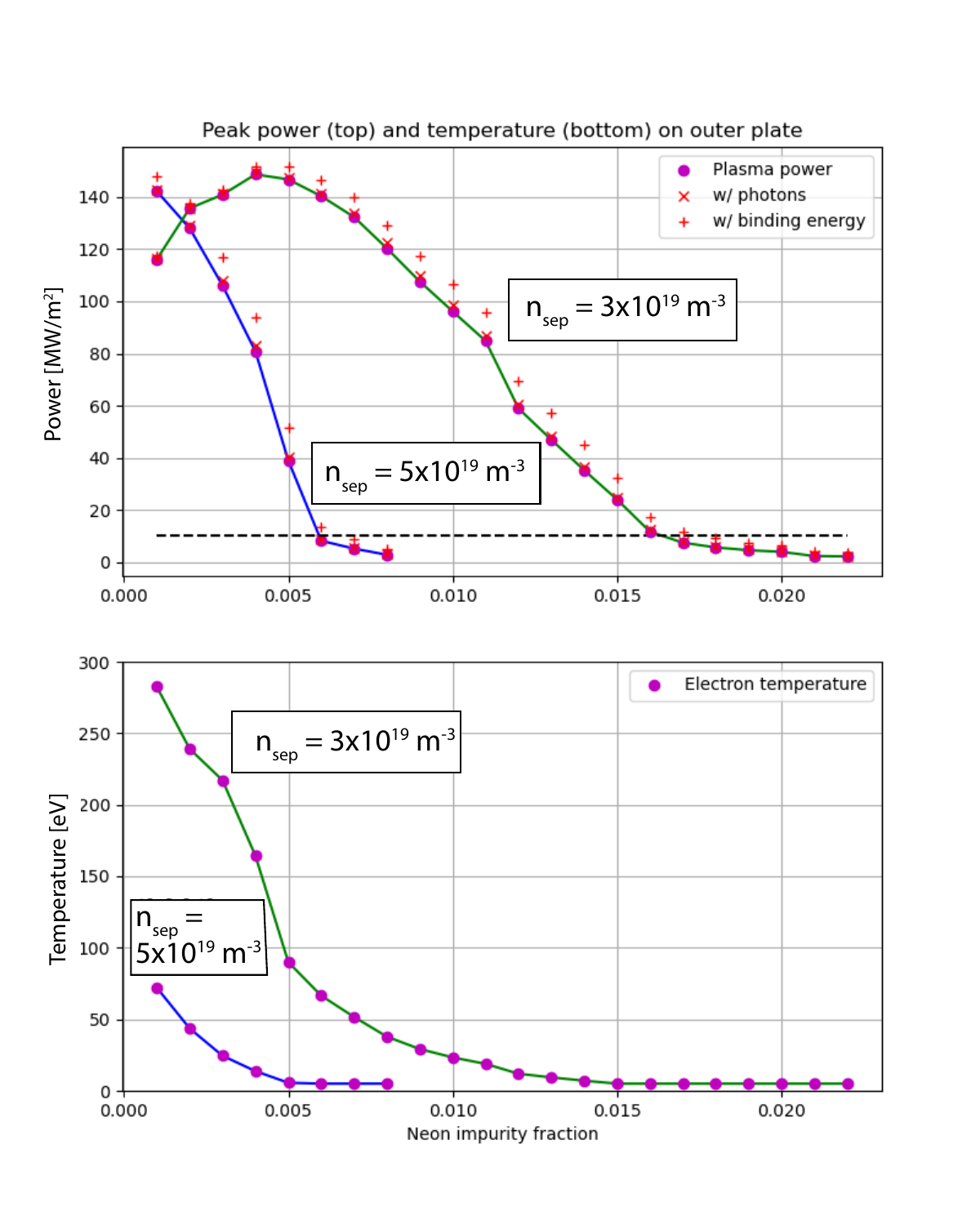}
    \caption{
      The peak power flux and peak temperature on the outer target
      plate vs. the power from the core $P_{core}$ and neon impurity
      fraction $\alpha_z$. The dashed line in the top plot shows the
      level 10 MW/m$^2$. Two cases are shown, with
      $n_{sep}$=3$\times$19$^{10}$ m$^{-3}$ and
      $n_{sep}$=5$\times$19$^{10}$ m$^{-3}$. For higher density, a
      lower impurity fraction is needed to achieve the detachment
      as the impurity radiation scales quadratically with the
      plasma density.  \hfill }
  \label{fig:ue_zimp_sweep}
\end{figure}
}

At some critical impurity fraction, which depends on both $n_{sep}$ and $P_{sol}$,
as well as the chosen impurity species, the divertor plasma
transitions to a state where the peak power flux drops below 10
MW/m$^2$ and the peak plasma temperature on the plate drops below 10
eV; at the same time, the neutral density einxceeds the plasma density
near the plate; these are signatures of a deeply detached divertor.
Plasma profiles for such deeply detached case corresponding to the base UEDGE
case with $P_{sol}$ = 27 MW, plasma density on the separatrix
$n_{sep}$ = 3 $\times$ 10$^{19}$, and neon impurity fraction $2.3\%$,
are shown in Fig. (\ref{fig:ue_detached_case}).
Because the high edge impurity concentrations required for strong
detachment can, in principle, poison the core plasma through fuel
dilution and enhanced core radiation, and in the limit drive
radiative collapse or loss of H-mode operating margin
\cite{Kallenbach2013Seeding,Pitts2019ITERDivertor}, we further study
the operating space in separatrix density, $P_{sol}$, and impurity
concentration to refine the accessible detached regime. One of the key benefits of higher separatrix density is that lower impurity concentration is required for detachment, and thus the risk of poisoning the core from impurity transport is reduced.

Note that in divertor modeling studies, it is generally asssumed that
$\sim$1 $\%$ of neon impurity in the divertor is compatible with a
high-performance burning plasma core \cite{Wigram2019,Lore2024}. In
part this is supported by long-pulse experiments on EAST with neon
injection in the divertor without deterioration of the core plasma
performance \cite{Gong2024}.

{
  \begin{figure}
    \includegraphics[scale=0.25]{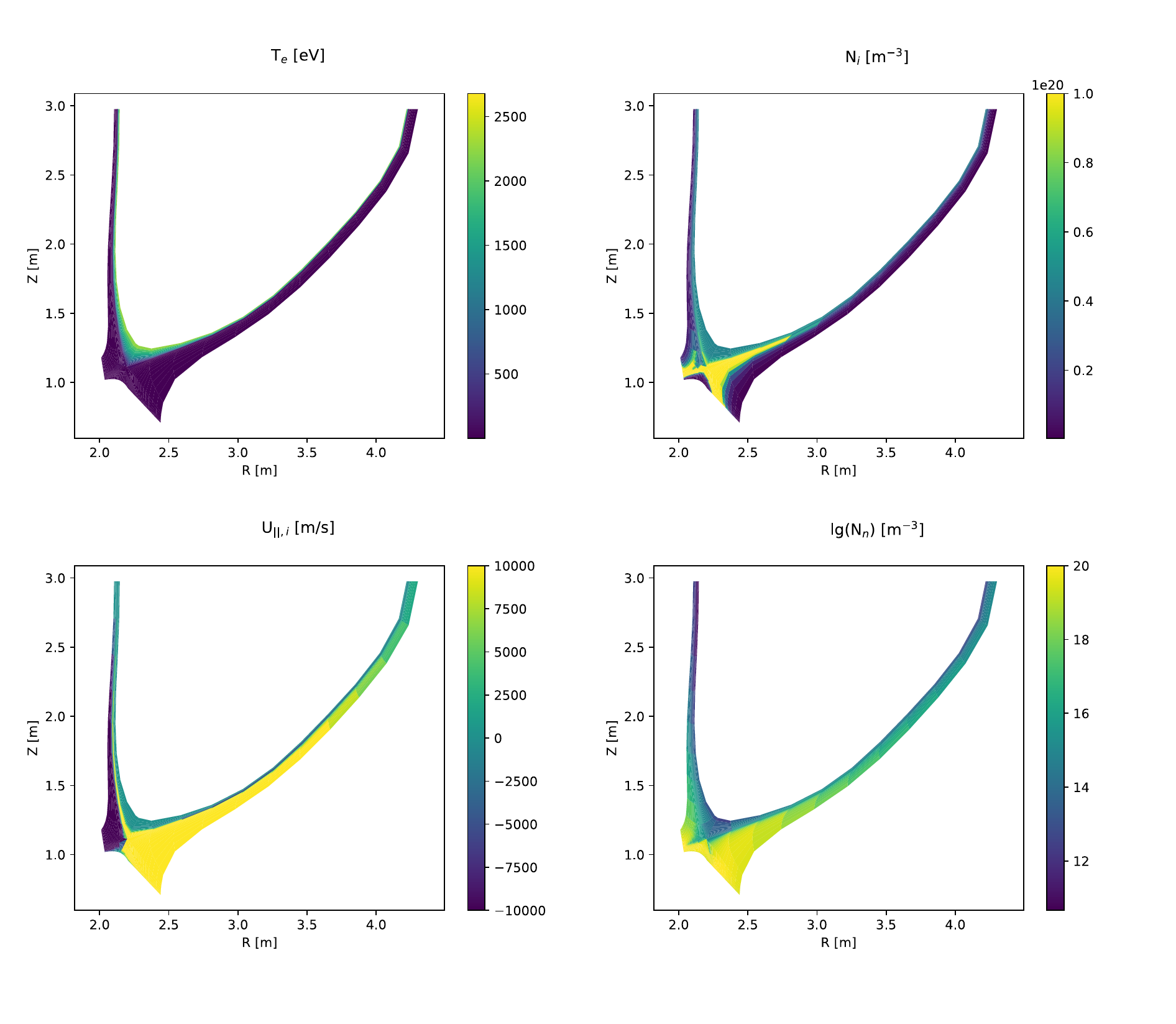}
    \caption{
      Plasma 2D profiles from the fully-detached UEDGE base
      case with $n_{sep}$ = 3 $\times$ 10$^{19}$, $P_{sol}$ = 27 MW,
      and 2.3$\%$ neon fraction is shown.  The peak power flux on
      the target plates is reduced to only $\sim$1 MW/m$^2$ due to the significant
      impurity radiation.  \hfill }
\label{fig:ue_detached_case}
\end{figure}
}

The threshold for detachment transition is a function of $P_{sol}$ and
the impurity fraction $\alpha_z$, given that other parameters, mainly $n_{sep}$  are
fixed. Scanning parameter $P_{sol}$ and the impurity fraction
$\alpha_z$, one can establish an approximate scaling for detachment
transition, as illustrated in Fig. (\ref{fig:ue_scan2D}).

The resolved target-plate heat-flux profiles also help clarify how the
detachment transition redistributes the local power load between the
inner and outer plates. In the lower-density branch the outer target
	remains the limiting location for the representative
	\SIlist{32.5;35}{MW} cases shown here, whereas in the higher-density
	branch the inner and outer peaks become more comparable as the neon
	fraction is increased, as shown in
	Fig. (\ref{fig:ue_target_plate_profiles}).
These resolved divertor heat-flux profiles are then used directly as
inputs to the target plate thermal model discussed in the following
subsection, where they are mapped onto a candidate FLiBe-cooled target
geometry.

{
  \begin{figure*}[!p]
    \centering
    \includegraphics[width=0.885\textwidth]{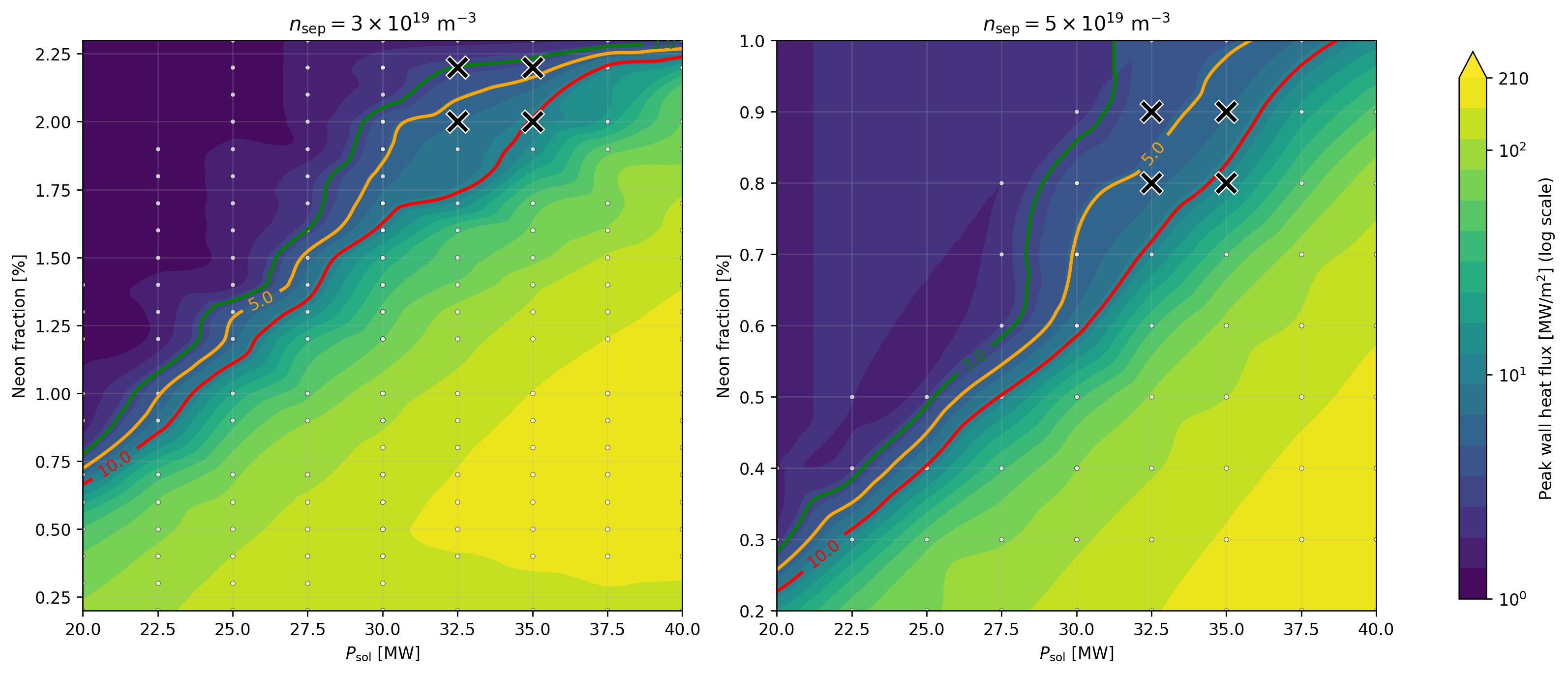}
    \caption{Contour plots of the peak wall heat flux extracted from the
      full poloidal $q_{\perp}$ profiles as a function of
      $P_{\mathrm{sol}}$ and neon impurity fraction $\alpha_z$ for
      $n_{\mathrm{sep}} = 3 \times 10^{19}\,\mathrm{m^{-3}}$ (left) and
      $n_{\mathrm{sep}} = 5 \times 10^{19}\,\mathrm{m^{-3}}$ (right).
      The green, orange, and red contours mark the
      \SI{3}{MW/m^2}, \SI{5}{MW/m^2}, and \SI{10}{MW/m^2} levels,
	      respectively. Black crosses mark the cases used for the resolved target-plate
	      profile comparisons: $P_{\mathrm{sol}}=\SIlist{32.5;35}{MW}$
	      with Ne$=2.0\%,2.2\%$ for the lower-density branch and
	      Ne$=0.8\%,0.9\%$ for the higher-density branch.\hfill }
    \label{fig:ue_scan2D}
    \vspace{0.5em}
    \includegraphics[width=0.885\textwidth]{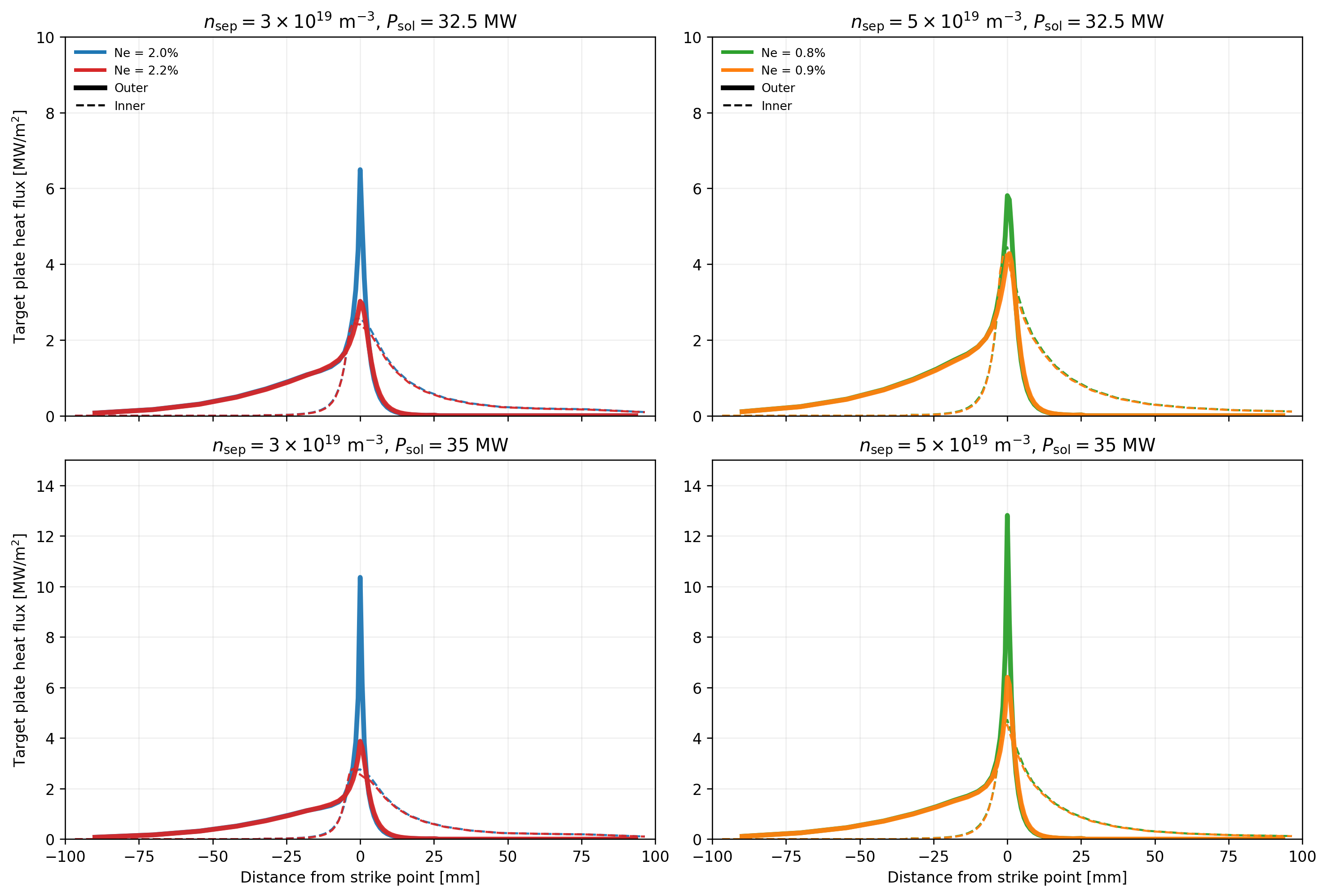}
    \caption{Resolved target-plate heat-flux profiles centered on the
      local strike points for representative cases near the detachment
	      threshold. The top row shows $P_{\mathrm{sol}} = \SI{32.5}{MW}$ and
	      the bottom row shows $P_{\mathrm{sol}} = \SI{35}{MW}$. The left
	      column corresponds to $n_{\mathrm{sep}} = 3 \times 10^{19}\,\mathrm{m^{-3}}$
	      with neon fractions of 2.0\% and 2.2\%, and the right column to
	      $n_{\mathrm{sep}} = 5 \times 10^{19}\,\mathrm{m^{-3}}$ with neon
	      fractions of 0.8\% and 0.9\%. Solid curves denote the outer
      target plate and dashed curves the inner target plate.\hfill }
    \label{fig:ue_target_plate_profiles}
  \end{figure*}
}

\input{first_wall}

%% file: first_wall.tex
\subsection{Thermal analysis of the target plate}

{
  \begin{figure*}[!p]
    \centering
    \includegraphics[width=0.94\textwidth]{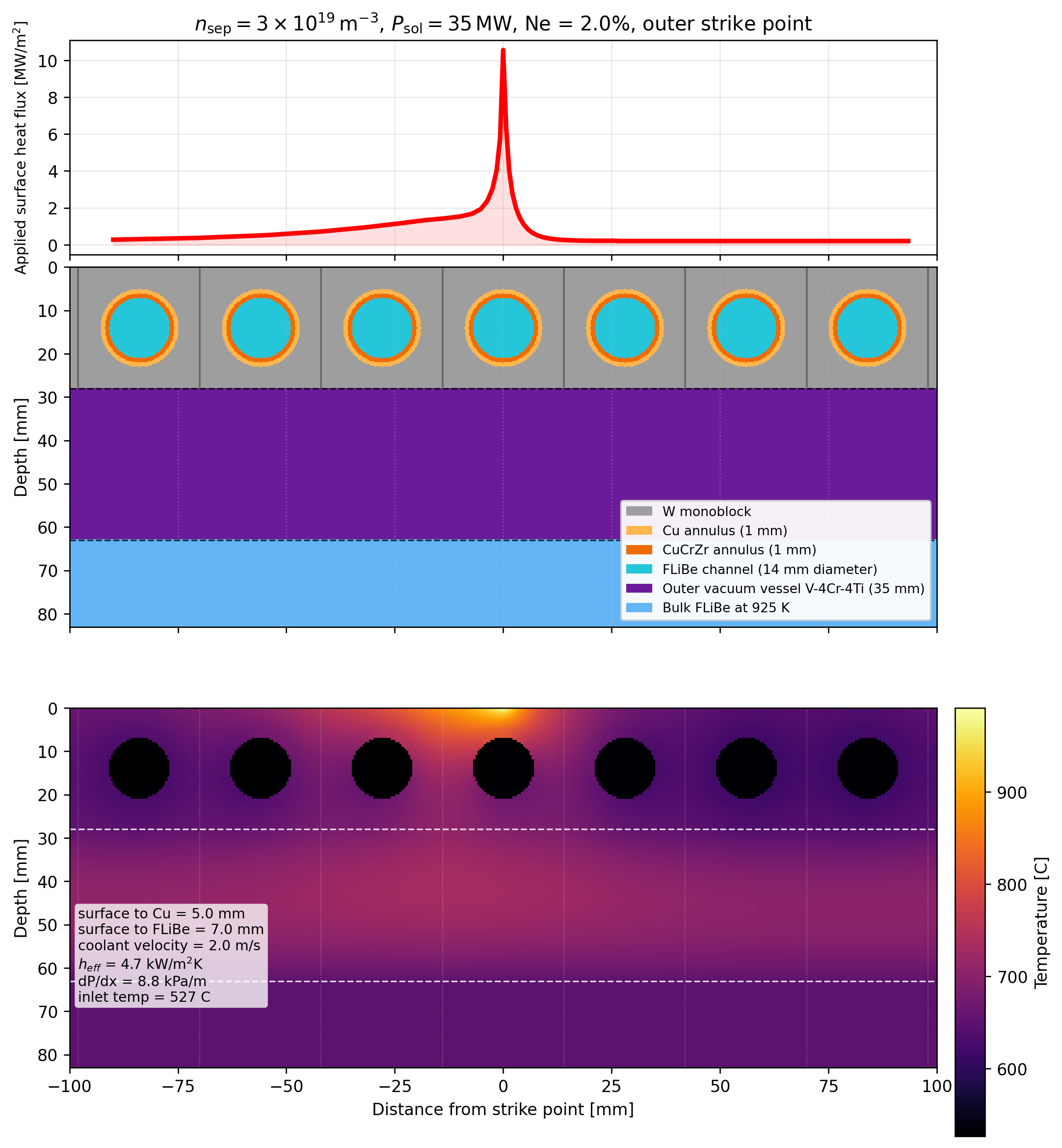}
	    \caption{ITER-inspired FLiBe monoblock target plate design. The upper panel shows the
	      applied outer-target surface heat-flux profile for the selected detached UEDGE case
	      $n_{\mathrm{sep}}=3\times10^{19}\,\mathrm{m^{-3}}$,
	      $P_{\mathrm{sol}}=\SI{35}{MW}$, Ne$=2.0\%$, including the
	      resolved UEDGE target heat flux and an additional
	      \SI{0.2}{MW/m^2} core-radiation load. The middle
	      panel shows the corresponding \SI{14}{mm}-diameter monoblock
	      cross-section, and the lower panel shows the associated 2D
	      temperature field for a nominal FLiBe velocity
      of \SI{2}{m/s}.\hfill}
    \label{fig:divertor_monoblock_geometry}
  \end{figure*}
}

The divertor target plates remain the hardest thermal challenge associated with power handling, even with deep detachment. Thus, a 2D thermal model is built here to understand heat transfer, maximum surface temperatures, and thermal-hydraulic performance. The
resolved UEDGE target heat-flux profiles from
Fig.~\ref{fig:ue_target_plate_profiles} are used directly as the
plasma-facing boundary condition for an ITER-inspired tungsten
monoblock design.

In addition to SOL power heating, volumetric nuclear heating is added
using parameters from the OpenMC simulations discussed in
Section~\ref{sec:neutronics}. Uniform volumetric source terms of
\SI{20}{MW/m^3} in the tungsten and \SI{10}{MW/m^3} in the outer
vacuum vessel V-4Cr-4Ti are included in the solved solid domain. A worst-case additional core-radiation
surface heat flux of \SI{0.2}{MW/m^2} is also added. These terms do
not change the conclusion that the selected strike-point case in
Fig.~\ref{fig:divertor_monoblock_geometry} remains below the tungsten
recrystallization limit, but they do affect the deeper solid-structure
temperature distribution and the FLiBe heat load.

As a starting point, we follow the geometric concept of the ITER
tungsten monoblock divertor discussed in the ITER IIS PFC
design note~\cite{ITERIISPFCDesignNote},
but replace the high-pressure water with FLiBe. The adopted
reference geometry is shown in
Fig.~\ref{fig:divertor_monoblock_geometry}. In this layout, a
\SI{28}{mm}-deep tungsten monoblock contains a centered
\SI{14}{mm}-diameter FLiBe channel surrounded by \SI{1}{mm} Cu and
\SI{1}{mm} CuCrZr annuli. On the other side, the monoblock is backed
by a \SI{3.5}{cm}-thick outer vacuum vessel V-4Cr-4Ti wall, followed
by the bulk FLiBe tank. Relative to the original ITER
\SI{12}{mm}-diameter channel concept, the \SI{14}{mm}-diameter FLiBe
channel reduces the minimum tungsten thickness between the
plasma-facing surface and the outer Cu annulus to \SI{0.5}{cm}.
The lower temperature-field portion of
Fig.~\ref{fig:divertor_monoblock_geometry} corresponds to the same
\SI{14}{mm} channel at a nominal FLiBe speed of \SI{2}{m/s}. For the
	$n_{\mathrm{sep}}=3\times10^{19}\,\mathrm{m^{-3}}$,
	$P_{\mathrm{sol}}=\SI{35}{MW}$, Ne$=2.0\%$ outer-target case, the
	conservative smooth-tube closure gives
	$h_{\mathrm{eff}}\approx\SI{4.7}{kW/m^2/K}$ and
	$dP/dx\approx\SI{8.8}{kPa/m}$; the peak tungsten surface temperature is
	approximately \SI{991}{C} with the added radiation and volumetric
	heating terms.

	A sweep of FLiBe channel diameter and coolant speed for the same
	selected detached outer-target case
	($n_{\mathrm{sep}}=3\times10^{19}\,\mathrm{m^{-3}}$,
	$P_{\mathrm{sol}}=\SI{35}{MW}$, Ne$=2.0\%$) is shown in
	Fig.~\ref{fig:divertor_monoblock_sweep}, including the same
	core-radiation and volumetric nuclear-heating terms. Over the
	\SIrange{9}{18}{mm} diameter and \SIrange{1}{5}{m/s} flow-rate space,
	the predicted peak tungsten surface temperature ranges from about
	\SI{1367}{C} down to about \SI{829}{C}. For the present heat-flux
range, a modestly larger coolant channel is beneficial, and the
\SI{14}{mm} design still preserves the \SI{0.5}{cm} minimum tungsten
thickness above the coolant channel.

{
  \begin{figure}[!t]
    \centering
    \includegraphics[width=\columnwidth]{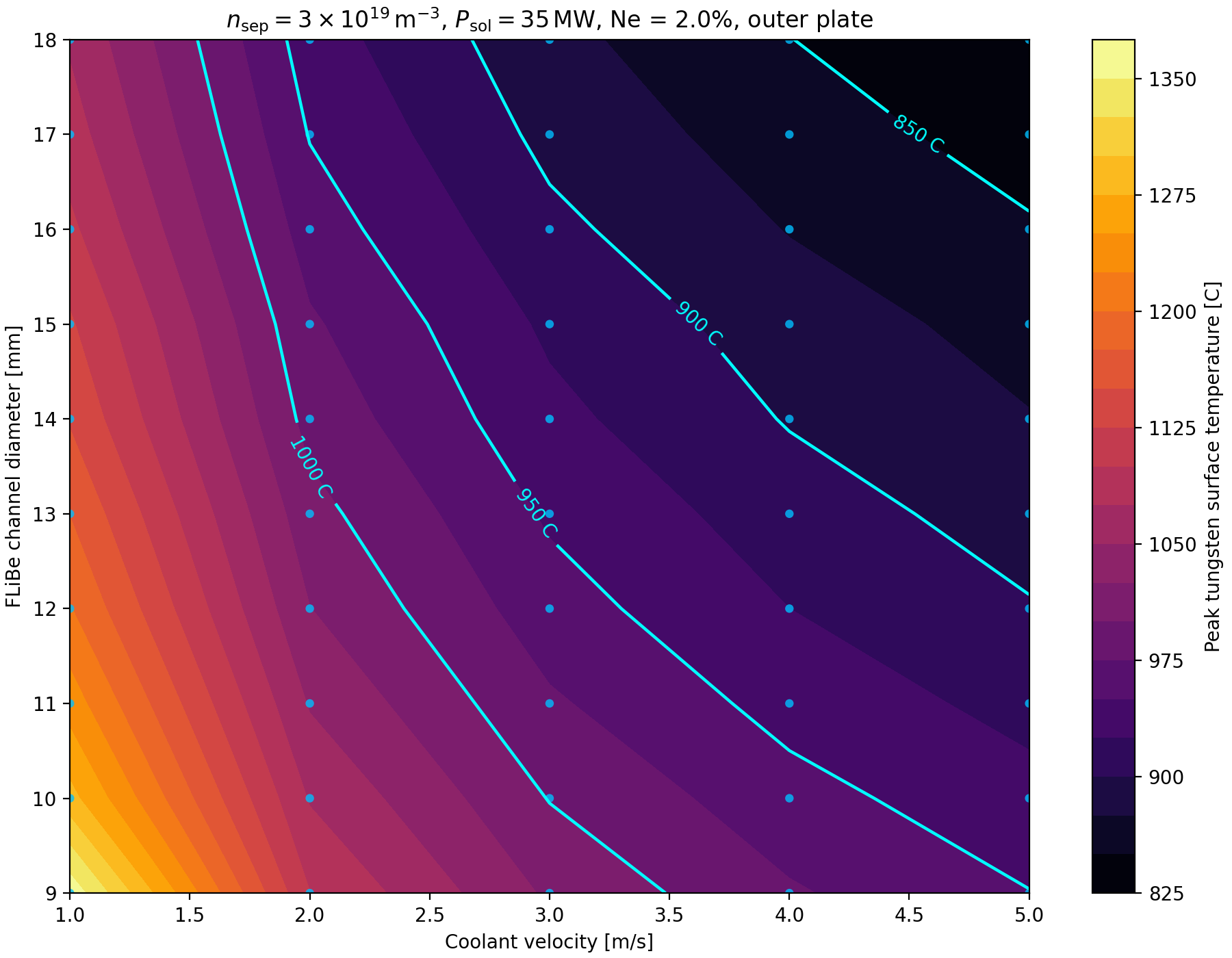}
	    \caption{Peak tungsten surface temperature as a function of FLiBe channel diameter and
	      coolant speed for the selected detached outer-target case
	      $n_{\mathrm{sep}}=3\times10^{19}\,\mathrm{m^{-3}}$,
	      $P_{\mathrm{sol}}=\SI{35}{MW}$, Ne$=2.0\%$, with the same
	      added core radiation and volumetric nuclear heating used in
	      Fig.~\ref{fig:divertor_monoblock_geometry}. Larger channel diameter and higher
      flow speed both reduce the peak tungsten temperature in this
      conservative smooth-tube model with no explicit
      swirl turbulence enhancement.\hfill }
    \label{fig:divertor_monoblock_sweep}
  \end{figure}
}

In each solid region, the steady temperature field
satisfies
\beq
\nabla \cdot \left( \kappa \nabla T \right) + q'''_{\mathrm{nuc}}(\mathbf{x}) = 0,
\eeq
with the plasma-facing boundary condition
\beq
-\hat{n}\cdot \kappa \nabla T = q_{\perp}(x) + q_{\mathrm{rad}},
\eeq
and the FLiBe-channel boundary condition
\beq
-\hat{n}\cdot \kappa \nabla T = h_{\mathrm{eff}}\!\left(D,v\right)
\left( T - T_c(x) \right),
\eeq
where $D$ is the FLiBe channel diameter, $v$ is the bulk channel
speed, $h_{\mathrm{eff}}$ is an effective coolant-side heat-transfer
coefficient, $q'''_{\mathrm{nuc}}(\mathbf{x})$ is the volumetric nuclear-heating source,
$q_{\mathrm{rad}}$ is the core-radiation heat flux, and $T_c(x)$ is the bulk FLiBe temperature marched along the
poloidal direction. The coolant heating is modeled with
\beq
\dot{m} c_p \frac{dT_c}{dx} = q'(x),
\eeq
where $\dot{m}$ is the FLiBe mass flow rate in one monoblock channel
and $q'(x)$ is the local heat removal per unit length inferred from
the imposed surface heat flux and the monoblock pitch. The coolant-side closure is treated conservatively in the present screening model by not explicitly crediting any swirl-tape heat-transfer
enhancement. Instead, the channel is modeled as a smooth circular tube
with hydraulic diameter equal to the FLiBe channel diameter. The local
flow state is characterized by the Reynolds and Prandtl numbers,
\beq
\mathrm{Re} = \frac{\rho v D}{\mu},
\eeq
\beq
\mathrm{Pr} = \frac{c_p \mu}{k_f},
\eeq
where $\rho$, $\mu$, $c_p$, and $k_f$ are the FLiBe density, dynamic
viscosity, specific heat, and thermal conductivity, respectively. A
standard smooth-tube circular-channel correlation is then used to obtain
the baseline Nusselt number $\mathrm{Nu}_{\mathrm{smooth}}$ and Darcy
friction factor $f$. The effective coolant-side heat-transfer coefficient
is written as
\beq
h_{\mathrm{eff}} = 0.7 \,\frac{\mathrm{Nu}_{\mathrm{smooth}} k_f}{D},
\eeq
with the prefactor 0.7 representing a conservative 30\% MHD penalty
on heat transfer, while the streamwise pressure gradient is
\beq
\frac{dP}{dx} = f \,\frac{\rho v^2}{2D}.
\eeq
Temperature and normal heat flux are taken to be continuous across the
tungsten, Cu, CuCrZr, and outer vacuum vessel V-4Cr-4Ti interfaces.
The back side of the outer vacuum vessel V-4Cr-4Ti wall is held at the
bulk FLiBe tank temperature of \SI{925}{K}, the side boundaries of the
local patch are treated as adiabatic so that the calculation
represents a repeated monoblock tiling around the strike-point region. This model is intended as a simplified
heat-transfer model rather than a detailed conjugate CFD
calculation, but it captures the lateral heat spreading in the armor
and the coolant-side trade between heat transfer and pumping penalty
that are central to monoblock performance. If explicit swirl-tape
turbulence enhancement is credited, FLiBe swirl-tube studies suggest
that $h_{\mathrm{eff}}$ can readily increase by more than an order of
magnitude relative to this conservative baseline, substantially
improving heat transfer\cite{Sorbom2020ARCRecent}.

{
  \begin{figure}[!t]
    \centering
    \includegraphics[width=\columnwidth]{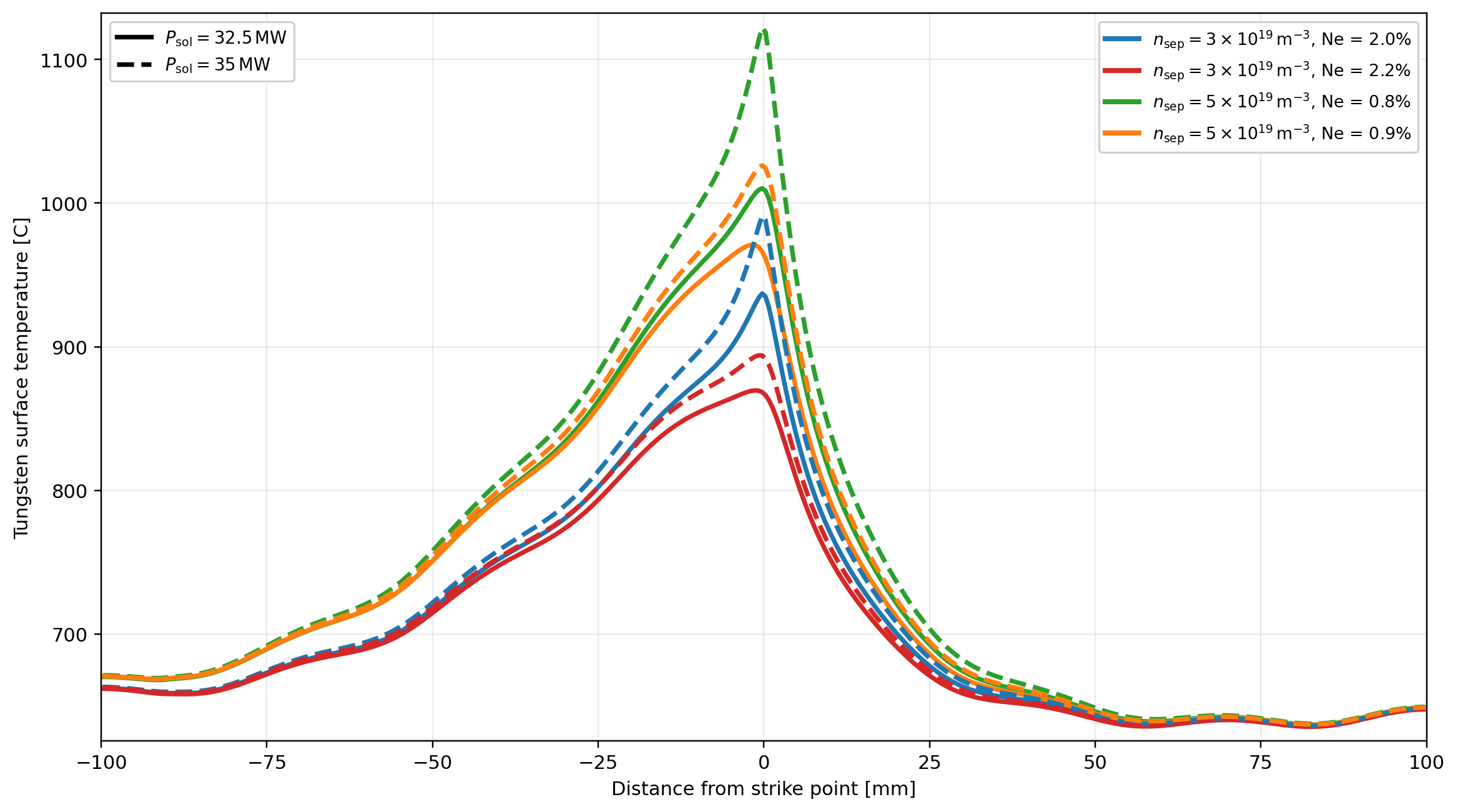}
    \caption{Outer-target tungsten surface temperature profiles for representative
      detached UEDGE cases. Color denotes the density branch and neon
	      fraction, while line style distinguishes
	      $P_{\mathrm{sol}}=\SI{32.5}{MW}$ (solid) from
	      $P_{\mathrm{sol}}=\SI{35}{MW}$ (dashed). All cases use the
      reference \SI{14}{mm}-diameter FLiBe monoblock, a nominal coolant
      speed of \SI{2}{m/s}, and the same conservative smooth-tube
      coolant closure with the same added core radiation and volumetric
      nuclear heating, but no explicit swirl-enhancement credit.\hfill }
    \label{fig:divertor_monoblock_surface_temp}
  \end{figure}
}

Finally, Fig.~\ref{fig:divertor_monoblock_surface_temp} compares the
outer-target tungsten surface temperature profiles for representative
detached cases in both density branches, using the same
	\SI{14}{mm}-diameter monoblock and \SI{2}{m/s} FLiBe speed. Across the
	cases shown, the peak outer-target temperatures remain below about
	\SI{1130}{C}, with the most demanding case among this set being the
	$n_{\mathrm{sep}}=5\times10^{19}\,\mathrm{m^{-3}}$,
	$P_{\mathrm{sol}}=\SI{35}{MW}$, Ne$=0.8\%$ case at approximately
	\SI{1122}{C}, slightly above the \SI{1100}{C} recrystallization
reference temperature for tungsten in this conservative no-swirl model
~\cite{Tsuchida2018TungstenRecrystallization}. The corresponding
Ne$=0.9\%$ case reduces the peak temperature to approximately
\SI{1026}{C}, indicating why the higher impurity content is important:
it drives deeper detachment, lowers the peak target heat flux, and
restores tungsten temperature margin.
\FloatBarrier

%% file: engineering.tex
\section{Magnet Engineering} 

\subsection{Magnet System Overview}

The tokamak magnet system is primarily comprised of high‑temperature superconducting (HTS) technology and is composed of three principal subsystems: toroidal field (TF) coils, a central solenoid (CS), and poloidal field (PF) coils. The main confinement and shaping systems (18 TF coils, 6 CS coils, and 8 PF coils) are HTS and arranged to provide magnetic confinement, plasma current drive, and equilibrium control. In addition, in-vessel resistive coils are considered for fast-response and X-point control in the high-radiation, high-heat-load environment near the plasma, but are not required to achieve the desired equilibrium. The TF coils are equipped with outer inter‑coil structures (OIS) positioned between the vacuum vessel ports and gravity supports. The CS stack includes a dedicated pre‑compression system and top‑mounted gravity supports. The PF coils are mechanically supported by the TF coil cases via flexible interfaces, with the exception of the innermost PF coils, which are clamped using sliding interfaces to accommodate radial motion.

\subsection{HTS Conductor and Cable Design}

All HTS coils employ a REBCO-based cable derived from an Insulated Stacked Tape CroCo architecture named SHIELD (Superconducting High Integrity Energy Link \& Distribution), developed by Maritime Fusion. For fusion applications, SHIELD is optimized for high-field operation using advanced-pinning HTS tape selections to preserve current density under strong background field and unfavorable field-angle regions. The same cable architecture is also compatible with non-advanced-pinning tape for lower-field power-distribution applications, enabling a shared manufacturing pathway across markets. This cross-market compatibility is intentional: the same manufacturing-line capacity can be shared between fusion magnet production and lower-field power-distribution cable markets, where demand is expected to grow with AI data-center and grid-infrastructure expansion.

\begin{figure}[!t]
    \centering
    \includegraphics[width=\linewidth]{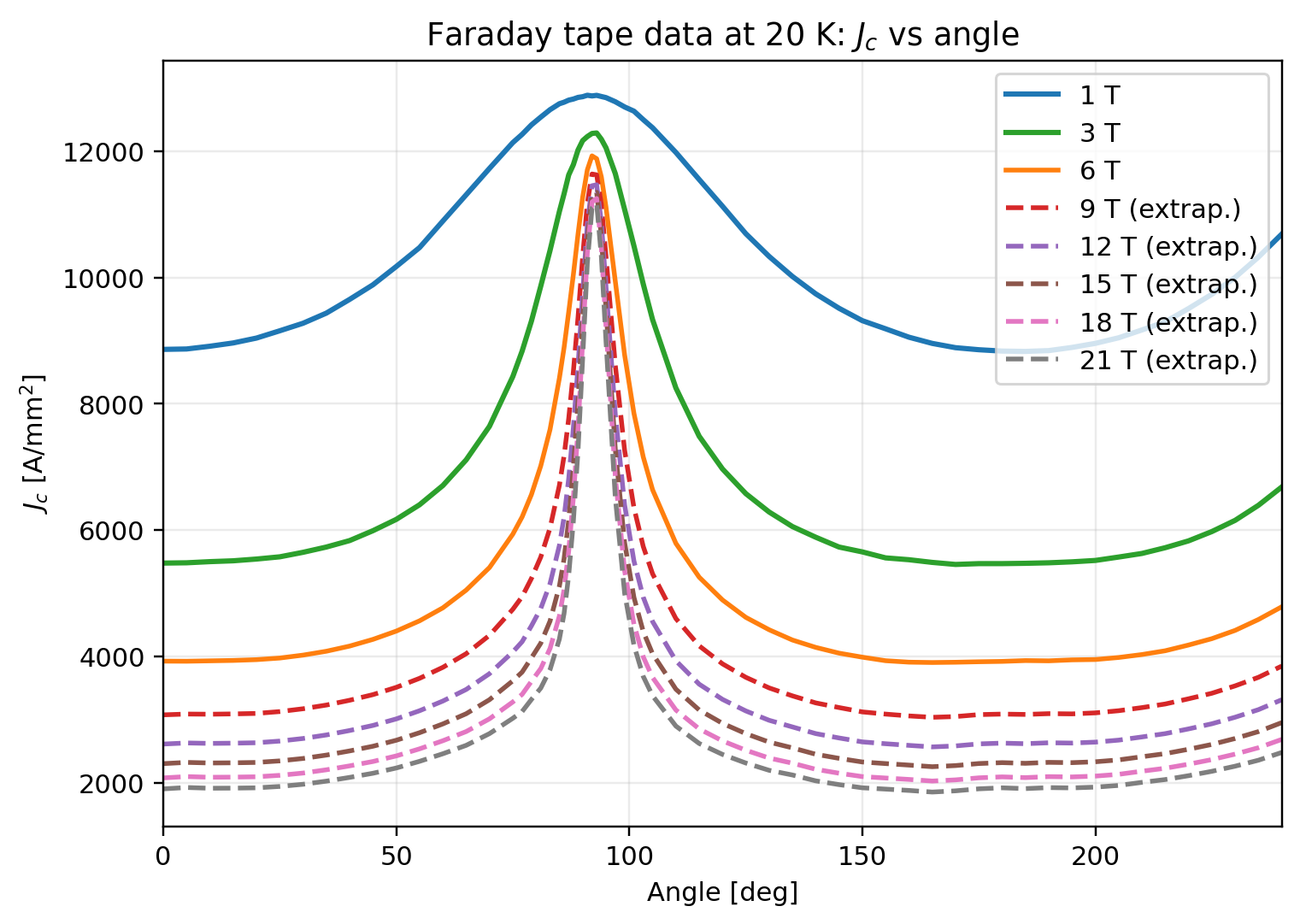}
    \par\medskip
    \includegraphics[width=\linewidth]{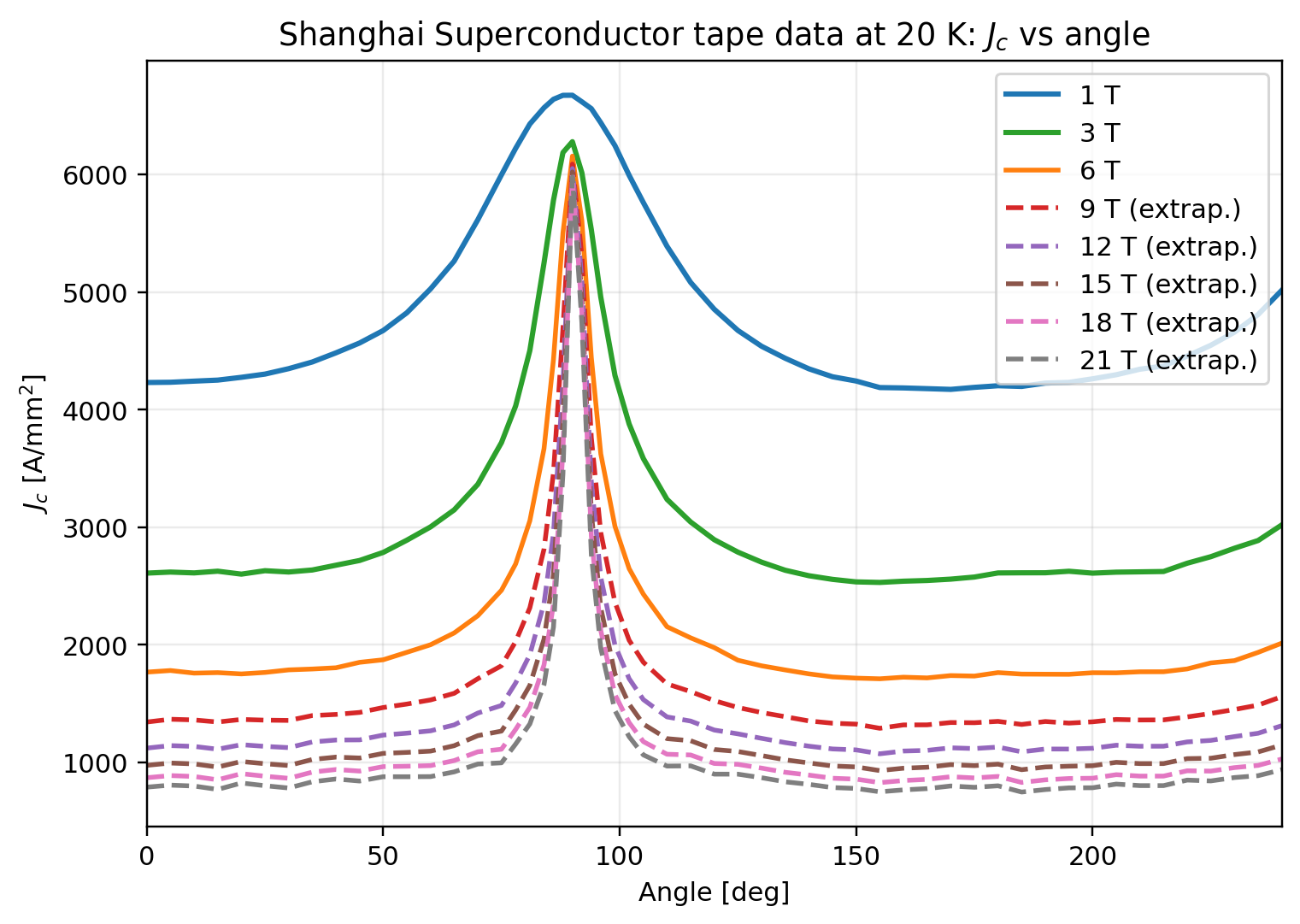}
    \caption{Measured HTS tape data at 20 K for Faraday Factory (top) and Shanghai Superconductor (bottom), taken from the Victoria University of Wellington HTS tape database \cite{VUWHTSTapeDatabase}. The grading calculations associated with these datasets assume a plated tape thickness of \SI{70}{\micro\meter}. The 6 T curves are interpolated within the measured window, while the 9--21 T curves are plotted as monotone high-field extensions anchored at the maximum measured field of \SI{8}{T} using $J_c(B,\theta)=J_c(B_{\max},\theta)\left(B/B_{\max}\right)^{m_{\mathrm{tail}}(\theta)}$ with $m_{\mathrm{tail}}(\theta)\le 0$.}
    \label{fig:hts_jc_angle_comparison}
\end{figure}

The cable uses a rectangular, highly modular cross-section that supports straightforward bundling of electrically insulated conductors for magnet-specific winding pack layouts. A key benefit of electrically insulated winding packs is that they permit fast charging and discharging without the large resistive losses associated with non-insulated magnets. In very large grid-scale non-insulated magnet systems, charge and discharge can extend to many days or even weeks unless cryogenic cooling capacity is drastically oversized to absorb the transient losses associated with higher dI/dt. That directly penalizes plant availability, since substantial time must be spent de-energizing the magnets before maintenance work can begin. In contrast, the insulated Yinsen magnet architecture is intended to permit full ramping on a timescale of 15 minutes. The high engineering current density available from REBCO more than compensates for the winding-pack current-density penalty introduced by turn insulation, and the slow quench-propagation characteristics of HTS keep the burden on the protection system manageable. This geometry is paired with a split-body manufacturing process in which the HTS tapes are first PbSn-plated, additional solder is applied to each conductor half during assembly, and the two halves are then brought together under mechanical preload before oven or inductive heating to produce a controlled, homogeneous solder bond throughout the stack in a way that is compatible with high-throughput mass manufacturing. Within each individual stack, the tape assembly is monolithic, so local dropouts are mitigated through current sharing inside the stack. Across the tokamak, conductor builds are designed to accept qualified tape from multiple manufacturers, and coil-by-coil grading is used to reduce tape count in regions with lower local field magnitude or more favorable field angle, while maintaining at least 50\% margin to the local critical current of the cable. As shown in Fig.~\ref{fig:hts_jc_angle_comparison}, the tape critical current depends strongly on the local field angle; for conservatism, and to preserve additional margin in the grading methodology described later, the worst-case field angle is assumed when assigning tape count. In depopulated regions, removed HTS tapes are replaced with copper tapes so that conductor stabilization and thermal mass are preserved while superconductor usage is reduced. Operating up to 50 kA, the conductor uses an 8 mm copper former radius with a CroCo-shaped groove in the copper former sized to accept both 12 mm and 8 mm tapes, achieving a tape fill factor of 79\%. The resulting square jacketed conductor has an overall cross section of 21 mm by 21 mm, including 1.5 mm of electrical insulation. The tape fill factor is intentionally not maximized so that sufficient copper cross section remains to provide thermal mass during a quench and limit hotspot temperature to below 200$^\circ$C.  Coolant channels, operable with either Helium or Nitrogen working fluids, are integrated into the stainless steel jacket at the interface with the copper former to improve both thermal-hydraulic performance and conductor manufacturability. To reduce AC losses, particularly in the CS modules where the wider tape selections are expected to be most penalizing, two mitigation strategies are proposed. First, conductor twist pitch can be optimized such that tape orientation remains as favorable as possible to the local magnetic-field direction, thereby reducing the perpendicular field component that drives hysteresis loss. Second, a novel lengthwise insulation concept is proposed in which the copper former and outer jacket are periodically segmented electrically along the cable length, while the soldered HTS tape stack remains continuous, in order to reduce the effective loop area available for recirculating eddy currents. This approach may require a thin layer of insulation between the copper former and jacket if only the copper former is segmented. Additional non-conductive coolant-channel jumpers are needed to bridge the insulated breaks. The overall trade space among critical-current margin, allowable CS power-supply voltage and ramp duration, installed cryogenic capacity to absorb AC losses, and hotspot detection remains an active area of conductor optimization.

\subsection{Toroidal Field Coil Design and Assembly}

The present TF layout uses 18 coils, each built from seven electrically insulated double-pancake pairs. The current packed cross section contains 188 turns per TF coil at a terminal current of 43.6 kA, corresponding to 8.20 MA-turn per coil. The inboard legs of the TF coils are wedged against one another, forming a self-supporting structure capable of reacting the dominant centering forces. Insulated shear keys are installed in the curved inboard regions to resist overturning moments during pulsed operation. Four outer inter-coil structures connect adjacent TF coils through insulated pins, providing global stiffness while electrically isolating the coils. Stainless steel 316LN was selected as the TF structural material because it combines extensive ITER-relevant characterization with a yield strength of 1122 MPa. The case material must carry the large Lorentz loads generated by the conductors without yielding, while also retaining sufficient ductility and fracture toughness for faulted and cyclic loading. It must also be stiff enough to limit strain in the REBCO tapes to below 0.4\%.


\begin{figure}[H]
    \centering
    \includegraphics[width=0.72\columnwidth]{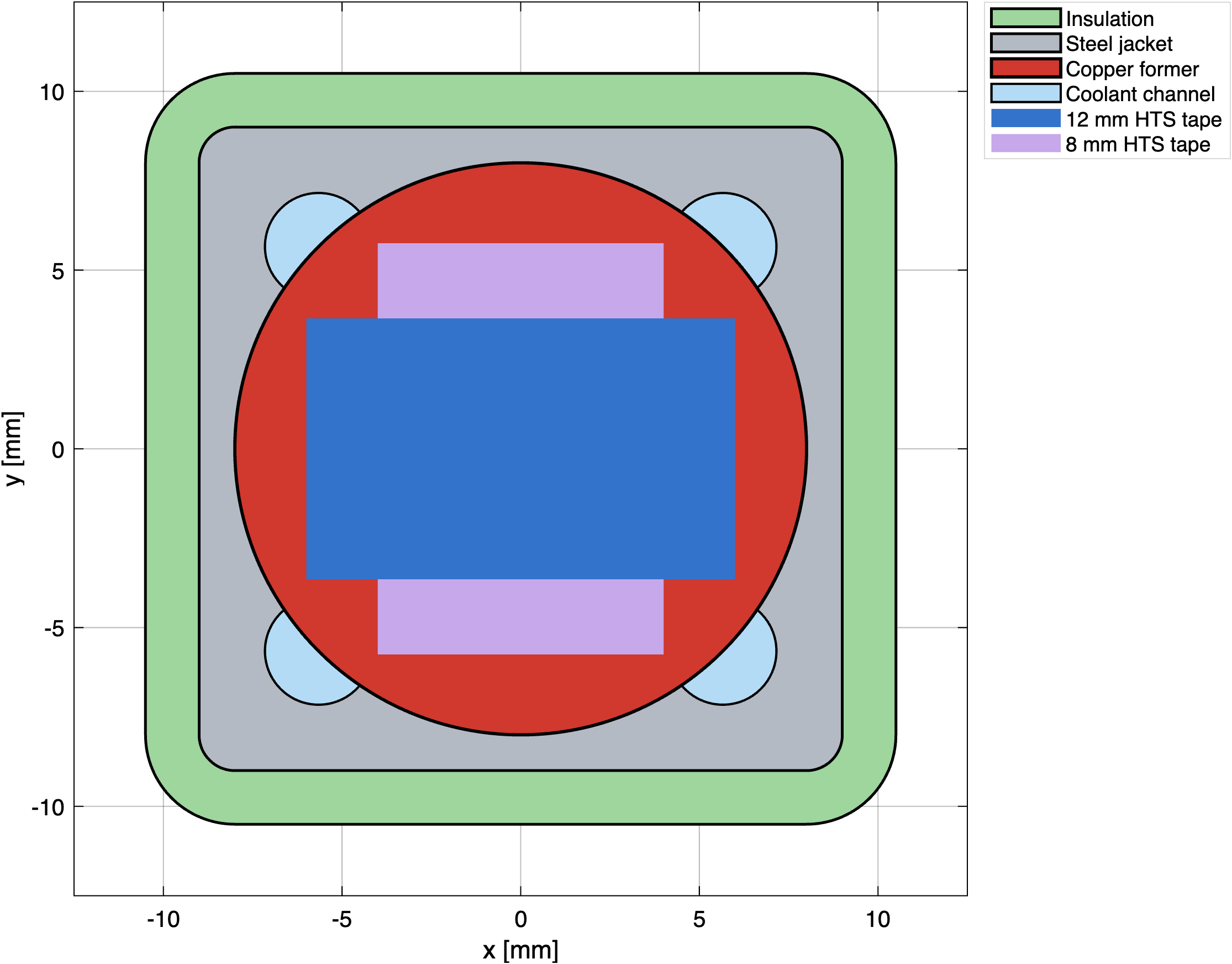}
    \caption{SHIELD CroCo cable cross section showing the insulated outer package, stainless jacket, copper former, integrated coolant channels, and combined 12 mm / 8 mm HTS tape stack.}
    \label{fig:shield_cable_design}
\end{figure}

\vspace{0.5cm}

\begin{table}[H]
    \centering
    \scriptsize
    \caption{\protect\raggedright SHIELD cable geometry}
    \begin{tabular}{l@{\hspace{0.8em}}l@{\hspace{-4.8em}}p{1.35cm}}
        \hline
        Parameter & Value &  \\
        \hline
        Copper former radius & 8.0 mm & \\
        Minimum copper thickness & 1.0 mm & \\
        Width $\times$ Height & 21.0 mm $\times$ 21.0 mm & \\
        Insulation thickness & 1.5 mm & \\
        Coolant channel radius & 1.5 mm & \\
        Total area & 436 mm$^2$ & \\
        Tape area & 121 mm$^2$ & (27.82\%) \\
        Coolant area & 15 mm$^2$ & (3.37\%) \\
        Steel area & 107 mm$^2$ & (24.65\%) \\
        Copper area & 80 mm$^2$ & (18.33\%) \\
        Insulation area & 112 mm$^2$ & (25.82\%) \\
        \hline
    \end{tabular}
    \label{tab:shield_cable_geometry}
\end{table}

\begin{table}[H]
\centering
\scriptsize
\caption{\raggedright TF Coil Design}
\label{tab:tf_coil_design}
\begin{tabular}{ll}
\hline
\textit{Parameter} & \textit{Value} \\ \hline
Number of TF coils & 18 \\
MA-turns & 8.20 MA-turn/TF \\
Double pancakes & 7 \\
Number of turns & 188 \\
Terminal current & 43.6 kA \\
WP current density & 86.2 A/mm$^2$ \\
Self inductance & 0.43 H \\
Stored energy single TF & 406 MJ \\
Total stored energy & 27 GJ \\
12mm REBCO tape length & 144 km \\
Radial case thickness & 80 mm \\
Peak field on conductor & 19.5 T \\ \hline
\end{tabular}
\end{table}

\begin{figure}[!b]
    \centering
    \includegraphics[width=0.86\columnwidth]{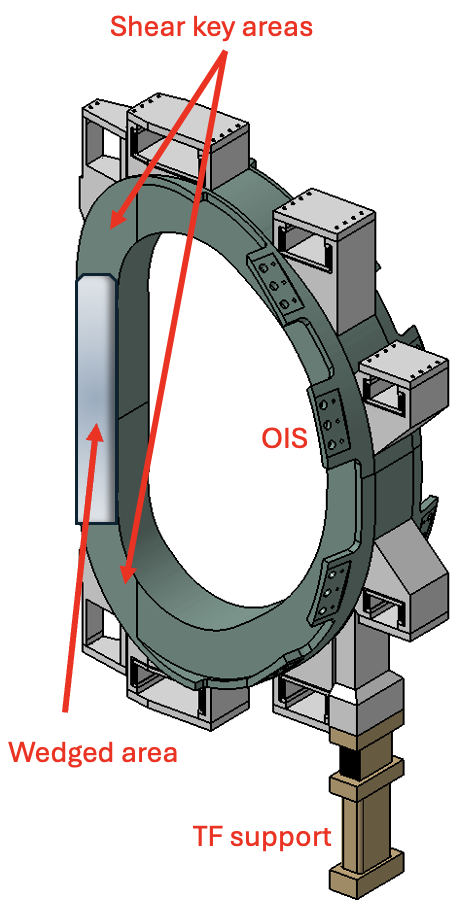}
    \caption{Simplified TF structural frame highlighting wedged area, shear-key regions, OIS interfaces, and representative PF/TF support attachment locations.}
    \label{fig:tf_structure_features}
\end{figure}

\begin{figure}[!t]
    \centering
    \includegraphics[width=\columnwidth]{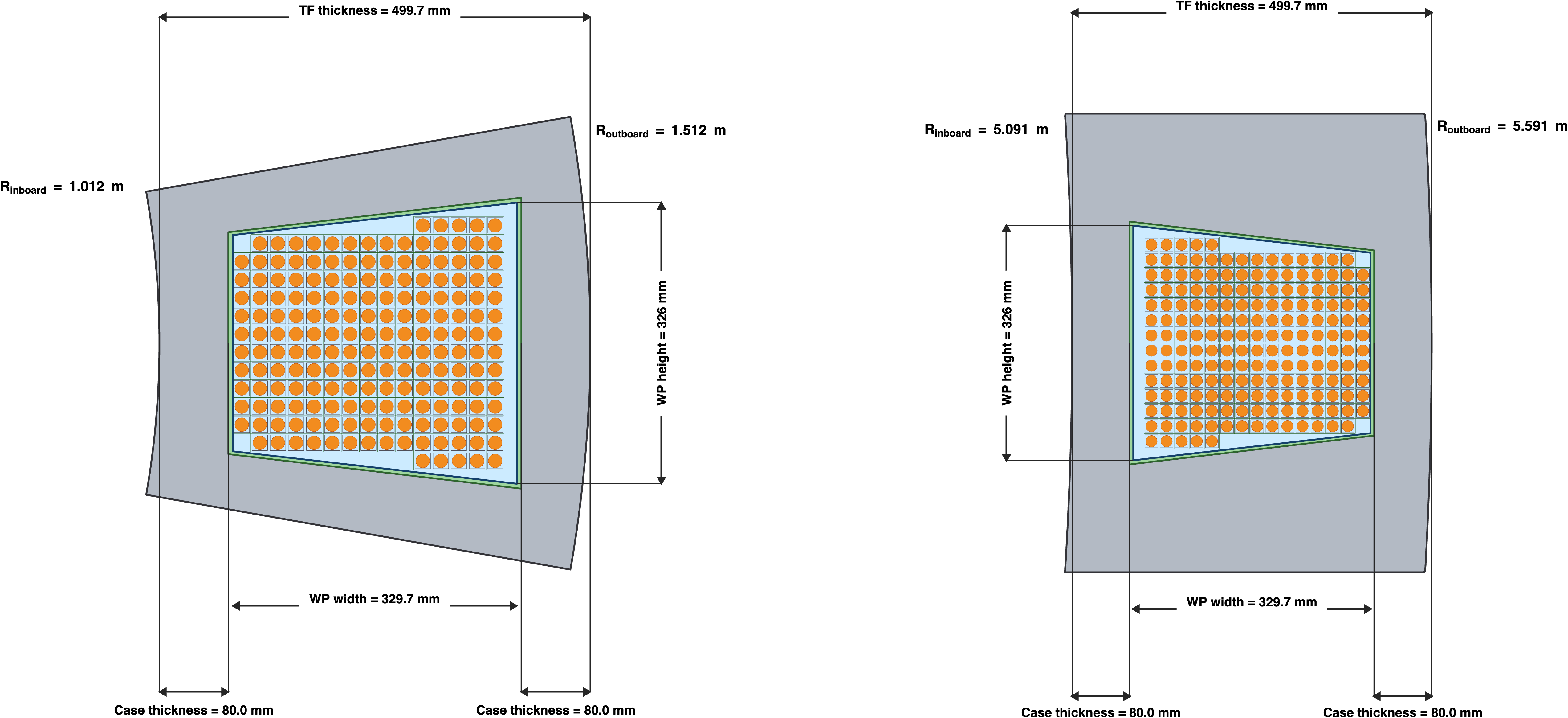}
    \caption{TF coil inboard and outboard cross sections depicting winding-pack configuration and structural case thicknesses.}
    \label{fig:tf_case_windingpack}
\end{figure}

\begin{figure*}[!t]
    \centering
    \includegraphics[width=0.95\textwidth]{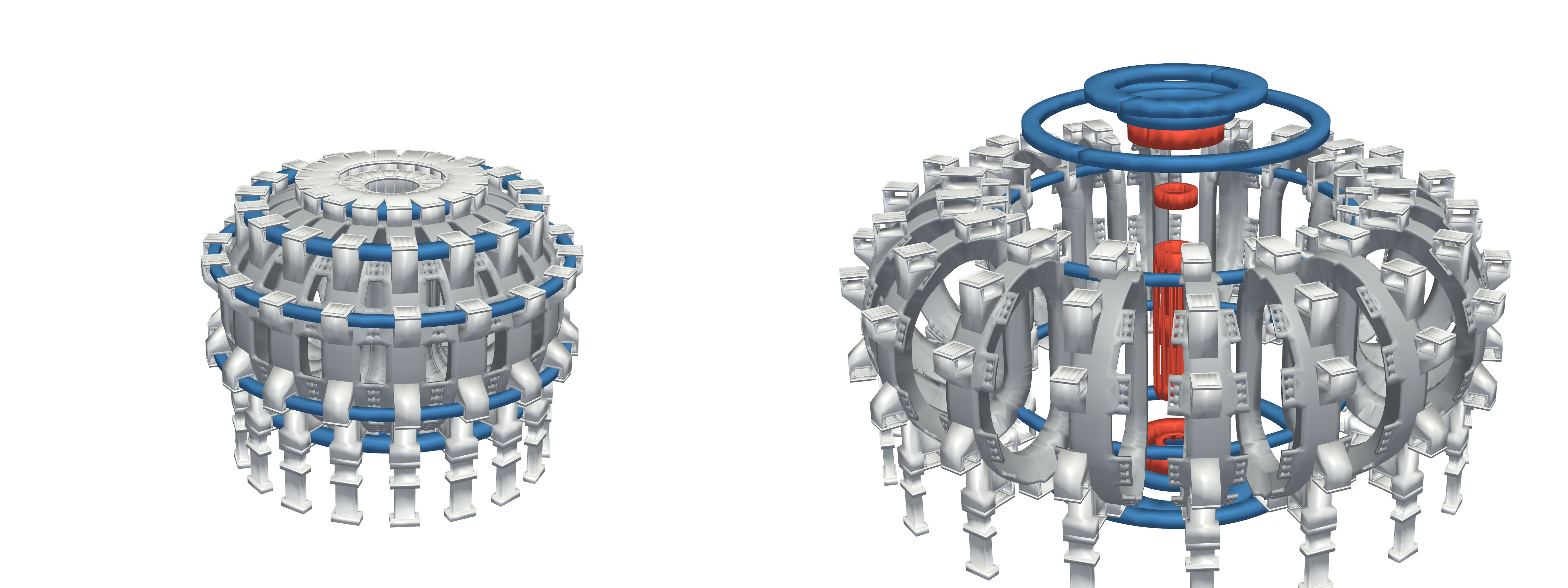}
    \caption{Full 18-coil TF cage rendered from the CAD master model, shown assembled (left) and exploded (right). The assembled view shows the toroidal wedging of the inboard legs, the structural case surrounding each D-shape winding pack, and the discrete outer-intercoil support (OIS) bolt-group locations on each toroidal cut face and the radial gravity supports are also visible.}
    \label{fig:tf_cad_assembly}
\end{figure*}

\subsection{Poloidal Field and Central Solenoid Assembly}

All superconducting PF coils are located outside of the TF to alleviate the nuclear heating and damage challenge. The PF coils are mechanically attached to the TF coil cases via welded flanges and clamped in place by cover plates following installation. Radial motion of PF coils is accommodated through flexible support elements and low‑friction interfaces (e.g., aluminum‑bronze pads). All PF coils, except PF1U and PF1L, are supported using thin flexible plates that permit relative radial displacement between the PF, CS, TF, and vacuum vessel structures. The CS stack consists of six independently powered coils, separated by interface plates that allow controlled axial and radial motion. The CS winding-pack current density is set to 40 A mm$^{-2}$, corresponding to a centerline bore field of approximately 24 T, to provide the required bipolar flux swing of approximately $+45.4$ Wb to $-45.4$ Wb. The implications for flat-top duration and flux consumption are discussed further in Section~\ref{sec:pulse} using time-dependent TokaMaker-informed pulse simulations, as their exists substantial margin on critical current density to further increase available flux swing. Vertical pre‑compression is applied using tie‑plate assemblies distributed along the inner and outer surfaces of the stack. This pre‑compression is established via thermal shrink‑fitting and serves to counteract tensile electromagnetic loads during operation. The CS assembly is suspended from eight flexible vertical gravity supports, which carry the full weight of the stack and react electromagnetic and seismic loads while permitting radial movement. Radial restrainers with spring elements are installed at the base of the CS to ensure centering without over‑constraining the structure. Each CS and PF coil is supplied by independent electrical‑cryogenic feeder lines.

\subsection{In-vessel Coil Candidates}
\label{sec:invessel_coil_candidates}

Because the main PF system is located outside the TF cage, Yinsen also considers a limited set of in-vessel normal-conducting coils for fast local control functions that benefit from reduced response time and closer magnetic coupling to the plasma boundary. The primary motivations are improved vertical stabilization authority and more direct control of the X-point and divertor topology, with additional future options including error-field correction and resistive-wall-mode control. In contrast with the primary TF, CS, and PF systems, these coils are not intended to be superconducting. The in-vessel environment is instead characterized by high nuclear heating, strong thermal coupling to hot FLiBe, and a chemically aggressive blanket environment that is not compatible with practical cryogenic operation.

\begin{figure}
    \centering
    \includegraphics[width=\linewidth]{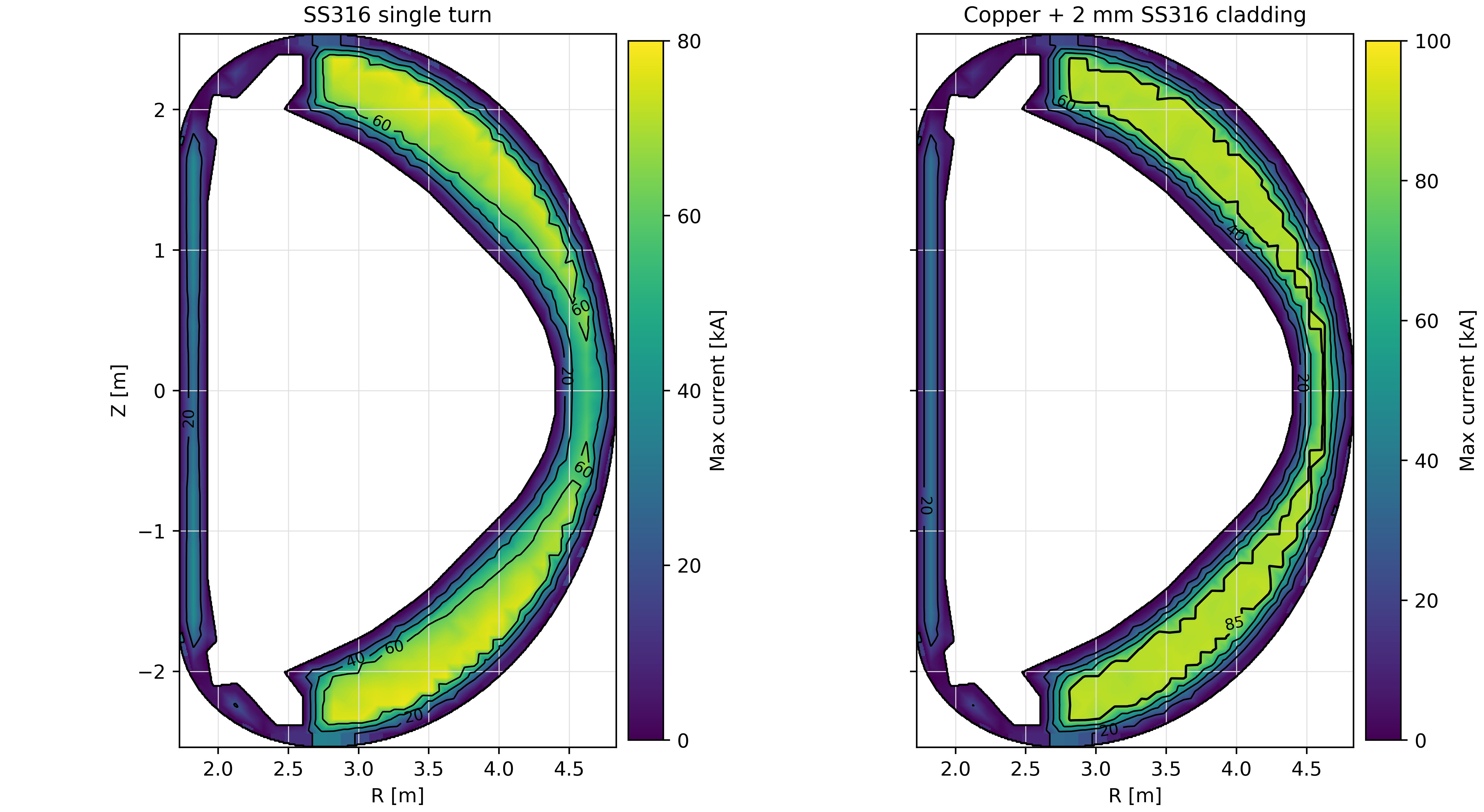}
    \caption{Maximum achievable current by location for single-turn in-vessel coils in the updated divertor-shaped blanket geometry, shown for SS316 and for copper with a 2 mm SS316 cladding layer at a fixed current density of 1 A/mm$^2$. Both panels use the same 0--100 kA color scale. The outboard crescent remains the most favorable region, while the high-field-side inner leg is still the most thermally constrained.}
    \label{fig:invessel_coil_ss316_current_map}
\end{figure}

A two-dimensional nonlinear thermal model was used to screen where resistive in-vessel coils could be placed inside the FLiBe blanket tank while accounting for both volumetric nuclear heating and ohmic dissipation. Candidate coils were evaluated at a fixed current density of $1 A/mm^2$, with feasible regions defined by keeping the peak conductor temperature below 60\% of the material melting temperature. Two material cases were considered: pure copper, and copper protected by a 2 mm SS316 cladding layer for improved chemical compatibility in hot FLiBe. Figure~\ref{fig:invessel_coil_ss316_current_map} shows that the outboard off-midplane region is the most favorable location, allowing approximately 75--80 kA with SS316 cladding and 80--90 kA for unclad copper. The high-field-side inner leg is more thermally constrained, but lower-current resistive coils of order 30 kA still appear feasible there and could potentially support divertor strike-point control. Despite this thermal feasibility, the practical challenges of maintenance, electrical isolation, and routing high-current leads through the blanket tank make in-vessel coils undesirable for the baseline design. Given the control authority already provided by the exterior PF coils, Yinsen does not adopt in-vessel coils in the baseline, but retains them as an option for future design iterations.

\subsection{Electromagnetic and Structural Analysis}

A detailed electromagnetic‑structural analysis was performed using COMSOL, incorporating all TF, CS, and PF coils as well as a simplified plasma current distribution. A 20$^\circ$ sector model with cyclic symmetry was employed to reduce computational cost while preserving global force balance. Uniform current density was applied to the TF coils, while realistic current distributions were assigned to the CS, PF coils, and plasma. The simulations show that the peak magnetic field reaches 19.5 T on the TF coils, while the on‑axis toroidal field achieves the design target of 9.29 T. The magnetic field decays rapidly outside the device, falling to near‑zero within approximately 0.5 m of the outer boundary. Electromagnetic force analysis shows that the dominant loads on all coils act in the radial direction, consistent with toroidal field–dominated operation. Force and torque magnitudes are comparable to those observed in ITER‑class devices, confirming the structural survivability of the HTS magnet system. Stress distributions remain within allowable limits for the conductor, insulation, and structural materials, validating the adopted support and pre‑compression strategies.

Structural response of the TF case is assessed with a separate CAD-mesh solid-mechanics model driven by the analytical TF centering body force $\mathbf{f} = -K\,\mathbf{r}/r^{4}$, with $K = (B_0 R_0)^2/\mu_0 = 6.93\times10^{8}$~N\,m$^{-1}$. This closed-form body force avoids the need to co-solve the full magnetostatic problem inside the structural domain and still captures the dominant inward loading seen by each coil to within a few percent of the cyclic-symmetric COMSOL result discussed above. The outer case is imported from the tessellated CAD geometry (Fig.~\ref{fig:tf_single_coil_cad}) and meshed with approximately $10^{5}$ second-order tetrahedra; refinement is concentrated on the inboard wedge land and on the OIS bolt bosses, where the largest gradients are expected. SS316LN linear-elastic material properties ($E = 200$~GPa, $\nu = 0.30$, $\rho = 7930$~kg\,m$^{-3}$) are used throughout, and the model is solved in COMSOL with the default direct sparse solver.

\begin{figure}[!b]
    \centering
    \includegraphics[width=0.95\columnwidth]{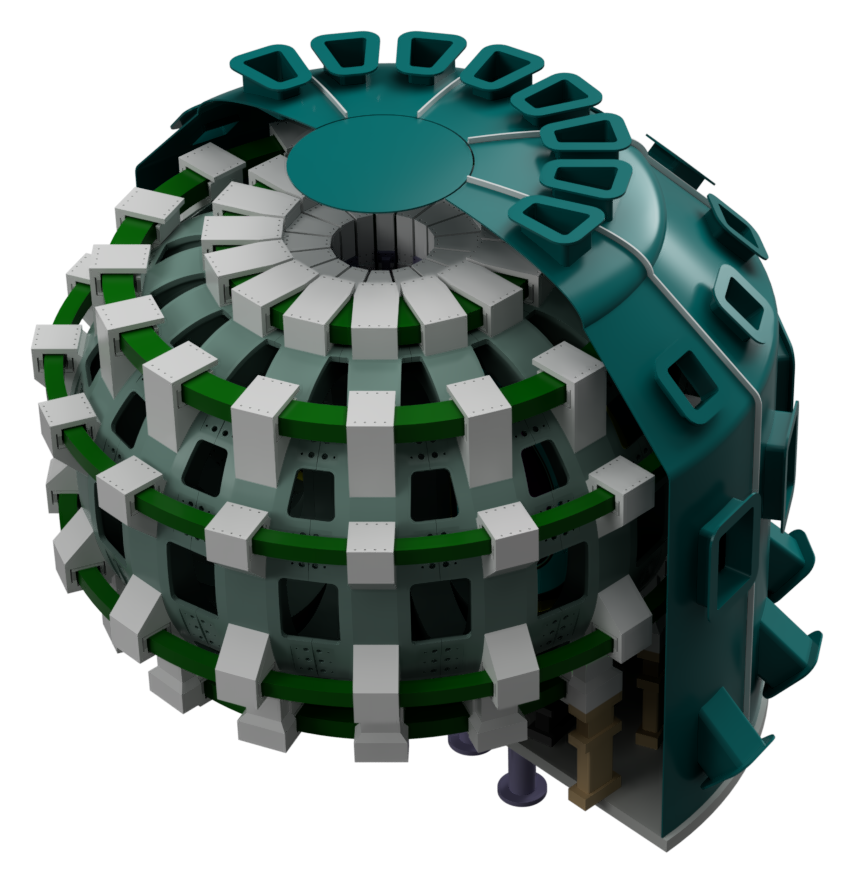}
    \caption{Isometric cutaway of the Yinsen tokamak assembly showing the integrated TF cage, vacuum vessel, blanket region, central solenoid, and gravity supports.}
    \label{fig:tokamak_assembly_cutaway}
\end{figure}

\begin{figure}[!t]
    \centering
    \includegraphics[width=0.9\linewidth]{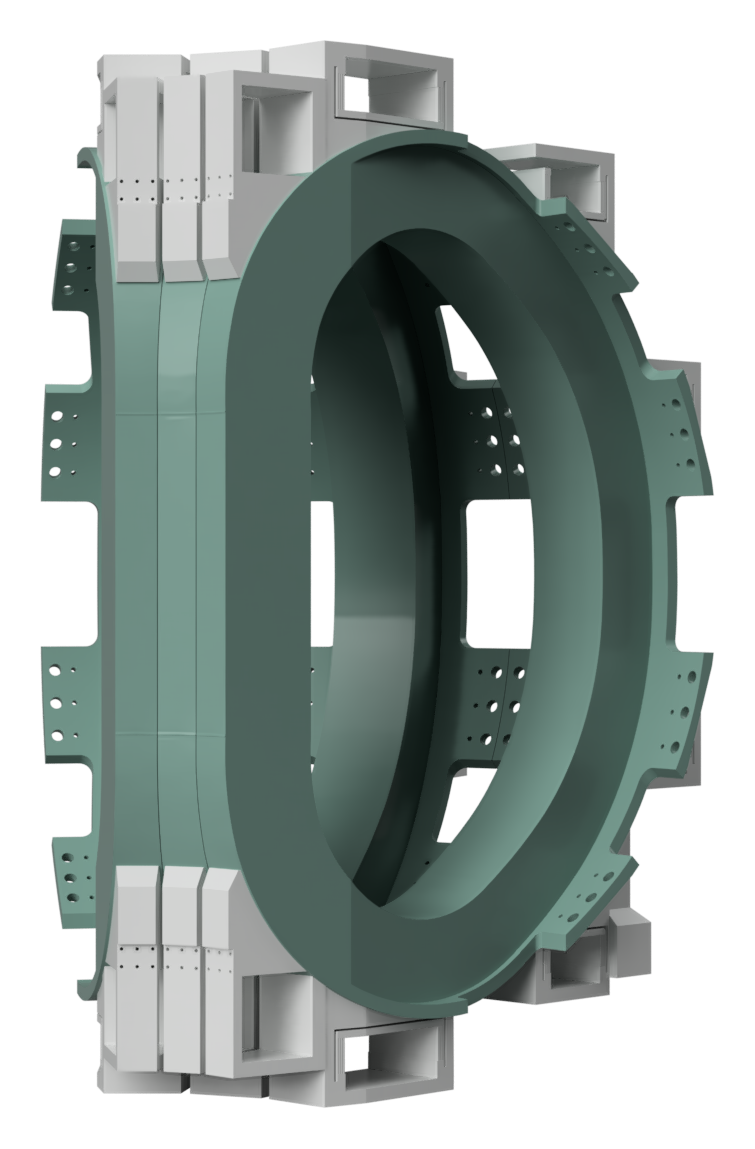}
    \caption{TF case CAD used in the structural analysis: three adjacent TF coils shown mated together, illustrating the toroidal wedging of the inboard legs, the D-shape winding pack nested inside each machined SS316LN case, and the paired OIS bolt bosses on each toroidal cut face that transfer shear between neighbouring coils.}
    \label{fig:tf_single_coil_cad}
\end{figure}

Mechanical boundary conditions reflect the actual inter-coil interfaces visible in Fig.~\ref{fig:tf_single_coil_cad}. The inboard wedge face is represented by a pair of narrow spring-foundation bands at $z \in [100, 280]$~mm and $z \in [-280, -100]$~mm, matching the geometric land where neighbouring coils bear against each other at the nominal $\pm 10^{\circ}$ toroidal wedge angle rather than the full inboard face. A roller constraint at the base represents the gravity support, and four discrete spring-foundation patches per toroidal cut face -- centred on each OIS bolt group at radii of 4818, 5358, 5357, and 4830~mm and at $y = \pm 2000, \pm 700$~mm -- model the shear reactions that pass between neighbouring coils through the OIS studs and shim stack. All springs use a constant normal stiffness of $k = 5 \times 10^{13}$~N\,m$^{-3}$, chosen so that the predicted stand-off displacement at the wedge land is consistent with the compressive travel of the nominal insulating shim stack under the expected contact pressure.

\begin{figure*}[!t]
    \centering
    \includegraphics[width=0.72\textwidth]{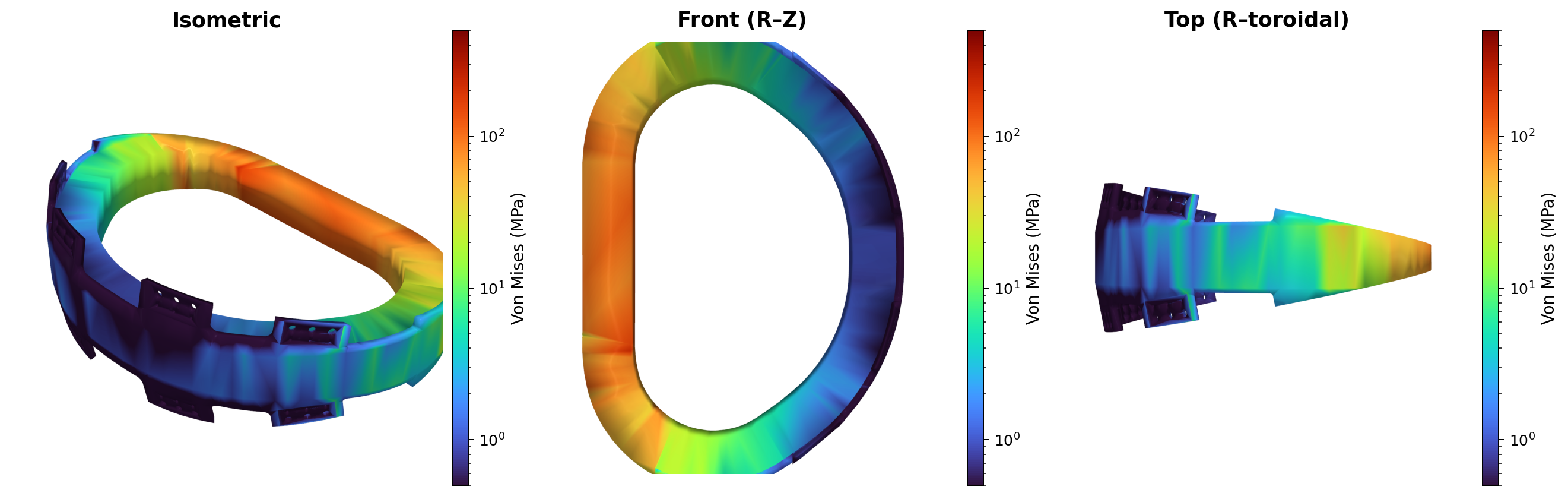}
    \caption{Von Mises stress on the TF case (logarithmic colour scale; gold markers indicate the four OIS contact centres). The analytical centering body force $\mathbf{f} = -K\,\mathbf{r}/r^{4}$ ($K = 6.93\times10^{8}$~N\,m$^{-1}$) drives the case inward; the inboard spring wedge and four discrete OIS patches react the load. The peak nodal stress is 464~MPa at the inboard midplane.}
    \label{fig:tf_vm_stress_log}

\end{figure*}

Under this loading, the centering body force drives each coil radially inward. Load is transferred primarily into the inboard wedge springs on the high-field side and secondarily into the eight OIS patches on the two toroidal cut faces, which together react both the net inward load and the out-of-plane bending induced by the asymmetric inboard--outboard force distribution. The resulting equivalent (von Mises) stress field is shown in Fig.~\ref{fig:tf_vm_stress_log}. The peak nodal stress is 464~MPa and occurs along the inboard midplane, well within the SS316LN allowable ($\sigma_y \approx 830$~MPa at 4~K) and consistent with the expected wedge-reaction behaviour, providing a safety margin of over 44\%. The four OIS patches on each cut face show localized stress increases where the spring foundations react shear between adjacent coils, but none approach the inboard peak. The outboard D-shape legs, away from any restrained boundary, carry membrane-level stresses that are roughly an order of magnitude lower than the inboard peak, indicating that the case is adequately sized everywhere outside the wedge. Future work will augment this analysis with an explicit winding-pack solid to capture interaction between the conductor stack and the structural case, and will resolve the local stress concentration around each OIS bolt hole with a sub-model.

\begin{table*}[!t]
\centering
\scriptsize
\caption{CS and PF coil design parameters.}
\label{tab:pf_cs_reduced}
\begingroup
\renewcommand{\arraystretch}{1.22}
\resizebox{\textwidth}{!}{%
\begin{tabular}{lcccccccccccccc}
\toprule
\textbf{Parameter} & \textbf{CS 1} & \textbf{CS 2} & \textbf{CS 3} & \textbf{CS 4} & \textbf{CS 5} & \textbf{CS 6} & \textbf{PF 1} & \textbf{PF 2} & \textbf{PF 3} & \textbf{PF 4} & \textbf{PF 5} & \textbf{PF 6} & \textbf{PF 7} & \textbf{PF 8} \\
\midrule
MA-turn & 16.4 & 16.4 & 16.4 & 16.4 & 16.4 & 16.4 & 6.2 & 3.1 & 6.6 & 3.0 & 3.0 & 6.6 & 3.1 & 6.2 \\
\addlinespace[1.5pt]
R (m) & 0.80 & 0.80 & 0.80 & 0.80 & 0.80 & 0.80 & 1.53 & 3.31 & 4.90 & 5.76 & 5.76 & 4.90 & 3.31 & 1.53 \\
\addlinespace[1.5pt]
Z (m) & -2.50 & -1.50 & -0.49 & 0.52 & 1.52 & 2.53 & 3.13 & 3.46 & 2.55 & 0.92 & -0.92 & -2.55 & -3.46 & -3.13 \\
\addlinespace[1.5pt]
Width (m) & 0.44 & 0.44 & 0.44 & 0.44 & 0.44 & 0.44 & 0.39 & 0.39 & 0.39 & 0.39 & 0.39 & 0.39 & 0.39 & 0.39 \\
\addlinespace[1.5pt]
Height (m) & 0.94 & 0.94 & 0.94 & 0.94 & 0.94 & 0.94 & 0.39 & 0.39 & 0.39 & 0.39 & 0.39 & 0.39 & 0.39 & 0.39 \\
\addlinespace[1.5pt]
J$_{\mathrm{WP}}$ (A/mm$^2$) & 47.7 & 47.7 & 47.7 & 47.7 & 47.7 & 47.7 & 52.5 & 25.8 & 55.6 & 25.0 & 25.0 & 55.6 & 25.8 & 52.5 \\
\addlinespace[1.5pt]
Turns radial & 19 & 19 & 19 & 19 & 19 & 19 & 16 & 16 & 16 & 16 & 16 & 16 & 16 & 16 \\
\addlinespace[1.5pt]
Turns vertical & 41 & 41 & 41 & 41 & 41 & 41 & 16 & 16 & 16 & 16 & 16 & 16 & 16 & 16 \\
\addlinespace[1.5pt]
Total turns & 774 & 774 & 774 & 774 & 774 & 774 & 256 & 256 & 256 & 256 & 256 & 256 & 256 & 256 \\
\addlinespace[1.5pt]
I$_{\mathrm{term}}$ (kA) & 21.2 & 21.2 & 21.2 & 21.2 & 21.2 & 21.2 & 24.3 & 12.0 & 25.8 & 11.7 & 11.7 & 25.8 & 12.0 & 24.3 \\
\addlinespace[1.5pt]
12mm REBCO tape length (km) & 81.8 & 82.1 & 84.1 & 84.2 & 82.0 & 81.8 & 38.5 & 32.2 & 123.6 & 49.8 & 49.8 & 123.6 & 32.2 & 38.5 \\
\bottomrule
\end{tabular}}
\endgroup
\end{table*}

\begin{figure}
    \centering
    \includegraphics[width=0.90\linewidth]{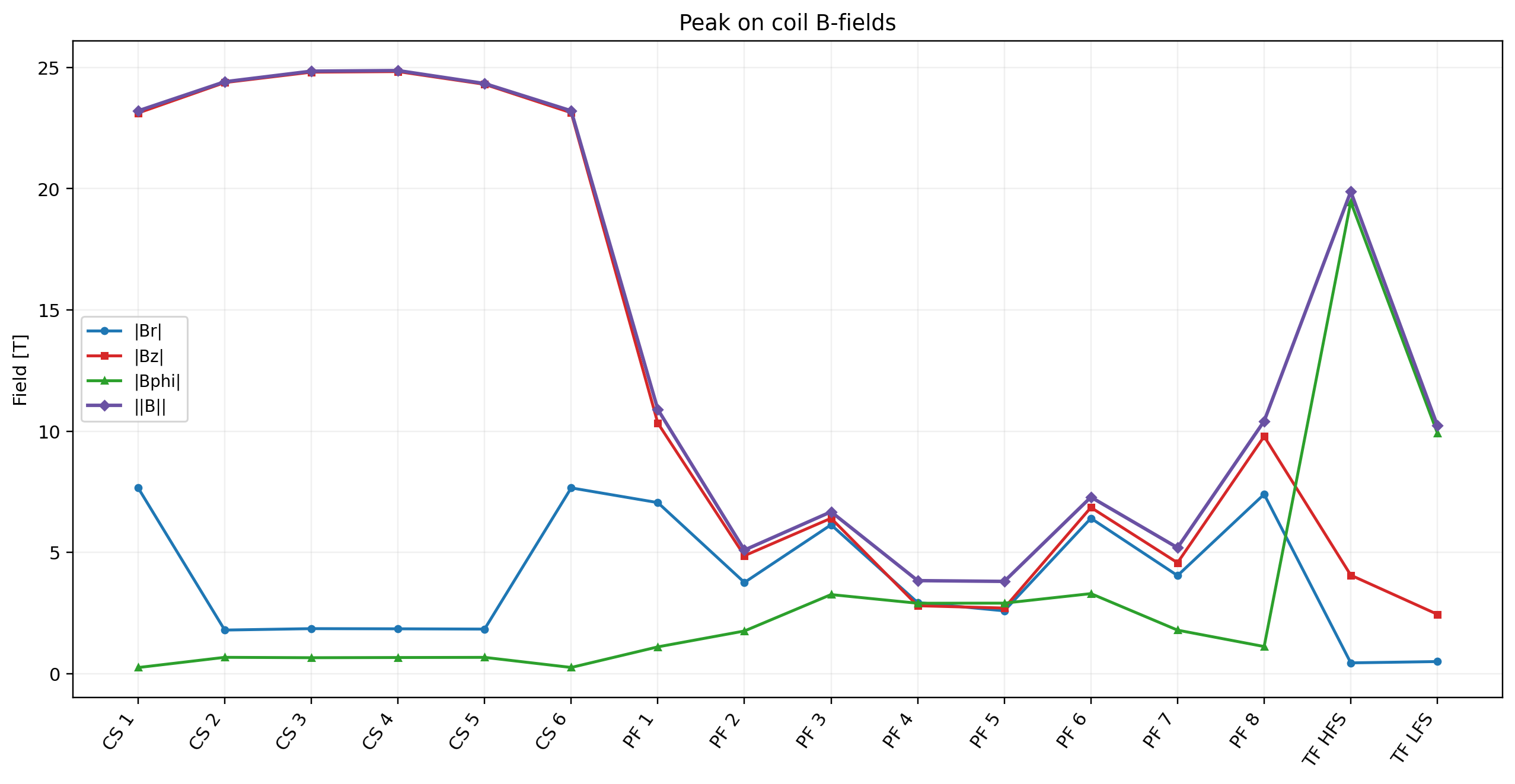}
    \caption{Peak magnetic field on conductor.}
    \label{fig:pf_cs_peak_field_on_conductor}
\end{figure}

\begin{figure}
    \centering
    \includegraphics[width=0.90\linewidth]{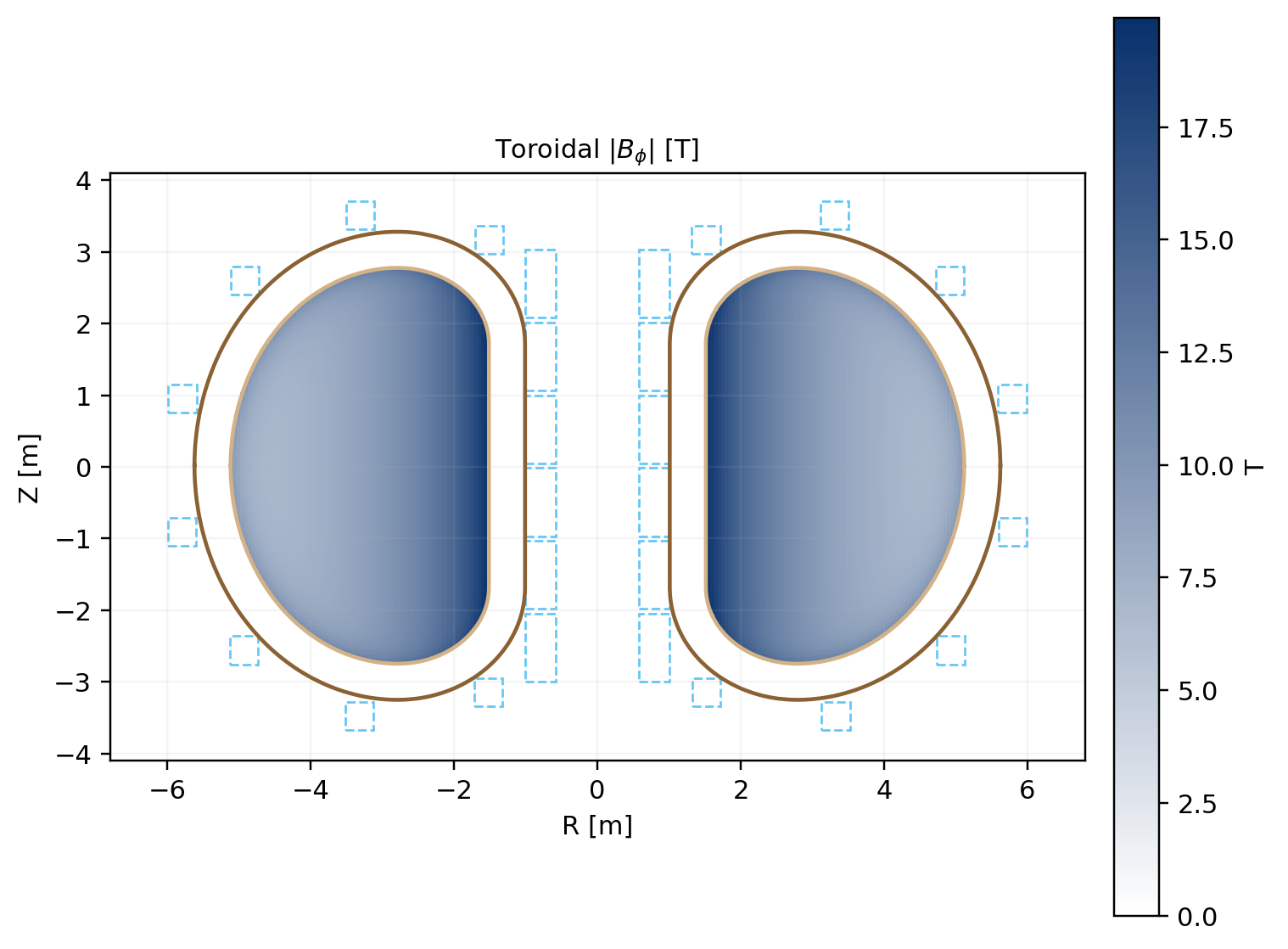}
    \caption{Toroidal field magnitude $|B_{\phi}|$.}
    \label{fig:pfcs_bphi_map}
\end{figure}

\begin{figure}
    \centering
    \includegraphics[width=0.82\linewidth]{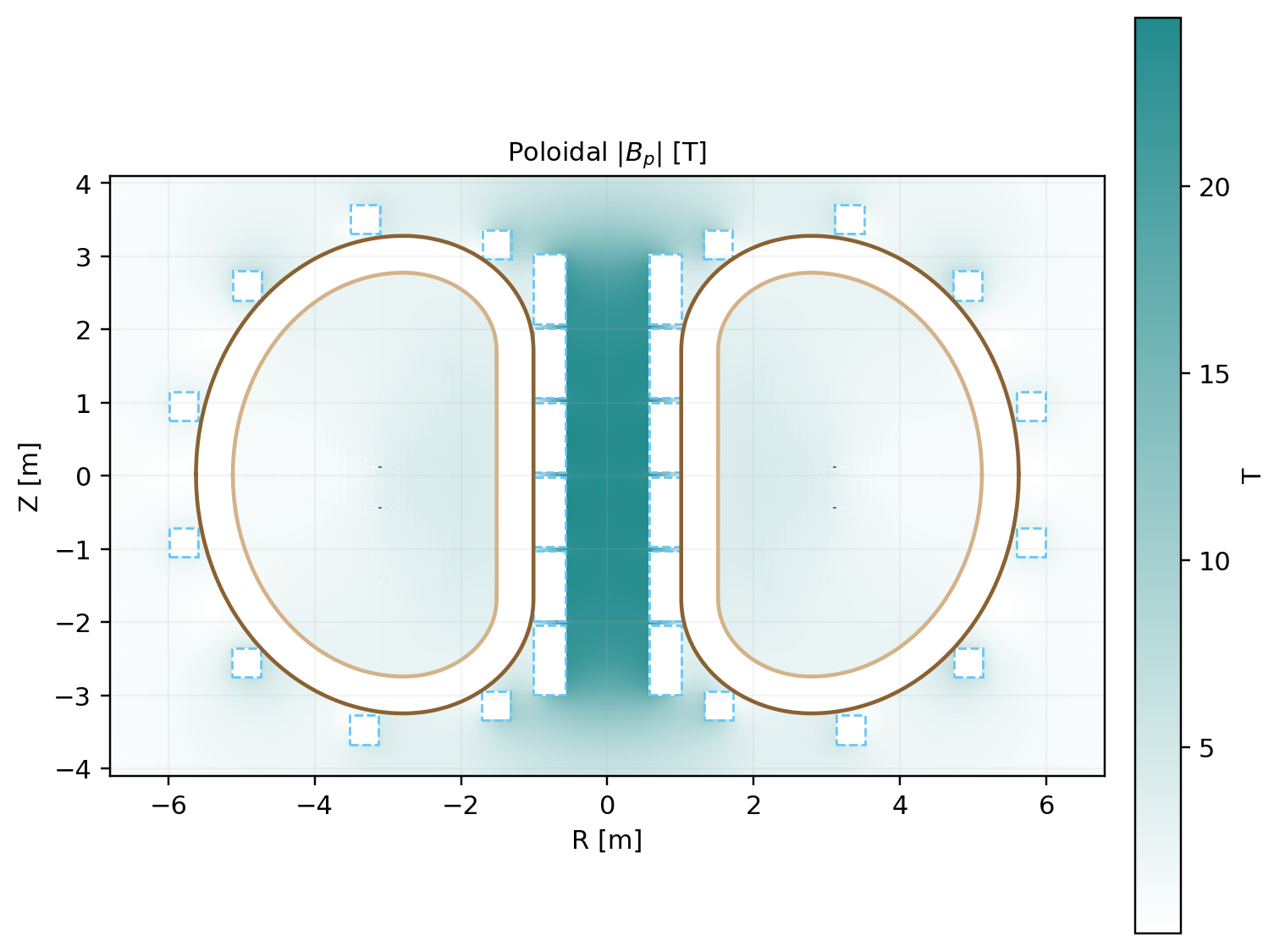}
    \caption{Poloidal field magnitude $|B_p|$.}
    \label{fig:pfcs_bp_map}
\end{figure}

\begin{figure}
    \centering
    \includegraphics[width=0.82\linewidth]{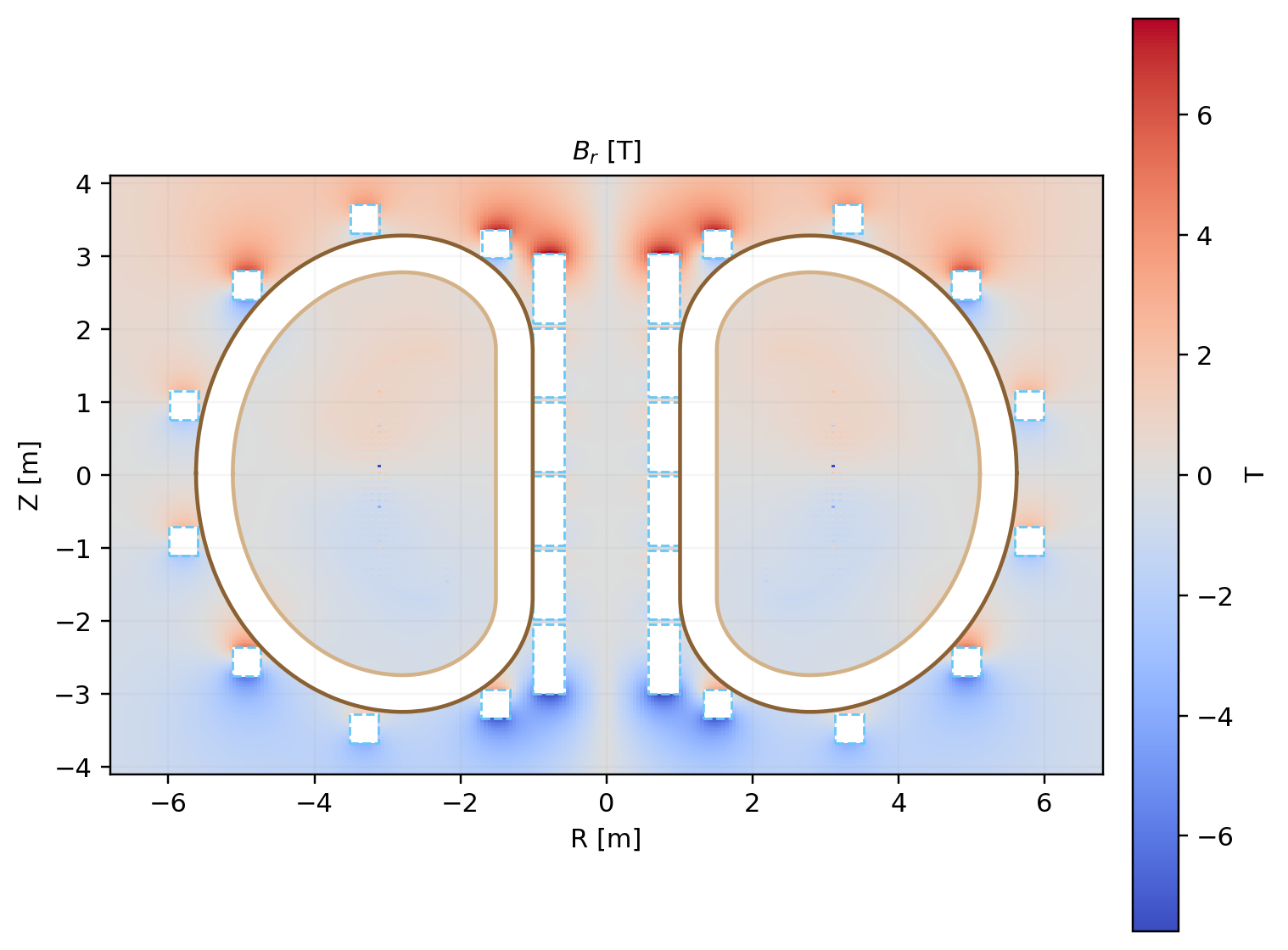}
    \caption{Radial field component $B_r$.}
    \label{fig:pfcs_br_map}
\end{figure}

\begin{figure}
    \centering
    \includegraphics[width=0.82\linewidth]{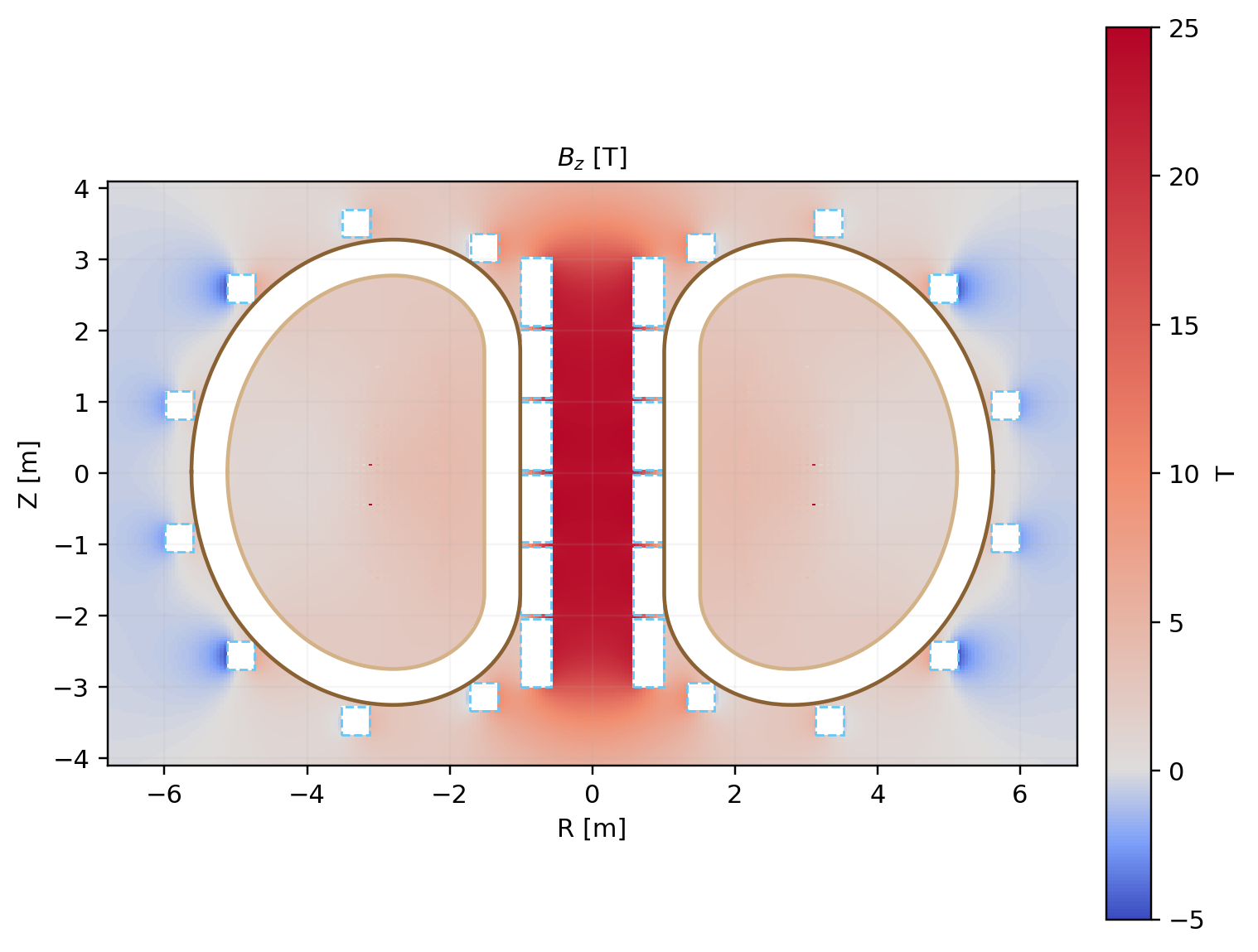}
    \caption{Vertical field component $B_z$.}
    \label{fig:pfcs_bz_map}
\end{figure}

\begin{figure}[!t]
    \centering
    \includegraphics[width=0.84\linewidth]{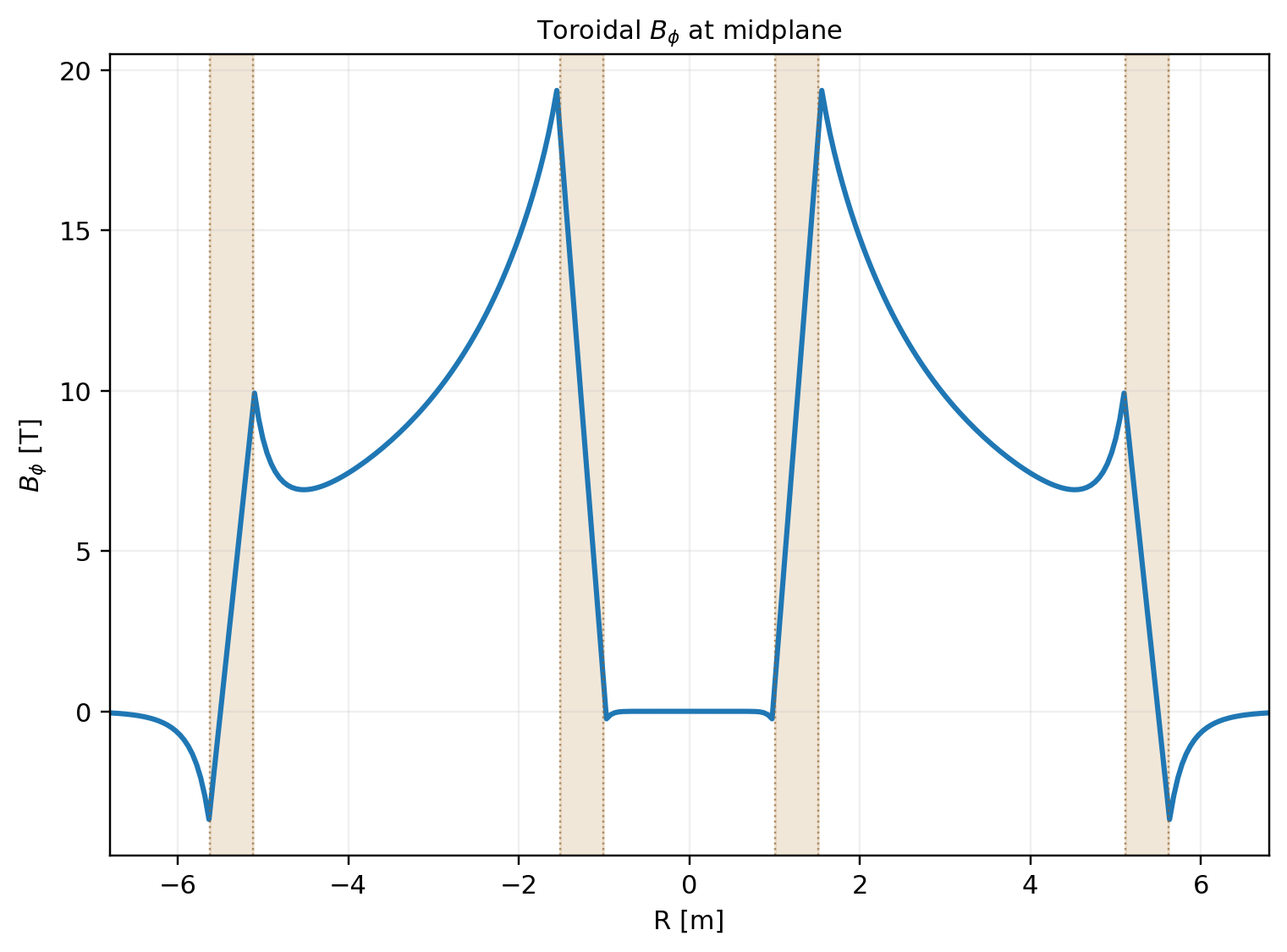}
    \caption{Midplane toroidal-field profile. The shaded regions denote the TF radial span.}
    \label{fig:pfcs_bphi_midplane}
\end{figure}

\begin{figure}[!t]
    \vspace{1.5cm}
    \centering
    \includegraphics[width=0.84\linewidth]{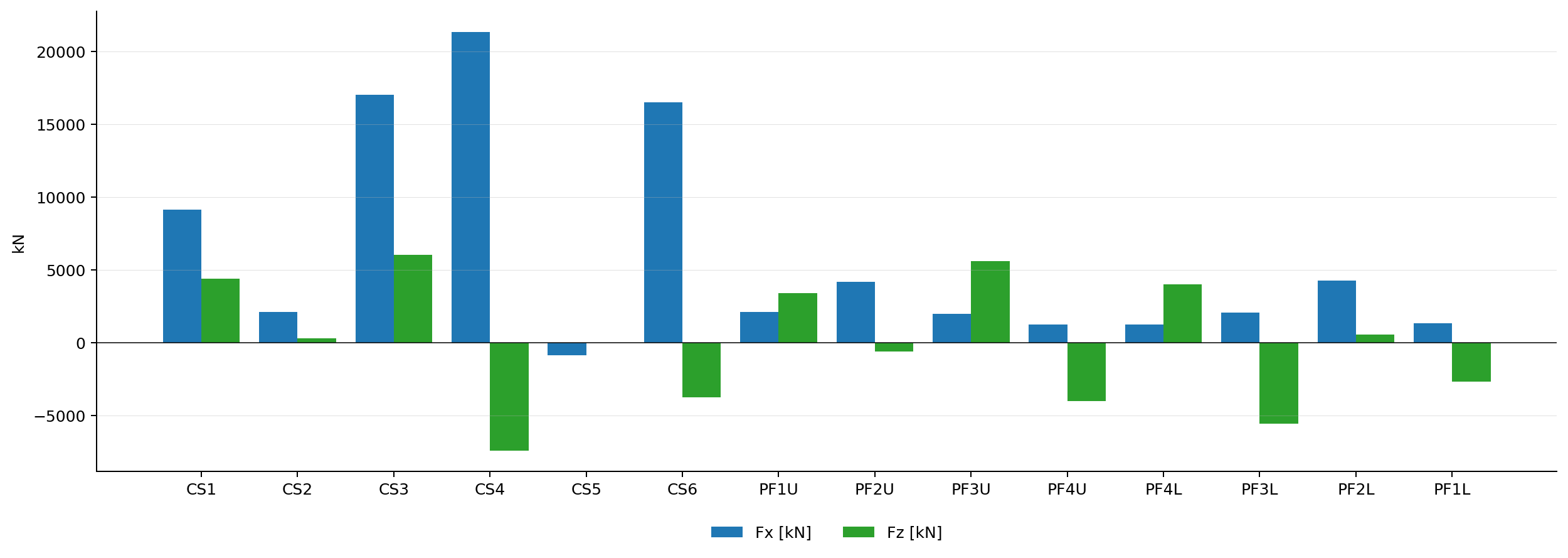}
    \caption{Resultant force components on the PF and CS coils from the axisymmetric filament electromagnetic model applied to the production six-coil CS stack (uniform $h=0.941$~m, $R=0.79$~m, $Z$ centers at $\pm 0.48$, $\pm 1.48$, $\pm 2.54$~m) and the eight PF coils of Table~\ref{tab:pf_cs_reduced}. Forces are reported per sector ($1/18$ of the axisymmetric total) to preserve the convention of the prior one-sector analysis.}
    \vspace{0.8cm}
    \label{fig:pf_cs_forces_one_sector}
\end{figure}

For the PF and CS system, the same electromagnetic model is also used to evaluate the local field on conductor across all magnets with the TF, PF, CS, and plasma current energized simultaneously to produce the worst-case field seen by the coils. For many of the equilibria studied, the coil currents for PF4 and PF5 remain low, $< 1MA$, however a minimum bound of 3MA was used for grading and structural analysis to cover operational flexibility regarding varying internal inductance, beta profiles, ramp up profiles etc. This provides the local field and current envelope used for conductor grading and for checking that the packed winding-current densities in Table \ref{tab:pf_cs_reduced} remain at or below 50\% of the critical-current limit when evaluated turn-by-turn for each coil. As expected for a TF-dominated machine, the radial force component is generally the largest term, while vertical load reversals between upper and lower coils reflect the corresponding support reactions that must be carried through the PF and CS attachment hardware.

Net force components on each PF and CS coil under the bounding current set are resolved with an axisymmetric filament model: each coil is discretised as a rectangular grid of circular toroidal-current filaments, the plasma is represented by the torodial current density output from FUSE with $I_p = 9.67$~MA, and the Biot--Savart poloidal field at every filament is summed over all other filaments via closed-form complete-elliptic-integral expressions. The radial force component is generally the largest term on the PF coils and on the central CS segments, while vertical load reversals between upper and lower coils reflect the up-down asymmetry of the shaping system and the corresponding support reactions that must be carried through the PF and CS attachment hardware. Per-sector values shown in Fig.~\ref{fig:pf_cs_forces_one_sector} are $1/18$ of the axisymmetric total, preserving the convention of the prior one-sector COMSOL analysis for direct comparison.

\begin{figure}[!t]
    \centering
    \includegraphics[width=\linewidth]{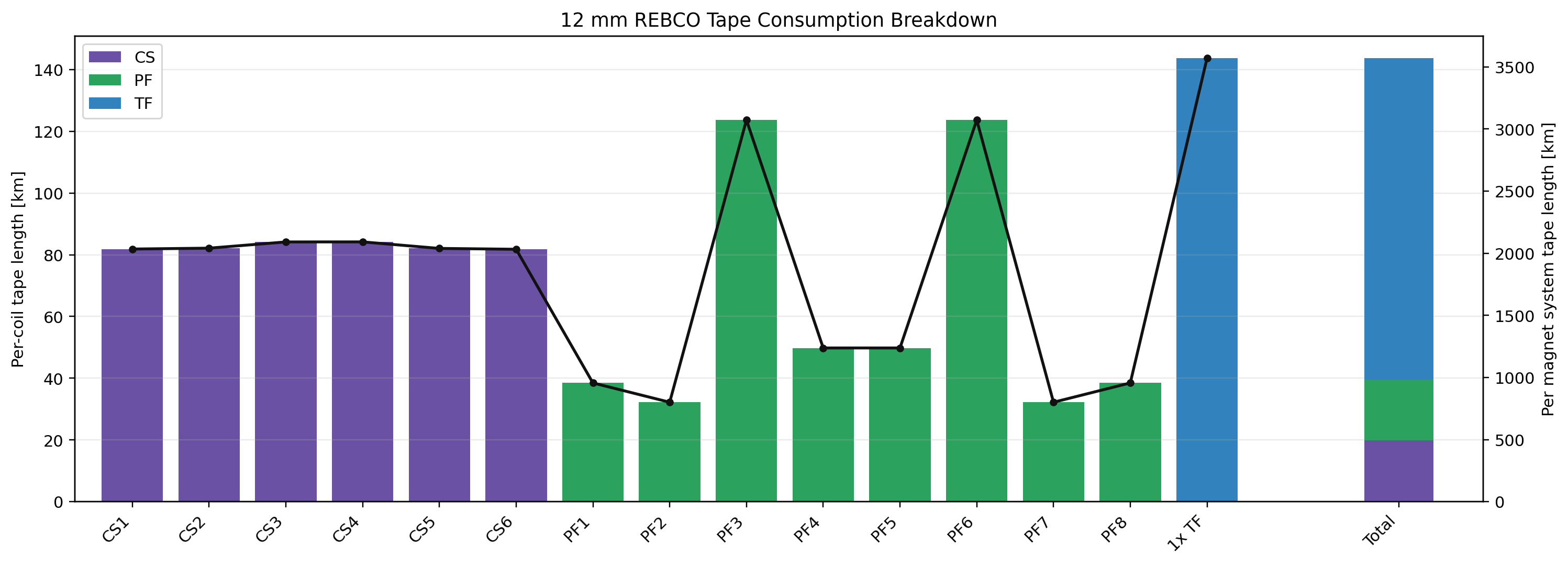}
    \caption{Faraday Factory tape-length summary for the bounding PF/CS/plasma operating point. The left y-axis shows coil-by-coil tape length, with the TF contribution reported for a single TF coil. The right y-axis shows the machine totals for the CS, PF, and full 18-coil TF systems.}
    \label{fig:faraday_tape_length_summary}
\end{figure}

Using the Faraday Factory tape dataset under the same bounding PF/CS/plasma operating point, the resulting conductor inventory remains concentrated in the TF system, while the six CS coils and eight PF coils together still represent a non-trivial fraction of total HTS consumption. Figure \ref{fig:faraday_tape_length_summary} summarizes this split both at machine level and coil-by-coil, which is useful for connecting the local grading results back to procurement-scale tape demand. With grading, including only the amount of tape needed to provide a 50\% margin to $I_c$ given the local $B$ field, the total 12\,mm REBCO tape required is \SI{3600}{km}.

\section{Neutronics}\label{sec:neutronics} 

Understanding where neutron power is deposited is one of the most restrictive constraints in the plant design space. For D--T operation, approximately $80\%$ of the fusion energy leaves the plasma as neutrons, so shielding, breeding, heating, activation, and lifetime are all tightly coupled design drivers. In high-power-density tokamaks, this neutron loading can produce severe nuclear heating in surrounding structures, high displacement damage, and high activation inventories. These effects are especially consequential for superconducting magnets: cryogenic operation penalizes every watt of parasitic heat with large wall-plug overhead (at $20~\mathrm{K}/300~\mathrm{K}$, Carnot COP is $\sim 7.1\%$ while practical cryocooler COP is $\sim 1.5\%$), and high-energy neutron fluence can degrade HTS critical current over mission life.

Yinsen uses a deliberately lower-power-density strategy to reduce these neutron-driven penalties while preserving mission-relevant performance. OpenMC was used extensively, to verify that candidate geometries remain feasible from shielding, nuclear heating, material damage, and tritium-breeding perspectives. This workflow was used to assess in-vessel loading, magnet dose/fluence, vacuum-vessel survivability, and plant-level power-balance implications of blanket energy multiplication.

The blanket/shield concept includes a tungsten first wall and an inner FLiBe (Li$_2$BeF$_4$) coolant channel~\cite{Bocci2020} that contribute to neutron multiplication and tritium production, while the primary FLiBe blanket channel provides most of the breeding volume. Vacuum-vessel structural components use V-4Cr-4Ti alloy~\cite{Muroga2002}, while the FLiBe-facing blanket tank is constructed of Hastelloy N (Ni-16Mo-7Cr-5Fe), the nickel-base alloy developed at ORNL for the Molten Salt Reactor Experiment to provide corrosion resistance against fluoride salts at elevated temperature~\cite{HastelloyN_MSRE}.

\begin{figure*}[!t]
\centering
\includegraphics[width=\textwidth]{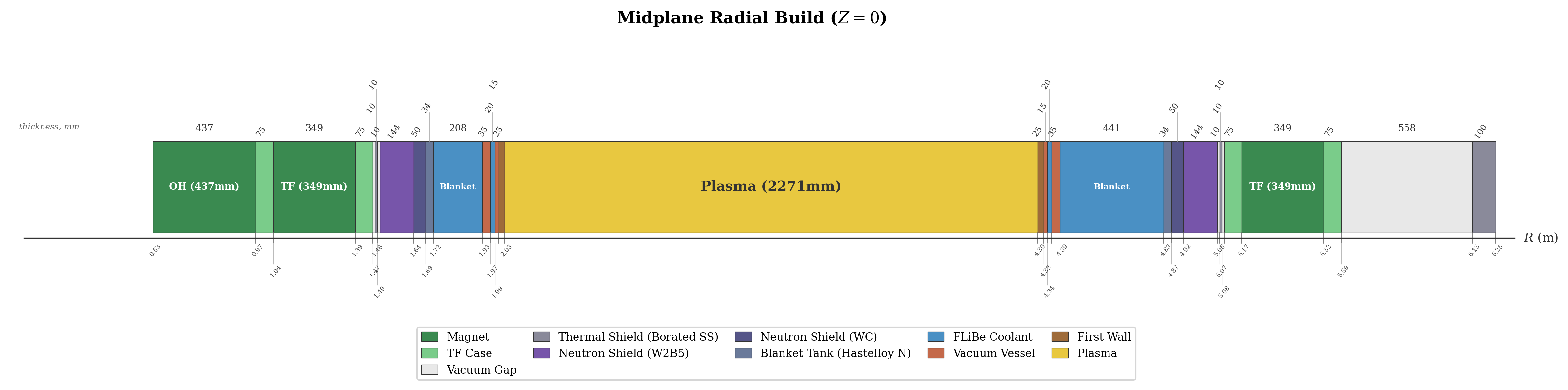}
\caption{Radial build at the midplane, showing the sequence of plasma-facing, breeding, shielding, and magnet-adjacent layers used throughout the neutronics analysis.}
\label{fig:neutronics_radial_build}
\end{figure*}

\begin{figure}[!t]
\centering
\includegraphics[width=\columnwidth]{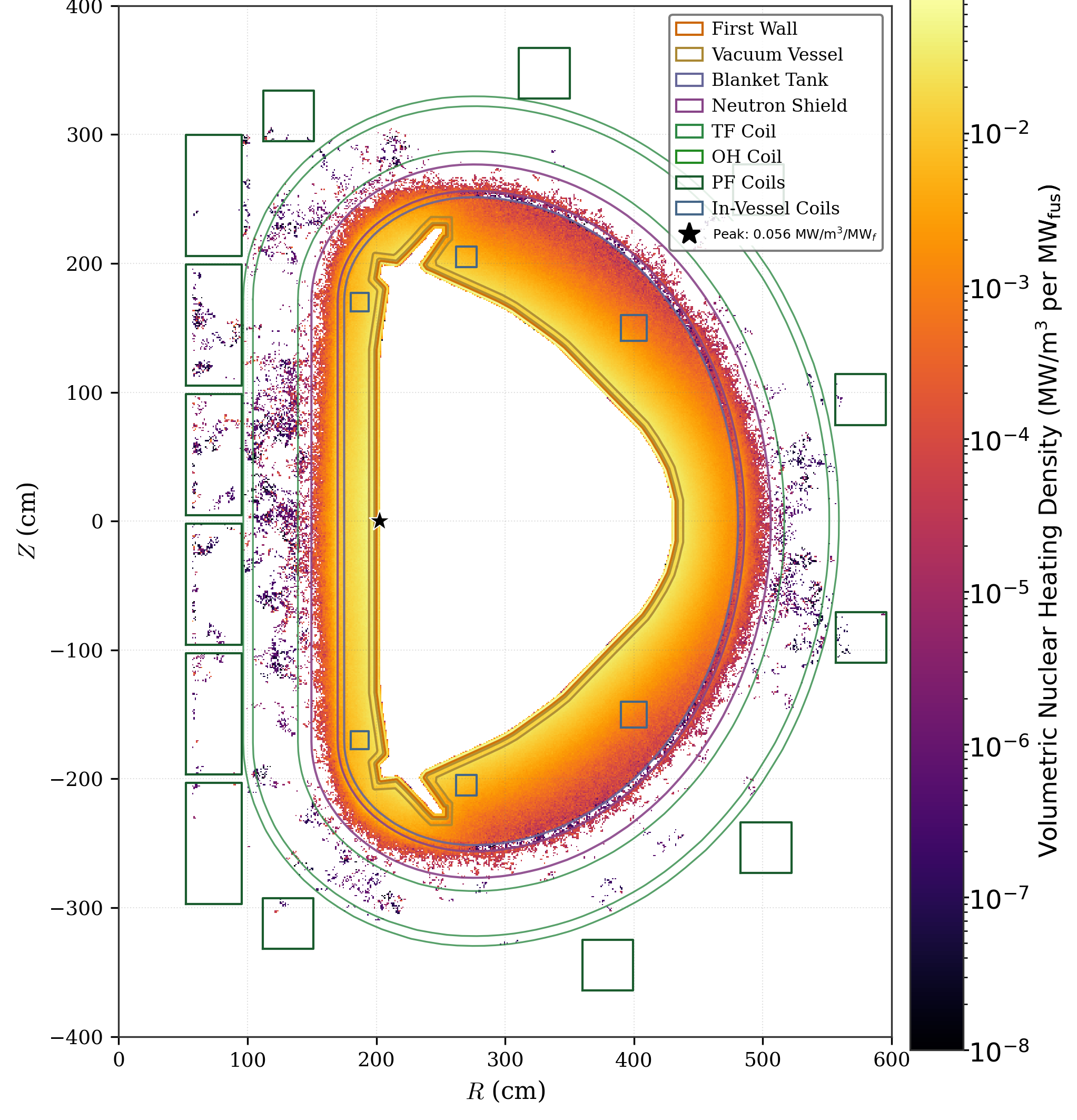}
\caption{Total volumetric nuclear heating density normalized per MW of fusion power, with component geometry outlines.}
\label{fig:neutronics_heating_map_perMW}
\end{figure}

\begin{figure}[!t]
\centering
\includegraphics[width=\columnwidth]{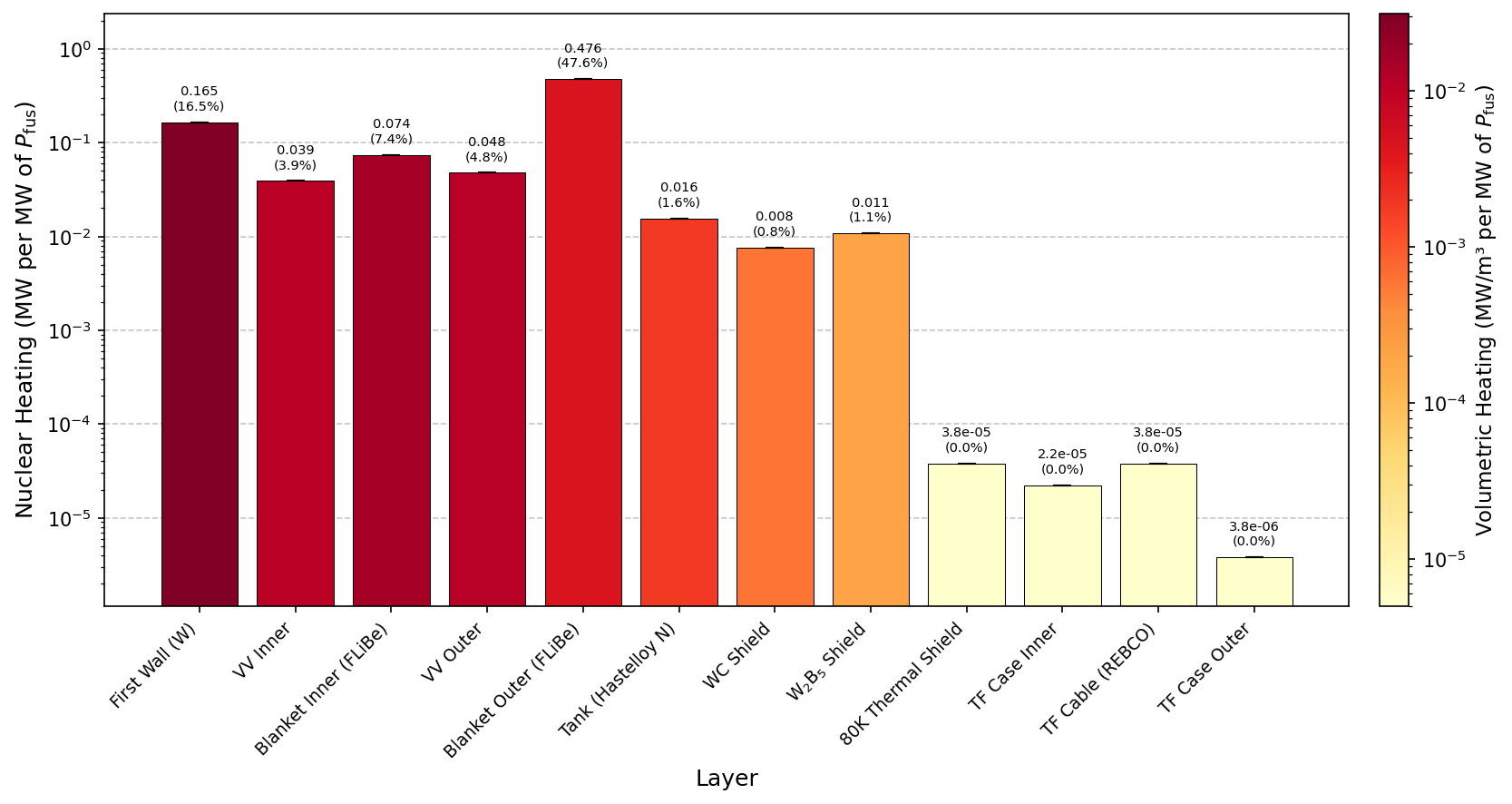}
\caption{Integrated nuclear heating by layer normalized per MW of fusion power, \% is relative to total fusion power.}
\label{fig:neutronics_heating_layers_perMW}
\end{figure}

\begin{figure}[!t]
\centering
\includegraphics[width=\columnwidth]{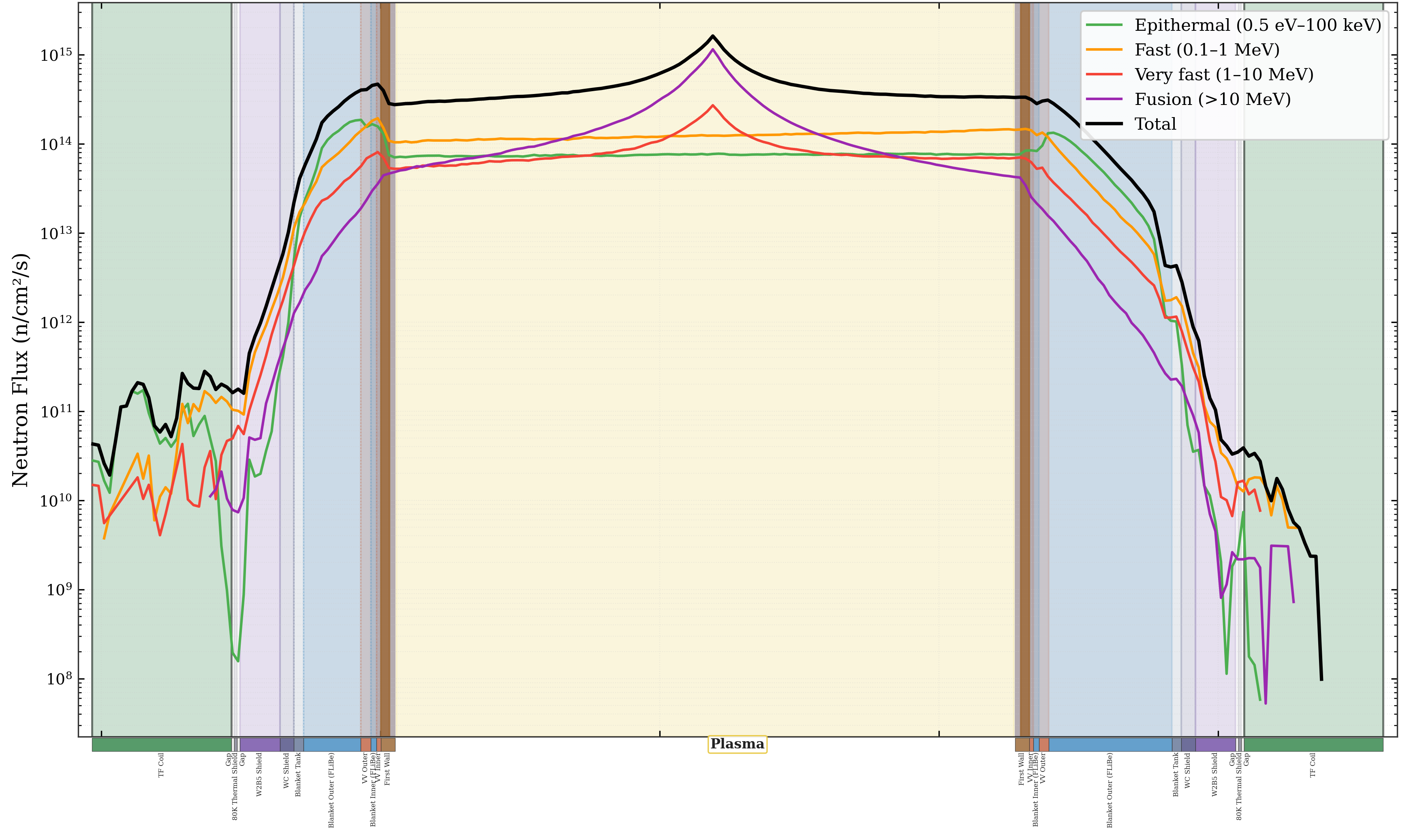}
\caption{Midplane neutron particle flux by energy group (n/cm$^2$/s), illustrating spectral slowing-down and attenuation through the radial build.}
\label{fig:neutronics_midplane_flux}
\end{figure}

\begin{figure}[!t]
\centering
\includegraphics[width=\columnwidth]{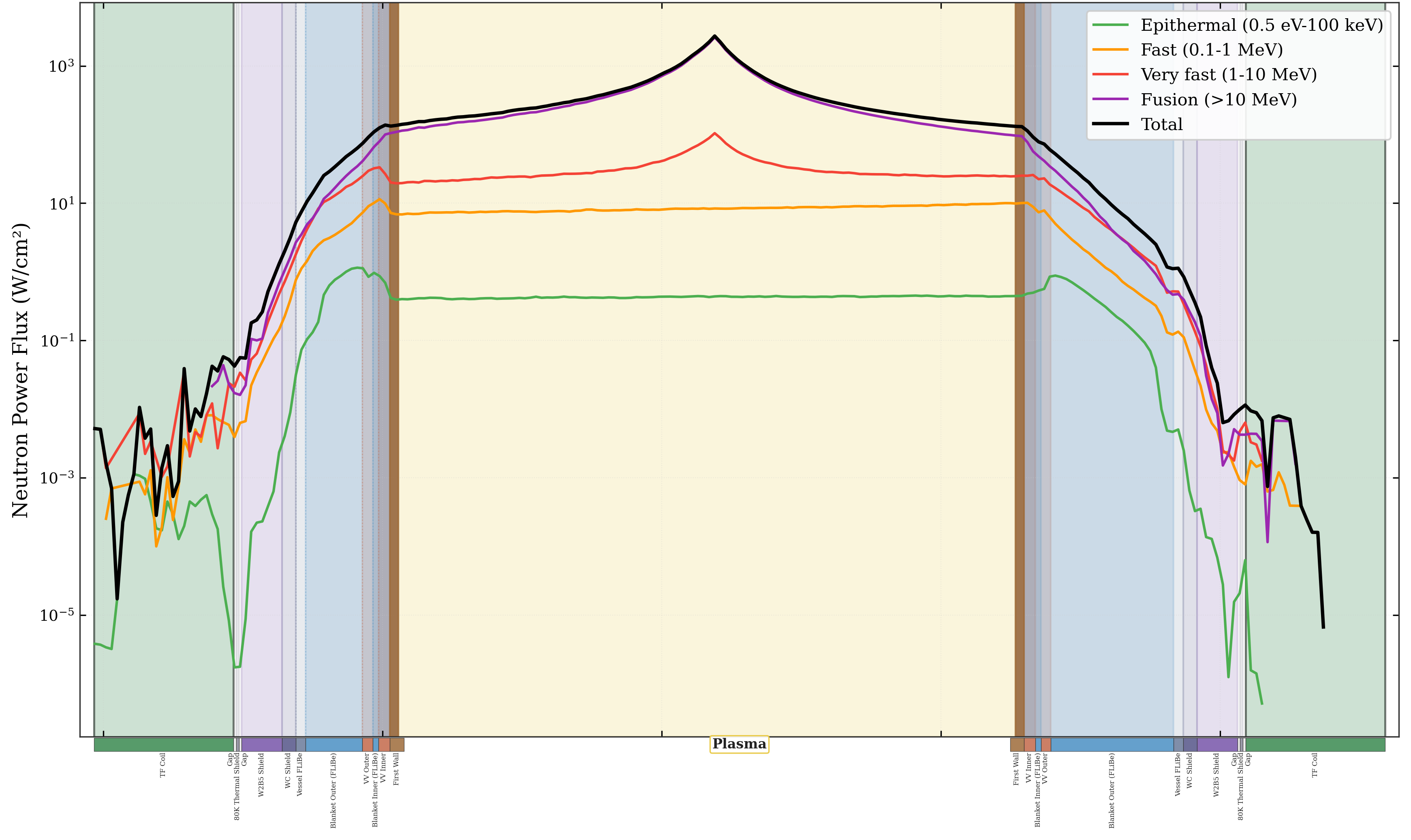}
\caption{Midplane energy-weighted neutron power flux by energy group (W/cm$^2$), showing where deposited power is concentrated radially and spectrally.}
\label{fig:neutronics_midplane_energy_flux}
\end{figure}

Outside the breeding region, Yinsen adopts a double-layer neutron-shield strategy using WC followed by W$_2$B$_5$: the dense WC layer is used first to reduce fast-neutron energy and associated volumetric heating, while the outer W$_2$B$_5$ layer is used to absorb the softened spectrum before it reaches the magnet system. The effectiveness of this moderation-and-absorption sequence is reflected directly in the radial-build, heating, and flux results summarized in Figures~\ref{fig:neutronics_radial_build}--\ref{fig:neutronics_midplane_energy_flux}.

Because Yinsen targets a lower fusion power and lower required availability than grid-scale baseload plants, the tritium-accounting burden is correspondingly reduced. Even so, the present neutronics design still imposes $TBR \geq 1.1$ as the baseline self-sufficiency requirement.

\begin{table}[htbp]
\centering
\scriptsize
\renewcommand{\arraystretch}{1.4}
\resizebox{\columnwidth}{!}{%
\begin{tabular}{m{5.6cm}cc}
\hline
\textit{Metric} & \textit{Symbol} & \textit{Value} \\ \hline
TBR ($^6$Li\,$\approx$\,30\,\%, no dedicated multiplier) & $TBR$ & 1.10 \\ \hline
Blanket energy mult. & $M_E$ & 1.04 \\ \hline
Peak DPA inner VV & $DPA_{\mathrm{VV,in}}^{\mathrm{pk}}$ & $0.0337$\,DPA/(MW$\cdot$yr) \\ \hline
Average DPA inner VV & $DPA_{\mathrm{VV,in}}^{\mathrm{avg}}$ & $0.025$\,DPA/(MW$\cdot$yr) \\ \hline
Peak DPA outer VV & $DPA_{\mathrm{VV,out}}^{\mathrm{pk}}$ & $0.0224$\,DPA/(MW$\cdot$yr) \\ \hline
Average DPA outer VV & $DPA_{\mathrm{VV,out}}^{\mathrm{avg}}$ & $0.019$\,DPA/(MW$\cdot$yr) \\ \hline
Peak He production inner VV & $\dot{c}_{\mathrm{He,in}}^{\mathrm{pk}}$ & $0.111$\,appm/(MW$\cdot$yr) \\ \hline
Average He production inner VV & $\dot{c}_{\mathrm{He,in}}^{\mathrm{avg}}$ & $0.082$\,appm/(MW$\cdot$yr) \\ \hline
Peak He production outer VV & $\dot{c}_{\mathrm{He,out}}^{\mathrm{pk}}$ & $0.0569$\,appm/(MW$\cdot$yr) \\ \hline
Average He production outer VV & $\dot{c}_{\mathrm{He,out}}^{\mathrm{avg}}$ & $0.048$\,appm/(MW$\cdot$yr) \\ \hline
Peak TF fast-neutron flux & $\dot\Phi_{n,\mathrm{TF}}$ & $4.0\!\times\!10^{7}$\,n/cm$^2$/s/MW(fus) \\ \hline
TF magnet lifetime & $\tau_{\mathrm{TF}}$ & $1.72\!\times\!10^{4}$\,MW$\cdot$yr \\ \hline
Peak inboard TF vol.\ heating & $\dot q_{\mathrm{TF,in}}$ & $1.25\!\times\!10^{-6}$\,W/cm$^3$/MW(fus) \\ \hline
Total TF nuclear heating & $P_{\mathrm{mag}}$ & $0.057$\,kW/MW(fus) \\ \hline
Total in-vessel coil nuclear heating & $P_{\mathrm{IVC}}$ & $3.12$\,kW/MW(fus) \\ \hline
Divertor nuclear heating & $P_{\mathrm{div}}$ & $0.013$\,MW/MW(fus) \\ \hline
\end{tabular}%
}
\caption{Neutronics summary metrics for the Yinsen baseline. Peak and average damage and gas-production rates are listed separately for the inner and outer vacuum vessel. All rates except TBR and the blanket energy multiplier are normalized per MW$\cdot$yr of fusion power.}
\label{tab:neutronics_summary}
\end{table}

After multiple iterations of material selection, layer ordering, and radial thickness, Figure~\ref{fig:neutronics_radial_build} shows the midplane build adopted for the present Yinsen baseline. This stack-up is the configuration that simultaneously satisfies the neutronics-driven requirements on tritium breeding, nuclear heating, shutdown dose, and magnet protection, while remaining compatible with the thermal constraints and through-thickness temperature gradients. The most difficult trade is on the high-field side, where blanket thickness must be balanced directly against shield thickness near the inner TF leg. The relatively low blanket energy multiplication of 1.04 is likely a direct consequence of this same geometry trade: total blanket volume is reduced, especially on the inboard side, additionally volume was reserved for potential in-vessel coils. Figure~\ref{fig:neutronics_radial_build} therefore serves both as the adopted design point and as the reference geometry for interpreting all subsequent heating, flux, damage, and breeding results.

Table~\ref{tab:neutronics_summary} shows that this same geometry also produces the intended distribution of component lifetimes within the plant. The reported displacement-damage rates are computed from the OpenMC damage-energy tally with the NRT model, using effective displacement thresholds of 40\,eV for V-4Cr-4Ti and 90\,eV for tungsten, while the hydrogen and helium production rates are derived from $(n,Xp)$ and $(n,X\alpha)$ reaction tallies. At the 130\,MW baseline and 40\% utilization, the peak HFS leg of the inner-vacuum-vessel damage rate corresponds to about 1.84\,DPA per calendar year, so the adopted 35\,DPA limit implies a vessel lifetime of 20 years, exactly as intended by the low-power-density design strategy. The total lifetime integrated fusion power to reach 35\,DPA on the HFS leg inner vacuum vessel region is 1040\,MW$\cdot$yr. The corresponding He/DPA values in the vessel remain in the expected fusion-materials range of a few appm/DPA, well below the regime where helium-driven grain-boundary degradation would become an independent dominant constraint. By contrast, the TF fast-neutron flux corresponds to a fast-flux-limited TF lifetime of $1.72\times 10^{4}$\,MW$\cdot$yr, or about 315 calendar years at the same operating point, roughly sixteen vacuum-vessel lifetimes. Because the TF inner leg is the most demanding superconducting location, this indicates that the HTS magnet system is comfortably a lifetime component in the present mission class. The PF and CS systems are less exposed than that bounding TF location, so their direct nuclear heating and irradiation burden are correspondingly negligible on plant-lifetime timescales, with effectively unbounded radiation lifetime relative to the vessel. The associated TF nuclear heating is also very small: at 130\,MW of fusion power it corresponds to only about 7.4\,kW total at 20\,K, which is readily accommodated by a compact cryoplant. Even at 500\,MW of fusion power, this scales to only about 28.5\,kW to the cold mass, which can be taken as a practical upper bound on the HTS-magnet nuclear-heating contribution to cryoplant sizing for the present reactor plant design.

The next question is whether that chosen stack-up actually puts the neutron power where the plant needs it. Figures~\ref{fig:neutronics_heating_map_perMW} and~\ref{fig:neutronics_heating_layers_perMW}, together with Table~\ref{tab:heating_by_layer}, show that it does. Most of the nuclear heating is deposited directly in the FLiBe blanket volume, where it can be usefully recovered, while the first wall and vacuum-vessel layers intercept a smaller but still important share that must be removed by the in-vessel cooling loops. Outboard of the blanket tank, the shield materials then drive the heating down rapidly before the thermal shield and magnet-adjacent regions are reached. This is the core systems result of the neutronics design: the FLiBe blanket volumes remain the dominant thermal sink, while the downstream structural and cryogenic-adjacent layers are pushed into a regime that is manageable for the thermal and magnet systems. That layer-wise breakdown is what allows the same neutronics solution to be carried consistently into the BOP heat-load model and the cryogenic design basis.

This breakdown also clarifies the thermal character of the in-vessel solid structures. For the first wall and adjacent interior structural layers, the dominant neutron-driven load is volumetric nuclear heating generated within the material itself, rather than a heat source applied only at the surface. That energy must therefore be conducted out through the local structural stack and then transferred into the FLiBe coolant channels in order to keep peak material temperatures and through-thickness thermal gradients within acceptable limits.

\begin{table}[!b]
\centering
\scriptsize
\caption{Nuclear heating by layer at the two reference power levels}
\label{tab:heating_by_layer}
\renewcommand{\arraystretch}{1.3}
\resizebox{\columnwidth}{!}{%
\begin{tabular}{m{2.8cm}cccc}
\hline
\textit{Layer} & \textit{130\,MW} & \textit{185\,MW} & \textit{Vol.\ Heat} & \textit{Share} \\
               & (MW)             & (MW)             & (kW/m$^3$/MW) &  (\%) \\ \hline
First Wall (W)            & 21.5 & 30.5 & 31.3  & 16.5 \\ \hline
VV Inner                  &  5.1 &  7.2 & 10.8  &  3.9 \\ \hline
Blanket Inner (FLiBe)     &  9.6 & 13.7 & 15.8  &  7.4 \\ \hline
VV Outer                  &  6.2 &  8.8 & 10.9  &  4.8 \\ \hline
Blanket Outer (FLiBe)     & 61.8 & 88.0 &  4.5  & 47.5 \\ \hline
Tank (Hastelloy N) &  2.0 &  2.9 &  1.8  &  1.5 \\ \hline
WC Shield                 &  1.0 &  1.4 &  0.6  &  0.8 \\ \hline
W$_2$B$_5$ Shield         &  1.4 &  2.0 &  0.2  &  1.1 \\ \hline
80K Thermal Shield        & 0.002 & 0.003 & $<$0.01 & $<$0.1 \\ \hline
TF Case Inner             & 0.003 & 0.004 & $<$0.01 & $<$0.1 \\ \hline
TF Cable (REBCO)      & 0.004 & 0.006 & $<$0.01 & $<$0.1 \\ \hline
TF Case Outer             & 0.000 & 0.001 & $<$0.01 & $<$0.1 \\ \hline
\end{tabular}%
}
\end{table}

Starting from the 14.1\,MeV DT fusion source, the midplane spectral results show how the neutron population is progressively scattered out of the fusion-energy group and redistributed into very-fast, fast, and then moderated energy groups as it traverses the first wall, FLiBe blanket, and shield stack. This slowing-down behavior is exactly what the adopted radial build is intended to produce: most of the neutron power is captured in the breeding blanket, while the WC and W$_2$B$_5$ shield sequence suppresses the residual high-energy tail before it reaches magnet-adjacent structures. In the process, the total neutron flux is reduced by nearly four orders of magnitude between the first wall and the TF conductor, dropping from about $2\times10^{14}$\,n/cm$^2$/s to about $4\times10^{10}$\,n/cm$^2$/s at the 130\,MW baseline.

The divertor benefits from a materially softer and less power-dense neutron spectrum than the inboard midplane path. That reduction in nuclear heating is important because the divertor already faces the most demanding local thermal environment in the machine from the SOL-driven surface heat flux; keeping the neutron contribution comparatively small helps keep the combined divertor heat-removal problem tractable.

\begin{figure}[!t]
\centering
\includegraphics[width=\columnwidth]{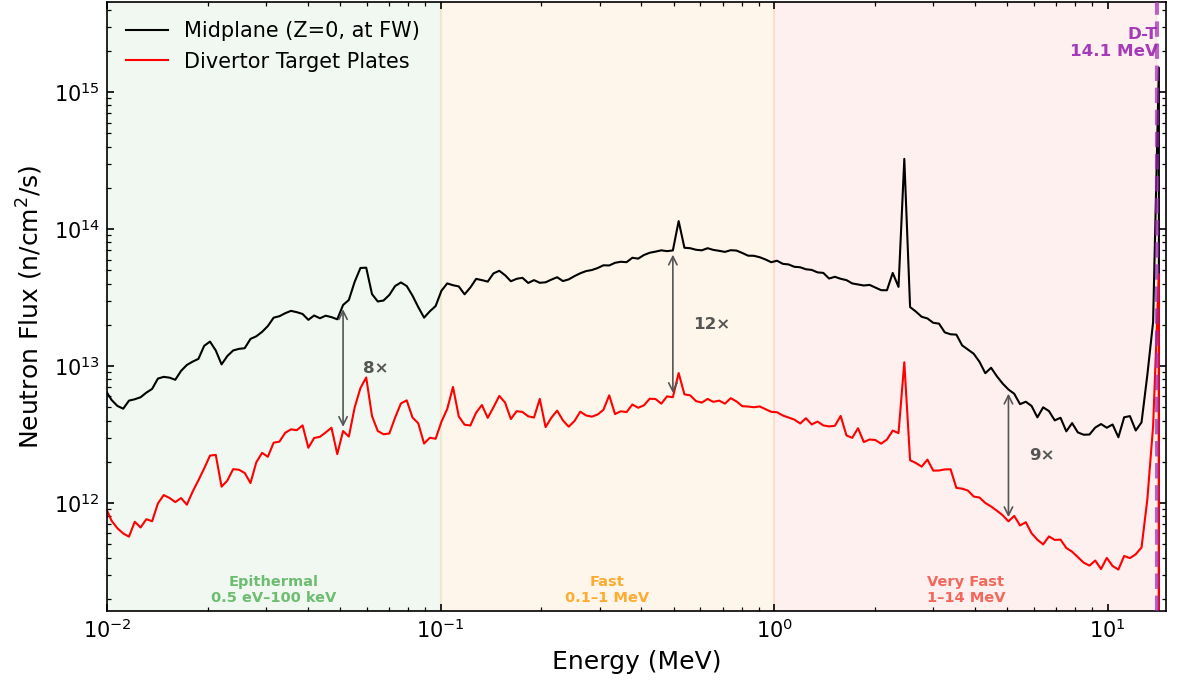}
\caption{Comparison of divertor-region and midplane neutron flux, showing the softer divertor spectrum.}
\label{fig:neutronics_divertor_midplane_flux}
\end{figure}

\begin{figure}[!t]
\centering
\includegraphics[width=\columnwidth]{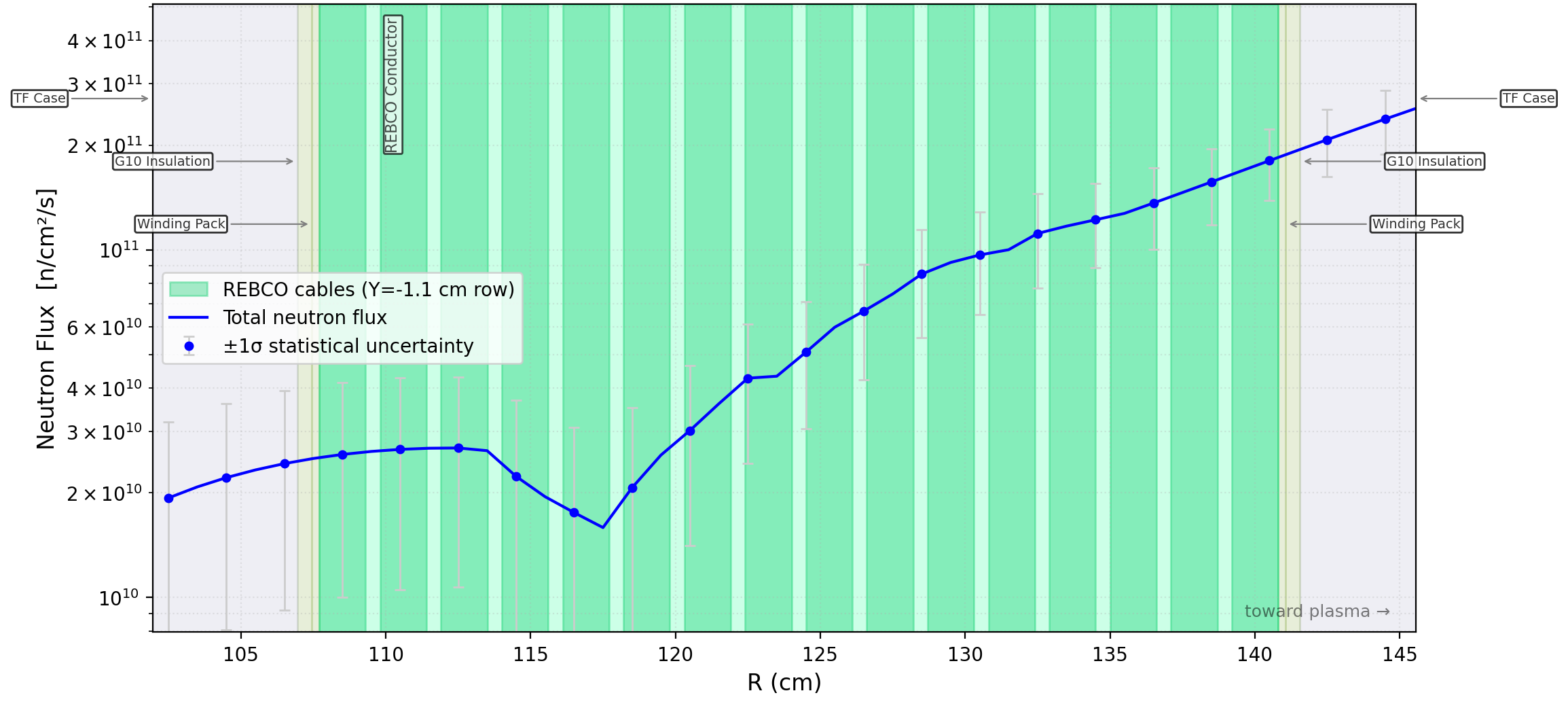}
\caption{Radial neutron-flux profile through the HFS TF coil at the midplane, showing flux attenuation across the TF case, G10 winding-pack insulation, and REBCO conductor regions. Green bands indicate individual REBCO cable positions.}
\label{fig:tf_fast_flux_radial}
\end{figure}

It is also useful to inspect the local flux environment that the TF conductor actually sees through the inboard shield stack. Figure~\ref{fig:tf_fast_flux_radial} shows the radial profile of total neutron flux through the HFS TF coil at the midplane ($|Z|\!<\!50$\,cm) from a standalone 250\,k-particle transport run. The profile resolves the individual sub-regions of the detailed TF model---outer steel case, G10 winding-pack insulation, REBCO conductor, and inner steel case---with the green bands marking individual REBCO cable positions. The flux drops by approximately one order of magnitude across the case and winding-pack stack, from ${\sim}10^{12}$\,n/cm$^2$/s at the outer case face to ${\sim}10^{11}$\,n/cm$^2$/s at the inner case. This attenuation profile sets the scale for the magnet-protection sensitivity study that follows.

Residual fast-neutron flux can be further attenuated by doping the magnet casing. Related magnet-adjacent engineered shielding approaches, including metal-hydride composite concepts, have also been proposed for compact fusion systems \cite{Fletcher2025MetalHydride}. A parametric study of hafnium doping (10\,wt\% Hf addition) was therefore carried out for the three material layers between the blanket and the TF conductor: the blanket tank, the thermal shield, and the TF case. All eight on/off combinations were evaluated at 300\,k source particles per configuration. Because hafnium has a large thermal-neutron absorption cross section, it offers a way to further suppress the residual neutron flux reaching the superconducting windings without requiring additional radial build.

The principal finding is that Hf doping in the TF case alone reduces the fast-neutron flux at the TF conductor by ${\sim}9\,\%$, while adding Hf to both the thermal shield and TF case yields a combined reduction of ${\sim}11\,\%$. Doping the blanket tank provides no additional benefit for magnet protection and can slightly worsen the result, likely because the spectrum at that location is still too fast for Hf absorption to be effective, so the few captures that do occur mainly add prompt-gamma heating rather than useful shielding. This makes Hf doping of the TF case a potentially attractive neutronics lever for reducing residual magnet flux, although its cost and impact on structural integrity would still need to be evaluated before adoption in the baseline design.

\begin{figure}[!t]
\centering
\includegraphics[width=\columnwidth]{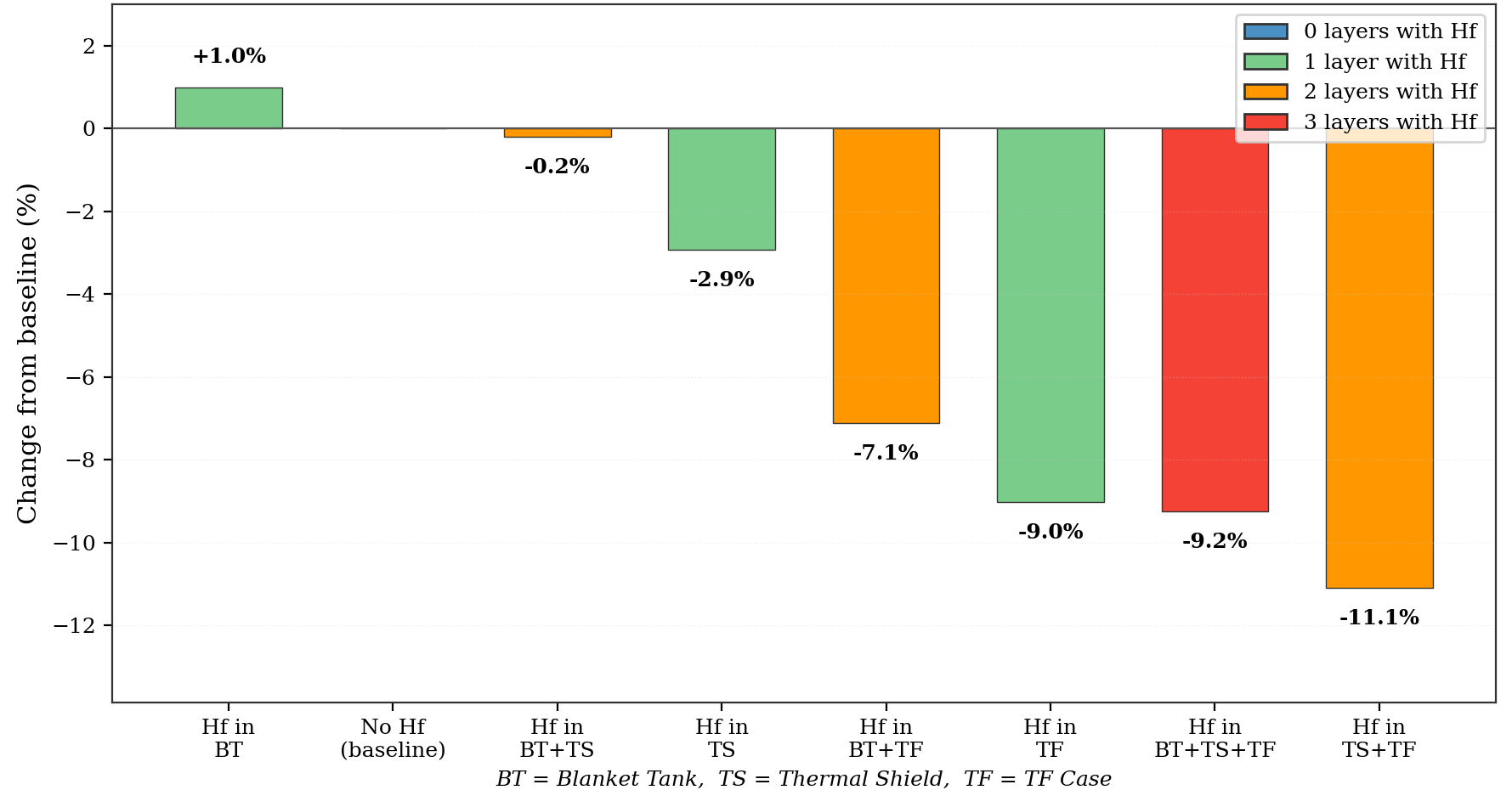}
\caption{Percentage change in TF conductor fast-neutron flux relative to the undoped baseline for all eight Hf-doping configurations (10\,wt\% Hf replacing Fe in each doped layer). Hf in the TF case alone reduces the conductor flux by ${\sim}9\,\%$; combined thermal-shield and TF-case doping yields ${\sim}11\,\%$ reduction.}
\label{fig:hf_doping_sweep}
\end{figure}

\begin{figure}[!t]
\centering
\includegraphics[width=\columnwidth]{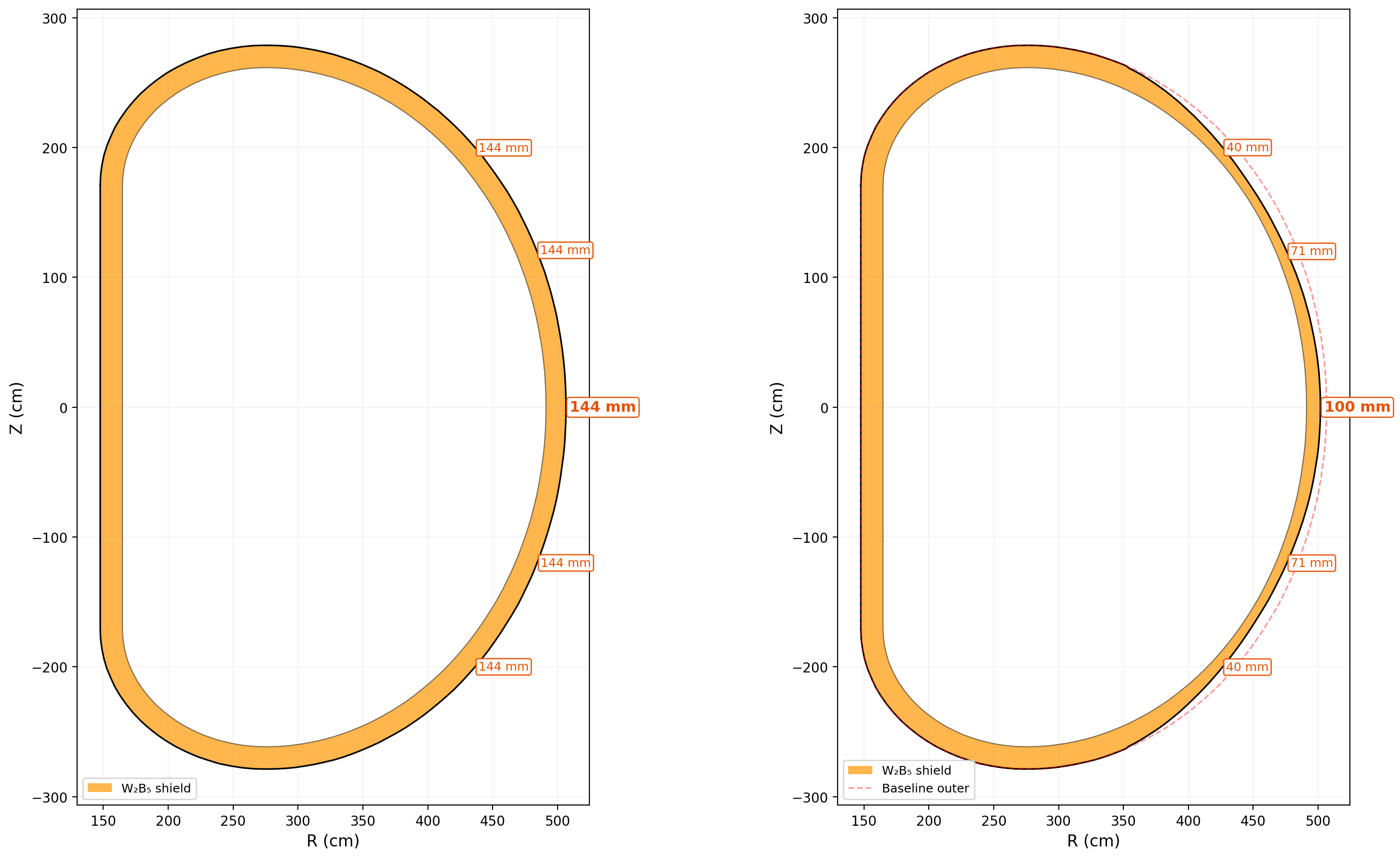}
\caption{Low-field-side W$_2$B$_5$ shield-thinning optimization relative to the poloidally uniform baseline shield build. The low-field-side shield is reduced away from the midplane until the total TF nuclear heating reaches the retained \SI{25}{kW} limit at \SI{300}{MW} $P_f$.}
\label{fig:neutronics_lfs_shield_thinning}
\end{figure}

Another optimization explored here is thinning the low-field-side W$_2$B$_5$ shield relative to the baseline geometry, in which the shield layers are taken to be poloidally uniform. The motivation comes directly from Section~\ref{sec:direct_capex}, where the W$_2$B$_5$ shield is one of the dominant cost rows before this refinement. Using an iterative OpenMC workflow, the low-field-side W$_2$B$_5$ layer was progressively reduced away from the midplane until the total allowed TF-coil nuclear-heating limit reached \SI{25}{kW} at \SI{300}{MW} of fusion power. The resulting slimmed-down geometry is shown in Figure~\ref{fig:neutronics_lfs_shield_thinning}. Because the high-field-side inner-leg shielding is unchanged, this modification affects the cryogenic load but not the limiting TF lifetime. The corresponding W$_2$B$_5$ shield mass falls from \SI{741}{t} to \SI{563}{t}, reducing the projected fabricated shield cost from about \SI{92}{MUSD} to about \SI{70}{MUSD}.

\subsection{Tritium Breeding}

\begin{figure}[!b]
\centering
\includegraphics[width=\columnwidth]{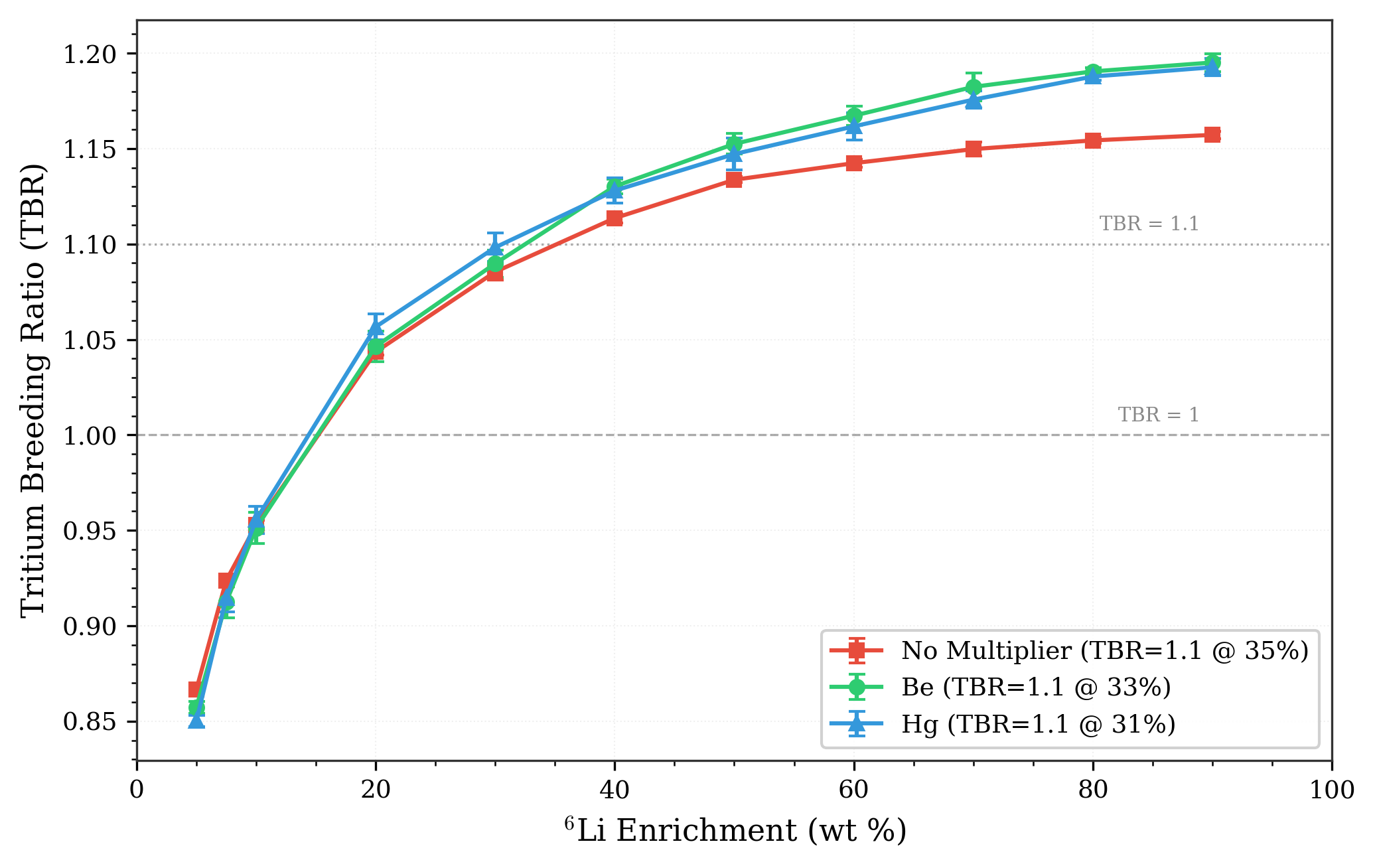}
\caption{Tritium breeding ratio as a function of $^6$Li enrichment in FLiBe for the present Yinsen baseline, showing that the self-sufficiency target can be reached without extreme enrichment or a dedicated multiplier layer.}
\label{fig:neutronics_tbr_enrichment}
\end{figure}

The $^6$Li enrichment sweep shows that Yinsen can close its tritium balance with only modest isotopic enrichment. A key reason is the use of V-4Cr-4Ti for the in-vessel structural material: vanadium alloys are comparatively transparent to fusion neutrons and impose much less parasitic absorption on the breeding blanket than high-nickel candidate structural alloys such as Inconel~\cite{PSFC20JA056}. In the present baseline, that transparency is sufficient to achieve approximately $TBR \approx 1.1$ with only about $30\%$ $^6$Li enrichment and no dedicated neutron multiplier layer, as summarized in Figure~\ref{fig:neutronics_tbr_enrichment}. Both Be and Hg were nevertheless considered as possible multiplier materials because of their large fusion-relevant $(n,2n)$ cross sections, and Hg remains of separate interest because the same reaction pathway could in principle support mercury-to-gold transmutation concepts in specialized blankets~\cite{Rutkowski2025Alchemy}. For the present Yinsen mission class, however, the breeding result itself indicates that a distinct multiplier layer is not required.

The enrichment sensitivity therefore defines a practical operating window that satisfies the Yinsen self-sufficiency target without requiring extreme isotopic enrichment or a dedicated multiplier region.

\subsection{Activation Analysis}

Activation behavior is what ultimately determines how quickly the machine can be approached after shutdown, how maintenance must be staged, and which components can plausibly be recycled or require long-term controlled handling. Both the activation timeseries and the shutdown-dose results in this section share their irradiation basis with the vacuum-vessel DPA analysis above: they are evaluated at the end-of-life fluence corresponding to the 35\,DPA vessel limit, i.e.\ approximately $1040$\,MW$\cdot$yr of integrated full-power operation ($20$ calendar years at the $130$\,MW, $40\%$ utilization baseline). The reported activated inventories therefore represent the conservative end-of-life bound for the power core, and any earlier point in the vessel life would yield correspondingly lower decay heat and dose. In the present Yinsen baseline, the activated structures remain close to their full-power source strength through roughly the first 24 hours after shutdown, with the first wall sustaining dose rates of order 100\,Sv/h. A more rapid cooldown phase begins only after the shorter-lived activation products start to decay, and by about 50 years most major structural components fall below the 10\,$\mu$Sv/h recycling threshold. For comparison, higher-power-density compact devices such as ARC have reported century-scale or longer cooldown times to approach the same hands-on recycling criterion for structural materials \cite{Bocci2020ARCActivation}. These timescales define the practical maintenance-access windows and waste-handling pathways for the power core.

\begin{figure}[!t]
\centering
\includegraphics[width=\columnwidth]{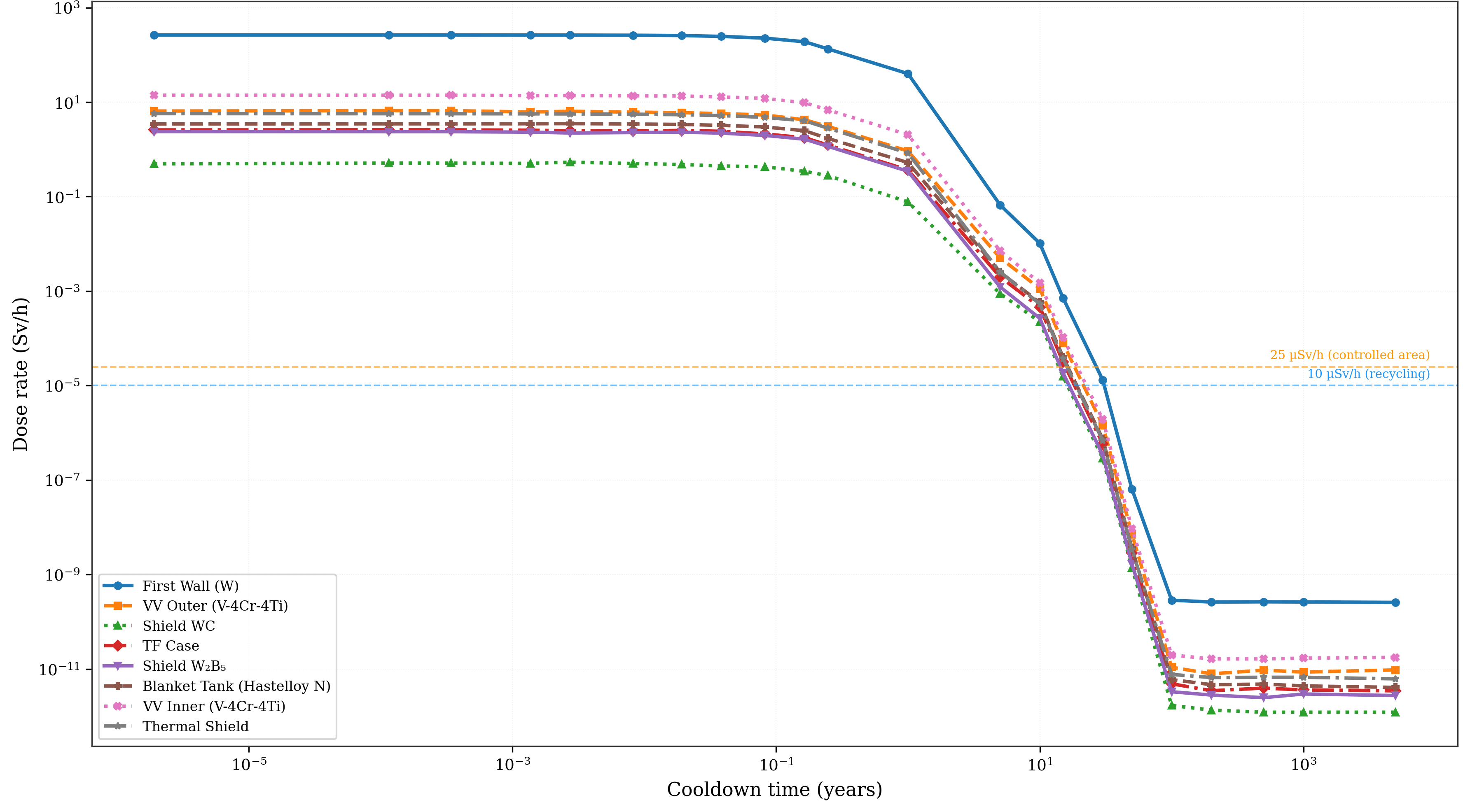}
\caption{Activation decay behavior for principal components after shutdown, used to define maintenance access windows and component handling pathways over post-operation cooldown time.}
\label{fig:neutronics_activation}
\end{figure}

The shutdown dose maps then show how that activation inventory translates into the external maintenance environment. Using ICRP-116 conversion coefficients~\cite{ICRP116}, the results track how the effective dose field around the cryostat and port regions relaxes in both space and time as short-lived isotopes decay. This is the systems-level information needed to define controlled-access zones, outage timing, and maintenance sequencing.

\begin{figure}[!t]
\centering
\includegraphics[width=\columnwidth]{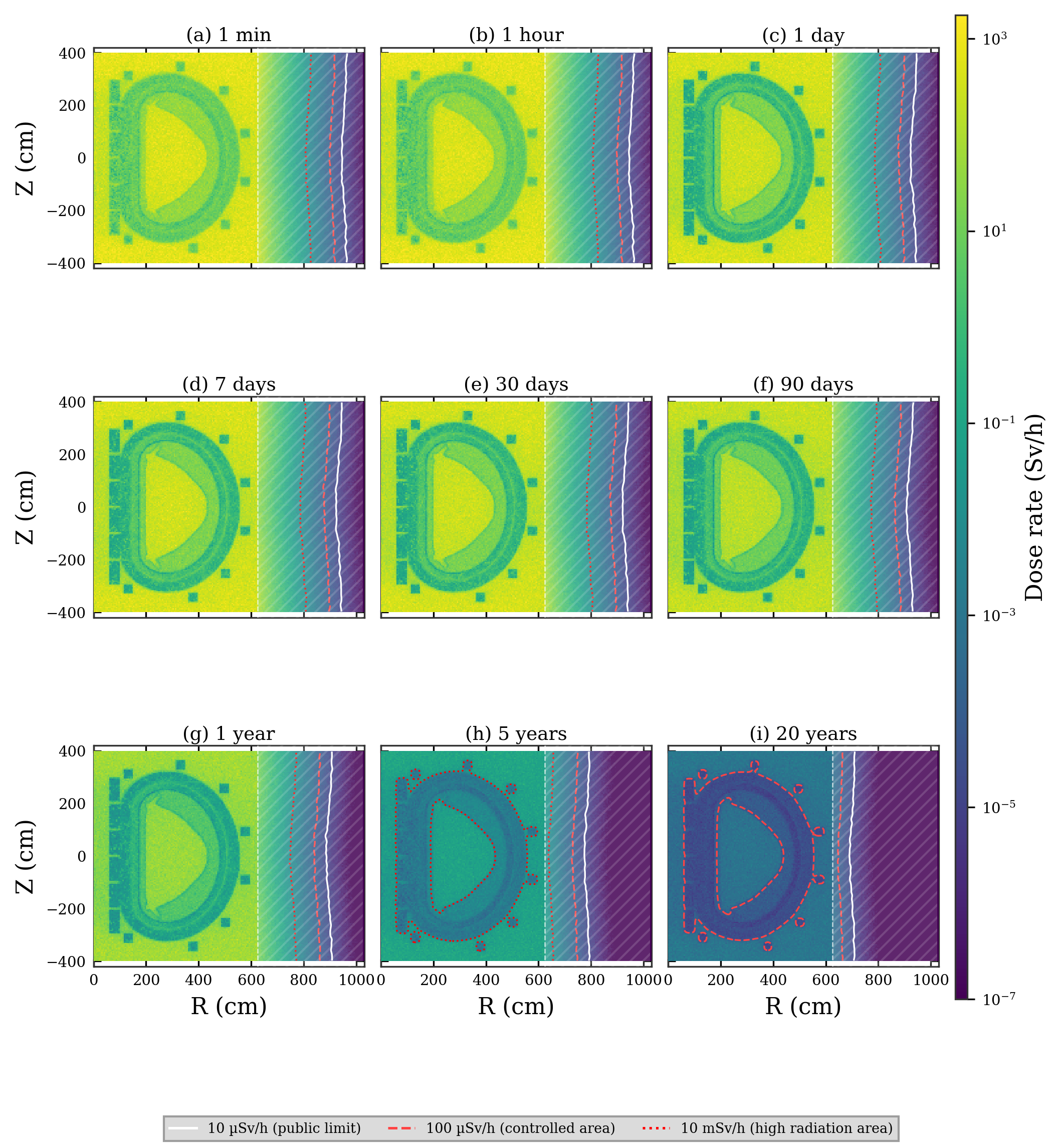}
\caption{Dose-rate maps after shutdown at multiple post-operation times (1\,min to 1000\,years), showing the temporal and spatial relaxation of activation-driven dose around the cryostat and port regions. Contour lines mark the 10\,\textmu Sv/h public limit (white solid), 100\,\textmu Sv/h controlled-area threshold (red dashed), and 10\,mSv/h high-radiation boundary (red dotted). The colorbar floor is set to the natural background dose rate ($\sim$0.1\,\textmu Sv/h).}
\label{fig:neutronics_external_dose}
\end{figure}

\begin{figure*}[!t]
\centering
\includegraphics[width=0.48\textwidth]{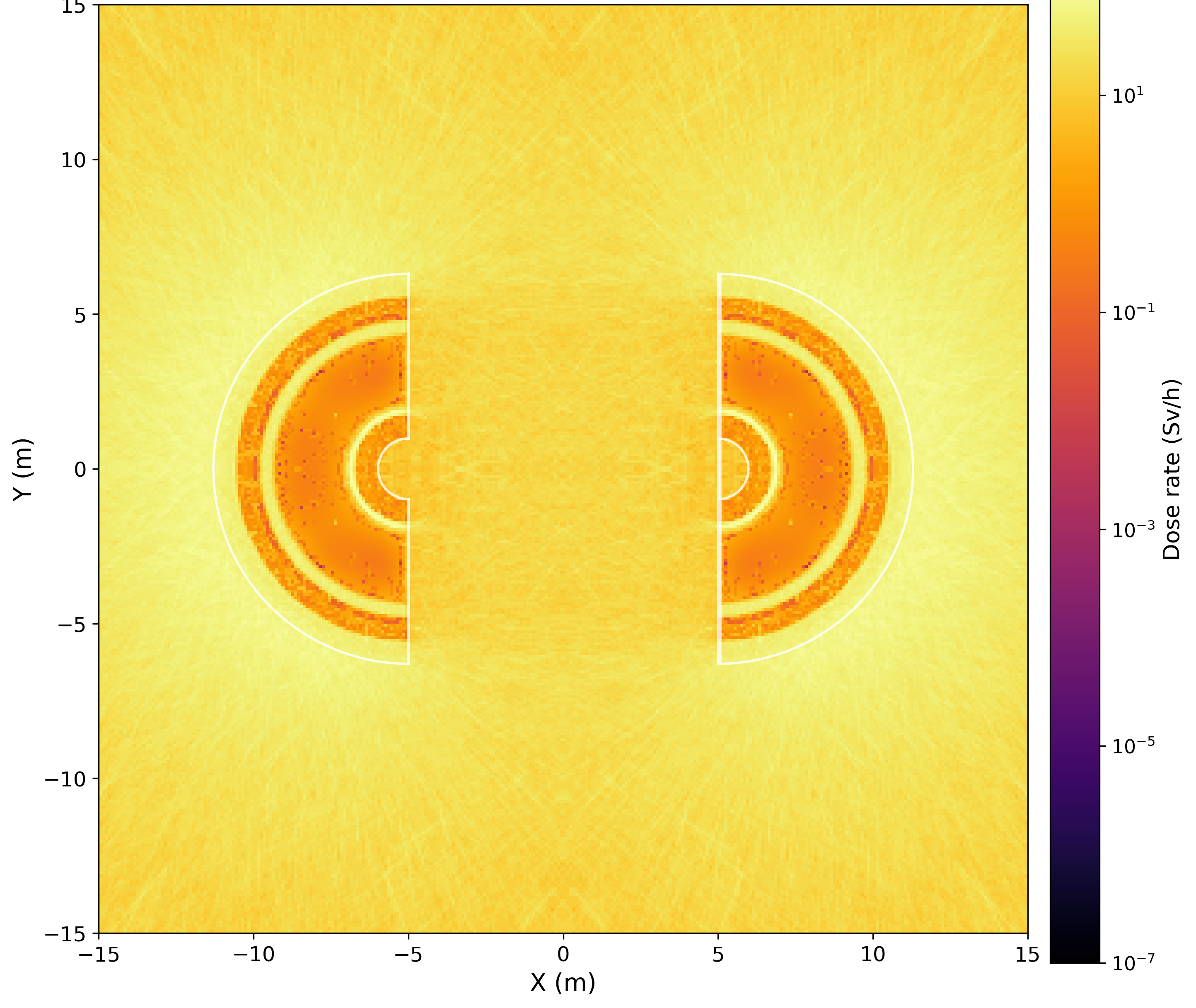}%
\hfill
\includegraphics[width=0.48\textwidth]{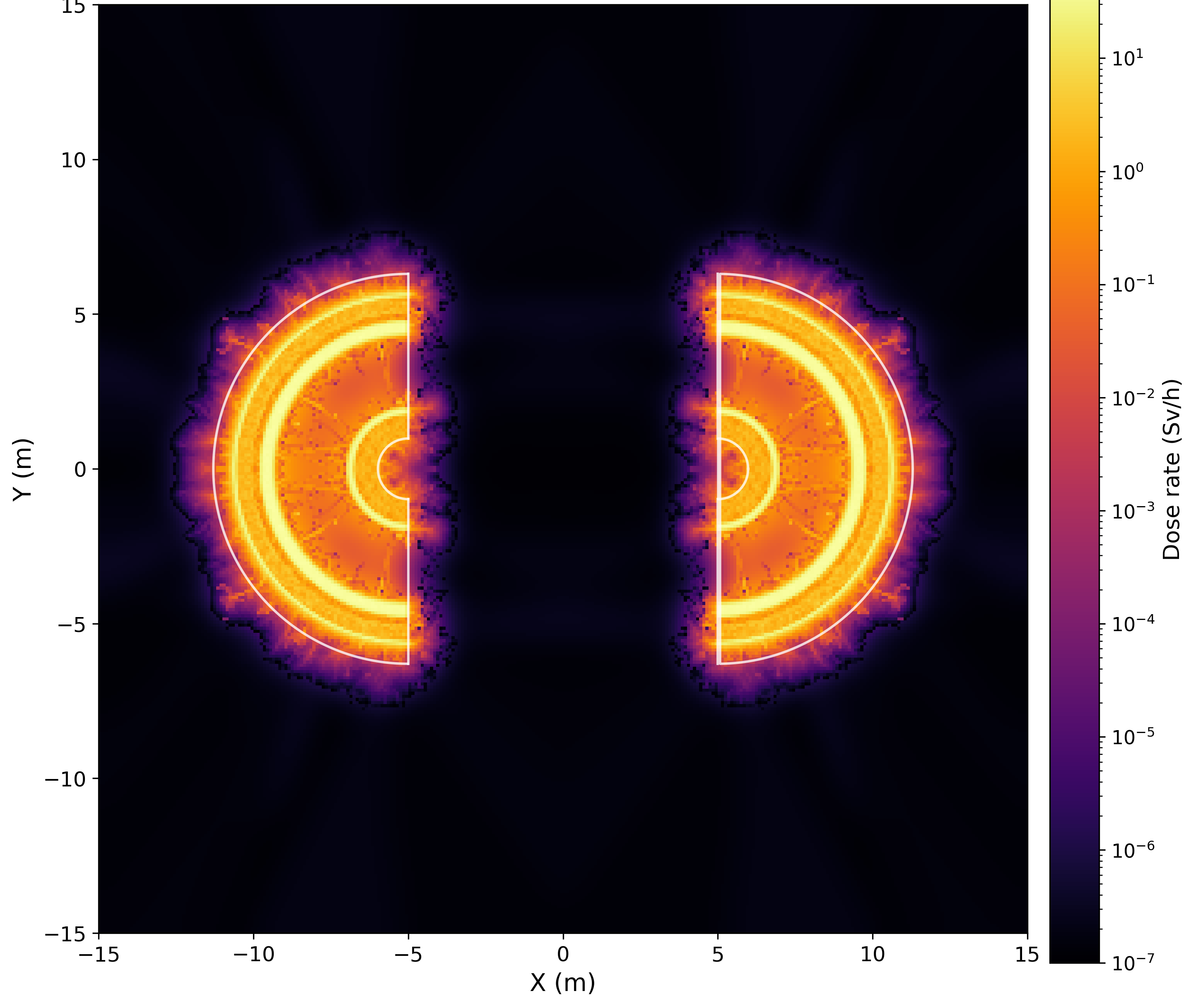}
\caption{Top-down shutdown dose-rate maps in the maintenance hall with the tokamak split into two halves separated by 10\,m (end-of-life irradiation, $\sim$1040\,MW$\cdot$yr at the 35\,DPA vessel limit, evaluated at 1~day cooldown). The dotted yellow iso-line marks the 10\,mSv/h high-radiation boundary. Left: air-filled hall---dose exceeds 10\,mSv/h throughout the room, with a peak of 151\,Sv/h. Right: water-flooded hall---the 10\,mSv/h contour collapses to the immediate vicinity of the tokamak halves, demonstrating that water shielding reduces the ambient dose rate by several orders of magnitude.}
\label{fig:maintenance_dose}
\end{figure*}

The same analysis also makes clear that post-shutdown maintenance conditions depend strongly on the surrounding hall environment. With the tokamak halves separated by 10\,m in air, essentially the entire maintenance volume remains in a high-radiation state and lethal dose rates persist several metres from the machine. Flooding the hall with water suppresses the far-field dose by several orders of magnitude, shrinking the 10\,mSv/h contour to the immediate vicinity of the machine halves and bringing the inter-half region below controlled-area thresholds. That result is what motivates the water-flooding maintenance strategy adopted for Yinsen in the event that major maintenance or vacuum-vessel replacement is required beyond what can be accomplished through the ports. It is proposed that the ICRH system can be used for drying out the vessel post submersion, in a procedure similar to baking and wall conditioning, although further analysis is needed on the details here.

\begin{figure}[!t]
\centering
\includegraphics[width=\columnwidth]{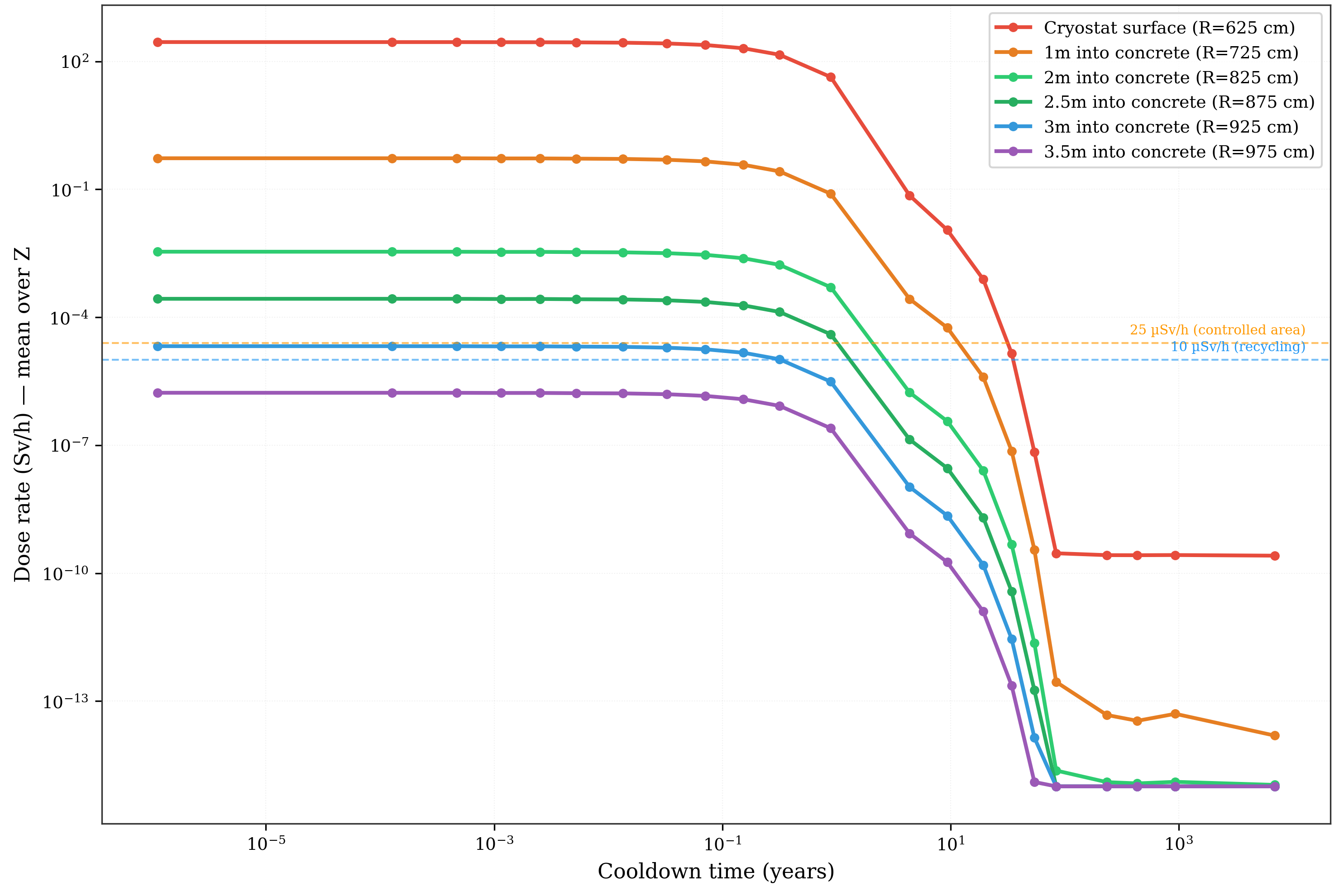}
\caption{Shutdown dose rate versus cooldown time at successive radial locations through the 4\,m ordinary-concrete biological shield wrapping the cryostat. Reference contours mark the 10\,$\mu$Sv/h public-recycling and 25\,$\mu$Sv/h controlled-area limits.}
\label{fig:neutronics_concrete_bioshield}
\end{figure}

The reference design includes a passive concrete biological shield wrapped around the cryostat. The neutronics model therefore includes a cylindrical shell of ordinary Portland-cement concrete (NBS~04 composition: H/C/O/Na/Mg/Al/Si/K/Ca/Fe at $2.3$\,g/cm$^3$) inserted as a post-DAGMC layer in the shutdown-dose transport step, with thickness and density set by configuration. The reference design uses a 4\,m thick shell extending from the cryostat outer radius ($R\!\approx\!6.26$\,m) to $R\!\approx\!10.26$\,m, and the shutdown-dose tally mesh is placed immediately outside the bioshield outer surface so that the reported far-field dose rates already include the concrete attenuation. Figure~\ref{fig:neutronics_concrete_bioshield} shows how the dose rate evolves with cooldown time at successive radial locations across the shield. Across all cooldown times, 3.5\,m of ordinary concrete reduces the post-cryostat dose by several orders of magnitude, dropping below the 25\,$\mu$Sv/h controlled-area threshold and approaching the 10\,$\mu$Sv/h public-recycling limit at the bioshield outer surface.

\subsection{Proliferation Resistance Assessment}\label{sec:proliferation}

A sometimes-raised concern for molten-salt blanket fusion reactors is the potential for clandestine breeding of weapons-usable fissile material by dissolving fertile isotopes in the blanket salt. In the uranium pathway, $^{238}$U captures a neutron and decays via $^{239}$Np to $^{239}$Pu; in the thorium pathway, $^{232}$Th captures a neutron and decays via $^{233}$Pa to $^{233}$U. Both $^{239}$Pu and $^{233}$U are weapons-usable, with an IAEA significant quantity (SQ) of 8\,kg each~\cite{IAEA2001}.

To quantify this risk for Yinsen, a parametric neutronics study was performed with OpenMC in which UF$_4$ or ThF$_4$ was dissolved in the FLiBe blanket salt at weight fractions from 0.5\% to 10\%. Two analyses were conducted: (i)~a transport-only sweep to assess the impact on TBR and blanket fission rate across the full loading range, and (ii)~coupled depletion calculations at selected compositions (2\% and 5\,wt\% for each fertile salt) to track fissile-material buildup over a 75\,MW$\cdot$yr irradiation campaign at 185\,MW fusion power.

\begin{figure}[!t]
\centering
\includegraphics[width=\columnwidth]{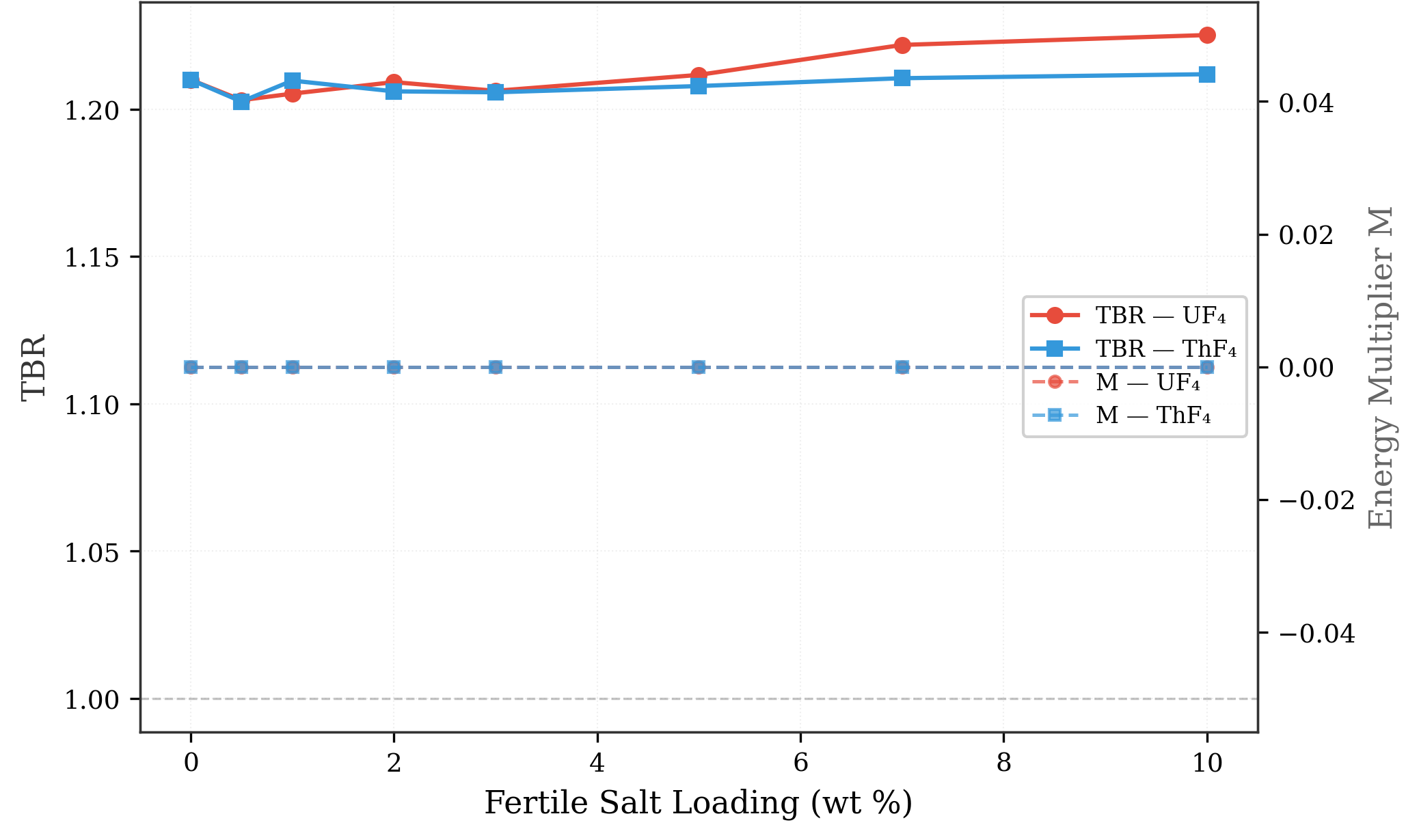}
\caption{Combined TBR and blanket energy multiplier as a function of fertile salt loading, illustrating the tradeoff between tritium breeding and thermal output from fission energy release.}
\label{fig:fertile_tradeoff}
\end{figure}

The first result is that fertile-salt loading has very little leverage on the tritium economy of the blanket. Even with up to 10\,wt\% UF$_4$ or ThF$_4$ added to the FLiBe, the tritium breeding ratio remains essentially unchanged, while the blanket energy multiplier increases only modestly as fission reactions begin to contribute. At 10\,wt\% UF$_4$, TBR even rises slightly (by ${\sim}2\,\%$) because $^{238}$U fast fission ($E_{\mathrm{threshold}}\sim 1$\,MeV) releases additional neutrons that partially compensate for parasitic absorption. ThF$_4$ shows the same qualitative behavior but with a weaker effect, keeping TBR within $\pm$1\,\% of the undoped baseline across the full loading range, as shown in Figure~\ref{fig:fertile_tradeoff}.

\begin{figure}[!t]
\centering
\includegraphics[width=\columnwidth]{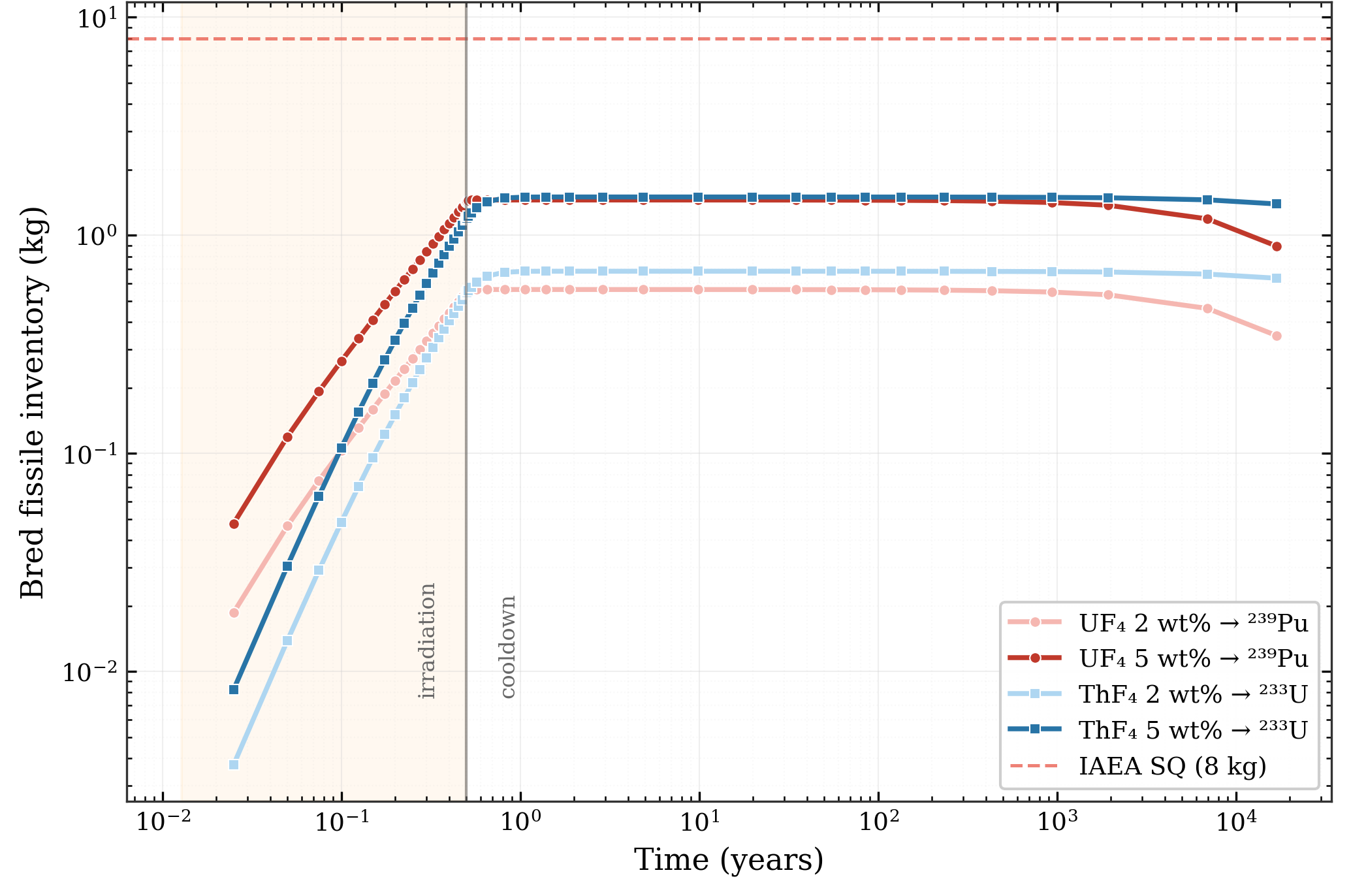}
\caption{Bred fissile-mass accumulation versus time for fertile-salt-doped FLiBe blankets: $^{239}$Pu from UF$_4$-doped salt and $^{233}$U from ThF$_4$-doped salt, each at 2\,wt\% and 5\,wt\% loading. The dashed red line marks the 8\,kg IAEA significant quantity. Even at 5\,wt\% loading, the bred inventory after ${\sim}0.41$\,FPY reaches only ${\sim}1.5$\,kg in either case, far below the SQ.}
\label{fig:fertile_buildup}
\end{figure}

\begin{figure}[!t]
\centering
\includegraphics[width=\columnwidth]{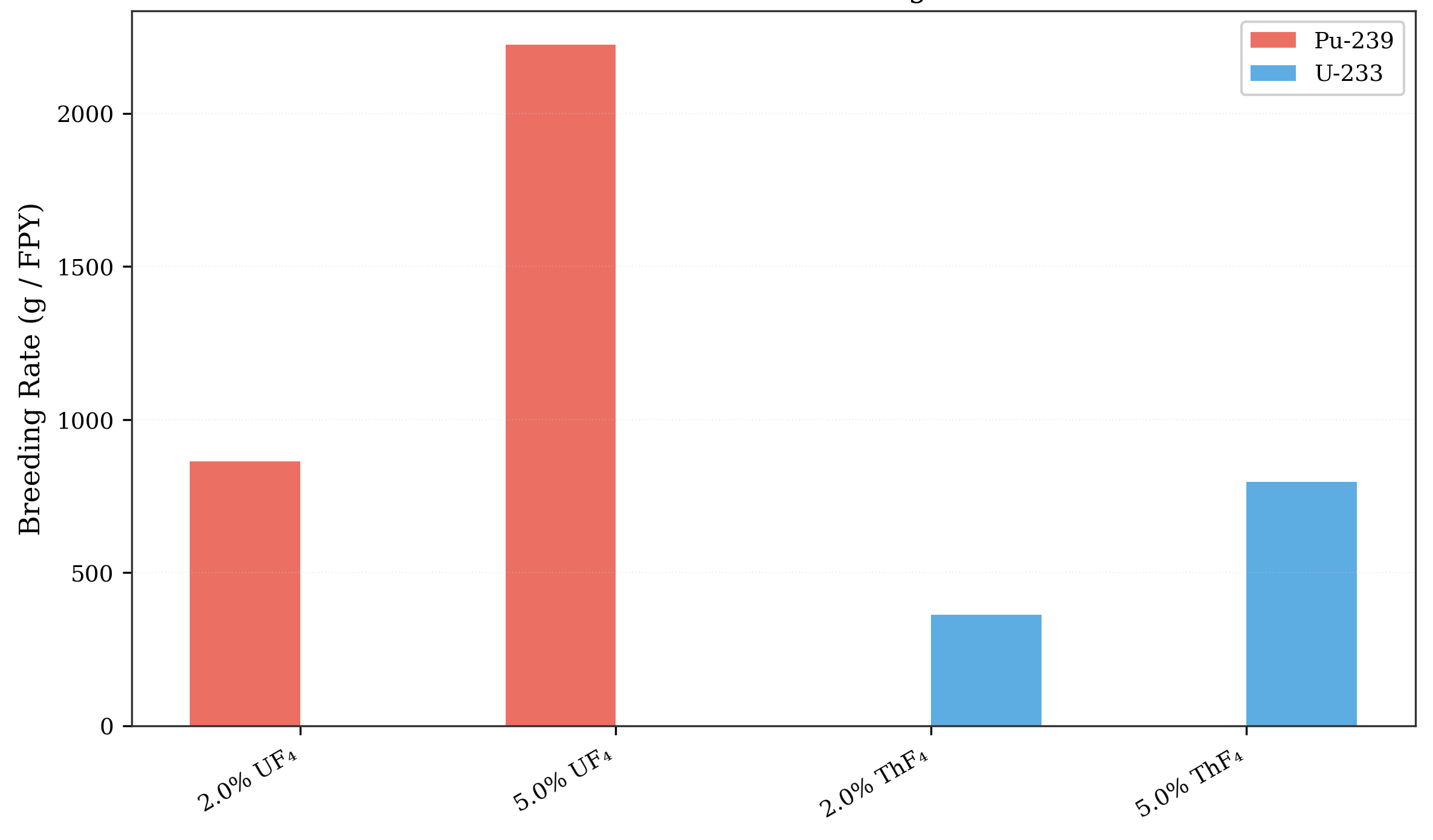}
\caption{Fissile material breeding rates in grams per full-power-year for each fertile salt composition, showing that even the most aggressive loading produces only $\sim$1.4--3.7\,kg of fissile material per FPY.}
\label{fig:fertile_rate}
\end{figure}

What ultimately matters for safeguards, however, is not the change in TBR but the absolute fissile inventory that can be accumulated in the salt. Over the 75\,MW$\cdot$yr depletion campaign (approximately 0.41\,FPY at 185\,MW), the peak bred inventory remains small in every case studied. Table~\ref{tab:fertile_breeding} and Figure~\ref{fig:fertile_buildup} show that even at the most aggressive 5\,wt\% loading, the accumulated $^{239}$Pu is only 1.45\,kg and the accumulated $^{233}$U is only 1.50\,kg, both well below the 8\,kg IAEA significant quantity. The corresponding breeding rates, summarized in Figure~\ref{fig:fertile_rate}, remain of order only 1.4--3.7\,kg per full-power-year. Reaching one significant quantity of either isotope would therefore require approximately 2.2\,FPY of continuous operation at 5\,wt\% loading, a scenario that would be operationally conspicuous and readily detectable through standard safeguards measures.

\begin{table}[H]
\centering
\scriptsize
\caption{Peak fissile material inventory after 75\,MW$\cdot$yr irradiation at 185\,MW, for selected fertile salt compositions dissolved in the FLiBe blanket.}
\label{tab:fertile_breeding}
\renewcommand{\arraystretch}{1.3}
\begin{tabular}{lccc}
\toprule
\textit{Additive} & \textit{wt\%} & \textit{Product} & \textit{Mass (kg)} \\
\midrule
UF$_4$  & 2.0 & $^{239}$Pu & 0.56 \\
UF$_4$  & 5.0 & $^{239}$Pu & 1.45 \\
ThF$_4$ & 2.0 & $^{233}$U  & 0.68 \\
ThF$_4$ & 5.0 & $^{233}$U  & 1.50 \\
\midrule
\multicolumn{3}{l}{IAEA significant quantity} & 8.0 \\
\bottomrule
\end{tabular}
\end{table}

Small bred inventories are only part of the barrier. Even granting the inventories above, converting dissolved actinide fluorides in FLiBe into separated weapons-usable metal is not a benchtop operation: it requires the same class of pyrochemical reprocessing line developed for the ORNL Molten Salt Reactor Experiment and Molten Salt Breeder Reactor programs~\cite{McNeeseFerris1971}. Two routes are conceivable. The first is fluoride volatility, in which elemental F$_2$ is sparged through the molten salt to convert UF$_4$ into volatile UF$_6$ for off-gas capture; this route works tolerably for uranium but is notoriously poor for plutonium because PuF$_6$ is thermally unstable and disproportionates back to non-volatile PuF$_4$ in the same temperature window, so the MSRE-era experience is that fluoride volatility cannot cleanly recover plutonium from a fluoride melt~\cite{SoodPatil1996}. The second is reductive extraction into liquid bismuth using a Li- or Th-metal reductant in a multistage countercurrent contactor, the route adopted in the MSBR chemical-plant flowsheet. That process demands tight control of the UF$_3$/UF$_4$ redox ratio, molybdenum-bonded contactor internals, freeze-valve salt isolation, and a dedicated HF/F$_2$ off-gas train. Either pathway is, in practice, an industrial-scale chemical facility rather than anything that could plausibly be concealed in normal plant operations.

The chemical barrier is compounded by the radiological and toxicological state of the salt itself. Operating FLiBe carries a continuous inventory of tritium together with short-lived neutron-activation products including $^{20}$F ($T_{1/2}\!\approx\!11$\,s, 5.4\,MeV $\beta/\gamma$), $^{16}$N, and $^{19}$O, and longer-lived $^{14}$C, all generated in situ from $(n,\gamma)$ and $(n,\alpha)$ reactions on the carrier salt~\cite{FlibeActivation2019}. In any fertile-doped scenario the salt would additionally carry the full fission-product spectrum from in-blanket fission of the bred material. Contact dose rates near freshly drained salt are accordingly in the Sv/h range, and even after weeks of cooldown the inventory remains far above hands-on limits. Any extraction campaign would therefore have to be performed entirely by remote manipulation inside a gamma-shielded hot cell with hermetic containment for both tritium and beryllium aerosols. Taken together, the low breeding rates, the need for sustained undeclared fertile-salt loading, and the requirement for an MSR-scale remote pyrochemical processing facility indicate that the proliferation pathway for Yinsen is minimal and manageable within normal safeguards and institutional-oversight frameworks.

\section{Energy Storage and Pulsed Power Operations}\label{sec:pulse}

Yinsen is intentionally operated in an inductively driven long pulsed regime, rather than attempting to design and operate a first-of-a-kind facility in a fully steady-state scenario where current drive efficiency and stability challenges remain significant. The primary consequence of inductive pulses is the need for periodic charging and discharging of the central solenoid and repeated plasma current ramp-up/ramp-down cycles. Additionally, the pulsed fusion power output, even with long flat-top durations, results in a pulsed generated electric power output.

For marine propulsion, pulsed export does not need to be perfectly flattened at the customer interface because vessel inertia provides a natural low-pass filter on shaft power. The cyclic stresses that this operating mode imposes on shafts, gearboxes, propulsors, and associated powertrain components still require dedicated follow-on study. By contrast, off-grid applications that cannot tolerate a pulsed output profile would require additional energy capacitance to buffer the fusion pulses before delivery to the load \cite{MANTA2024}.

Time-resolved plant power-flow analysis was therefore developed around a 34 kV AC backbone bus connecting the turbine-generator output, energy storage, magnet power supplies, RF systems, cryogenic plant, tritium fuel cycle facility, and customer load. For the marine mission case, the customer load is represented by a 25--70 MWe permanent-magnet propulsion motor. In this architecture, short high-power intervals during plasma initiation, ramp, and inter-pulse recharge must be absorbed without violating converter, feeder, protection, or bus-voltage limits.

To make that architecture physically realizable without resorting to very large line-frequency transformers and one-off converter stacks, the detailed concept uses solid-state transformers (SSTs) between the 34 kV medium-voltage backbone and a shared 1.5 kV DC magnet bus. That choice is driven less by novelty than by packaging and controllability. Conventional MV transformers remain attractive for fixed industrial duty, but they are bulky and inflexible once a particular voltage ratio and feeder topology are frozen. By contrast, SST architectures now being pushed by data-center distribution, microgrids, and naval power systems offer precisely the features needed here: modular galvanic isolation, controllable bidirectional power flow, and the ability to interface a common MV AC source with a tightly regulated low-voltage DC bus that can also host energy storage. In the Yinsen plant this DC node is where the battery and capacitor systems discussed later in the section are tied in, so the storage plant can absorb the fast CS/PF charge--discharge pulses locally rather than forcing those transients onto the ship service bus or turbine-generator terminals.

\begin{figure}[htbp]
    \centering
    \includegraphics[width=\linewidth]{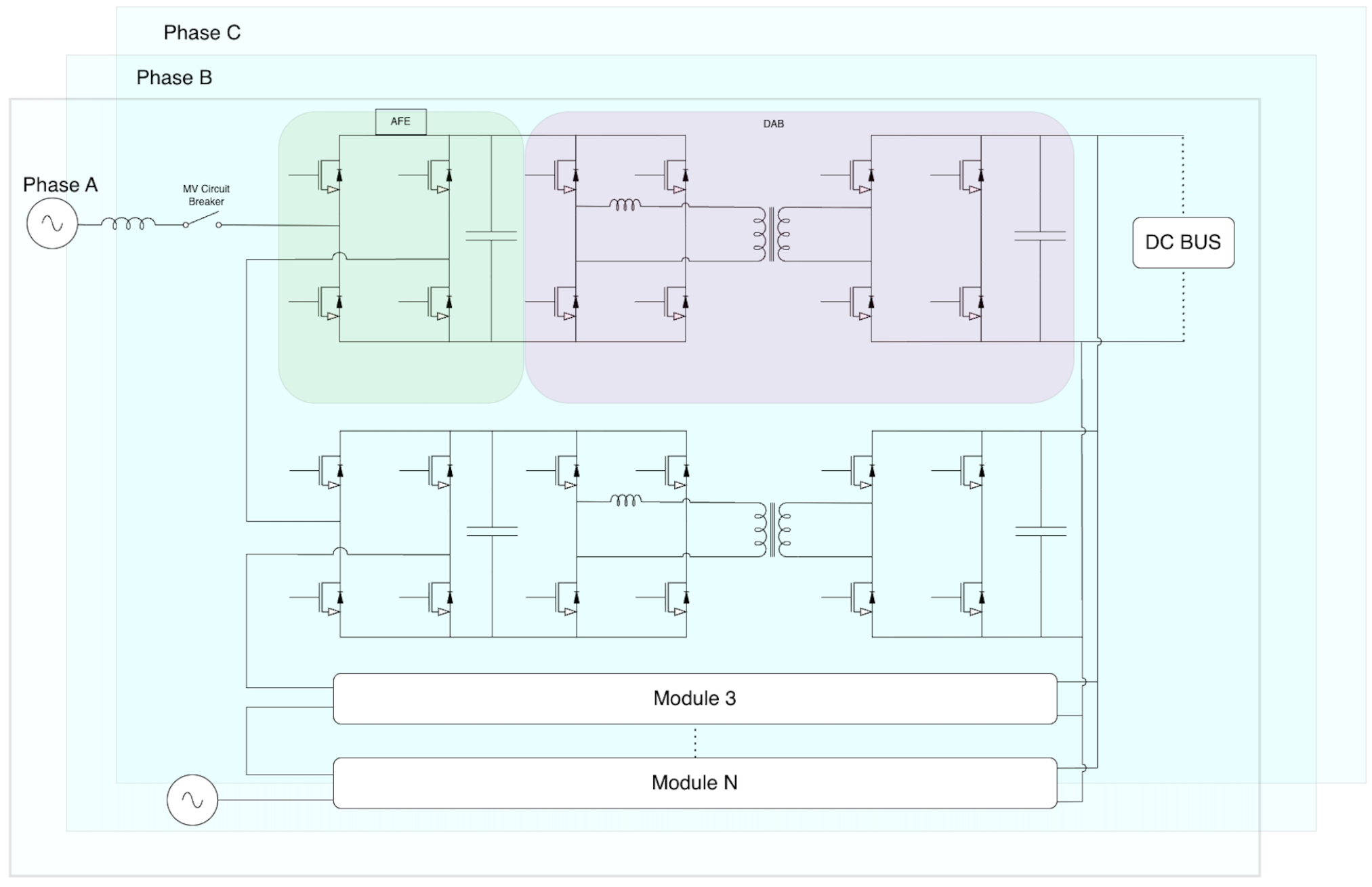}
    \caption{Conceptual solid-state-transformer interface between the 34 kV three-phase backbone and the shared 1500 V DC magnet bus. A repeated 3.3 kV SiC conversion cell is used to build the medium-voltage stack, provide galvanic isolation through a high-frequency link, and regulate the common DC bus to which storage and magnet supplies are attached.}
    \label{fig:sst_mv_bus}
\end{figure}

\begin{figure}[htbp]
    \centering
    \includegraphics[width=\linewidth]{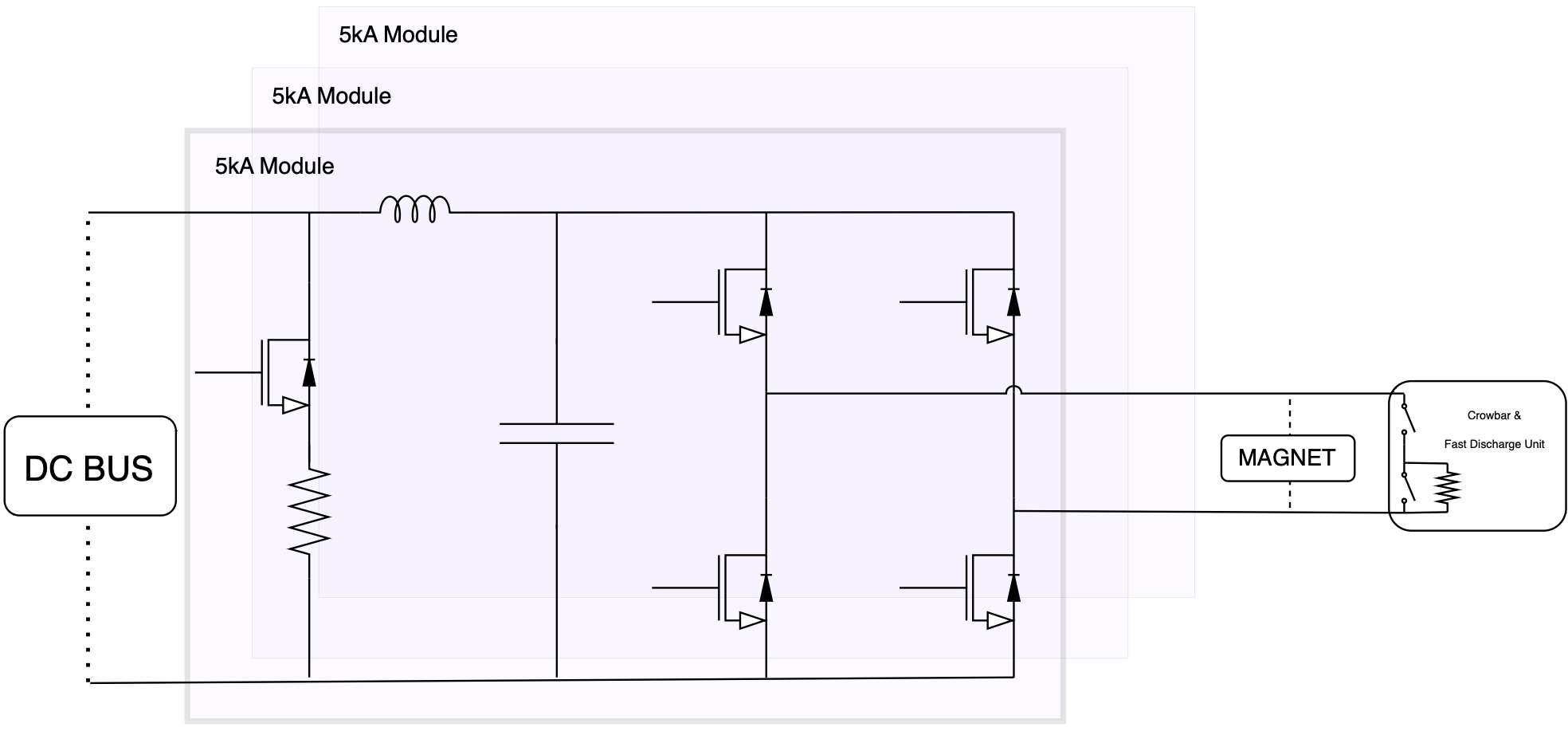}
    \caption{Modular magnet power-supply unit tied to the shared 1500 V DC bus. Each 5 kA module combines a regulated front end, DC-link energy buffering, an H-bridge output stage, and a dedicated crowbar/fast-discharge branch so higher-current feeders can be built by straightforward paralleling of identical modules.}
    \label{fig:sst_magnet_module}
\end{figure}

The proposed SST implementation is built around a repeated 3.3 kV SiC conversion cell. On a 34 kV line-to-line backbone, the corresponding line-to-neutral peak voltage is approximately 27.8 kV, so a 17-cell series stack limits the average blocked voltage per cell to about 1.6 kV and leaves substantial margin below the device class rating for fault transients, unequal sharing, and insulation design reserve. The same device family can then be reused on the low-voltage side, but now arranged in parallel to build current capacity on the 1.5kV DC bus. That reuse matters because it keeps the architecture manufacturable: the same gate drives, thermal hardware, qualification basis, and spare parts pool can serve both the medium-voltage SST front end and the coil power-supply back end. Figure \ref{fig:sst_mv_bus} shows the three-phase SST concept used to step the 34 kV backbone down to the shared DC bus, while Figure \ref{fig:sst_magnet_module} shows the modular magnet feeder built from repeated 5 kA units.

Downstream of the shared DC bus, each magnet supply is treated as its own current-programmed conversion chain. Each CS and PF coil is assigned an H-bridge supply with its own output filter and chopper stage, together with a dedicated quench-protection and crowbar circuit. The H-bridge building block is taken to be a 5 kA, 1.5 kV module, corresponding to 7.5 MW per module at full DC utilization; multiple units are then paralleled to achieve the required current capacity for each individual coil circuit. For the CS and PF systems, this low-voltage high-current architecture couples naturally to the hybrid storage plant described earlier. The relevant pulse-energy cases in the present study are in the 1-3 GJ range, so the storage system is sized to support the CS and PF feeders directly at the DC bus, where battery energy capacity and capacitor pulse-power capability can be blended without repeatedly cycling the full plant AC distribution system. The requirement for stored energy is derived from integrated time-dependent Tokamaker simulations of coil currents and voltages discussed later in this section. Additional solid-state switch network units (SNUs) are retained on the CS feeders to provide the higher transient-voltage capability needed for plasma breakdown and other fast waveform excursions. In the present concept these SNUs are implemented with back-to-back IGCT-based switching elements so that large bipolar voltage transients can be imposed on the CS circuit for startup assistance without relying on slow mechanical switching or overstressing the steady-state H-bridge stack.

The TF coil power-supply requirement is different because its nominal voltage is set primarily by the desired charge and discharge schedule rather than by fast shaping transients. Here the design target is a 15 minute ramp-up and 15 minute ramp-down, so the steady TF converter voltage is selected from that inductive time-base and then supplemented with a dedicated fast-discharge path. Using the present TF design point of 43.6 kA terminal current, the nominal charging voltage for a 15 minute ramp is about 872 V. That operating point fits naturally on the 1500 V DC bus: a TF feeder assembled from nine parallel 5 kA modules can supply the required 43.6 kA while still retaining current margin and modular redundancy. Unlike the CS and PF feeders, the TF supply does not require large local energy storage because the toroidal field is not repeatedly pulsed on the same timescale as plasma initiation and shaping. Instead, the TF converter is sized primarily around current capability and controlled charge up duration.

Each TF protection branch uses a hybrid DC interruption path made up of a normally closed mechanical bypass switch operating in parallel with a solid-state breaker branch. A dedicated dump resistor is connected so that, once the coil group is disconnected from its main energization path, its stored magnetic energy is forced into a controlled resistive discharge rather than an uncontrolled arc path. A pyro-breaker is retained as a backup sacrificial interrupter, and an earthing resistor is included to control floating-potential behavior during abnormal events. In normal operation the mechanical bypass carries the current with minimal loss and the static breaker remains blocked. When a fast discharge is commanded, the solid-state branch is first turned on so current can be commutated away from the opening mechanical path with minimal arcing. After the bypass switch has opened fully, the static branch is then turned off on a controlled timescale, forcing current into the dump resistor and establishing the intended discharge trajectory. This sequence preserves low-loss steady operation while still providing a fast and well-defined interruption mechanism for quench, fault, or emergency ramp-down scenarios.

This protection architecture is particularly important for the TF system because the full string stores 27 GJ. In the present partitioning, the six FDUs each protect a three-coil group, so the nominal magnetic energy handled by one protection branch is reduced to 4.5 GJ before additional current-sharing and discharge-shaping margins are considered. That is still a large energy inventory, but it is much more tractable than forcing a single interrupter to control the full toroidal-field system. The SST-fed DC architecture and the grouped FDU concept therefore work together: the former provides a compact, modular means of charging and regulating the magnets from the plant backbone, while the latter ensures that quench protection and emergency discharge remain localized, well defined, and compatible with practical maintenance boundaries.

The storage split and sizing are computed with a time-dependent plant model that evaluates battery-only, capacitor-only, and hybrid storage on a common lifetime-equivalent basis rather than fixed, arbitrary DoD or C-rate assumptions.

In this section, pulse energy is defined from the CS/PF magnetic inventory:
\begin{equation}
E_{\mathrm{pulse}} \equiv E_{\mathrm{CS,0}} + E_{\mathrm{PF,0}},
\end{equation}
where $E_{\mathrm{CS,0}}$ and $E_{\mathrm{PF,0}}$ are the stored CS and PF energies at the start of a pulse. The phase-wise exchanged energies are then written as
\begin{equation}
\begin{aligned}
\Delta E_{\mathrm{RU}} &= E_{\mathrm{RU}} - E_{\mathrm{0}},\\
\Delta E_{\mathrm{RD}} &= E_{\mathrm{end}} - E_{\mathrm{FT}},\\
\Delta E_{\mathrm{DW}} &= E_{\mathrm{0}} - E_{\mathrm{end}},
\end{aligned}
\end{equation}
with corresponding average powers
\begin{equation}
\bar{P}_{j} = \frac{\Delta E_{j}}{t_{j}},\qquad
j\in\{\mathrm{RU},\mathrm{RD},\mathrm{DW}\}.
\end{equation}
Here RU is ramp-up, FT denotes the flat-top state at the end of burn, RD is ramp-down (reverse direction of ramp-up transfer), and DW is dwell/recharge. We also note that there exists the opportunity with 4 quadrant power supplies to re-capture the inductive energy transfer during ramp down from the plasma through the magnets and power supplies back to the energy storage. However, for the sake of being conservative and the requirement of reliable remote start-up capability, we do not take advantage of this feature for sizing the storage elements.

For any selected operating scenario, these relations map the pulse program into phase durations, exchanged energies, and average powers for RU, FT, RD, and DW. Lifetime duty is represented by the required pulse count $N_{\mathrm{req}}$, and feasibility is evaluated against technology-specific lifetime models (battery: DoD/C-rate/temperature dependent; capacitor: usable-voltage-swing/temperature dependent).

For batteries, the lifetime model used in the time-dependent plant model is
\begin{equation}
\begin{aligned}
N_{\mathrm{life,bat}} &=
N_{\mathrm{ref,bat}}\,
\Phi_{\mathrm{DoD}}\,
\Phi_{\mathrm{Crate}}\,
\Phi_{T,\mathrm{bat}},\\
\Phi_{\mathrm{DoD}} &=
\left(\frac{\mathrm{DoD}_{\mathrm{ref}}}{\mathrm{DoD}}\right)^{a_{\mathrm{bat}}},\\
\Phi_{\mathrm{Crate}} &=
\left(\frac{C_{\mathrm{rate,ref}}}{C_{\mathrm{rate}}}\right)^{b_{\mathrm{bat}}},\\
\Phi_{T,\mathrm{bat}} &=
\exp\!\left[-k_{T,\mathrm{bat}}(T-T_{\mathrm{ref}})\right].
\end{aligned}
\end{equation}
where DoD is depth of discharge (fraction of nameplate energy used per pulse) and $C_{\mathrm{rate}}$ is the normalized charge/discharge power relative to energy capacity. From a battery-longevity perspective, this favors shallow DoD operation (energy oversizing to remain near mid-state-of-charge), lower $C_{\mathrm{rate}}$ operation (reduced pulse-power stress), and avoidance of elevated temperature operation. Cost uses $\mathcal{C}$ notation below to avoid conflict with $C_{\mathrm{rate}}$. For capacitors, we use
\begin{equation}
\begin{aligned}
N_{\mathrm{life,cap}} &=
N_{\mathrm{ref,cap}}\,
\Phi_{u}\,
\Phi_{T,\mathrm{cap}},\\
\Phi_{u} &=
\left(\frac{u_{\mathrm{ref}}}{u}\right)^{n_{\mathrm{cap}}},\\
\Phi_{T,\mathrm{cap}} &=
\exp\!\left[-k_{T,\mathrm{cap}}(T-T_{\mathrm{ref}})\right].
\end{aligned}
\end{equation}
where $u$ is usable energy fraction set by voltage swing. Note that this capacitor damage model does not include a battery-like C-rate dependence; instead, it is governed by the allowable voltage window $u$, since capacitors generally tolerate pulsed operation without equivalent cycle-damage penalty provided minimum and maximum voltage bounds are respected. Appendix A includes reference values, exponents, cost and sizing assumptions used for the energy storage model.

Given a candidate hybrid split $\alpha$ (capacitor share of pulse energy), we set
\begin{equation}
E_{\mathrm{cap,usable}}=\alpha E_{\mathrm{pulse}},
\end{equation}
\begin{equation}
E_{\mathrm{bat,usable}}=(1-\alpha)E_{\mathrm{pulse}}.
\end{equation}
then compute required nameplate energy for each side as the maximum of lifetime-driven and power-driven constraints. Total hybrid metrics are
\begin{equation}
E_{\mathrm{name,total}}=E_{\mathrm{name,bat}}+E_{\mathrm{name,cap}},
\end{equation}
\begin{equation}
\mathcal{C}_{\mathrm{total}}=\mathcal{C}_{\mathrm{bat}}+\mathcal{C}_{\mathrm{cap}},
\end{equation}
with $\mathcal{C}_{\mathrm{bat}}$ and $\mathcal{C}_{\mathrm{cap}}$ derived from nameplate energy and technology-specific USD/GJ assumptions.

For the baseline storage-sizing case ($E_{\mathrm{req}}=1.5$ GJ pulse-usable demand envelope and $N_{\mathrm{req}}=10^6$), the model predicts a hybrid cost-optimum at $\alpha=0.79$ with $E_{\mathrm{name,bat}}=31.50$ GJ, $E_{\mathrm{name,cap}}=2.38$ GJ, and total cost 6.71 MUSD.
For the selected hybrid point, this corresponds to approximately 8.75 MWh of battery nameplate and 0.66 MWh of capacitor nameplate. In battery terms, that is roughly two to three Tesla Megapack-class units, based on current per-unit Megapack energy ratings of 3.854--3.916 MWh depending on duration configuration \cite{TeslaMegapack2026}, while the capacitor side corresponds to about five 0.5 GJ pulse-capacitor container-equivalents under the present packaging assumptions.
The underlying lifetime, cost, packing, and optimization assumptions used in this hybrid-storage screening model are collected in Appendix~\ref{app:pulsed_power_assumptions}. Battery cycle-life behavior is anchored to DOE grid-storage benchmark data, while capacitor lifetime assumptions follow Eaton supercapacitor guidance; the remaining coefficients are retained as engineering placeholders intended for replacement once vendor-specific battery and pulse-capacitor hardware data become available.
\addtocounter{table}{1}

\begin{figure}[!t]
    \centering
    \includegraphics[width=\columnwidth]{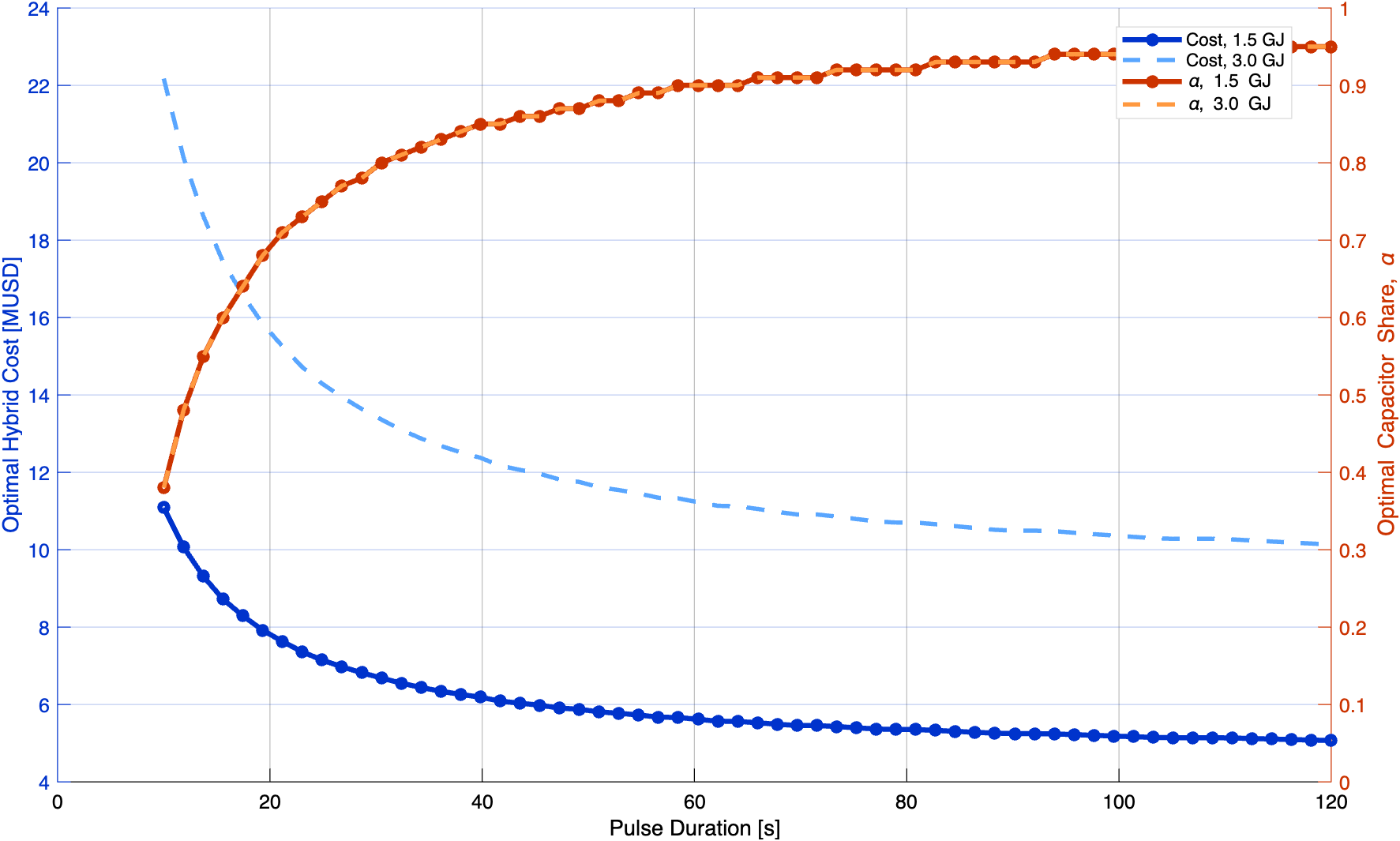}
    \caption{Cost-optimal hybrid storage cost and capacitor share $\alpha$ as a function of pulse duration for $E_{\mathrm{pulse}}=1.5$ GJ (solid) and $3.0$ GJ (dashed).}
    \label{fig:storage_energy_sweep}
\end{figure}

Parametric sweeps were then performed to quantify how pulse demand and duty requirements move the optimum split, as shown in Figure~\ref{fig:storage_energy_sweep}. Sweeping pulse duration shows that the minimum hybrid cost falls as longer-duration pulses allow greater capacitor participation, while the cost-optimal capacitor share $\alpha$ rises accordingly. The same trend is shown for both $E_{\mathrm{pulse}}=1.5$ and $3.0$ GJ cases.

A two-dimensional sweep over $(t_{\mathrm{pulse}},N_{\mathrm{req}})$ at fixed $E_{\mathrm{pulse}}$ maps the transition between capacitor-heavy and battery-heavy solutions. Short pulses and high lifetime-pulse requirements push the optimum toward low $\alpha$, while longer pulse durations permit substantially larger capacitor participation. Here pulse refers to the transient ramp-up demand rather than the flattop interval, which is comparatively quasi-steady from the standpoint of storage sizing.

\begin{figure}[!t]
    \centering
    \includegraphics[width=\columnwidth]{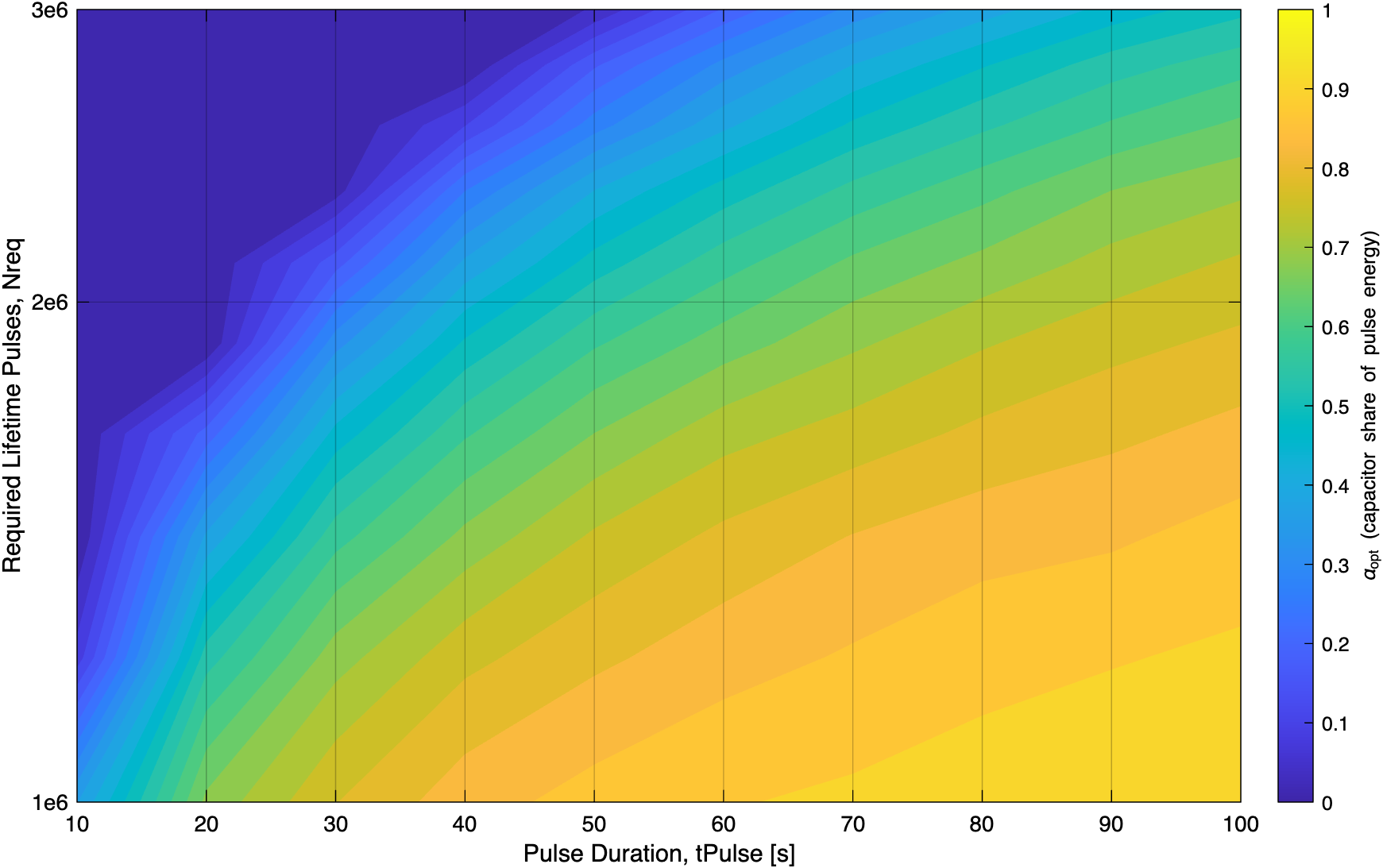}
    \caption{Cost-optimal capacitor share $\alpha_{\mathrm{opt}}$ from the time-dependent model across pulse duration and required lifetime pulse count at fixed pulse energy. This map is used to select hybrid splits consistent with duty-cycle targets and storage lifetime limits.}
    \label{fig:storage_alpha_2d}
\end{figure}

This equivalent-sizing framework makes the trade study explicit: batteries carry lower cost per unit stored energy (USD/GJ) but incur higher cycling damage under fast pulsed duty, while capacitors tolerate pulse power and cycling with lower damage per pulse but at higher USD/GJ. The selected hybrid architecture is therefore chosen where total cost and cumulative damage are jointly minimized for the Yinsen mission pulse schedule.

Flux consumption is a first-order driver of this optimization because central-solenoid volt-seconds set the achievable ramp trajectory, flat-top duration, and dwell recharge time. Breakdown and startup flux consumption are estimated using FUSE and TokaMaker with evolving conductivity and profile assumptions; flat-top flux consumption is comparatively low because high electron temperature implies low loop voltage. Consequently, achievable pulse length for fixed CS build is highly sensitive to startup/ramp flux use, motivating future trajectory-refinement studies with TORAX.

\begin{figure}[!t]
    \centering
    \includegraphics[width=\columnwidth]{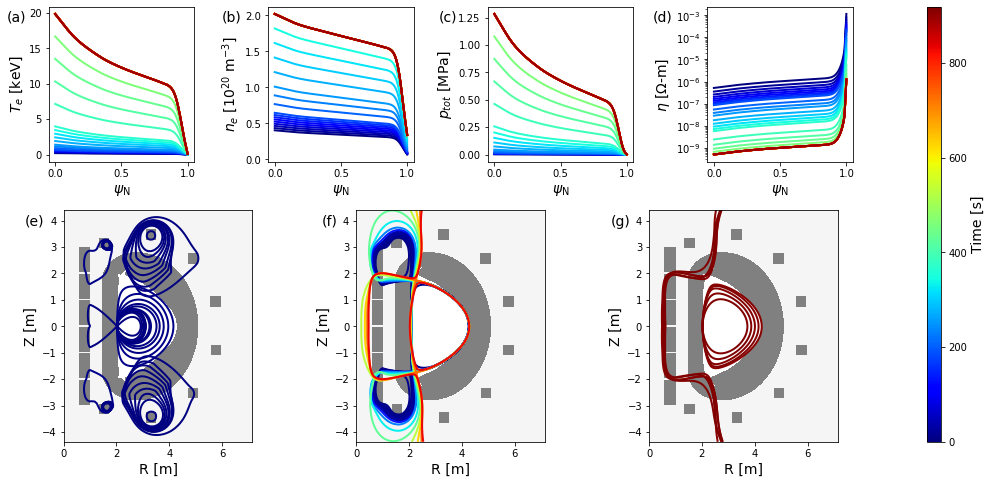}
    \caption{(a) Electron temperature, (b) electron density, (c) total pressure and (d) resistivity profiles as a function of the pulse. The corresponding shape evolution is shown for the (e) ramp-up, (f) flattop and (g) ramp-down sections of the pulse.}
    \label{fig:pulse_shapes}
\end{figure}

For the present work, the TokaMaker code \cite{hansen_tokamaker_2024} is used to model the evolution of coil currents and consumed flux over the course of a template discharge. TokaMaker captures the evolution of flux throughout the ramp-up, flattop, and ramp-down phases of the discharge, but requires assumptions on the evolution of kinetic profiles in order to self-consistently capture the impact of plasma resistivity. As such, kinetic profiles from FUSE are adopted for the flattop phase of the discharge. As shown in figure~\ref{fig:pulse_shapes}, as simple evolution of these profiles is enforced during the ramp-up phase, leading to evolution of the plasma resistivity. $Z_\mathrm{eff}=1.5$ is held constant during this calculation. The loop voltage $V_\mathrm{loop}$ also evolves over the course of the discharge, dropping from $\sim5-6\,$V just after breakdown to $V_\mathrm{loop}=0.04\,$V during the current flattop. This low loop voltage and low resistance plasma, result in very little ohmic heating during flattop, approximately 0.5 MW. This assumed profile evolution leads to an estimate of $\sim20\,$Wb of total flux consumption during the ramp-up phase and $\sim40\,$Wb of total flux consumption during the flattop phase, pointing to a total flux consumption of  $\sim60\,$Wb. While predicting the precise ramp up flux consumption remains a challenge with large error bars, given the 90Wb of flux available from the CS, long-pulse discharges exceeding 15 minutes remain possible with continued discharge optimization. Additionally, it is proposed that instead of ramping down and fully terminating the plasma at end of pulse, instead the plasma is ramped down to a low current and cold state while the CS coils are being individually charged up from their perspective power supplies. Thus avoiding flux consumption associated with breakdown between each pulse. This requires further study particularly regarding the ability to maintain shape control and stability while coils are being charged in preparation of the next pulse.

\begin{figure}[!t]
    \centering
    \includegraphics[width=\columnwidth]{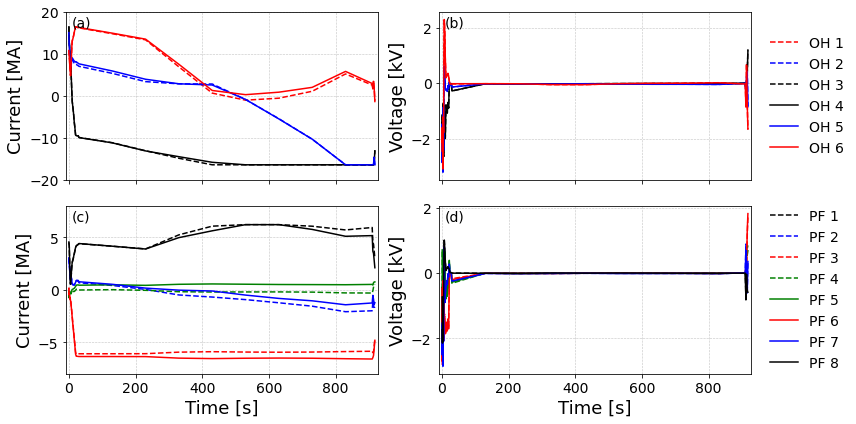}
    \caption{Coil currents (a, c) and associated power supply voltages (b, d) for the CS (a, b) and PF (c, d) coils.}
    \label{fig:pulse_currents}
\end{figure}

Figure~\ref{fig:pulse_shapes} also shows the evolution of the plasma shape throughout this nominal Yinsen pulse, which is achieved via the current and voltage trajectories shown in figure~\ref{fig:pulse_currents} for each coil. No effort is made at this time to optimize coil trajectories, though a loose $2.5\,$kV limit is imposed on the power supply for each coil throughout the pulse. Notably, this simulation only accounts for plasma evolution after breakdown and while the Ohmic coils are ramped sequentially in figure~\ref{fig:pulse_currents}, the role of driving flux will be equally shared among all CS coils  in the final Yinsen design. The integration of the (terminal) coil currents and voltages from figure~\ref{fig:pulse_currents} yields a usable stored energy requirement of 1.5GJ and a maximum cumulative instantaneous power draw of 300MW during ramp up, used in later sections for sizing the power supply facility. Lengthening the ramp up duration longer than 30 seconds along with better coil regulation could significantly reduce this.

Power-supply limits (maximum voltage, current, and slew rate) then define the admissible operating envelope for CS/PF trajectories. The time-dependent plant model is used to quantify coupled tradeoffs between these electrical limits, thermal inertia in the blanket and power conversion train, and constant parasitic demand (dominated by cryogenic load). For the baseline case, gross turbine electric output exceeds the constant recirculating power loads over the entire pulse cycle, including the dwell period between pulses, and dynamic FLiBe flow control maintains adequate blanket temperature margin above the adopted \SI{800}{K} lower operating bound, discussed later in Section~\ref{sec:bop}.

The same model is used to estimate ride-through capability for always-on house loads. With baseline assumptions of approximately 10 MW of recirculating power loads, not including ICRH heating since that is only required during active fusion-power operation, the selected hybrid battery nameplate (about 8.75 MWh) provides roughly 50 minutes of coverage for constant-load-only operation if dispatched from full charge. In contingency scenarios, a dedicated 10 MW diesel generator can be used as a backup source to carry the baseline recirculating load directly; under the same 10 MW constant-load condition, this reduces the net battery draw to approximately zero and in principle allows the battery state of charge to be preserved while the plant is held in a controlled standby or shutdown state. Importantly, during non-fusion operation the cryogenic plant no longer has to remove neutron-induced and high operational thermal loads; power demand trends toward heat-leak compensation with comparatively small residual heating, which can substantially increase cold-hold duration for the magnet system beyond the conservative constant-load estimate above. This backup path is intended for resilience and controlled shutdown/transition support rather than routine pulse support.

\begin{figure}[!t]
    \centering
    \includegraphics[width=\columnwidth]{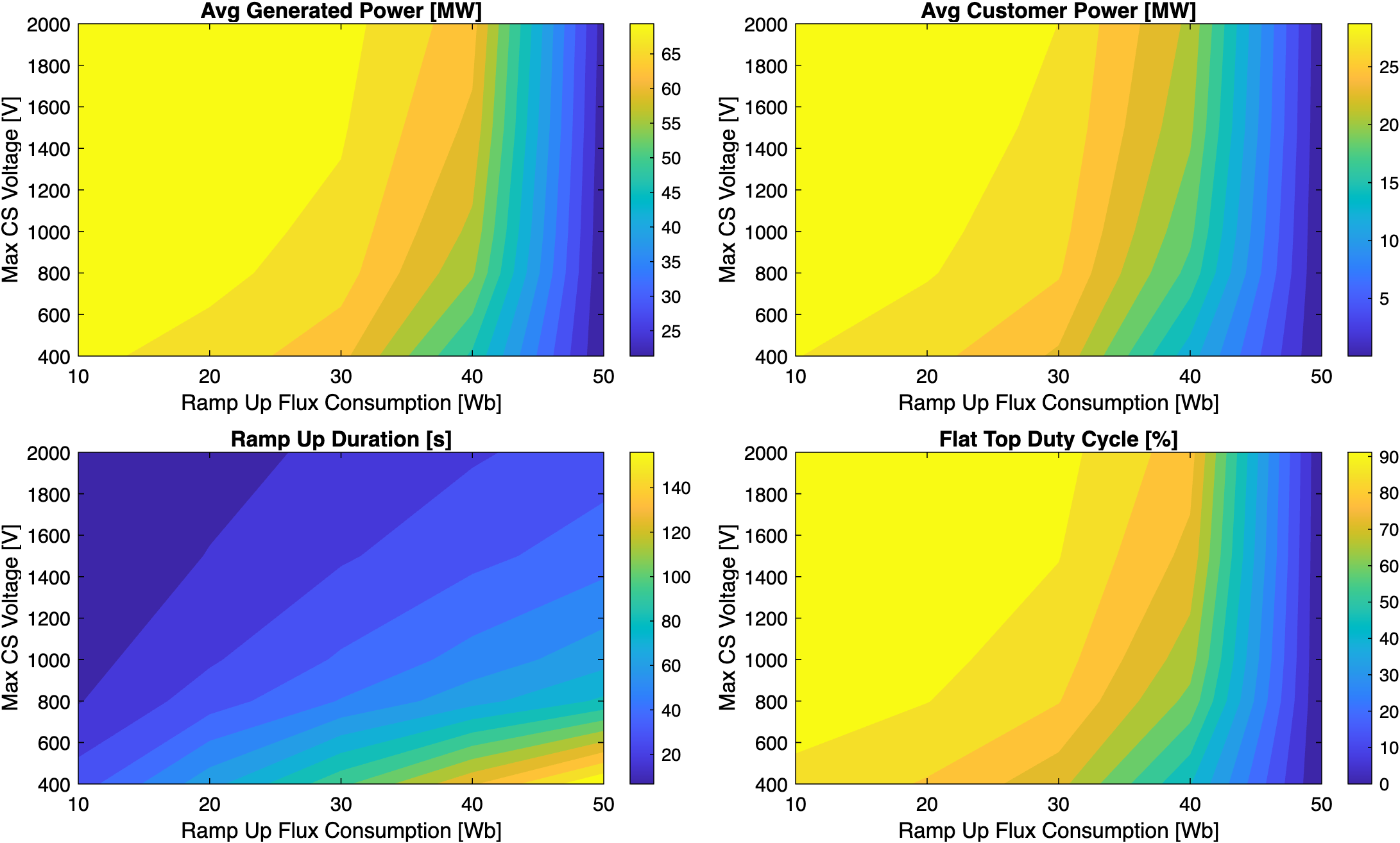}
    \caption{Two-parameter sensitivity from the time-dependent plant model for $V_{\mathrm{CS,max}}$ and ramp-up CS flux use. Panels show average generated power, average customer power, ramp-up duration, and flat-top duty cycle. The duty cycle is defined as $100\times t_{\mathrm{flat}}/(t_{\mathrm{dwell}}+t_{\mathrm{rampup}}+t_{\mathrm{flat}}+t_{\mathrm{rampdown}})$.}
    \label{fig:storage_sweep_yinsen}
\end{figure}

This sensitivity map, shown in Figure~\ref{fig:storage_sweep_yinsen}, shows that CS ramp-up flux allocation is the dominant lever on mission-level pulse utility: increasing ramp-up flux consumption sharply reduces flat-top duration and duty cycle, with the highest-flux cases collapsing to near-zero flat-top. Increasing allowable CS voltage primarily shortens ramp and dwell intervals, which increases duty cycle and average delivered power, but also shifts transient burden toward higher short-duration storage power demand. These trends are consistent with the lifetime-equivalent sizing results and motivate co-optimization of flux trajectory and converter limits rather than treating them independently.

\begin{figure*}[!t]
    \centering
    \includegraphics[width=0.92\textwidth]{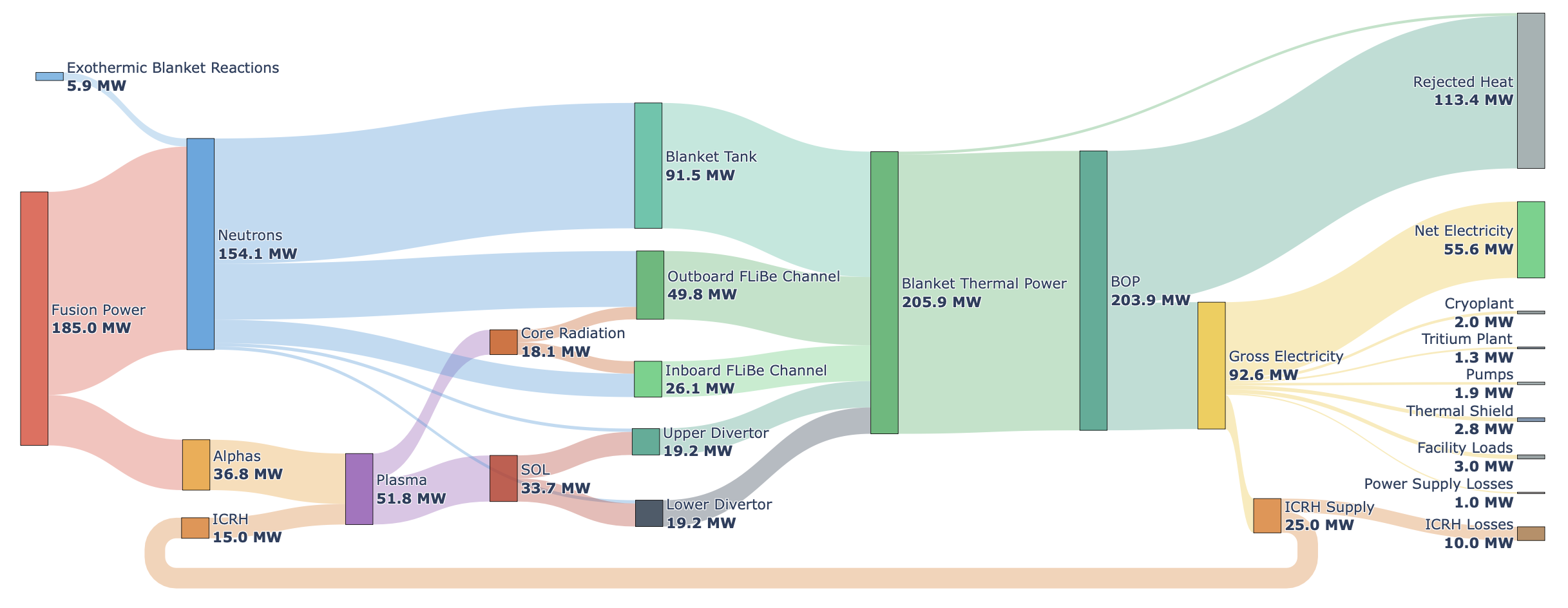}
    \caption{Plant power-flow summary for Yinsen at the higher-power case B operating point, showing the partition of fusion power into blanket and divertor heat loads, gross electric generation, recirculating loads, rejected heat, and net electric output.}
    \label{fig:bop-power-balance}
\end{figure*}

\section{Balance of plant}\label{sec:bop}
The balance of plant (BOP) converts thermal power extracted from the first wall, blanket, and divertors into usable electricity while also setting important constraints on plant layout, startup behavior, and overall efficiency. For Yinsen, the primary coolant is FLiBe, the eutectic molten-salt mixture LiF--BeF$_2$, which serves simultaneously as the tritium-breeding medium and the main heat-transport fluid. That choice is attractive because FLiBe remains liquid over a wide operating window, is chemically stable at reactor-relevant temperatures~\cite{Forsberg2020,Muramatsu2002}, and enables single-phase, low-pressure heat removal with correspondingly modest pumping-power requirements relative to high-pressure water/steam systems. Its electrical conductivity is best interpreted as modest rather than negligible: tabulated LiF--BeF$_2$ values near the operating range are about \SI{250}{S/m}~\cite{Janz1988MoltenSalts}, roughly 50 times higher than seawater at ambient conditions~\cite{NOAAPMELConductivity}, but still \(10^4\)--\(10^5\) times lower than structural metals such as stainless steel or copper. This comparatively low conductivity relative to metals helps limit, rather than eliminate, MHD pressure-drop penalties in the high background magnetic fields of a compact tokamak. From a blanket perspective, lithium provides tritium breeding while beryllium contributes neutron multiplication and moderation. Taken together, these features make FLiBe a natural thermal interface between the fusion core and a compact high-temperature secondary power-conversion system.

\subsection{Primary FLiBe loop and secondary-cycle selection}
The primary-side FLiBe operating window strongly shapes the BOP architecture and is in many ways well matched to supercritical CO$_2$ conversion. The freezing point of FLiBe is \SI{732}{K}, so the primary loop must remain comfortably above this threshold; in practice, a working lower bound of roughly \SI{800}{K} is imposed to preserve margin against local freezing during transients and distribution nonuniformity. At the same time, present high-temperature structural and heat-exchanger materials place a practical upper operating limit near \SI{973}{K}~\cite{Forsberg2020}. Within that range, Yinsen adopts a conservative FLiBe hot-leg / blanket-tank outlet target of approximately \SI{925}{K}. Yinsen therefore operates in a relatively narrow but still attractive high-temperature window that naturally favors compact Brayton-cycle conversion rather than lower-temperature steam systems. This is consistent with prior FLiBe blanket studies, which find sCO$_2$ to compare favorably with steam and helium when both thermal efficiency and compact turbomachinery are considered together~\cite{Segantin2020}. For Yinsen, that compactness matters as much for plant layout as for cycle efficiency: CO$_2$ at supercritical conditions permits substantially denser turbomachinery than helium and therefore a more compact balance-of-plant footprint for marine and distributed-deployment applications. Cycle design and analysis were carried out using an extension of the ThermalSystemModels.jl (TSM) Julia package built on ModelingToolkit~\cite{Ma2021}. Initially developed for lead-lithium blanket studies, TSM was extended here to support FLiBe as both the blanket and divertor coolant. A steady-state parameter sweep over the secondary-side pressure range, compressor inlet temperature, and mass flow rate was then used to select cycle parameters with high efficiency in the allowed thermal window.

\begin{figure}[!t]
    \centering
    \includegraphics[width=\columnwidth]{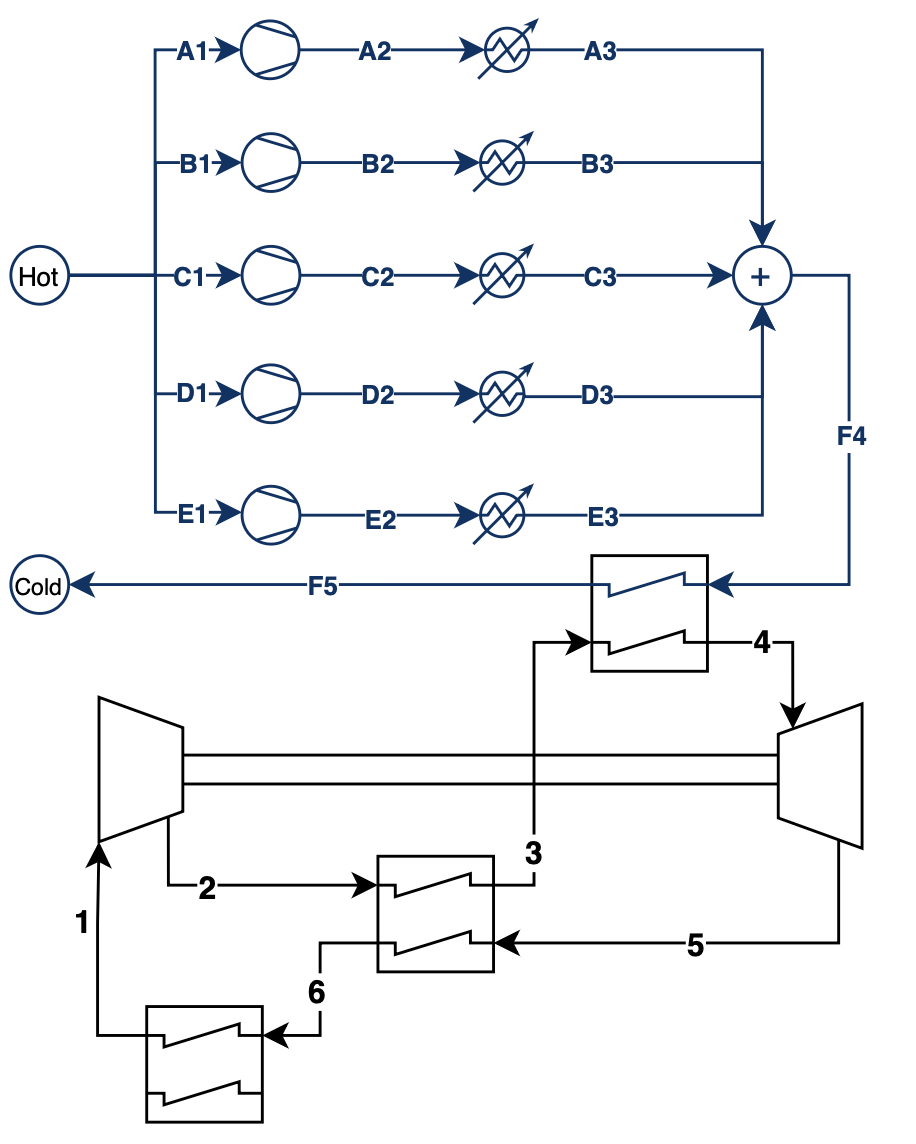}
    \caption{BOP schematic. Five primary FLiBe loops discharge into the blanket tank before heat is transferred to the sCO\textsubscript{2} secondary side (nodes 1--6).}
    \label{fig:bop}
\end{figure}

The baseline BOP system design for Yinsen is shown in Figure~\ref{fig:bop}. The cycle of choice is a simple Brayton cycle with regeneration (the recirculating of waste heat from the turbine outlet back into the secondary side loop). The primary-side FLiBe flow and pumping architecture is similar to the conceptual design for ARC shown in Figure 11 of \cite{Kuang2018}, with five principal loops in the present Yinsen implementation: a high-field-side first-wall/vacuum-vessel loop, a low-field-side first-wall/vacuum-vessel loop, an upper divertor loop, a lower divertor loop, and the blanket-tank loop. The four interior coolant branches are intended to remove the dominant surface heat fluxes so that the plasma-facing components and vacuum-vessel structures remain within their allowable operating temperatures. These interior branches discharge into the blanket tank, which acts as the main, more slowly circulating reservoir for extracting volumetric nuclear heating from the surrounding blanket and shield structures before feeding the primary-to-secondary heat exchanger. Each of the five loops is equipped with its own pump so that branch mass flow rates can be controlled independently.

The thermal inputs to the primary FLiBe system are assembled from both plasma and neutronics calculations. The Case B analysis point at 185~MW of fusion power was selected because, as discussed in Section~\ref{sec:direct_capex}, modest oversizing of the balance-of-plant equipment adds relatively little cost compared with the tokamak core. In practice, the Yinsen BOP would likely be sized closer to \(300~\mathrm{MW}_{\mathrm{th}}\) in order to preserve upgrade margin and operating flexibility. At the 185~MW fusion-power point, FUSE simulations were used to provide the core-radiation surface heat flux incident on the relevant first-wall loops through ray tracing, and to determine the divertor surface heat flux through the SOL model. Separately, OpenMC neutronics results were used to determine the volumetric nuclear heating deposited in the radial layers associated with each of the five primary FLiBe loops. Convective and conductive heat transport was then modeled from the tungsten first wall outward through the cooled structures and into the blanket tank and neutron shields. Table~\ref{tab:flibe-loops} gives the corresponding FLiBe design mass-flow targets required through each coolant loop to extract these loads, consistent with the loop-wise thermal inputs shown in Figure~\ref{fig:bop-inputs-profile}. At the outer edge of the neutron-shield stack, an aluminum-silicate insulating layer together with a \SI{1}{cm}-thick water-cooled layer defines an approximately room-temperature thermal boundary. This provides an intermediate thermal staging point between the $\sim$\SI{900}{K} blanket tank and the \SI{20}{K} HTS magnet system; between that room-temperature boundary and the magnets sits the \SI{80}{K} thermal shield, which further suppresses radiative heat loading on the HTS magnet system. In the baseline design, the room-temperature neutron-shield water loop intercepts about \(0.025~\mathrm{MW}_{\mathrm{th}}/\mathrm{MW}_{\mathrm{fusion}}\), while the \SI{80}{K} thermal shield is budgeted to intercept about \(0.001~\mathrm{MW}_{\mathrm{th}}/\mathrm{MW}_{\mathrm{fusion}}\); these correspond to roughly \(0.003~\mathrm{MW}_{\mathrm{e}}/\mathrm{MW}_{\mathrm{fusion}}\) of wall-plug power at the \SI{80}{K} stage and \(0.015~\mathrm{MW}_{\mathrm{e}}/\mathrm{MW}_{\mathrm{fusion}}\) for the remaining \SI{20}{K} magnet refrigeration allowance. The resulting loop-wise thermal inputs and representative through-thickness thermal staging are illustrated in Figure~\ref{fig:bop-inputs-profile}, while the representative primary-to-secondary state points are summarized in Tables~\ref{tab:flibe-hx-state}--\ref{tab:secondary}. For context, the selected CO$_2$ state points span roughly \SIrange{8}{25}{MPa}, with the compressor inlet near \SI{305}{K}, so the cycle remains above the CO$_2$ critical point (\SI{7.38}{MPa}, \SI{304.13}{K}) and avoids crossing into two-phase operation while still benefiting from the high fluid density and low compression work available near the critical region~\cite{NISTCO2WebBook,Dostal2004}. In that sense, the \SI{25}{MPa} high side should be understood not as unusually aggressive, but as consistent with standard compact recuperated sCO$_2$ Brayton-cycle practice.

\begin{figure}[!t]
    \centering
    \includegraphics[width=\columnwidth]{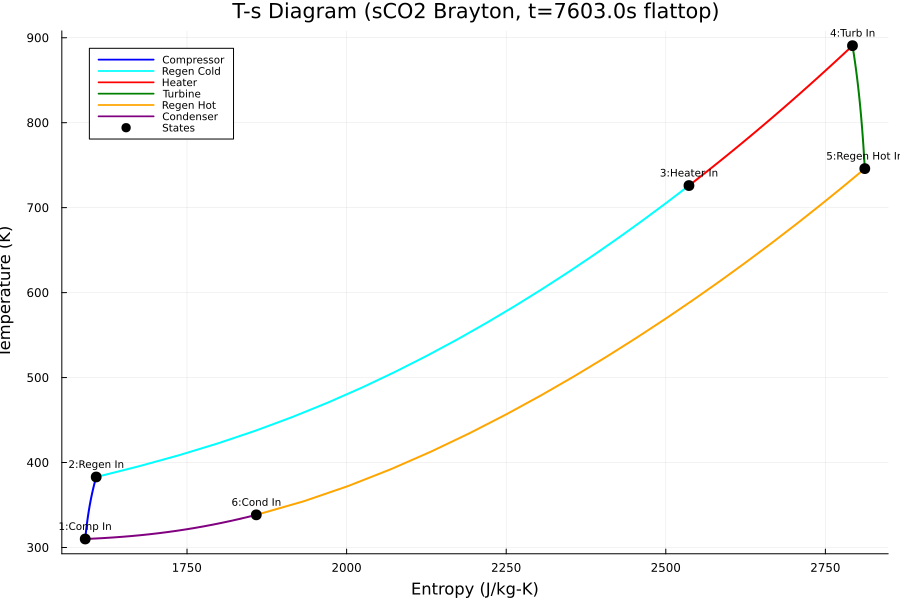}
    \caption{T--s diagram for the baseline regenerated sCO\textsubscript{2} Brayton cycle used in the transient BOP analysis.}
    \label{fig:Ts}
\end{figure}

\begin{figure}[!t]
    \centering
    \includegraphics[width=\columnwidth]{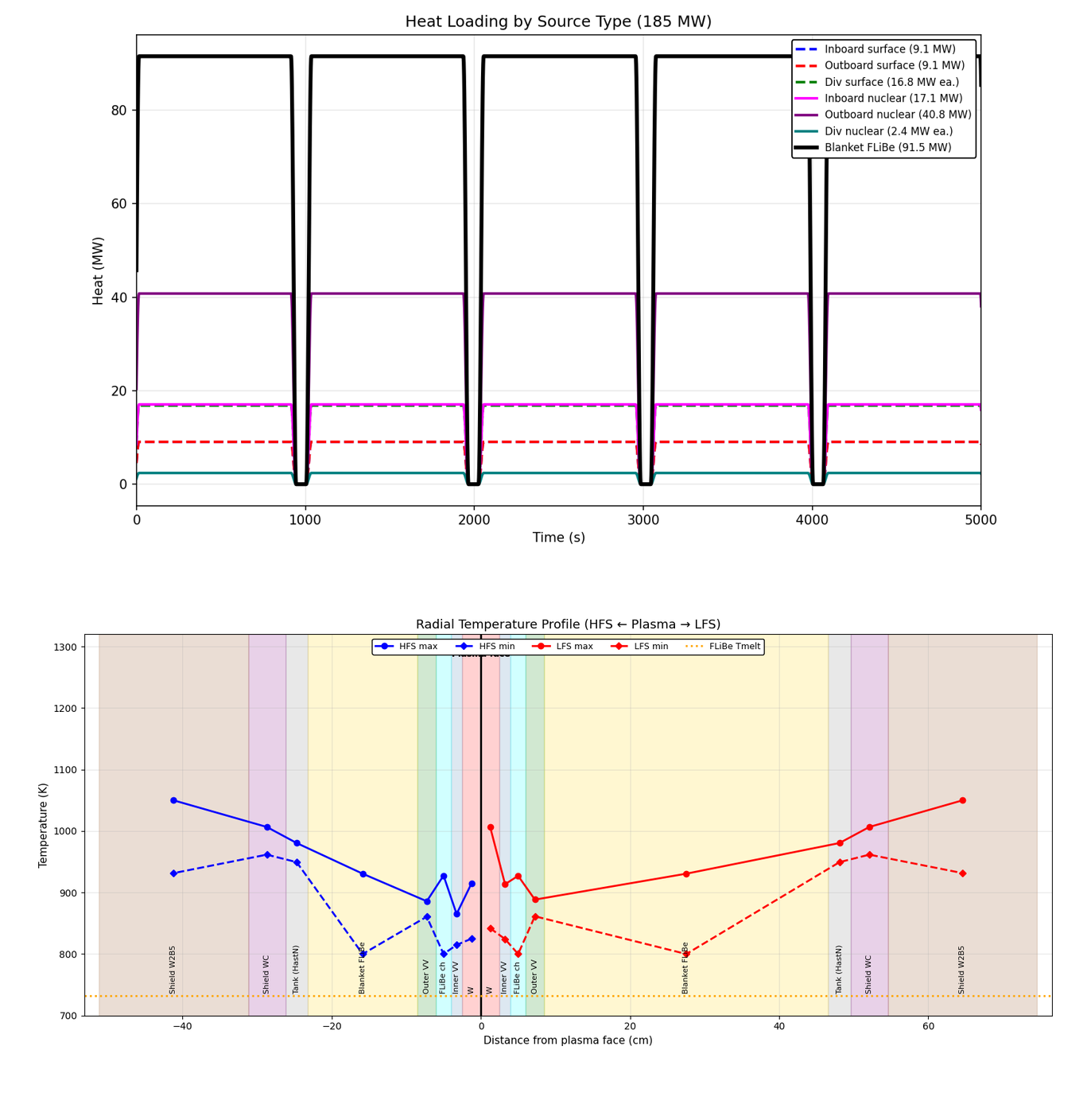}
    \caption{Thermal inputs used in the reduced-order BOP model. The upper panel shows the imposed pulsed heat-loading schedule by source type for the five-loop FLiBe architecture, including surface heating on the first-wall and divertor branches together with volumetric nuclear heating in the blanket and shield structures. The lower panel shows a representative through-thickness temperature profile from the tungsten first wall outward through the cooled structures, blanket tank, and neutron-shield stack.}
    \label{fig:bop-inputs-profile}
\end{figure}

The secondary-side loop consists of a compression and expansion phase, with turbine exhaust heat recuperated back into the cycle through regeneration. A small steady-state parameter study was carried out over the minimum and maximum pressure of the CO$_2$ loop, the compressor inlet temperature, and the cycle mass flow rate in order to identify a compact high-efficiency operating point. In addition to the selected Brayton cycle with regeneration, a second configuration including intercooling was also examined. The regeneration+intercool cycle improved the overall efficiency only marginally, by about 0.3 percentage points, while requiring an additional component and an extra intermediate-pressure design variable at the compressor. Given the volumetric space constraints of the intended applications and the advantage of keeping the BOP as simple as possible, the regeneration-only cycle is selected as the baseline design. Representative design inputs and flattop operating states for the primary and secondary cycles are collected in Tables~\ref{tab:flibe-loops}--\ref{tab:secondary}, and the T-s diagram for the secondary side cycle is shown in Figure \ref{fig:Ts}.

\input{generated/bop_tables_generated.tex}

\subsection{Transient loading under pulsed operation conditions}

The transient startup behavior of the FLiBe primary system was investigated to evaluate its response to cyclical thermal loading during pulsed operation, together with the resulting thermal and power-conversion response of the secondary CO$_2$ side. The imposed pulse schedule uses a smoothed pulse waveform with a \SI{30}{s} rise, \SI{900}{s} flat-top, \SI{30}{s} fall, and \SI{60}{s} dwell, as shown in the upper panel of Figure~\ref{fig:bop-inputs-profile}. Within the reduced-order model, the four principal poloidal sectors that receive the dominant pulsed loads; the HFS first wall, LFS first wall, upper divertor, and lower divertor, are each represented as a one-dimensional radial stack of lumped thermal masses connected by series thermal resistances. Plasma surface flux is applied at the tungsten armor face, while volumetric nuclear heating is deposited directly into each solid layer and into the bulk FLiBe according to fractions frozen from the baseline neutronics solution and scaled linearly with fusion power. The lower panel of Figure~\ref{fig:bop-inputs-profile} therefore represents not just a schematic thermal profile, but the radial staging path through which heat is conducted out of the plasma-facing structures, through the vessel and shield materials, and ultimately toward the FLiBe and the room-temperature shield boundary.

\begin{figure}[!t]
    \centering
    \includegraphics[width=\columnwidth]{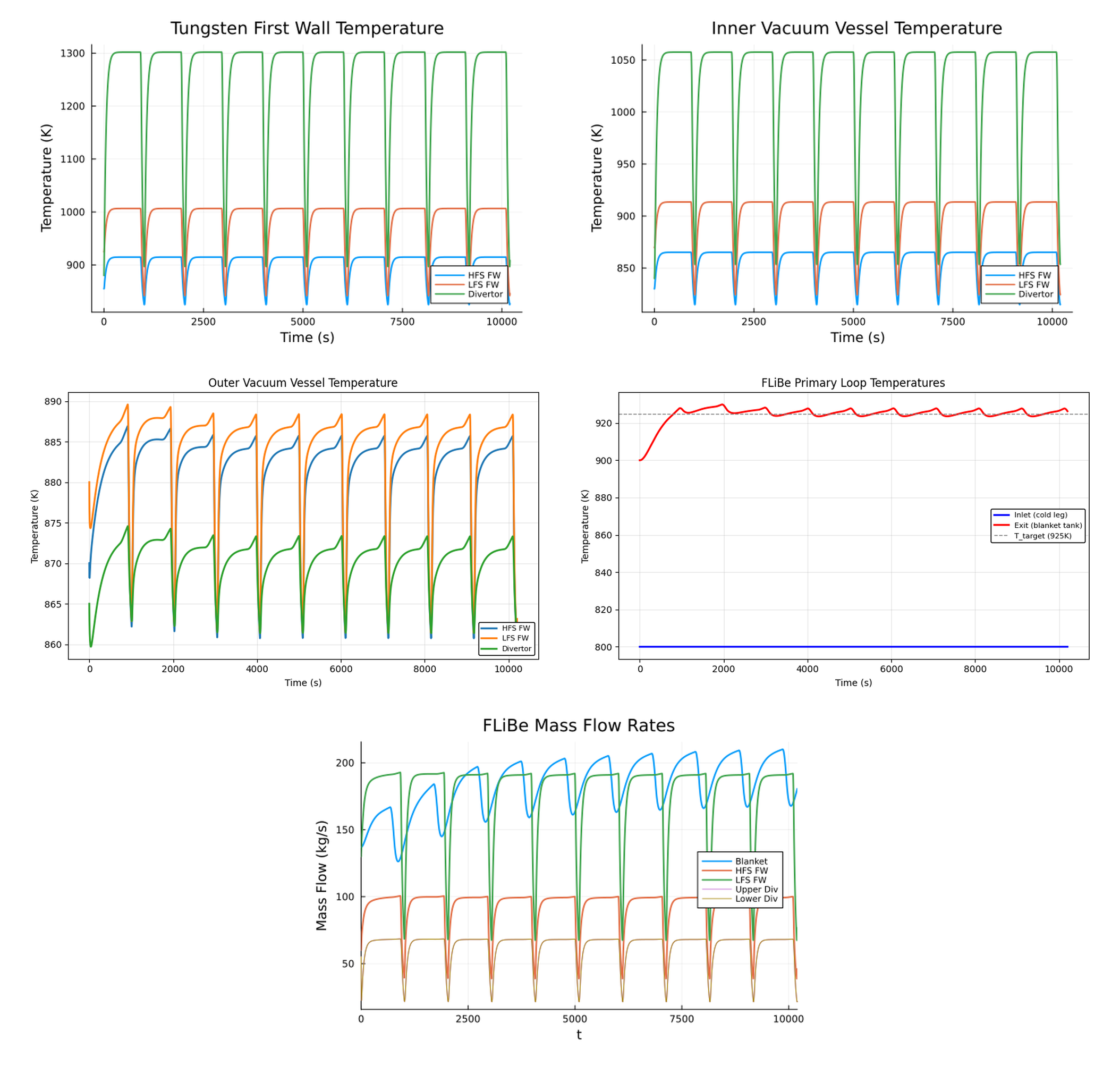}
    \caption{Primary-side transient response during repeated pulsed operation. The four upper panels show tungsten first-wall, inner-vacuum-vessel, outer-vacuum-vessel, and FLiBe outlet temperatures. The lower panel shows the independently commanded mass-flow histories of the five FLiBe loops.}
    \label{fig:bop-primary-response}
\end{figure}

Each FLiBe branch is then treated as a moving thermal capacitance: \SI{800}{K} cold-leg FLiBe is drawn from a common \SI{90}{m^3} bulk-mixing tank, the branch absorbs both wall heat and its own direct nuclear deposition, and the heated return discharges back to that tank before the primary-side FLiBe-to-CO$_2$ heat exchanger. A small poloidal-conduction coupling is retained between neighboring sectors to represent the continuity of the surrounding shells, while the secondary side is modeled as a closed regenerated supercritical CO$_2$ Brayton cycle coupled to the primary through an effectiveness-based heat-exchanger element. Independent pump control is applied to the FLiBe loops so that the four interior branches track the pulsed surface loading, while the blanket loop uses pulse-tracking feedforward with a slow integral correction to hold the tank outlet near the intended operating point of approximately \SI{925}{K}. The primary-side response is summarized in Figure~\ref{fig:bop-primary-response}.

After the initial startup transient, the tungsten and vacuum-vessel temperatures settle into repeatable pulse-to-pulse excursions, while the FLiBe outlet temperatures and independently controlled loop mass flows remain within an envelope consistent with the imposed surface and volumetric heating. For the divertor branch in particular, the applied surface heat load was distributed across the divertor wetted area so that the transient model captures the correct integrated thermal power entering that loop. In reality, the divertor heat flux is strongly concentrated near the strike region as discussed in Section~\ref{sec:divertor_power_exhaust}; however, this assumption is implemented for power accounting in the transient BOP model. This assumption also gains validity as deep detachment is achieved. Over the same repeated pulse sequence, the blanket-loop hot side remains near \SI{925}{K}, while the primary heat exchanger returns FLiBe to approximately \SI{800}{K}.

\begin{figure}[!t]
    \centering
    \includegraphics[width=\columnwidth]{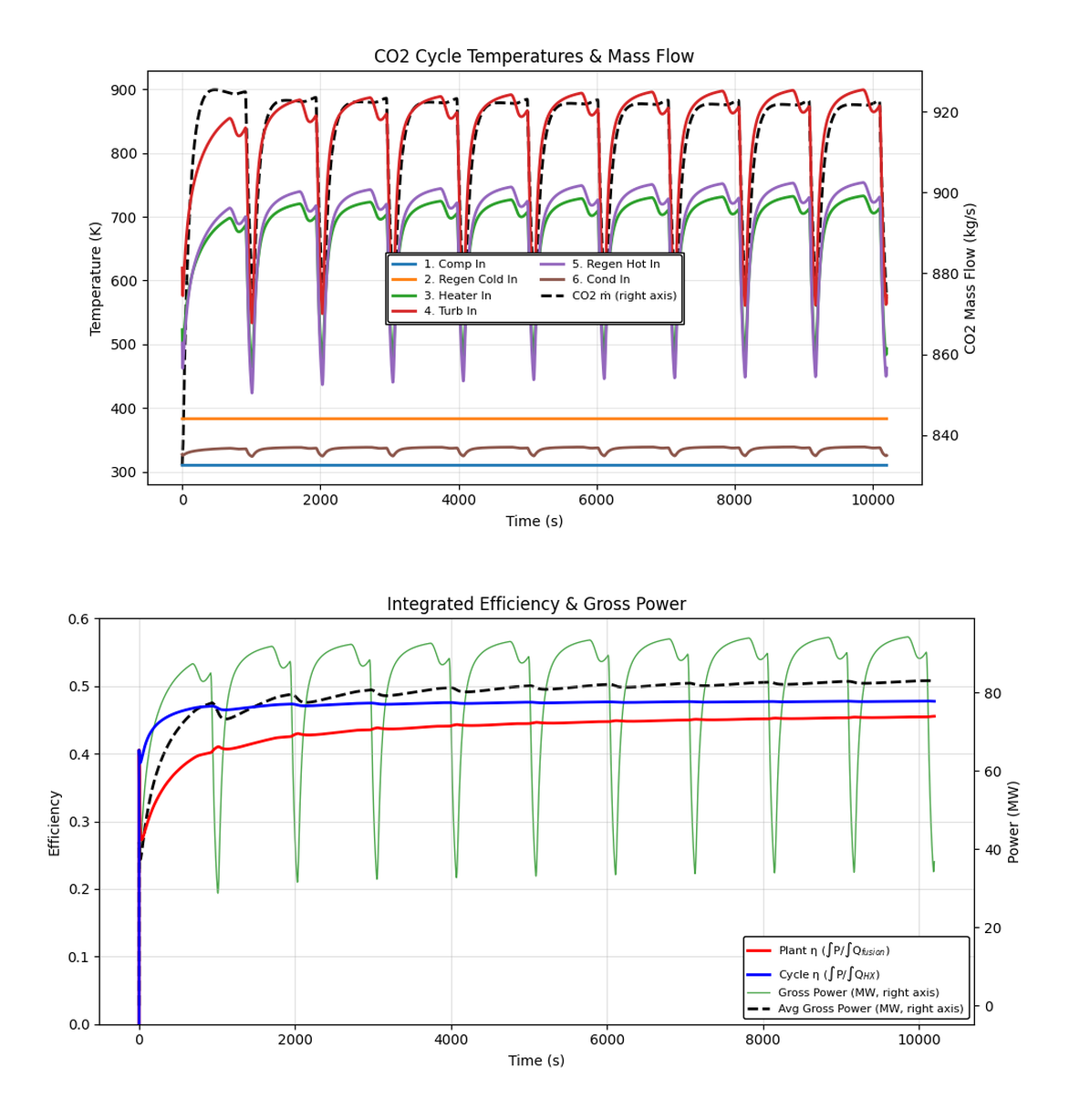}
    \caption{Secondary-side and plant-level transient response during repeated pulsed operation. The upper panel shows the CO\textsubscript{2} cycle-state temperatures, while the lower panel shows the corresponding plant and cycle efficiency histories over the same pulse train.}
    \label{fig:bop-secondary-response}
\end{figure}

The corresponding secondary-side and plant-level response is summarized in Figure~\ref{fig:bop-secondary-response}. Gross plant power varies from about \SI{30}{MW} during dwell to nearly \SI{95}{MW} during flattop and the primary heat-exchanger duty follows the pulsed thermal input between roughly \SIrange{90}{193}{MW}, while the commanded CO\textsubscript{2} flow stabilizes near \SIrange{872}{923}{kg/s} and the plant efficiency settles into a repeatable operating band after the initial startup transient. Because of the thermal inertia of the working fluids, the plant can still dispatch tens of MWe during dwell, more than enough to cover the recirculating loads, which are largely parasitic once ICRH is turned off.

\begin{figure}[!htbp]
    \centering
    \includegraphics[width=0.95\columnwidth]{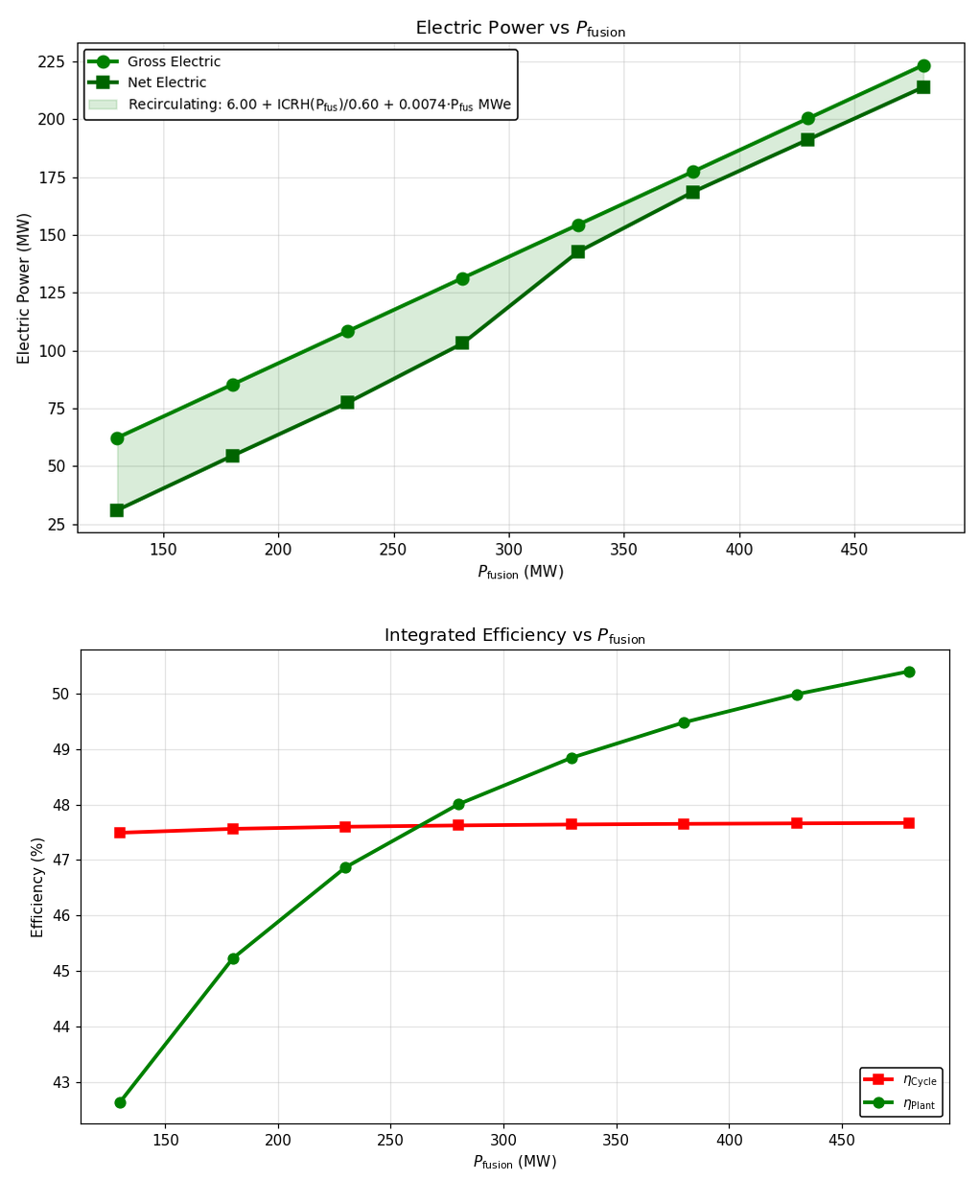}
    \caption{Fusion-power sweep trends for the baseline five-loop FLiBe primary system and regenerated sCO\textsubscript{2} secondary cycle. The two panels show how flattop and run-averaged gross plant power, together with integrated plant and cycle efficiency, scale as the fusion power is increased across the Yinsen-relevant operating range.}
    \label{fig:bop-pfusion-sweep-power}
\end{figure}

\begin{figure}[!htbp]
    \centering
    \includegraphics[width=0.95\columnwidth]{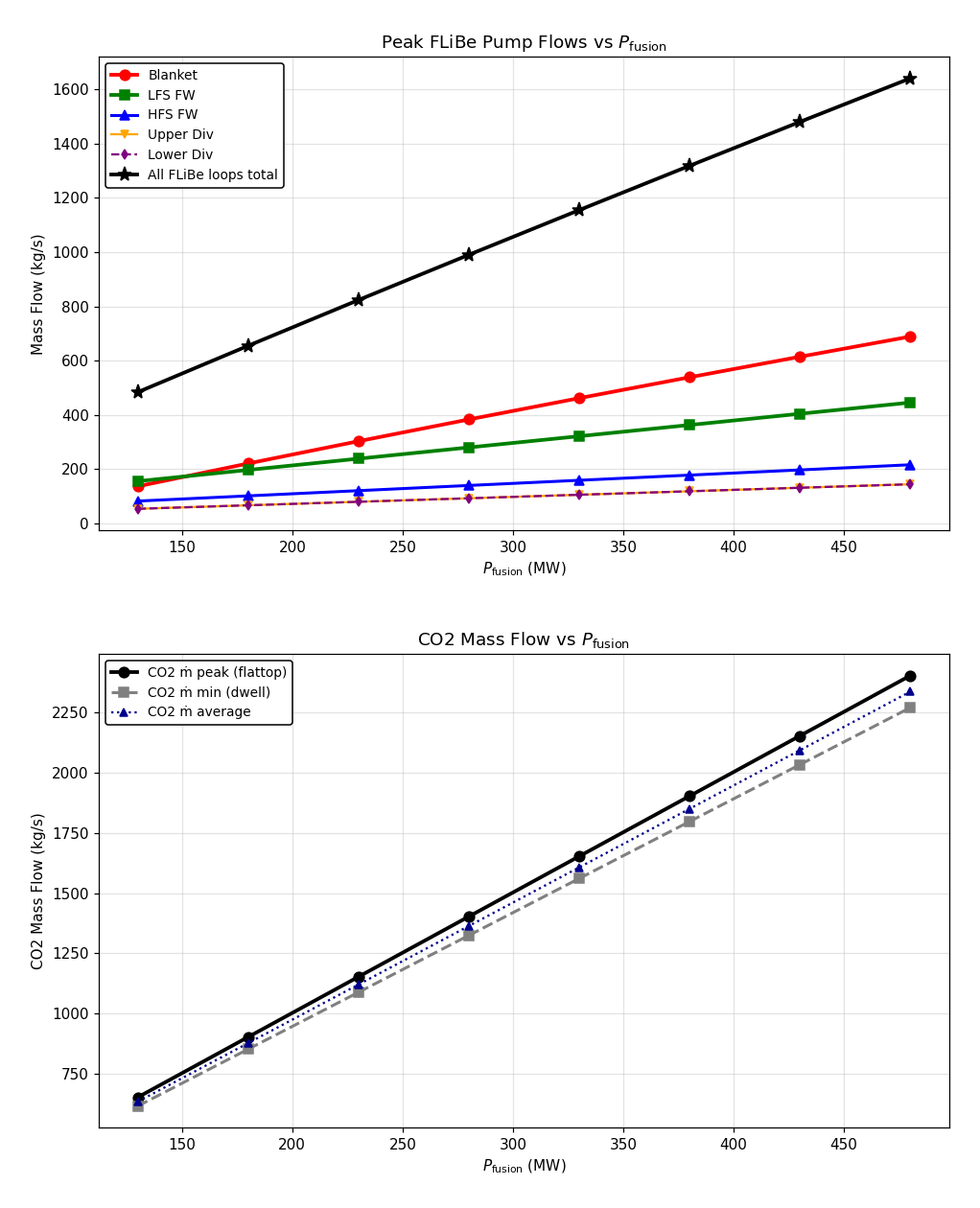}
    \caption{Primary- and secondary-side flow scaling over the same fusion-power sweep. The upper panel shows the required FLiBe loop mass-flow rates, while the lower panel shows the corresponding CO\textsubscript{2} mass-flow requirement on the secondary side.}
    \label{fig:bop-pfusion-sweep-flow}
\end{figure}

To assess how the same BOP architecture scales beyond the baseline case, an additional sweep was carried out from \SI{130}{MW} to \SI{480}{MW} fusion power using the same five-loop FLiBe layout, regenerated sCO\textsubscript{2} cycle, and control structure. Across that sweep, the ICRH heating input was scaled using the worst-case $Q$ predicted by the POPCON analysis at each $P_{\mathrm{fusion}}$ until $P_{\mathrm{SOL}}$ reached \SI{40}{MW}; above roughly \SI{330}{MW} fusion power, the plasma is predicted to be effectively ignited and therefore no longer require sustained external heating power. Figures~\ref{fig:bop-pfusion-sweep-power} and~\ref{fig:bop-pfusion-sweep-flow} show that gross plant power rises linearly across this range: flattop gross plant output increases from about \SI{70}{MW} at \SI{130}{MW} fusion power to about \SI{253}{MW} at \SI{480}{MW}, while pulsed average gross plant output increases from about \SI{62}{MW} to \SI{223}{MW}. At the same time, the integrated cycle efficiency remains nearly unchanged at about 47.5--47.7\%, whereas the integrated plant efficiency improves from about 42.6\% to 50.4\% as fixed parasitic loads are diluted at higher power. The required primary and secondary mass-flow rates also increase smoothly and monotonically, with the blanket-loop peak FLiBe flow rising from about \SI{137}{kg/s} to \SI{689}{kg/s} and the peak CO\textsubscript{2} flow rising from about \SI{651}{kg/s} to \SI{2404}{kg/s}. In that sense, the BOP does not exhibit a sharp systems-level scaling cliff over this sweep; the more constraining limits at high power remain the coupled first-wall and divertor thermal loads together with the broader materials-limited reactor envelope discussed earlier.

\section{Fuel Cycle Engineering}

The Yinsen fuel cycle is evaluated here with a 0-D residence-time model built using the open-source PathSim/PathView workflow of Delaporte-Mathurin et al.~\cite{DelaporteMathurin2026} and the same ARC-class topology analyzed by Meschini et al.~\cite{Meschini2023FuelCycle}, but with Yinsen-specific parameters. The baseline case uses $P_{\mathrm{fus}} = 130~\mathrm{MW_{th}}$, $TBE = 0.02$, $TBR = 1.10$, a $900~\mathrm{s}$ burn plus $60~\mathrm{s}$ dwell pulse window, and an effective availability factor $AF = 0.403125$ from $43\%$ macro availability and $0.9375$ pulse duty. At this stage the model is used as a systems-level tool to determine the minimum startup inventory, tritium doubling time, and the residence-time sensitivities that matter most for Yinsen.

\begin{figure}[!b]
    \centering
    \includegraphics[width=0.98\columnwidth]{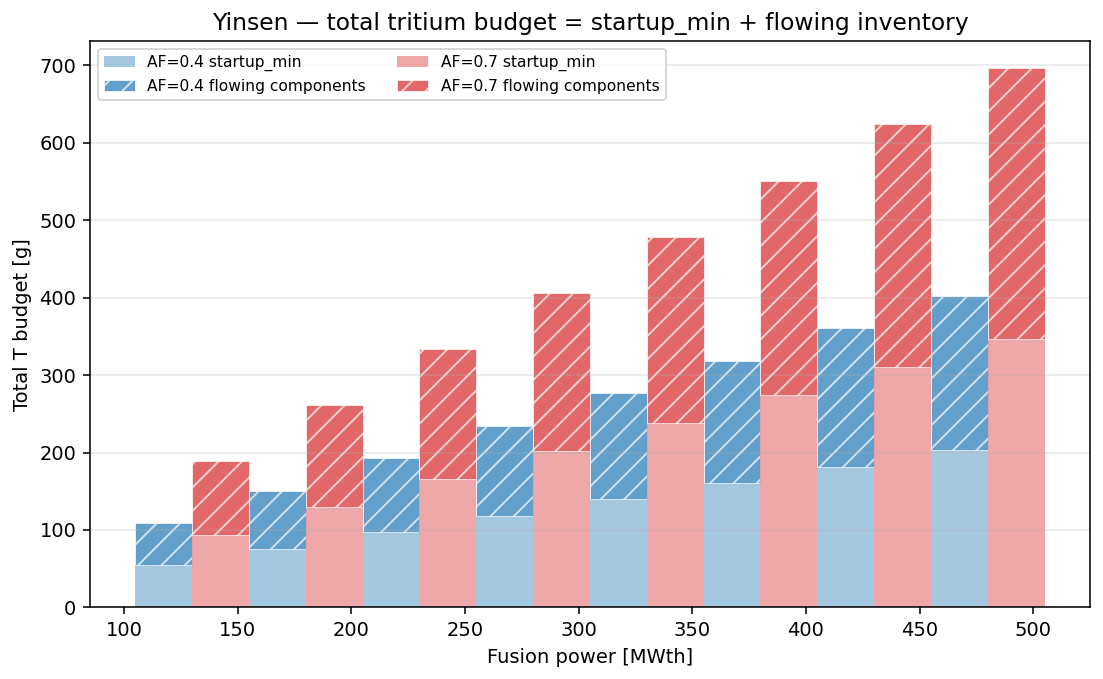}
    \caption{Total in-system tritium budget over the fusion-power and effective-availability sweep. Higher $AF$ increases the required startup budget almost in proportion to annual throughput, while the doubling time changes only weakly.}
    \label{fig:fuel_cycle_total_budget}
\end{figure}

\begin{figure*}[!t]
    \centering
    \includegraphics[width=0.98\textwidth]{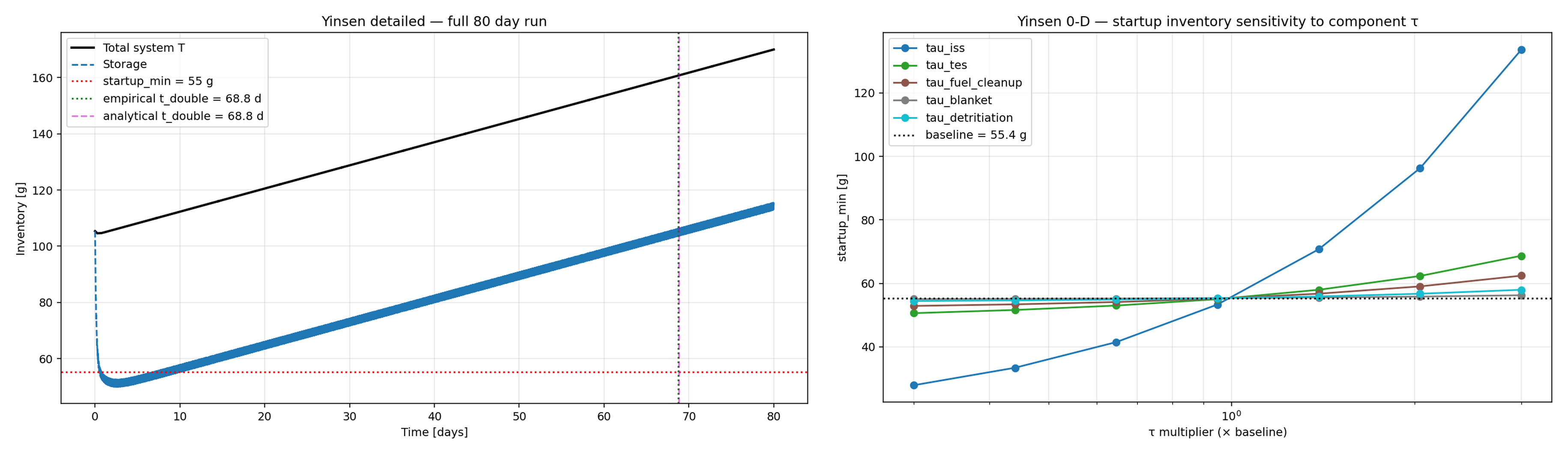}
    \caption{Fuel-cycle baseline and dominant startup-inventory sensitivity. Left: 80-day pulse-resolved inventory evolution for the $130~\mathrm{MW_{th}}$ baseline, showing the startup minimum together with the empirical and analytical doubling times. Right: startup-inventory sensitivity to dominant residence times. The ISS residence time is the dominant lever, while the blanket residence time has only a weak effect at the present Yinsen baseline.}
    \label{fig:fuel_cycle_baseline_sensitivity}
\end{figure*}

At the baseline point, the minimum startup inventory from the LTI shift solve is $55.35~\mathrm{g}$. An 80-day pulse-resolved run gives analytical and empirical doubling times of $68.83$ and $68.76$ days, respectively, showing that the closed-form doubling-time estimate is already accurate at this fidelity. To visualize the transient with operating margin, the pulse-resolved run is initialized with $105.35~\mathrm{g}$ in storage, i.e. the startup minimum plus $50~\mathrm{g}$, which yields a minimum storage inventory of about $50~\mathrm{g}$ during the early transient. The steady non-storage inventory is about $54~\mathrm{g}$ and is dominated by the isotope separation system (ISS, $39.3~\mathrm{g}$) and the tritium extraction system (TES, $9.6~\mathrm{g}$), while the blanket itself contains only about $0.47~\mathrm{g}$. In other words, the minimum tritium budget required to operate the baseline plant is already about $109~\mathrm{g}$ before any additional site reserve is added. Figure~\ref{fig:fuel_cycle_baseline_sensitivity} summarizes both the pulse-resolved baseline and the dominant startup sensitivity.

A $0.3\times$--$3\times$ sweep of the dominant component residence times shows that $\tau_{\mathrm{ISS}}$ is the main lever on both startup inventory and doubling time. Over that range, varying $\tau_{\mathrm{ISS}}$ changes the startup minimum by $105.8~\mathrm{g}$, compared with $18.1~\mathrm{g}$ for $\tau_{\mathrm{TES}}$ and only $1.2~\mathrm{g}$ for $\tau_{\mathrm{blanket}}$. The same ranking appears in doubling time. For Yinsen, therefore, the next fidelity target should be the ISS and recycle train rather than a more elaborate blanket-holdup model. For example, halving $\tau_{\mathrm{ISS}}$ from $3~\mathrm{h}$ to $1.5~\mathrm{h}$, the kind of reduction plasma centrifuges or superpermeable pumps could enable, drops the analytical doubling time from $68.8$ days to $44.4$ days and the startup minimum from $55.35~\mathrm{g}$ to $35.7~\mathrm{g}$, while doubling $\tau_{\mathrm{ISS}}$ to $6~\mathrm{h}$ pushes the doubling time out to $117.6$ days. Both bracket cases were confirmed by full pulse-resolved 80-day simulations, with empirical and analytical doubling times agreeing to better than $1\%$.

At fixed $TBR$, $TBE$, and $AF$, the startup minimum scales almost perfectly linearly with fusion power, $startup_{\min} \approx 0.426~\mathrm{g/MW} \cdot P_{\mathrm{fus}}$, rising from $55.35~\mathrm{g}$ at $130~\mathrm{MW}$ to $204.37~\mathrm{g}$ at $480~\mathrm{MW}$, while the doubling time remains essentially unchanged at $68.8$ days because both startup inventory and breeding surplus scale linearly with burn rate. Changing $AF$ has a different effect. Raising the effective availability factor from $0.40$ to $0.70$ increases the total tritium budget strongly but only weakly improves doubling time, so $AF$ behaves mainly as a revenue lever rather than a breeding lever. Figure~\ref{fig:fuel_cycle_total_budget} shows the resulting total in-system budget, defined here as the startup minimum plus the steady flowing inventory: it rises from about $109~\mathrm{g}$ to $189~\mathrm{g}$ at $130~\mathrm{MW}$ when $AF$ goes from $0.40$ to $0.70$, from $235~\mathrm{g}$ to $406~\mathrm{g}$ at $280~\mathrm{MW}$, and from $402~\mathrm{g}$ to $696~\mathrm{g}$ at $480~\mathrm{MW}$. For Yinsen, the practical implication is clear: the blanket remains important for achieving the required breeding ratio, but the next fidelity step for startup inventory and doubling time should target ISS holdup and recycle time first.

\FloatBarrier
\section{Facility Layout}\label{sec:facility_layout}

The facility layout includes the major plant systems required by the present Yinsen concept: the tokamak hall and bioshield, RF plant, cryoplant, vacuum pumping and tritium plant, heat-extraction and conversion equipment, magnet power supplies, solid-state transformers, and pulsed energy storage for rampup and inductive current drive. The aim is to determine whether the selected operating point can be accommodated within a compact, maintainable plant arrangement once those support systems are explicitly sized and accounted for.

\begin{figure*}[!t]
    \centering
    \begin{minipage}[c]{0.56\textwidth}
        \centering
        \includegraphics[width=\textwidth]{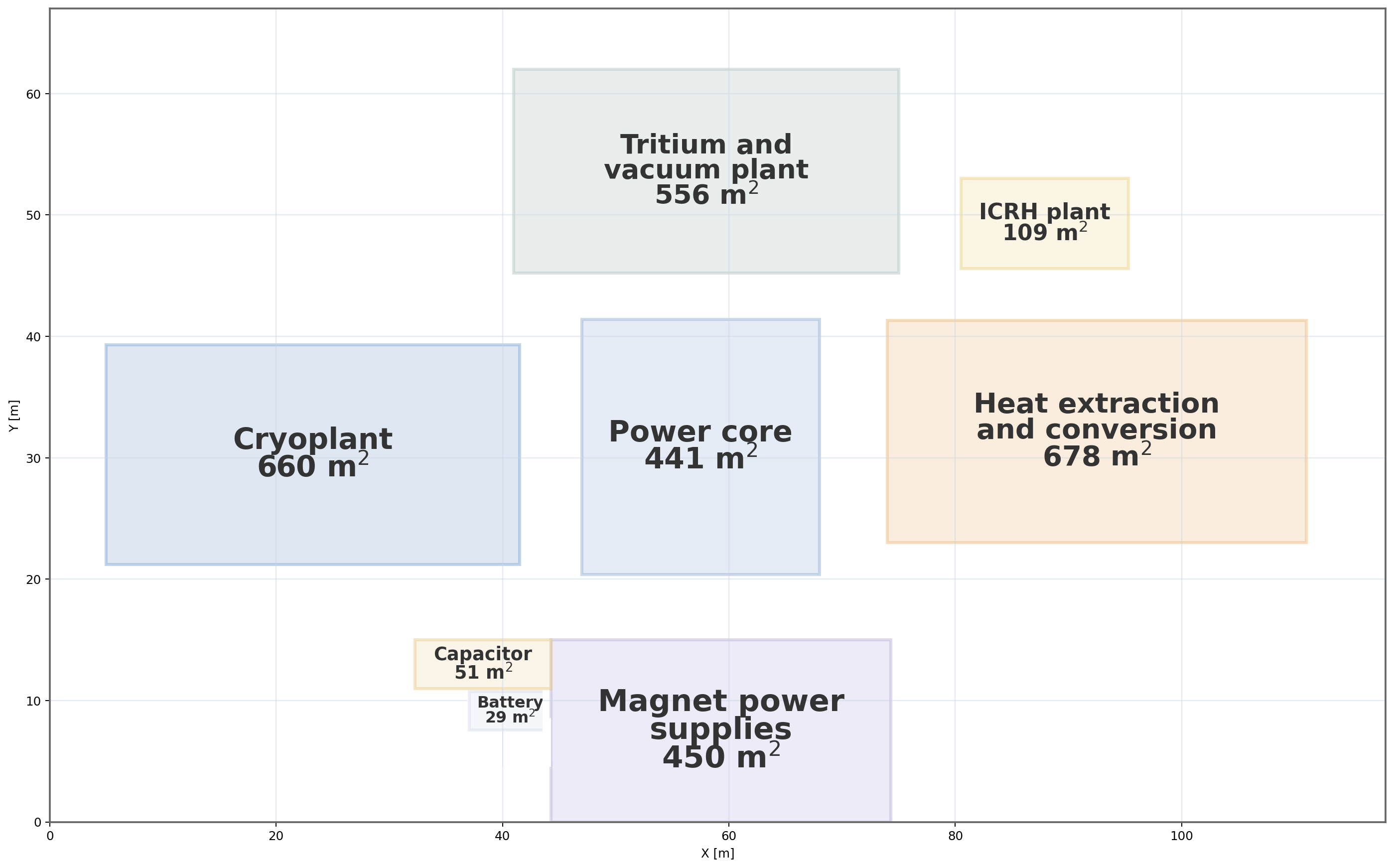}
    \end{minipage}\hfill
    \begin{minipage}[c]{0.36\textwidth}
        \centering
        \includegraphics[width=\textwidth]{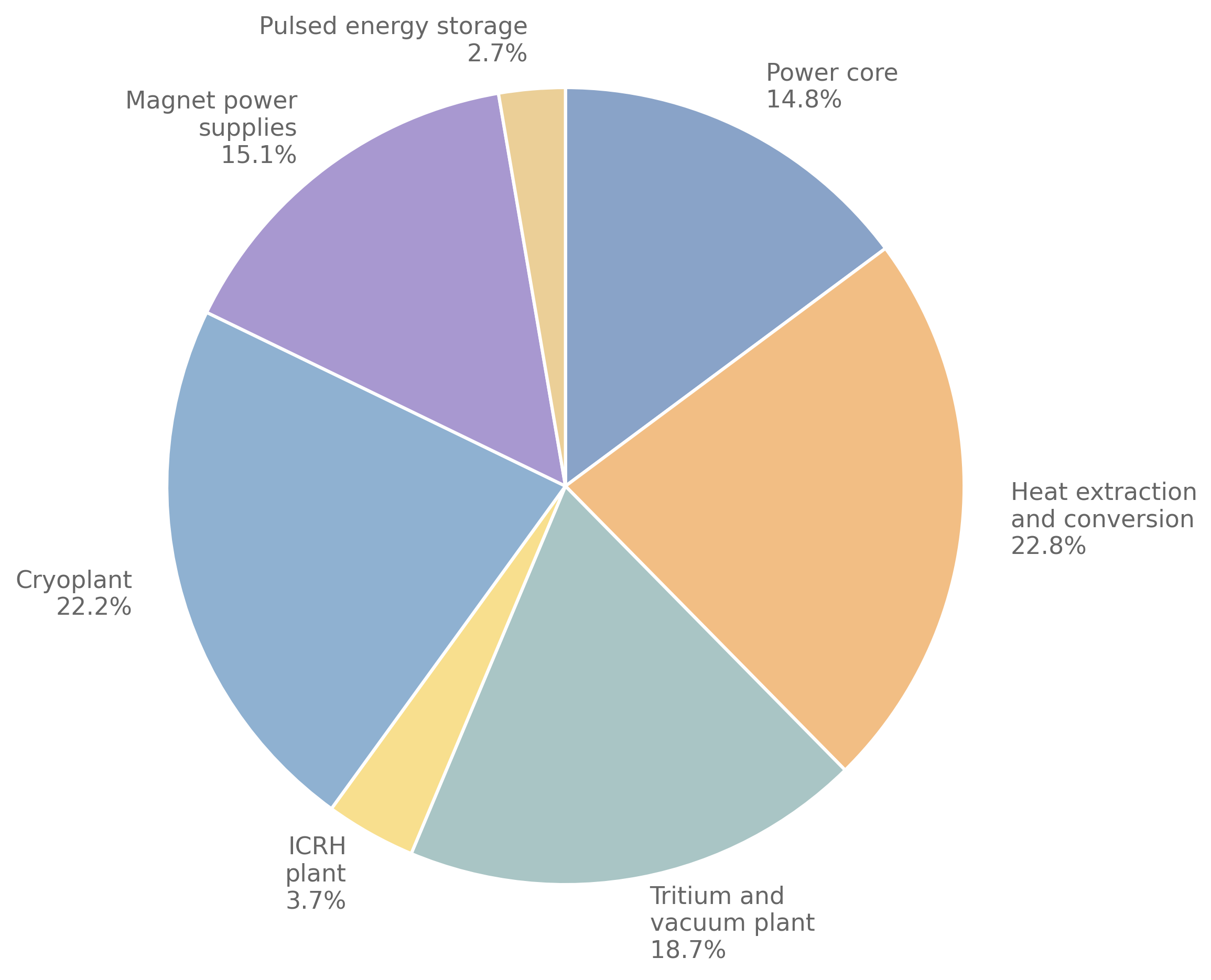}
    \end{minipage}
    \caption{Facility layout and subsystem footprint split from the unified layout and direct-capital model. Left: representative subsystem blocks arranged around the centered power core. Right: footprint share by major subsystem. The support plant occupies a footprint comparable to that of the tokamak hall and bioshield.}
    \label{fig:facility_layout_direct_capex}
\end{figure*}

The layout model is a block-level equipment representation and not intended to be an architectural design. Each major subsystem is represented by a rectangular block with an assigned plan area and installed height. The power-core envelope is derived using the cryostat radius and height together with the assumed \SI{4}{m} concrete bioshield. The remaining blocks are sized with linear screening laws using the relevant subsystem driver: component-specific heat-conversion drivers for the FLiBe pumps, PCHE, turbomachinery skid, and heat-rejection cooler; Tritium inventory and pumped torus volume for tritium and vacuum systems; installed auxiliary-heating power for ICRH; required \SI{20}{K} cooling duty for the cryoplant; installed peak magnet power for the magnet-supply block; and installed nameplate energy for battery and capacitor storage. Fixed aspect-ratio assumptions then convert those areas into packable rectangles. This is sufficient for first-order site closure, even though detailed internal equipment arrangement is not yet resolved. The main assumptions are summarized in Table~\ref{tab:facility_layout_assumptions}, and Appendix~\ref{app:facility_assumptions} records the full basis.

The reference arrangement shown in Figure~\ref{fig:facility_layout_direct_capex} uses the power-core block as the geometric and functional center of the site. Cryogenic support is grouped on one side of the tokamak to limit cold-distribution runs, while the heat-extraction and conversion island is placed on the opposite side to receive the hot FLiBe outlet directly. The magnet power supplies and pulsed-energy storage are kept adjacent to the tokamak hall to minimize high-current bus length, and the tritium, vacuum, and ICRH systems are placed nearby subject to \SI{6}{m} maintenance and access aisles. Because the heat-extraction and conversion block also contains the generator, keeping the principal electrical loads like the magnet power supplies and ICRH plant close to that block likewise reduces the length of high-power distribution runs.

\begin{table*}[!t]
    \centering
    \caption{Facility-layout assumptions used for the Yinsen block-packing estimate; see Appendix~\ref{app:facility_assumptions} for more details.}
    \label{tab:facility_layout_assumptions}
    \scriptsize
    \setlength{\tabcolsep}{2.5pt}
    \renewcommand{\arraystretch}{1.08}
    \begin{tabular}{>{\raggedright\arraybackslash}m{2.9cm} >{\raggedright\arraybackslash}m{5.0cm} >{\raggedright\arraybackslash}m{4.2cm} >{\raggedright\arraybackslash}m{3.2cm}}
        \toprule
        Subsystem & Area scaling law & Inputs used & Basis \\
        \midrule
        Power-core block & $A_{\mathrm{core}}=[2(R_{\mathrm{cryostat}}+t_{\mathrm{shield}})]^2$ & $R_{\mathrm{cryostat}}=\SI{6.5}{m}$, $t_{\mathrm{shield}}=\SI{4.0}{m}$ & Geometry-derived nuclear-island envelope \\
        \midrule
        Heat extraction and conversion & $A=0.0168\,\dot{V}_{\mathrm{FLiBe}}+0.0167\,Q_{\mathrm{HX}}+4.48\,P_{\mathrm{gross}}+0.2255\,Q_{\mathrm{reject}}$ & $\dot{V}_{\mathrm{FLiBe}}=\SI{1856}{m^3/h}$, $Q_{\mathrm{HX}}=\SI{300}{MWth}$, $P_{\mathrm{gross}}=\SI{135}{MWe}$, $Q_{\mathrm{reject}}=\SI{165}{MWth}$ & Five FLiBe pumps, PCHE, turbine, generator, recuperator, and marine seawater heat-rejection exchanger \\
        \midrule
        Tritium and vacuum plant & $A=520\,M_{\mathrm{T}}+N_{\mathrm{pump}}A_{\mathrm{pump,unit}}$ & $M_{\mathrm{T}}=\SI{0.70}{kg}$, $N_{\mathrm{pump}}=48$, $A_{\mathrm{pump,unit}}=\SI{2}{m}\times\SI{2}{m}$ & ITER scaled tritium-plant footprint using maximum inventory, plus doubled KSTAR major-pump inventory assuming \SI{2}{m} by \SI{2}{m} envelope per pump \\
        \midrule
        ICRH plant & $A=(5.33+1.20+0.76)\,P_{\mathrm{aux}}$ & $P_{\mathrm{aux}}=\SI{15}{MW}$ & Standalone RF electrical-conversion package: HVPS cabinets, amplifier cabinets, and RF combiner; launcher hardware and line routing excluded \\
        \midrule
        Cryoplant & $A=16.5\,Q_{20\mathrm{K}}$ & $Q_{20\mathrm{K}}=\SI{40}{kW}$ & Cryogenic package footprint scaled directly from the Absolut Systems Cryo-Cool 1000 product-page package at \SI{1}{kW}@\SI{20}{K} \\
        \midrule
        Magnet power supplies & $A=1.5\,P_{\mathrm{inst}}$ & $P_{\mathrm{inst}}=\SI{300}{MW}$ & Heron Link converter cabinet: \SI{1.5}{m} by \SI{5}{m} at \SI{5}{MW}, giving \SI{1.5}{m^2/MW} \cite{HeronLink2025} \\
        \midrule
        Battery storage & $A=3.47\,E_{\mathrm{bat}}$ & $E_{\mathrm{bat}}=\SI{8.33}{MWh}$ & Tesla Megapack scaled installed battery-storage block \\
        \midrule
        Capacitor storage & $A=73.3\,E_{\mathrm{cap}}$ & $E_{\mathrm{cap}}=\SI{0.69}{MWh}$ & Eaton XLHV scaled installed capacitor-bank block \\
        \bottomrule
    \end{tabular}
\end{table*}

Under these assumptions, the power-core envelope is \SI{21.0}{m} by \SI{21.0}{m}; the heat-extraction and conversion block is \SI{36.8}{m} by \SI{18.4}{m}; the tritium and vacuum block is \SI{33.3}{m} by \SI{16.7}{m}; the ICRH block is \SI{14.8}{m} by \SI{7.4}{m}; the cryoplant block is \SI{36.3}{m} by \SI{18.2}{m}; and the magnet power-supplies block is \SI{30.0}{m} by \SI{15.0}{m}. The battery and capacitor blocks contribute \SI{29}{m^2} and \SI{51}{m^2}, respectively. The tokamak hall is therefore not the sole site driver. The auxiliary facilities occupy a similarly important share of the footprint.

The largest area drivers in the reference arrangement are the heat-extraction block at \SI{678}{m^2}, the cryoplant block at \SI{660}{m^2}, the tritium and vacuum block at \SI{556}{m^2}, the magnet power-supplies block at \SI{450}{m^2}, and the power-core envelope at \SI{441}{m^2}. The heat-extraction block is sized here at $Q_{\mathrm{HX}}=\SI{300}{MWth}$, with gross electric output taken as $45\%$ of that duty and the balance rejected to seawater. That reference point is intentionally above the conservative vacuum-vessel lifetime design point discussed earlier, which lies near \SI{130}{MW} of fusion power. The reasoning is that once the nuclear island, shielding, magnets, and tritium systems are being built at all, sizing the surrounding conversion and support plant for the larger operating envelope adds a relatively modest increment in capital while preserving useful operating flexibility. The cryoplant footprint was similarly oversized to provide \SI{40}{kW}@\SI{20}{K}, and it was scaled directly from the Absolut Cryo-Cool 1000 packaging basis \cite{AbsolutCryoCool1000}. The tritium and vacuum block reflects a tritium inventory of \SI{700}{g}, together with a vacuum pump count taken as twice KSTAR's major-pump count \cite{Cristescu2007TritiumInventory,Yoshida2002TritiumPlant,Kim2009KSTARVacuum}. The magnet power-supplies block is scaled from the Heron Link converter-cabinet footprint, using a \SI{1.5}{m} by \SI{5}{m} cabinet at \SI{5}{MW} as the packaging basis \cite{HeronLink2025}. The standalone ICRH electrical-conversion package is smaller, at \SI{109}{m^2}; its present basis uses ITER-style tetrode-chain hardware \cite{Patel2024HVPS,Anand2025Combiner,Athaullah2018HPA}. However, we would likely opt for a solid-state implementation, which could reduce that footprint by more than a factor of two while also improving wall-plug efficiency, but that reduction is not credited here. If a smaller overall site becomes important, the strongest levers are heat-conversion packaging, cryogenic packaging density, tritium inventory, and vacuum package simplification rather than further reductions in the reactor envelope alone.

For marine use, this largely planar layout should be read as conservative. Large vessels normally distribute machinery across multiple decks rather than requiring everything on a single planar level. The cryoplant and RF systems could therefore be vertically integrated, and the electrical systems, including battery storage, capacitor storage, and magnet power supplies, could likewise be stacked on adjacent decks. For a shipboard version, the more consequential question is therefore not two-dimensional footprint alone, but three-dimensional routing of high-current electrical paths, cryogenic services, and maintenance access within the hull.

Routine in-vessel maintenance is instead assumed to follow a JET-like remote-handling philosophy, with first-wall tiles, diagnostics, and antenna hardware exchanged through dedicated tooling and remote-manipulator access rather than by routine vessel replacement \cite{JETRemoteHandling1999,JETSykes2011}. The vacuum vessel is therefore not treated as a regularly replaced component. If vessel replacement, or another exceptional intervention of similar scale, were ever required, the fallback approach remains the one discussed in the neutronics section: the vessel would be separated into two halves, the halves translated apart, and each section withdrawn through the TF-coil cage. The layout must therefore preserve crane coverage, flooded-maintenance clearances, extraction corridors, and short-term laydown space near the tokamak hall. Figure~\ref{fig:facility_layout_direct_capex} should accordingly be read as a maintenance-aware equipment arrangement rather than as a minimum-area packing exercise.

\section{Direct Overnight Capital Cost}\label{sec:direct_capex}

\begin{figure*}[!t]
    \centering
    \includegraphics[width=0.78\textwidth]{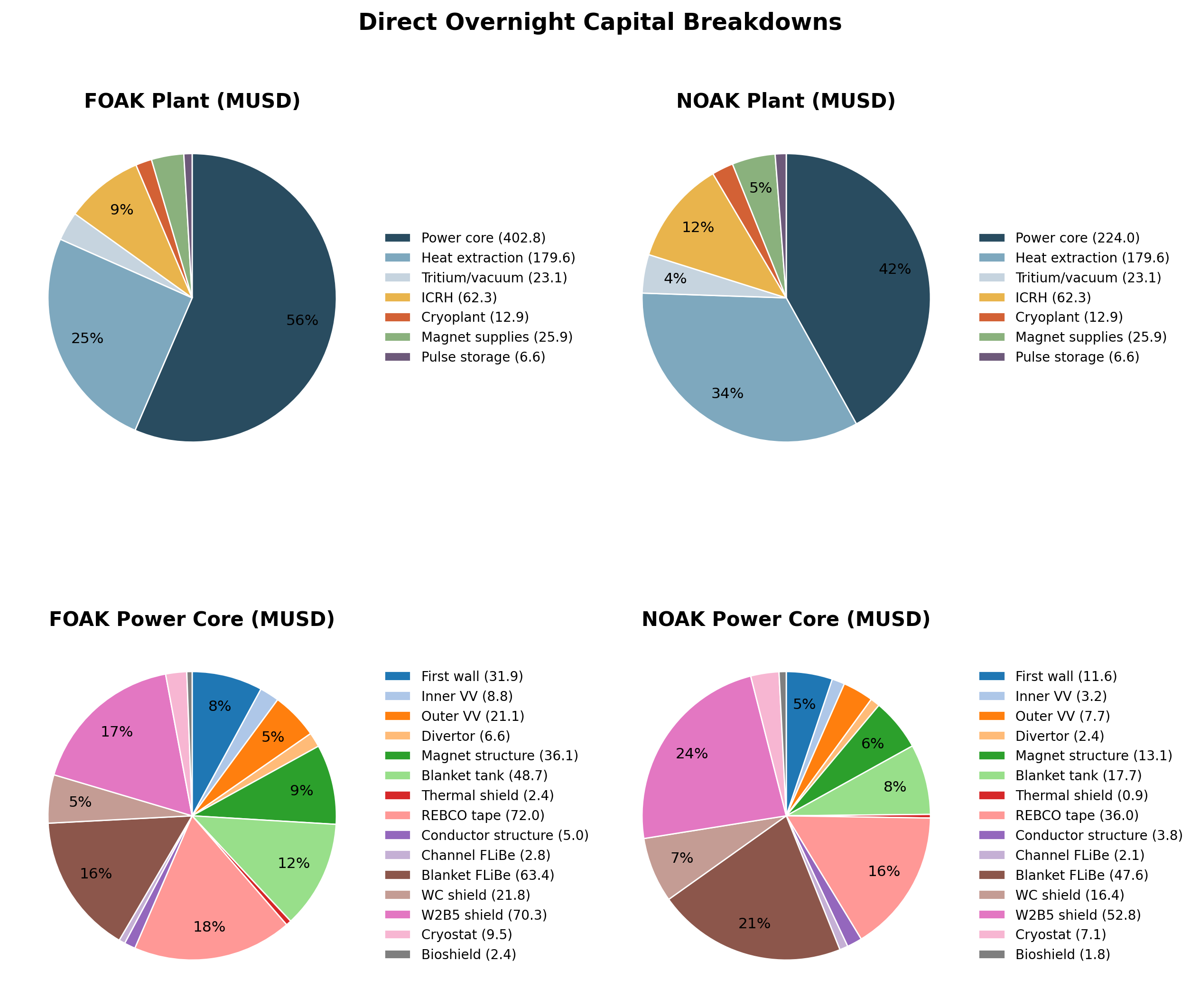}
    \caption{FOAK and NOAK direct overnight-capital breakdowns for the full plant (top) and power core (bottom).}
    \label{fig:direct_capex_breakdowns}
    \vspace{\baselineskip}
    \refstepcounter{table}
    {\footnotesize\raggedright\textbf{Table \thetable.} Yinsen power-core direct overnight-capital basis.\par}
    \label{tab:yinsen_direct_capex_core}
    \tiny
    \setlength{\tabcolsep}{1.6pt}
    \renewcommand{\arraystretch}{0.90}
    \resizebox{\textwidth}{!}{%
    \begin{tabular}{>{\raggedright\arraybackslash}m{2.35cm} >{\raggedright\arraybackslash}m{1.8cm} c c c >{\raggedright\arraybackslash}m{2.3cm} c c}
        \toprule
        Component & Material & Volume (m$^3$) & Mass (tonnes) & Material Cost & Fabricated Cost & Built Cost FOAK & Built Cost NOAK \\
        \midrule
        First wall & Tungsten & 5.27 & 101.5 & \SI{2.9}{MUSD} & 10$\times$ material cost & \SI{31.9}{MUSD} & \SI{11.6}{MUSD} \\
        \midrule
        Inner VV & V--4Cr--4Ti & 3.60 & 22.0 & \SI{0.8}{MUSD} & 10$\times$ material cost & \SI{8.8}{MUSD} & \SI{3.2}{MUSD} \\
        \midrule
        Outer VV & V--4Cr--4Ti & 8.50 & 51.9 & \SI{1.9}{MUSD} & 10$\times$ material cost & \SI{21.1}{MUSD} & \SI{7.7}{MUSD} \\
        \midrule
        Divertor & Tungsten & 1.05 & 20.3 & \SI{0.6}{MUSD} & 10$\times$ material cost & \SI{6.6}{MUSD} & \SI{2.4}{MUSD} \\
        \midrule
        Magnet structure & Stainless steel & 40.76 & 328.0 & \SI{3.3}{MUSD} & 10$\times$ material cost & \SI{36.1}{MUSD} & \SI{13.1}{MUSD} \\
        \midrule
        Blanket tank & Hastelloy N & 8.79 & 77.9 & \SI{4.4}{MUSD} & 10$\times$ material cost & \SI{48.7}{MUSD} & \SI{17.7}{MUSD} \\
        \midrule
        Thermal Shield & Stainless steel & 2.94 & 22.0 & \SI{0.2}{MUSD} & 10$\times$ material cost & \SI{2.4}{MUSD} & \SI{0.9}{MUSD} \\
        \midrule
        REBCO tape & 12 mm REBCO & 3600 [km] & 45.0 & \SI{72.0}{MUSD} & \SI{20}{USD/m} & \SI{72.0}{MUSD} & \SI{36.0}{MUSD} \\
        \midrule
        Conductor structure & 40\% Cu / 60\% steel & 32.61 & 262.4 & \SI{2.5}{MUSD} & Same as material cost & \SI{5.0}{MUSD} & \SI{3.8}{MUSD} \\
        \midrule
        Channel FLiBe & FLiBe & 4.70 & 9.1 & \SI{1.4}{MUSD} & Same as material cost & \SI{2.8}{MUSD} & \SI{2.1}{MUSD} \\
        \midrule
        Blanket tank FLiBe & FLiBe & 106.16 & 206.0 & \SI{31.7}{MUSD} & Same as material cost & \SI{63.4}{MUSD} & \SI{47.6}{MUSD} \\
        \midrule
        WC Shield & Tungsten carbide & 13.32 & 208.2 & \SI{10.9}{MUSD} & Same as material cost & \SI{21.8}{MUSD} & \SI{16.4}{MUSD} \\
        \midrule
        W$_2$B$_5$ Shield & Tungsten boride & 43.31 & 563.0 & \SI{35.2}{MUSD} & Same as material cost & \SI{70.3}{MUSD} & \SI{52.8}{MUSD} \\
        \midrule
        Cryostat & Steel & 59.88 & 473.0 & \SI{4.7}{MUSD} & Same as material cost & \SI{9.5}{MUSD} & \SI{7.1}{MUSD} \\
        \midrule
        Bioshield & Concrete & 4689.0 & 10784.0 & \SI{1.2}{MUSD} & Same as material cost & \SI{2.4}{MUSD} & \SI{1.8}{MUSD} \\
        \midrule
        Total power core & -- & -- & 13174.3 & \SI{173.7}{MUSD} & -- & \SI{402.8}{MUSD} & \SI{224.0}{MUSD} \\
        \bottomrule
    \end{tabular}
    }
    \vspace{-2pt}
\end{figure*}

In the absence of a validated FOAK fusion EPC cost model, and because studies such as ARIES and Sheffield are framed as top-down plant estimates rather than Yinsen-specific hardware inventories \cite{ARIESCostAccount2013,Sheffield1986Cost}, the present section uses a bottom-up direct overnight-capital estimate. The power core is costed from the current reactor-CAD inventory, with Table~\ref{tab:yinsen_direct_capex_core} reporting raw material costs, the fabrication cost basis applied to each component, and the resulting built costs for FOAK and NOAK power cores. For the first wall, vacuum-vessel shells, divertors, thermal shield, and mechanical load structures, the FOAK case takes fabrication to be ten times the material cost, while the NOAK case reduces that fabrication cost to three times the material cost. For comparison, PyFECONS \cite{WoodruffScientificPyFECONS2026} applies material multipliers of three to tungsten first-wall and magnet hardware rather than the stronger fabrication allowance adopted here. For the conductor structure (SHIELD cable), channel FLiBe, blanket-tank FLiBe, WC shield, W$_2$B$_5$ shield, cryostat, and bioshield, the FOAK case takes fabrication equal to the material cost, so the built cost is twice the material cost. In the NOAK case, those same simpler assembly rows are reduced to a built cost of 1.5 times the material cost. The shield rows are treated this way because bulk radiation shielding is mechanically simpler to assemble than vacuum-vessel shells, divertors, and superconducting-magnet structures. REBCO tape is treated separately as a purchased product, priced at \SI{20}{USD/m} in the FOAK case and \SI{10}{USD/m} in the NOAK case. Table~\ref{tab:yinsen_direct_capex_aux} gives the corresponding auxiliary-system costs as functions of the dominant design variables. The result is a direct overnight-capital screening estimate, not a bankable EPC number: contingency, owner's cost, interest during construction, licensing, O\&M, replacement campaigns, marine repackaging, and additional tokamak hardware such as ICRH antennas, waveguides, diagnostics, and similar systems are excluded.

The total FOAK direct overnight capital cost for Yinsen is estimated to be \SI{713.2}{MUSD}, of which \SI{402.8}{MUSD} sits in the power core and \SI{310.4}{MUSD} in the auxiliary systems. With the revised low-field-side shield treatment and the assumed reduction in fabricated cost for the NOAK case, the power-core total falls to \SI{224.0}{MUSD}, while the auxiliary block remains unchanged. Even with mass production and a strong learning rate for tokamak-core fabrication, the auxiliary plant still sets a substantial cost floor and the power core remains materially significant because the breeder and shielding inventories are now dominated more by raw material cost than by fabrication.

In the FOAK case, REBCO tape, the optimized W$_2$B$_5$ shield, and the blanket-salt inventory remain the dominant rows. The low-field-side shield thinning discussed earlier reduces the W$_2$B$_5$ burden materially, but the shield row still remains large because the shielding inventory itself is still substantial. The first-wall row is also elevated because the present geometry carries a thick \SI{2.5}{cm} tungsten layer, which produces a large tungsten mass and then applies a strong fabrication allowance to that mass. A tile-based tungsten first wall would likely come in lower than this screening treatment. There is also a clear opportunity to reduce shielding mass with more advanced neutronics materials, including hafnium-doped concepts explored earlier in this study. In a world where tokamak cores containing vacuum vessels, divertors, and HTS magnets become commoditized, the molten-salt blanket and shielding-material inventories can themselves become major capital-cost drivers.

With the shield rows treated as simple assemblies, the strongest sensitivities are now the adopted W$_2$B$_5$ bulk-procurement basis and the REBCO tape price. After that, the main drivers are the \SI{1330}{USD/kWe} heat-extraction rule, the \SI{4.15}{USD/W} ICRH assumption, the \SI{86.2}{USD/kVA} power-electronics basis for magnet power supplies, and the ITER-scaled tritium-plant cost basis. The conductor-structure row has a lower fabrication cost than other cryostat-inward components due to the simplified SHIELD cable manufacturing process, and the same simple-assembly rows are taken at 1.5 times material cost in the NOAK case rather than 2 times material cost in the FOAK case. Even with those uncertainties, the present estimate is still useful for showing how strongly the overall plant total depends on material procurement and fabrication maturity together.

It is also important that Yinsen is not being laid out as a conventional large-campus fusion facility with many separate support buildings and long utility runs. Because the intended applications are off-grid and marine-adjacent, much of the auxiliary hardware can be co-located tightly: the magnet power supplies, ICRH plant, and cryoplant can be grouped together, and the compact \SI{34}{kV} microgrid architecture built around solid-state transformers reduces the extent of high-voltage distribution and switchyard infrastructure that would otherwise add materially to site cost and size.

As a rough external benchmark, the DTT facility now under construction at ENEA Frascati reports an investment cost of about \SI{614}{MEUR}; it is a roughly \SI{2.1}{m}-major-radius, \SI{6}{T} device with \SI{45}{MW} of auxiliary heating and Nb$_3$Sn-based superconducting coils \cite{DTTBrochure2024,DTTInterim2019}. That is not an apples-to-apples comparison, since DTT is an experimental divertor test device rather than a breeding power-core concept, but it is still a useful scale anchor. Given Yinsen's larger overall machine size together with its HTS-based magnet system, breeding blanket, shielding inventory, and power-production balance of plant, a FOAK direct-capital estimate of about \SI{713}{MUSD} appears realistic as a first-pass screening value rather than obviously out of range with other modern fusion-class facilities.

\section{Fusion Power Shipping Economics}

To test whether Yinsen-like fusion systems could matter economically in maritime service rather than only as stationary plants, a separate containership propulsion study was carried out by recreating the LCOT model of Locatelli \emph{et al.} for nuclear propulsion of large container ships \cite{Locatelli2026}. That model computes levelized cost of transport (LCOT) from discounted variable OPEX, discounted fixed OPEX, and initial CAPEX while holding ship size, speed, trip length, utilization, and carbon-tax assumptions fixed across competing propulsion options. Before introducing the fusion case, the recreated fission cases were checked against the published results. For the cases of interest, the recreation reproduces the published values to within roughly 1--3\% for the average and unfavorable cases, and it reproduces the same technology ranking across the speed and carbon-tax scenarios reported in the paper. The combustion cases retain a modest absolute LCOT offset, but the analysis below is used mainly as a break-even CAPEX study rather than as a claim of exact absolute transport cost, so that residual calibration error is not central to the fusion comparison. Appendix~\ref{app:shipping_assumptions} gives the detailed shipping-economics basis, including the LCOT equations, mission derivation, propulsion-to-electric reinterpretation, fusion-case assumptions, and the break-even interpolation procedure. The shipping assumptions begin from the Locatelli baseline so that the fusion case remains anchored to the published nuclear-shipping comparison. The reference vessel is still a \num{12000} TEU Neo-Panamax containership, 70\% space utilization, and a port productivity of 210 container moves per hour. Two operational variables are then changed for the fusion case to reflect the deployment profile that is more plausible for a first-of-a-kind fusion-propulsion vessel. The representative trip length is reduced from \SI{8174.8}{nmi} to \SI{2500}{nmi}, corresponding to a dedicated transatlantic liner service linking a small number of certified home ports on the US East Coast, Western Europe, or West Africa. The annual time utilization is also reduced from 98.4\% to 70\% to account for longer commissioning periods, more frequent inspection windows for blanket and tritium systems, and longer drydock cycles than would be acceptable for a mature propulsion technology. Note that utilization in this context is more representative of availability in power plant context, meaning the portion of days per year in operation; it is not capacity factor nor representative of power throughput.

\twocolumn[{%
\begin{center}
\refstepcounter{table}
{\footnotesize\raggedright\textbf{Table \thetable.} Yinsen auxiliary-system direct overnight-capital estimate.\par}
\label{tab:yinsen_direct_capex_aux}
\vspace{3pt}
\footnotesize
\setlength{\tabcolsep}{4pt}
\renewcommand{\arraystretch}{1.08}
\resizebox{\textwidth}{!}{%
\begin{tabular}{>{\raggedright\arraybackslash}m{3.25cm} >{\raggedright\arraybackslash}m{3.1cm} >{\raggedright\arraybackslash}m{3.3cm} c >{\raggedright\arraybackslash}m{5.8cm}}
    \toprule
    Subsystem & Cost scaling law & Value & Cost & Basis \\
    \midrule
    Heat extraction and conversion & \SI{1330}{USD/kWe} & $P_{\mathrm{gross}}=\SI{135}{MWe}$ & \SI{179.6}{MUSD} & NREL concentrated-solar molten-salt power tower with sCO$_2$ cycle, Table~4 \cite{Augustine2023CSP}. \\
    \midrule
    Tritium and vacuum plant & \SI{32000000}{USD/kg} + \SI{14000}{USD/pump} & $M_{\mathrm{T}}=\SI{700}{g}$, $N_{\mathrm{pump}}=48$ & \SI{23.1}{MUSD} & ITER tritium-plant cost of \SI{96}{MUSD} scaled by \SI{3}{kg} maximum inventory. Doubled KSTAR vacuum pump count, \SI{14000}{USD/pump} \cite{Araiinejad2025AppliedEnergy}. \\
    \midrule
    ICRH plant & \SI{4.15}{USD/W} & $P_{\mathrm{aux}}=\SI{15}{MW}$ & \SI{62.3}{MUSD} & PyFECONS average ICRF-heating cost basis \cite{WoodruffScientificPyFECONS2026}. \\
    \midrule
    Cryoplant & \SI{322000}{USD/(kW@20K)} & $Q_{20\mathrm{K}}=\SI{40}{kW}$ & \SI{12.9}{MUSD} & Sumitomo RDK-500B2 quote converted to 2026 USD and normalized by a \SI{45}{W@20K} unit rating \cite{LabbaseRDK500B2Quote,FedH10USDCNY2026}. \\
    \midrule
    Magnet power supplies & \SI{86.2}{USD/kVA} & $P_{\mathrm{inst}}=\SI{300}{MW}$ & \SI{25.9}{MUSD} & Scaling from a \SI{3.3}{kV} SiC-based converter technoeconomics study \cite{Zheng2022SST}. \\
    \midrule
    Pulsed energy storage & \SI{360000}{USD/MWh} (battery) + \SI{5136000}{USD/MWh} (capacitor) & $E_{\mathrm{bat}}=\SI{8.33}{MWh}$, $E_{\mathrm{cap}}=\SI{0.694}{MWh}$ & \SI{6.6}{MUSD} & Tesla Megapack battery-storage basis together with Eaton XLHV capacitor-bank basis \cite{TeslaMegapack2026,NRELATB2025Battery,EatonXLHV2023,EatonSupercapacitorGuidelines}. \\
    \midrule
    Total auxiliary systems & -- & -- & \SI{310.4}{MUSD} & -- \\
    \bottomrule
\end{tabular}
}
\end{center}
}]

  These concessions make the shipping comparison more conservative, but they are a more credible FOAK operating profile. The financial basis still uses a 25-year amortization period and a 9.3\% discount rate. Operating points are evaluated at 17, 19, 22, and 25 kn under both today's ETS carbon price of \SI{74}{USD/tCO2} and a 2035 forecast case of \SI{220}{USD/tCO2}. For the fusion case, the reactor fixed-OPEX adder is initially held equal to the fission case, the startup tritium inventory is taken as the \SI{55.4}{g} value from the Yinsen fuel-cycle baseline, tritium is priced at \SI{30000}{USD/g}, annual tritium make-up is conservatively set to zero despite the modeled breeding surplus which could be another revenue stream, and the engine learning rate is set equal to the fission case. The same mission assumptions imply a specific electrical-demand envelope for a fusion-electric ship. Sea days are obtained from the shortened \SI{2500}{nmi} route and ship speed, while port days are obtained from unloading and reloading the full \num{16800} container moves at the stated port productivity. Shaft power is then back-calculated from the Locatelli fuel-consumption regression using VLSFO calorific value and a representative 50\% large-diesel efficiency, after which the fusion-electric sea-load case is obtained by dividing shaft power by a 95\% electric-drive efficiency and adding a \SI{3}{MWe} hotel load. During port stays, cold ironing is assumed, so hotel loads are supplied from shore power rather than from the reactor, which is beneficial because very low net power still incurs a high recirculating-power burden that must also be supplied by fusion power and would therefore consume substantial vacuum-vessel life for minimal delivered power. The resulting electrical demand is summarized in Table~\ref{tab:shipping_power_requirements}. The corresponding electric power rises from \SI{27.9}{MW} at 17 kn to \SI{78.6}{MW} at 25 kn, implying vacuum-vessel lifetimes from \SI{21.1}{yr} down to \SI{12.2}{yr} when considering vessel availability and cold ironing.

\begin{table}[H]
    \centering
    \caption{Mission power requirements for the revised FOAK Atlantic deployment interpreted as a fusion-electric vessel. $P_{\mathrm{shaft}}$ is mechanical shaft power [MW], $P_{\mathrm{elec,sea}}$ is the net electric power required during steaming [MWe], and the last two columns give the corresponding fusion-duty integral over 25 years and the implied vacuum-vessel lifetime.}
    \label{tab:shipping_power_requirements}
    \scriptsize
    \setlength{\tabcolsep}{0pt}
    \renewcommand{\arraystretch}{1.04}
    \begin{tabular*}{\columnwidth}{@{\extracolsep{\fill}}lcccccc@{}}
        \toprule
        \shortstack[c]{Speed} &
        \shortstack[c]{Sea\\d/yr} &
        \shortstack[c]{Port\\d/yr} &
        \shortstack[c]{$P_{\mathrm{shaft}}$} &
        \shortstack[c]{$P_{\mathrm{elec,sea}}$} &
        \shortstack[c]{MW$_{\mathrm{fusion}}$\\$\cdot$yr} &
        \shortstack[c]{VV lifetime\\{[}yr{]}} \\
        \midrule
        17 kn & 166 & 90 & 23.6 & 27.9 & 1233.4 & 21.1 \\
        19 kn & 159 & 97 & 32.6 & 37.3 & 1403.9 & 18.5 \\
        22 kn & 150 & 106 & 49.7 & 55.3 & 1726.5 & 15.1 \\
        25 kn & 142 & 114 & 71.9 & 78.6 & 2127.0 & 12.2 \\
        \bottomrule
    \end{tabular*}
\end{table}

That break-even framing is more useful here than a single headline LCOT because it turns the shipping problem directly into an engineering target. Table~\ref{tab:fusion_shipping_breakeven} gives the fusion-engine CAPEX required to match the main combustion-based competing propulsion options under the revised FOAK Atlantic deployment. The allowable fusion CAPEX rises with speed and with carbon price in the VLSFO and methanol cases, while the ammonia threshold is unchanged because ammonia is treated as zero direct CO$_2$ at the point of use.

\begin{figure}[!t]
    \centering
    \includegraphics[width=\columnwidth]{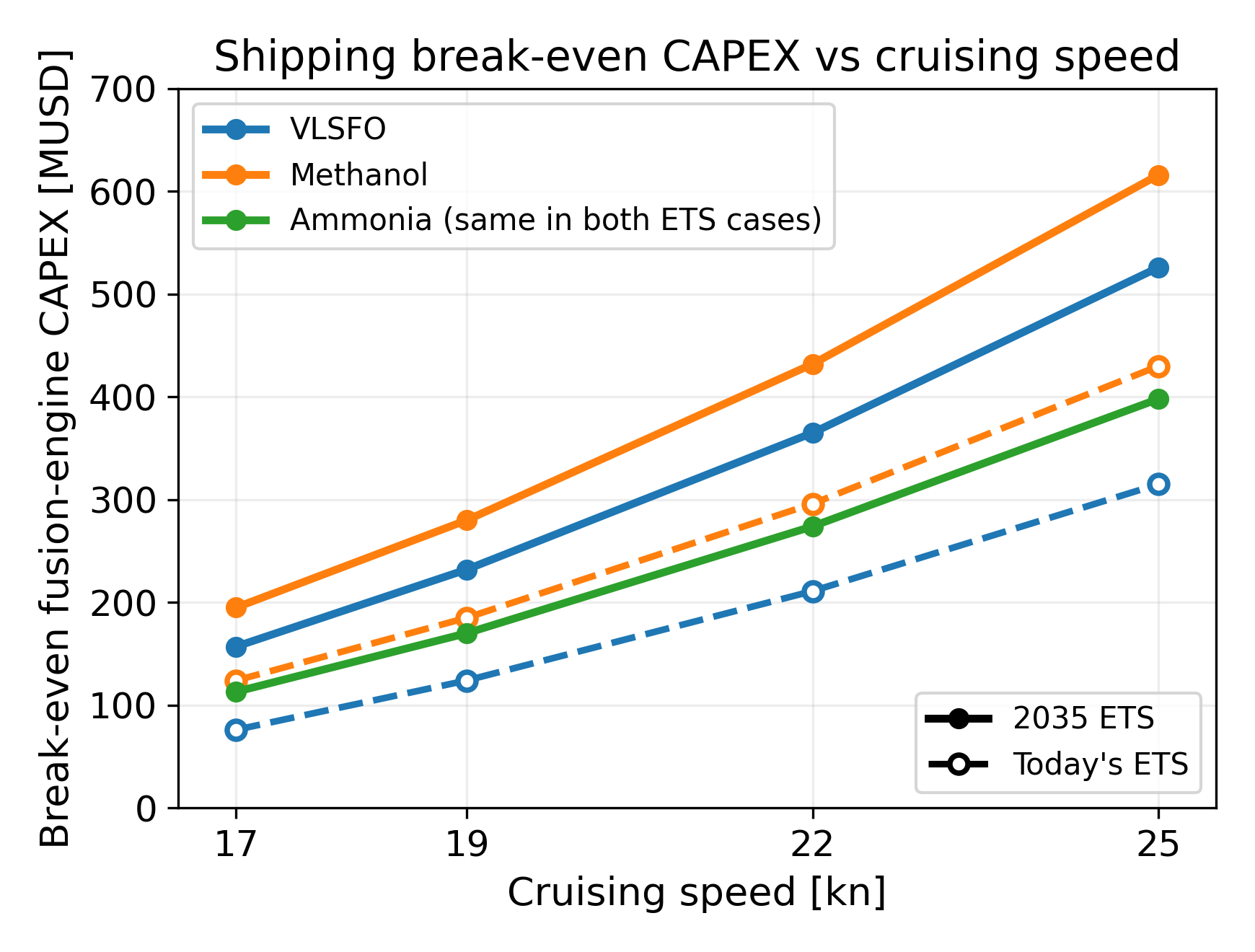}
    \caption{Marine fusion-engine CAPEX break-even versus cruising speed for the revised FOAK Atlantic deployment. Solid lines show the 2035 ETS case for VLSFO and methanol, dashed lines show today's ETS for those same fuels, and the ammonia line is unchanged because ammonia is treated here as zero direct CO$_2$ at the point of use.}
    \label{fig:shipping_breakeven_cases}
\end{figure}

Figure~\ref{fig:shipping_breakeven_cases} shows the 2035 ETS thresholds against ship speed. In that case, the allowable fusion-engine CAPEX rises to \SI{526}{MUSD} against VLSFO and \SI{616}{MUSD} against methanol at 25 kn. The ammonia threshold ranges from \SI{113}{MUSD} at 17 kn to \SI{398}{MUSD} at 25 kn because it is driven mainly by capital and fixed OPEX rather than carbon exposure. Taken together, the figure and Table~\ref{tab:fusion_shipping_breakeven} show that marine fusion economics remain strongly mission-dependent, with speed and carbon price still setting the most favorable conditions. These numbers have to be read with the right boundary in mind. 

\begin{table}[H]
    \centering
    \caption{Marine fusion-engine CAPEX break-even thresholds for the revised FOAK Atlantic deployment [MUSD, 2024 USD].}
    \label{tab:fusion_shipping_breakeven}
    \scriptsize
    \setlength{\tabcolsep}{3.2pt}
    \renewcommand{\arraystretch}{1.02}
    \begin{tabular}{lcccc}
        \toprule
        \multicolumn{5}{c}{Today's ETS, \SI{74}{USD/tCO2}} \\
        \cmidrule(lr){1-5}
        Beat... & 17 kn & 19 kn & 22 kn & 25 kn \\
        \midrule
        VLSFO & 76 & 124 & 211 & 315 \\
        Methanol & 124 & 185 & 296 & 430 \\
        Ammonia & 113 & 170 & 274 & 398 \\
        \midrule
        \addlinespace[1.6ex]
        \multicolumn{5}{c}{2035 ETS, \SI{220}{USD/tCO2}} \\
        \cmidrule(lr){1-5}
        Beat... & 17 kn & 19 kn & 22 kn & 25 kn \\
        \midrule
        VLSFO & 157 & 232 & 365 & 526 \\
        Methanol & 195 & 280 & 432 & 616 \\
        Ammonia & 113 & 170 & 274 & 398 \\
        \bottomrule
    \end{tabular}
\end{table}

The present Yinsen reference direct-capital estimate of about \SI{0.71}{BUSD} is not itself a marine-engine CAPEX, so it cannot be inserted directly into Table~\ref{tab:fusion_shipping_breakeven}; the power level, packaging, shielding arrangement, balance-of-plant boundary, and regulatory envelope would all change in a shipboard design. Even so, the table is useful because it identifies the cost band a marine fusion system would need to reach under a conservative FOAK deployment profile. Under today's ETS, fusion would need to fall to roughly \SIrange{76}{315}{MUSD} to beat VLSFO across the speed range and \SIrange{124}{430}{MUSD} to beat methanol. In the more favorable 2035 ETS case, the allowable CAPEX rises to about \SI{526}{MUSD} against VLSFO and \SI{616}{MUSD} against methanol at 25 kn. Framed that way, the present FOAK estimate of about \SI{713}{MUSD} is still above the immediate head-to-head range, but it is not an order of magnitude away from competitiveness; the remaining gap is concentrated in a handful of dominant cost rows rather than spread uniformly across the plant.

That path becomes clearer in the learning case. In the present Yinsen screening model, the FOAK power core is about \SI{402.8}{MUSD}, while the NOAK case reduces that portion to about \SI{224.0}{MUSD} through lower fabrication cost in the more intricate hardware and a lighter 1.5$\times$ built-cost treatment for the simpler assembly rows. With the auxiliary block still at about \SI{310.4}{MUSD}, that corresponds to a total reference-plant cost of about \SI{534}{MUSD}. On the break-even chart, that already places the NOAK case directly in the competitive range for high-speed methanol service under the 2035 ETS case, and very close to parity with VLSFO in the same regime. Thus, while immediate competitiveness remains challenging, this marine-entry target is closer than the grid case, where FOAK tokamaks may have an LCOE several times higher than competing energy sources such as solar even after production learning \cite{Lindley2023FusionEconomics}. That conclusion is strengthened further by several operational advantages that are not yet credited in the simple LCOT screen. A fusion ship does not need to bunker conventional fuel, which gives the vessel more routing and scheduling flexibility and reduces dependence on fuel-supply infrastructure that can constrain commercial operations. It also makes sustained high-speed operation much more natural, because the usual penalty that pushes combustion ships toward slow steaming is largely removed once propulsion is no longer dominated by bunker-fuel cost. Fusion therefore retains one of the main strategic attractions often assigned to fission shipping, sustained high-speed transit without a fuel-cost penalty, while avoiding the separate port-access and regulatory complications associated with bringing enriched uranium, and in some cases HEU-based systems, into commercial harbors. In that sense, manufacturing learning in cryostat-inward hardware, paired with lower shield-material procurement, lower REBCO pricing, and marine-specific auxiliary packaging, points to a credible commercial entry path led by higher-speed service rather than by direct competition in the lowest-cost bulk-shipping regime. Reaching the maximum speed and power regime with a practical ship reactor would still either require advanced structural materials that can preserve lifetime in-vessel components, or acceptance of a once-per-lifetime vacuum-vessel replacement procedure as part of major midlife maintenance. But for the moderate-speed cases, a vacuum-vessel lifetime of approximately 20 years remains a reasonable minimum threshold for viability of the fusion-powered ship concept.

%% file: generated/bop_tables_generated.tex
\begin{table}[htbp]
\centering
\caption{Peak heat loads and corresponding mass-flow targets for the five primary FLiBe loops in the reduced-order BOP model.}
\scriptsize
\begin{tabular}{l l S[table-format=2.1] S[table-format=3.1]}
\toprule
{Loop} & {Description} & {$Q$ (\si{MW})} & {$\dot{m}_{\mathrm{peak}}$ (\si{kg/s})} \\
\midrule
A & Blanket tank                  & 91.5 & 314.9 \\
B & Inboard FLiBe channel         & 26.1 & 89.0 \\
C & Outboard FLiBe channel        & 49.8 & 170.2 \\
D & Upper divertor                & 19.2 & 60.7 \\
E & Lower divertor                & 19.2 & 58.7 \\
\midrule
  & \textbf{Total}                & \textbf{205.8} & \textbf{693.5} \\
\bottomrule
\end{tabular}
\label{tab:flibe-loops}
\end{table}

\begin{table*}[htbp]
\centering
\caption{Representative FLiBe heat-exchanger state points, averaged over the steady flattop portion of the transient run.}
\scriptsize
\begin{tabular}{c l S[table-format=1.2] S[table-format=3.1] S[table-format=4.0] S[table-format=4.0] S[table-format=3.0]}
\toprule
{Node} & {Location} & {$P$ (\si{MPa})} & {$T$ (\si{K})} & {$h$ (\si{kJ/kg})} & {$s$ (\si{J/kg\text{-}K})} & {$\dot{m}$ (\si{kg/s})} \\
\midrule
F4 & HX hot-side inlet  & 2.80 & 925.5 & 2210 & 560 & 608.9 \\
F5 & HX hot-side outlet & 2.80 & 800.0 & 1910 & 212 & 608.9 \\
\bottomrule
\end{tabular}
\label{tab:flibe-hx-state}
\end{table*}

\begin{table*}[htbp]
\centering
\caption{sCO\textsubscript{2} state points in cycle order, averaged over the steady flattop portion of the transient run.}
\scriptsize
\begin{tabular}{c l S[table-format=2.1] S[table-format=3.1] S[table-format=4.0] S[table-format=4.0] S[table-format=2.2] S[table-format=3.0]}
\toprule
{Node} & {Component} & {$P$ (\si{MPa})} & {$T$ (\si{K})} & {$h$ (\si{kJ/kg})} & {$s$ (\si{J/kg\text{-}K})} & {$v$ ($10^{-3}$\,\si{m^3/kg})} & {$\dot{m}$ (\si{kg/s})} \\
\midrule
1 & Compressor inlet & 8.0 & 310.0 & 382 & 1591 & 3.05 & 917.4 \\
2 & Regen. cold-side inlet & 25.0 & 383.0 & 426 & 1608 & 1.83 & 917.4 \\
3 & Heater inlet & 25.0 & 706.1 & 886 & 2502 & 5.46 & 917.4 \\
4 & Turbine inlet & 25.0 & 864.9 & 1084 & 2755 & 6.86 & 917.4 \\
5 & Regen. hot-side inlet & 8.0 & 722.8 & 925 & 2775 & 17.12 & 917.4 \\
6 & Condenser inlet & 8.0 & 337.2 & 466 & 1852 & 5.45 & 917.4 \\
\bottomrule
\end{tabular}
\label{tab:secondary}
\end{table*}

%% file: conclusions.tex
\section{Conclusion}

This paper has presented the physics and engineering basis for \textit{Yinsen}, a low-power-density HTS tokamak reactor concept aimed at off-grid applications such as maritime propulsion, remote power, and industrial energy supply. The central design choice is to anchor the reactor not to an aggressive grid-competitive power density, but to a materials-limited blanket-area-normalized fusion power density of \(P_f/S_b = 0.7~\mathrm{MW/m^2}\). With the present assumptions of a \(35~\mathrm{DPA}\) structural limit, a \(20\)-year plant lifetime, \(40\%\) utilization, and a first-order geometric damage-peaking factor, that choice leads naturally to a reactor class near \(130~\mathrm{MW}\) of fusion power, corresponding to a lifetime-integrated fusion throughput of about \(1040~\mathrm{MW\cdot yr}\). In return, the design avoids making routine replacement of the vacuum vessel and primary in-vessel structures a baseline operating requirement, which is one of the most important distinctions between Yinsen and more aggressive compact grid-fusion concepts.

Within that reactor class, integrated \textsc{FUSE} studies produce a self-consistent high-field baseline with a shaped \(9.29~\mathrm{T}\), \(9.67~\mathrm{MA}\) plasma sustained with \(15~\mathrm{MW}\) of ICRH at 140MHz. The resulting plasma gain is \(Q=9.6\) for the \(130~\mathrm{MW}\) case and \(Q=13.7\) for the \(185~\mathrm{MW}\) case. The broader value of this result is not that Yinsen pushes to the edge of plasma performance, but that it does not need to. The selected operating point remains compatible with H-mode, vertical controllability, and a practical off-grid recirculating-power budget while still providing a useful net-electric power. In that sense, the plasma results support the main thesis of the paper: once the plant is deliberately pulled back from the highest power-density regime, compact HTS tokamaks still retain enough performance headroom to form a credible reactor design.

The same pattern appears in the power-exhaust and neutronics results. Without impurity radiation, the UEDGE base case produces an outer-target peak heat flux close to \(100~\mathrm{MW/m^2}\), but neon-seeded detachment can reduce the target loading by nearly two orders of magnitude; for example, the \(P_{\mathrm{sol}}=27~\mathrm{MW}\), \(n_{\mathrm{sep}}=3\times10^{19}~\mathrm{m^{-3}}\), \(2.3\%\) neon case reduces the target heat flux to roughly \(1~\mathrm{MW/m^2}\). The resolved detached target profiles were also mapped onto an ITER-inspired FLiBe-cooled tungsten monoblock. In the most demanding selected case, \(P_{\mathrm{sol}}=35~\mathrm{MW}\), \(n_{\mathrm{sep}}=5\times10^{19}~\mathrm{m^{-3}}\), and \(0.8\%\) neon, the conservative smooth-tube \(14~\mathrm{mm}\)-diameter FLiBe channel model with added core radiation and volumetric nuclear heating gives a peak tungsten surface temperature of about \(1122^\circ\mathrm{C}\), while increasing the neon fraction to \(0.9\%\) restores margin by reducing the peak to about \(1026^\circ\mathrm{C}\). OpenMC neutronics calculations further show that the low-power-density strategy succeeds at its intended purpose: the vacuum vessel remains the lifetime-limiting solid structure, with a peak damage rate corresponding to \(20\) years at the \(130~\mathrm{MW}\) baseline, while the HTS magnet system remains comfortably a lifetime component. The TF inner leg fast-flux limit corresponds to roughly sixteen vacuum-vessel lifetimes, and total TF nuclear heating is only about \(7.4~\mathrm{kW}\) at \(20~\mathrm{K}\) at the baseline, rising to only about \(28.5~\mathrm{kW}\) even at the \(\sim 500~\mathrm{MW}\) upper operating envelope. In parallel, the blanket design still achieves \(TBR \approx 1.1\) with only about \(30\%\) \(^{6}\mathrm{Li}\) enrichment and no dedicated neutron multiplier layer, which materially simplifies the breeding concept.

At the plant level, the same baseline remains internally consistent. The balance-of-plant studies support a supercritical CO$_2$ cycle well matched to the FLiBe temperature window, and the transient primary/secondary-side modeling shows that repeated pulsed operation can be accommodated without driving the coolant loops or power-conversion system outside a repeatable operating envelope. Across the \(130\)--\(480~\mathrm{MW}\) fusion-power sweep, the integrated cycle efficiency remains near \(47.5\%\)--\(47.7\%\), while integrated plant efficiency improves from about \(42.6\%\) to \(50.4\%\). The pulsed-power architecture, based on a medium-voltage AC backbone, local storage, and solid-state interfaces, is therefore not a side issue but part of the core viability argument for an off-grid machine in this power class. The fuel-cycle analysis reaches a similar conclusion: the tritium challenge remains tractable, but the dominant inventory is not in the blanket. The minimum in-system tritium budget is about \(109~\mathrm{g}\) for the \(130~\mathrm{MW}\) baseline, and the dominant lever is the isotope separation and recycle system rather than blanket holdup. That is an important design clarification for future work, because it shifts the next fidelity target from blanket-centric tritium accounting toward ISS residence time and processing architecture.

Taken together, these results support a different commercialization path for tokamak fusion than the one usually implied by pilot-plant and DEMO discussions. Yinsen does not argue that low-power-density fusion is the final economic optimum. It argues that there is a credible intermediate regime in which an HTS tokamak can still achieve useful net electric output while materially reducing first-generation penalties in structural lifetime, shielding burden, power exhaust, tritium inventory, and plant complexity. In that sense, Yinsen should be viewed as a reactor class designed to make the first practical fusion systems easier to build and operate, not as a permanently capped endpoint for the technology.

For that reason, a FOAK tokamak is unlikely to find its first market in parts of the grid that are already served by the lowest-cost bulk electricity. Those markets reward the lowest capital cost per delivered megawatt-hour and can lean on large transmission networks to absorb variability, share reserves, and spread infrastructure risk. The better early fit is in the applications considered here: marine propulsion, islanded or remote grids, and industrial sites where fuel delivery, logistics, and on-site autonomy matter as much as headline LCOE. In those cases the comparison is usually against expensive transported fuels and operational constraints, not against the cheapest onshore electricity. That is why the marine and off-grid cases are central to the argument of this paper rather than side examples. They are where a first-generation fusion system has the clearest chance of being useful before it is cheap.

The lower power and lower power density also help with a more basic problem: getting from \(Q>1\) to a machine that produces enough net power to matter. Fusion studies often jump quickly from burning-plasma performance to pilot-plant scale, because high recirculating power makes small plants look unattractive. The Yinsen results suggest that there is still a worthwhile middle ground. By backing away from the most aggressive blanket loading, the design relaxes vacuum-vessel lifetime, shielding thickness, divertor heat flux, cryogenic duty, tritium inventory, and balance-of-plant transients at the same time, while still producing net electric power in the tens-of-megawatts range. That does not make it the right answer for the main grid, but it does make it relevant for ships, remote infrastructure, and off-grid industrial service. The value of the low-power-density strategy is therefore not simply that each individual constraint becomes easier; it is that the whole reactor begins to look more buildable.

The study also points to where the remaining hard problems actually sit. Many are not limits of plasma performance, but ordinary, if still demanding, system-engineering problems: pulsed-power packaging, tritium-processing residence time, cryogenic plant sizing, shielding procurement, maintenance access, and the cost of otherwise conventional balance-of-plant equipment. That is useful, because these are areas where progress can come from iteration, standardization, supply-chain maturity, and borrowing from adjacent industries, rather than from waiting on a new confinement breakthrough. The hybrid battery-capacitor architecture and the thermal inertia of the working fluids both suggest that pulsed operation can be handled in a practical way. Likewise, the fuel-cycle work indicates that the next important step is tighter control of isotope-separation and recycle holdup rather than a radically different blanket concept. Those are real engineering challenges, but they are also the kind that can plausibly improve from one design generation to the next.

Several follow-on tasks remain important. The most immediate are H-mode access, thermomechanical analysis of the FLiBe-cooled divertor, refinement of impurity transport and core-edge coupling in the detached-divertor regime, refinement of ISS-dominated fuel-cycle residence times, continued development of the pulsed-power and maintenance architecture, and more detailed cost and layout studies. However the main conclusion of this work remains that a compact HTS tokamak operating at deliberately reduced power density can preserve plasma performance while relaxing some of the hardest engineering constraints that confront more aggressive reactor concepts. For applications that value autonomy and fuel security, as much as absolute electrical power, that is already a significant and useful result.

%% file: appendix_pulsed_power.tex
\section{Pulsed-Power and Hybrid-Storage Assumption Basis}
\label{app:pulsed_power_assumptions}

Tables~\ref{tab:appendix_pulse_assumptions} and~\ref{tab:appendix_storage_assumptions} collect the pulse-definition, optimization, lifetime, and packaging assumptions used in Section~\ref{sec:pulse} for the battery/capacitor sizing model.

\vspace{-0.6em}
\begin{center}
\refstepcounter{table}
{\footnotesize\raggedright\textbf{Table \thetable.} Pulse assumptions.\par}
\label{tab:appendix_pulse_assumptions}
\vspace{2pt}
\tiny
\setlength{\tabcolsep}{2pt}
\renewcommand{\arraystretch}{0.94}
\begin{tabular}{>{\raggedright\arraybackslash}p{3.2cm} >{\raggedright\arraybackslash}p{2.9cm} >{\raggedright\arraybackslash}p{9.4cm}}
        \toprule
        Assumption & Value & Basis \\
        \midrule
        Pulse-usable demand envelope & $E_{\mathrm{req}}=\SI{1.5}{GJ}$ & Reference Yinsen CS/PF pulse-demand case from Section~\ref{sec:pulse}, used as the central hybrid-storage sizing point. \\
        Required lifetime pulse count & $N_{\mathrm{req}}=10^6$ & Design-study high-cycle pulse-service requirement used to distinguish battery- and capacitor-dominant solutions. \\
        Hybrid split sweep & $\alpha\in[0,1]$ with 101 points & Uniform capacitor-share sweep used to locate the minimum-cost hybrid dispatch. \\
        Battery DoD search window & $\mathrm{DoD}\in[0.01,0.40]$ & Conservative lithium-ion high-cycle pulse-service window informed by DOE grid-storage cycle-life guidance \cite{DOEESGC2022}. \\
        Battery C-rate search window & $C_{\mathrm{rate}}\in[0.2,6.0]$ & Search range spanning moderate to aggressive lithium-ion pulse service while letting the lifetime model choose the optimum \cite{DOEESGC2022}. \\
        Capacitor usable-fraction window & $u\in[0.05,0.80]$ & Electrostatic-storage voltage-window sweep, with $u_{\mathrm{ref}}=0.50$ retained as the reference life anchor following Eaton guidance \cite{EatonSupercapacitorGuidelines}. \\
        Reference temperatures & $T_{\mathrm{bat}}=T_{\mathrm{ref,bat}}=\SI{25}{\degreeCelsius}$; $T_{\mathrm{cap}}=T_{\mathrm{ref,cap}}=\SI{35}{\degreeCelsius}$ & Screening ambient temperatures used in the battery and capacitor life-derate models \cite{DOEESGC2022,EatonSupercapacitorGuidelines}. \\
        Capacitor power-density floor & $600~\si{MW/GJ}$ & Minimum capacitor nameplate enforced from pulse-power capability, consistent with high-voltage capacitor-bank packaging \cite{EatonXLHV2023,EatonSupercapacitorGuidelines}. \\
        Nameplate sizing margin & $1.05$ & Small post-optimization design margin so the selected nameplate does not sit exactly on the active constraint. \\
        Battery round-trip components & $\eta_{\mathrm{chg}}=\eta_{\mathrm{dis}}=0.96$ & Representative utility-scale battery charge/discharge efficiencies used in the dispatch model \cite{NRELATB2025Battery}. \\
        Capacitor round-trip components & $\eta_{\mathrm{chg}}=\eta_{\mathrm{dis}}=0.985$ & Representative high-voltage capacitor-bank efficiencies used in the same dispatch model \cite{EatonXLHV2023}. \\
        \bottomrule
    \end{tabular}
\end{center}

\vspace{-0.8em}
\begin{center}
\refstepcounter{table}
{\footnotesize\raggedright\textbf{Table \thetable.} Storage assumptions.\par}
\label{tab:appendix_storage_assumptions}
\vspace{2pt}
\tiny
\setlength{\tabcolsep}{2pt}
\renewcommand{\arraystretch}{0.94}
\begin{tabular}{>{\raggedright\arraybackslash}p{2.8cm} >{\raggedright\arraybackslash}p{2.6cm} >{\raggedright\arraybackslash}p{2.6cm} >{\raggedright\arraybackslash}p{7.8cm}}
        \toprule
        Assumption & Battery value & Capacitor value & Basis \\
        \midrule
        Reference life anchor & $0.80\!\to\!2400$, $0.70\!\to\!4500$, $0.60\!\to\!8000$, $0.30\!\to\!32000$, $0.05\!\to\!192000$ cycles & $N_{\mathrm{ref}}=10^6$ cycles at $u_{\mathrm{ref}}=0.50$ & DOE ESGC 2022 lithium-ion cycle-life anchors for depth-of-discharge dependence \cite{DOEESGC2022}; Eaton high-voltage capacitor lifetime guidance referenced to usable voltage swing \cite{EatonSupercapacitorGuidelines}. \\
        Life exponents used & $a_{\mathrm{bat}}=1.48$, $b_{\mathrm{bat}}=0.55$ & $n_{\mathrm{cap}}=1.40$ & $a_{\mathrm{bat}}$ is fitted to the DOE anchors; $b_{\mathrm{bat}}$ and $n_{\mathrm{cap}}$ are retained reduced-order screening exponents for hybrid optimization. \\
        Temperature derate & $k_{T,\mathrm{bat}}=\ln(2)/10~^\circ\mathrm{C}^{-1}$ & $k_{T,\mathrm{cap}}=\ln(2)/10~^\circ\mathrm{C}^{-1}$ & Common screening derate corresponding to factor-of-two life reduction per $+10^\circ$C, consistent with Eaton guidance and standard battery practice \cite{EatonSupercapacitorGuidelines,DOEESGC2022}. \\
        Cost per nameplate energy & $1.0\times10^5$ USD/GJ ($360$ USD/kWh) & $1.5\times10^6$ USD/GJ ($5400$ USD/kWh) & NREL utility-scale battery benchmark \cite{NRELATB2025Battery}; Eaton XLHV-derived capacitor-bank screening value \cite{EatonXLHV2023}. \\
        Effective installed volumetric density & $0.347$ GJ/m$^3$ ($96$ kWh/m$^3$) & $0.014$ GJ/m$^3$ ($3.9$ kWh/m$^3$) & Installed packaging densities back-calculated from Tesla Megapack and Eaton XLHV product dimensions and stored energy \cite{TeslaMegapack2026,EatonXLHV2023}. \\
        Installed mass assumption & $7.5$ t/GJ & $22$ t/GJ & Conservative installed-mass screening values for plant-layout and marine-packaging estimates relative to Tesla Megapack and Eaton XLHV hardware \cite{TeslaMegapack2026,EatonXLHV2023}. \\
        Container-class packaging & $18$ GJ/container & $0.5$ GJ/container & Representative package classes used to translate optimized storage nameplate into container-equivalent counts for layout and marine integration \cite{TeslaMegapack2026,EatonXLHV2023}. \\
        \bottomrule
    \end{tabular}
\end{center}

\clearpage

%% file: appendix_facility.tex
\section{Facility Layout Scaling-Law Basis}
\label{app:facility_assumptions}

Tables~\ref{tab:appendix_facility_power_core}--\ref{tab:appendix_facility_electrical_storage} collect the source basis for the facility scaling laws used in Section~\ref{sec:facility_layout}.

\begin{table}[H]
    \centering
    \caption{Power-core and packing basis.}
    \label{tab:appendix_facility_power_core}
    \footnotesize
    \setlength{\tabcolsep}{4pt}
    \renewcommand{\arraystretch}{1.06}
    \resizebox{\textwidth}{!}{%
    \begin{tabular}{>{\raggedright\arraybackslash}p{2.8cm} >{\raggedright\arraybackslash}p{3.4cm} >{\raggedright\arraybackslash}p{9.3cm}}
        \hline
        Item & Scaling law & Source \\
        \hline
        Cryostat envelope & $R_{\mathrm{cryostat}}, H_{\mathrm{cryostat}}$ from imported geometry & Current Yinsen cryostat envelope, with $R_{\mathrm{cryostat}}=\SI{6.5}{m}$ and $H_{\mathrm{cryostat}}\approx\SI{8.52}{m}$, taken directly from reactor CAD. \\
        \hline
        Concrete bioshield & $t_{\mathrm{shield}}=\SI{4}{m}$ & Present design-basis concrete shell thickness, carried consistently through Section~\ref{sec:facility_layout}, Table~\ref{tab:facility_layout_assumptions}, and the facility model. \\
        \hline
        Power-core block area & $A_{\mathrm{core}}=[2(R_{\mathrm{cryostat}}+t_{\mathrm{shield}})]^2$ & Shared tokamak-hall and bioshield envelope formed directly from the cryostat dimensions and the retained concrete-shield thickness. \\
        \hline
    \end{tabular}
    }
\end{table}

\begin{table}[H]
    \centering
    \caption{Heat-extraction and conversion scaling laws.}
    \label{tab:appendix_facility_heat_conversion}
    \footnotesize
    \setlength{\tabcolsep}{4pt}
    \renewcommand{\arraystretch}{1.06}
    \resizebox{\textwidth}{!}{%
    \begin{tabular}{>{\raggedright\arraybackslash}p{3.1cm} >{\raggedright\arraybackslash}p{4.2cm} >{\raggedright\arraybackslash}p{8.0cm}}
        \hline
        Item & Scaling law & Source \\
        \hline
        FLiBe pump footprint term & $A_{\mathrm{pump}}=0.0168\,\dot{V}_{\mathrm{FLiBe}}$ & Copenhagen Atomics molten-salt pump datasheet: \SI{300}{L/min}=\SI{18}{m^3/h} maximum flow and \SI{55}{cm} worst-case published dimension. Applying \SI{55}{cm} to both horizontal axes gives \SI{0.3025}{m^2} per \SI{18}{m^3/h} reference pump and therefore $0.0168\,\si{m^2/(m^3/h)}$ \cite{CopenhagenAtomicsPump2024}. \\
        \hline
        PCHE footprint term & $A_{\mathrm{PCHE}}=0.0167\,Q_{\mathrm{HX}}$ & Representative flattop state gives $Q_{\mathrm{HX}}\approx\SI{202}{MW}$ and counterflow $\Delta T_{\mathrm{lm}}\approx\SI{80}{K}$, so $UA\approx\SI{2.5}{MW/K}$. Ji \emph{et al.} report $U=\SIrange{1.10}{2.53}{kW/m^2.K}$ for an sCO$_2$ PCHE, and Tano \emph{et al.} report $>\SI{10}{MW/m^3}$ installed power density for a molten-salt-to-sCO$_2$ primary HX. With $q'''_{\mathrm{inst}}=\SI{10}{MW/m^3}$ and $H=\SI{6}{m}$, this gives $A_{\mathrm{PCHE}}=Q_{\mathrm{HX}}/(q'''_{\mathrm{inst}}H)=0.0167\,Q_{\mathrm{HX}}$ \cite{Ji2022PCHE,Tano2022AMPHE,Dostal2004}. \\
        \hline
        Turbomachinery skid term & $A_{\mathrm{turbo}}=4.48\,P_{\mathrm{gross}}$ & Published \SI{10}{MWe} WHRS core skid of about \SI{3.2}{m} by \SI{14}{m}, scaled linearly with gross electric output \cite{SwRI2020WHRS}. \\
        \hline
        Heat-rejection exchanger term & $A_{\mathrm{cool}}=0.2255\,Q_{\mathrm{reject}}$ & Bowman RK400-5883-6 tubular marine exchanger: \SI{2.5}{MW} typical duty and bounding rectangular plan area $1.392\times0.405=\SI{0.564}{m^2}$, giving $0.564/2.5=0.2255\,\si{m^2/MW}$ \cite{BowmanMarine2021}. \\
        \hline
        Combined heat-conversion block & $A_{\mathrm{HX}}=A_{\mathrm{pump}}+A_{\mathrm{PCHE}}+A_{\mathrm{turbo}}+A_{\mathrm{cool}}$ & Summed component footprint model using separate terms for FLiBe flow, exchanger duty, gross electric output, and rejected heat. \\
        \hline
    \end{tabular}
    }
\end{table}

\begin{table}[H]
    \centering
    \caption{Tritium and vacuum scaling laws.}
    \label{tab:appendix_facility_tritium_vacuum}
    \footnotesize
    \setlength{\tabcolsep}{4pt}
    \renewcommand{\arraystretch}{1.06}
    \resizebox{\textwidth}{!}{%
    \begin{tabular}{>{\raggedright\arraybackslash}p{3.1cm} >{\raggedright\arraybackslash}p{4.2cm} >{\raggedright\arraybackslash}p{8.0cm}}
        \hline
        Item & Scaling law & Source \\
        \hline
        Tritium-processing term & $A_{\mathrm{tri}}=(78\times20/3)\,M_{\mathrm{T}}=520\,M_{\mathrm{T}}$ & ITER tritium-plant building footprint of \SI{78}{m} by \SI{20}{m} for about \SI{3}{kg} tritium inventory, scaled directly with total tritium inventory \cite{Cristescu2007TritiumInventory,Yoshida2002TritiumPlant}. \\
        \hline
        Vacuum term & $A_{\mathrm{vac}}=N_{\mathrm{pump}}A_{\mathrm{pump,unit}}=48\times(2\times2)=192$ & Kim et al. list 24 major pumps across the KSTAR vacuum-vessel and cryostat pumping systems. Yinsen retains twice that inventory, and each major pump is assigned a conservative \SI{2}{m} by \SI{2}{m} plan envelope \cite{Kim2009KSTARVacuum}. \\
        \hline
        Combined tritium and vacuum block & $A_{\mathrm{tri+vac}}=520\,M_{\mathrm{T}}+192$ & Combined near-nuclear process block formed by adding the ITER-scaled tritium area to the retained KSTAR-derived vacuum package. \\
        \hline
    \end{tabular}
    }
\end{table}

\begin{table}[H]
    \centering
    \caption{KSTAR major-pump inventory retained as the reference vacuum-system package for Yinsen. Counts are taken from Kim et al. Table~1 and then doubled for the present Yinsen screening basis. Each retained pump is assigned a \SI{2}{m} by \SI{2}{m} plan envelope, giving \SI{4}{m^2} per pump.}
    \label{tab:appendix_facility_vacuum_inventory}
    \footnotesize
    \setlength{\tabcolsep}{3.5pt}
    \renewcommand{\arraystretch}{1.02}
    \begin{tabular*}{\textwidth}{@{\extracolsep{\fill}}>{\raggedright\arraybackslash}p{2.2cm} >{\raggedright\arraybackslash}p{5.0cm} >{\centering\arraybackslash}p{1.5cm} >{\centering\arraybackslash}p{1.7cm} >{\centering\arraybackslash}p{1.9cm}@{}}
        \hline
        System & Pump class & KSTAR count & Yinsen count & Yinsen area (\si{m^2}) \\
        \hline
        Vacuum vessel & Main pumps: 8 $\times$ OSAKA TG2810 TMP & 8 & 16 & 64 \\
        \hline
        Vacuum vessel & Subsidiary pumps: 2 $\times$ ULVAC U20HLK cryo-pump & 2 & 4 & 16 \\
        \hline
        Vacuum vessel & TMP backing pumps: 1 $\times$ EDWARDS iF1800 Dry & 1 & 2 & 8 \\
        \hline
        Vacuum vessel & Roughing pumps: 1 $\times$ EDWARDS iF1800 Dry & 1 & 2 & 8 \\
        \hline
        Cryostat & Main pumps: 4 $\times$ OSAKA TG2810 TMP + 3 $\times$ PFEIFFER TPH2301 TMP & 7 & 14 & 56 \\
        \hline
        Cryostat & Subsidiary pumps: 2 $\times$ ULVAC U20HLK cryo-pump & 2 & 4 & 16 \\
        \hline
        Cryostat & TMP backing pumps: 1 $\times$ EDWARDS iF1800 Dry & 1 & 2 & 8 \\
        \hline
        Cryostat & Roughing pumps: 2 $\times$ EDWARDS iF1800 Dry & 2 & 4 & 16 \\
        \hline
        Total & All major pumps & 24 & 48 & 192 \\
        \hline
    \end{tabular*}
\end{table}

\begin{table}[H]
    \centering
    \caption{ICRH scaling laws.}
    \label{tab:appendix_facility_icrh}
    \footnotesize
    \setlength{\tabcolsep}{4pt}
    \renewcommand{\arraystretch}{1.06}
    \resizebox{\textwidth}{!}{%
    \begin{tabular}{>{\raggedright\arraybackslash}p{3.1cm} >{\raggedright\arraybackslash}p{4.2cm} >{\raggedright\arraybackslash}p{8.0cm}}
        \hline
        Item & Scaling law & Source \\
        \hline
        ICRH HVPS term & $A_{\mathrm{HVPS}}=\frac{2(8\times1)}{3}\,P_{\mathrm{aux}}=5.33\,P_{\mathrm{aux}}$ & Patel et al. report one dual-output HVPS cabinet of size \SI{8}{m} by \SI{1}{m} by \SI{2.4}{m} for each \SI{1.5}{MW} ITER ICRH chain; two chains are combined for one \SI{3}{MW} source \cite{Patel2024HVPS}. \\
        \hline
        ICRH amplifier-cabinet term & $A_{\mathrm{amp}}=\frac{4(0.79\times1.14)}{3}\,P_{\mathrm{aux}}=1.20\,P_{\mathrm{aux}}$ & The ITER-India source architecture uses two parallel chains combined to one \SI{3}{MW} source; the FEC amplifier preprint reports an HPA cabinet envelope of \SI{790}{mm} by \SI{1140}{mm} by \SI{2673}{mm} \cite{Athaullah2018HPA,Patel2024HVPS}. \\
        \hline
        ICRH RF-combiner term & $A_{\mathrm{comb}}=\frac{2.4\times0.95}{3}\,P_{\mathrm{aux}}=0.76\,P_{\mathrm{aux}}$ & Anand et al. report a fixed broadside-stripline RF combiner with overall dimensions \SI{2400}{mm} by \SI{950}{mm} by \SI{600}{mm} for the ITER ICRH source \cite{Anand2025Combiner}. \\
        \hline
        Combined ICRH block & $A_{\mathrm{ICRH}}=A_{\mathrm{HVPS}}+A_{\mathrm{amp}}+A_{\mathrm{comb}}=7.29\,P_{\mathrm{aux}}$ & Summed standalone RF electrical-conversion package; launcher hardware and transmission-line routing are excluded. \\
        \hline
    \end{tabular}
    }
\end{table}

\begin{table}[H]
    \centering
    \caption{Cryoplant scaling law.}
    \label{tab:appendix_facility_cryo}
    \footnotesize
    \setlength{\tabcolsep}{4pt}
    \renewcommand{\arraystretch}{1.06}
    \resizebox{\textwidth}{!}{%
    \begin{tabular}{>{\raggedright\arraybackslash}p{3.1cm} >{\raggedright\arraybackslash}p{4.2cm} >{\raggedright\arraybackslash}p{8.0cm}}
        \hline
        Item & Scaling law & Source \\
        \hline
        Cryoplant block & $A_{\mathrm{cryo}}=(3\times5.5)\,Q_{20\mathrm{K}}=16.5\,Q_{20\mathrm{K}}$ & Absolut System Cryo-Cool 1000 product page: \SI{1}{kW}@\SI{20}{K}, dimensions \SI{3}{m} by \SI{5.5}{m}, and mass about \SI{6}{t}, giving \SI{16.5}{m^2/kW} at \SI{20}{K} \cite{AbsolutCryoCool1000}. \\
        \hline
    \end{tabular}
    }
\end{table}

\begin{table}[H]
    \centering
    \caption{Magnet-power and storage scaling laws.}
    \label{tab:appendix_facility_electrical_storage}
    \footnotesize
    \setlength{\tabcolsep}{4pt}
    \renewcommand{\arraystretch}{1.06}
    \resizebox{\textwidth}{!}{%
    \begin{tabular}{>{\raggedright\arraybackslash}p{3.1cm} >{\raggedright\arraybackslash}p{4.2cm} >{\raggedright\arraybackslash}p{8.0cm}}
        \hline
        Item & Scaling law & Source \\
        \hline
        Magnet power supplies & $A_{\mathrm{mag}}=1.5\,P_{\mathrm{inst}}$ & Heron Link converter cabinet: \SI{1.5}{m} by \SI{5}{m} footprint at \SI{5}{MW}, giving \SI{1.5}{m^2/MW} for packaged medium-voltage power electronics \cite{HeronLink2025}. \\
        \hline
        Battery storage & $A_{\mathrm{bat}}=3.47\,E_{\mathrm{bat}}$ & Tesla Megapack package basis, corresponding to \SI{3.47}{m^2/MWh} for installed battery storage \cite{TeslaMegapack2026,NRELATB2025Battery}. \\
        \hline
        Capacitor storage & $A_{\mathrm{cap}}=73.3\,E_{\mathrm{cap}}$ & Eaton XLHV package basis, corresponding to \SI{73.3}{m^2/MWh} for installed capacitor storage \cite{EatonXLHV2023,EatonSupercapacitorGuidelines}. \\
        \hline
    \end{tabular}
    }
\end{table}

\FloatBarrier

%% file: appendix_direct_capex.tex
\clearpage
\section{Direct Overnight Capital Assumption Basis}
\label{app:direct_capex_assumptions}

Table~\ref{tab:appendix_direct_capex_basis} collects the unit-cost anchors used in the direct overnight-capital tables in Section~\ref{sec:direct_capex}.

\begin{table}[H]
    \centering
    \caption{Unit-cost basis used in Tables~\ref{tab:yinsen_direct_capex_core} and~\ref{tab:yinsen_direct_capex_aux}.}
    \label{tab:appendix_direct_capex_basis}
    \footnotesize
    \setlength{\tabcolsep}{3pt}
    \renewcommand{\arraystretch}{1.08}
    \resizebox{\textwidth}{!}{%
    \begin{tabular}{>{\raggedright\arraybackslash}p{2.4cm} >{\raggedright\arraybackslash}p{3.0cm} >{\raggedright\arraybackslash}p{10.2cm}}
        \hline
        Item & Unit cost or basis & Source \\
        \hline
        Tungsten & \SI{29000}{USD/t} & Araiinejad and Shirvan, Table~4 \cite{Araiinejad2025AppliedEnergy}. \\
        \hline
        V--4Cr--4Ti & \SI{37000}{USD/t} & Araiinejad and Shirvan, Table~4 \cite{Araiinejad2025AppliedEnergy}. \\
        \hline
        FLiBe & \SI{154000}{USD/t} & Araiinejad and Shirvan, Table~4 \cite{Araiinejad2025AppliedEnergy}. \\
        \hline
        Hastelloy N & \SI{56800}{USD/t} & AM Material reference-price table: Hastelloy N at \SI{56.8}{USD/kg} \cite{AMMaterialReferencePrice}. \\
        \hline
        Tungsten carbide & \SI{52400}{USD/t} & AM Material reference-price table: irregular tungsten carbide at \SI{52.4}{USD/kg} \cite{AMMaterialReferencePrice}. \\
        \hline
        Tungsten boride & \SI{62500}{USD/t} & Chinatungsten News Center lists W$_2$B$_5$ at \SI{250}{USD/kg} EXW Xiamen, China for a \SI{100}{kg} order; Section~\ref{sec:direct_capex} assumes the much larger shielding purchase pays 25\% of that quoted cost, giving \SI{62.5}{USD/kg} \cite{CTIAW2B52025}. \\
        \hline
        Stainless steel & \SI{10000}{USD/t} & Araiinejad and Shirvan, Table~5 \cite{Araiinejad2025AppliedEnergy}. \\
        \hline
        Concrete & \SI{112}{USD/t} & Allegany County 2025 ready-mix bid: \SI{205}{USD/yd^3} delivered, converted using \SI{2.4}{t/m^3} \cite{AlleganyReadyMix2025}. \\
        \hline
        REBCO tape, FOAK & \SI{20}{USD/m} & Quote from vendor. \\
        \hline
        REBCO tape, NOAK & \SI{10}{USD/m} & Quote from vendor. \\
        \hline
        Heat extraction and conversion & \SI{1330}{USD/kWe} & NREL concentrated-solar molten-salt power tower with sCO$_2$ cycle, Table~4 \cite{Augustine2023CSP}. \\
        \hline
        Tritium plant & \SI{32000000}{USD/kg} & ITER tritium-plant cost of \SI{96}{MUSD} scaled by \SI{3}{kg} maximum inventory \cite{Araiinejad2025AppliedEnergy}. \\
        \hline
        Vacuum pumping & \SI{14000}{USD/pump} & Doubled KSTAR vacuum pump count with \SI{14000}{USD/pump} from Araiinejad and Shirvan \cite{Araiinejad2025AppliedEnergy}. \\
        \hline
        ICRH & \SI{4.15}{USD/W} & PyFECONS average ICRF-heating cost basis \cite{WoodruffScientificPyFECONS2026}. \\
        \hline
        Cryoplant & \SI{322000}{USD/(kW@20K)} & RDK-500B2 quote normalized to \SI{45}{W@20K} \cite{LabbaseRDK500B2Quote,FedH10USDCNY2026}. \\
        \hline
        Magnet power supplies & \SI{86.2}{USD/kVA} & Scaling from a \SI{3.3}{kV} SiC-based converter technoeconomics study \cite{Zheng2022SST}. \\
        \hline
        Battery storage & \SI{360000}{USD/MWh} & Megapack / NREL basis \cite{TeslaMegapack2026,NRELATB2025Battery}. \\
        \hline
        Capacitor storage & \SI{5136000}{USD/MWh} & Eaton XLHV basis \cite{EatonXLHV2023,EatonSupercapacitorGuidelines}. \\
        \hline
    \end{tabular}
    }
\end{table}

%% file: appendix_shipping.tex
\section{Shipping Economics Assumption Basis}
\label{app:shipping_assumptions}

This appendix states the reduced shipping-economics basis used for the marine fusion CAPEX break-even study. It is intended as a comparative LCOT framework rather than as a full commercial ship-cost prediction.

\subsection{LCOT framework and symbols}

The recreated containership model follows the LCOT structure of Locatelli \emph{et al.} \cite{Locatelli2026}. The total levelized cost of transport is written as
\begin{equation}
\mathrm{LCOT} = w + f + c,
\end{equation}
where $w$ is the discounted variable-OPEX contribution, $f$ is the discounted fixed-OPEX contribution, and $c$ is the CAPEX contribution divided by discounted transport output. With the same notation as the source paper,
\begin{equation}
\begin{aligned}
w &= \frac{\sum_{t=0}^{T-1} w_t(1+r)^{-t}}{L},\\
f &= \frac{\sum_{t=1}^{T} F_t(1+r)^{-t}}{L(k)},\\
c &= \frac{v}{L(k)},
\end{aligned}
\end{equation}
with
\begin{equation}
L = \sum_{t=0}^{T-1}(1+r)^{-t}, \qquad L(k)=L\,k.
\end{equation}
Here $T$ is the economic life, $r$ is the discount rate, $v$ is initial CAPEX, and $k$ is annual non-discounted transport output in TEU$\cdot$nmi per year.

\begin{table*}[!t]
    \centering
    \caption{Symbols used in the shipping LCOT appendix.}
    \label{tab:shipping_symbol_table}
    \scriptsize
    \setlength{\tabcolsep}{4pt}
    \renewcommand{\arraystretch}{1.04}
    \begin{tabular*}{\textwidth}{@{\extracolsep{\fill}}>{\raggedright\arraybackslash}p{1.8cm} >{\raggedright\arraybackslash}p{2.2cm} >{\raggedright\arraybackslash}p{8.8cm}@{}}
        \toprule
        Symbol & Units & Meaning \\
        \midrule
        $k$ & TEU$\cdot$nmi/yr & Annual transport work \\
        $L$ & -- & Discount-factor sum over the economic life \\
        $T$ & yr & Economic life used in the LCOT calculation \\
        $r$ & -- & Discount rate \\
        $w_t$ & USD/(TEU$\cdot$nmi) & Variable-OPEX term in year $t$ \\
        $F_t$ & USD/yr & Fixed-OPEX term in year $t$ \\
        $v$ & USD & Initial CAPEX \\
        $N_{\mathrm{trip}}$ & trips/yr & One-way trips completed per year \\
        $d_{\mathrm{sea}}$ & d/yr & Annual steaming days \\
        $d_{\mathrm{port}}$ & d/yr & Annual port days \\
        $P_{\mathrm{shaft}}$ & MW & Mechanical shaft power in the propulsion baseline \\
        $P_{\mathrm{elec,sea}}$ & MWe & Net electric power required during steaming in the fusion-electric reinterpretation \\
        $P_{\mathrm{avg}}$ & MWe & Calendar-year average net electric delivery \\
        \bottomrule
    \end{tabular*}
\end{table*}

\subsection{Mission model and annual transport work}

The reference ship remains the Locatelli baseline: a \num{12000} TEU Neo-Panamax containership, 70\% space utilization, and 210 container moves per hour at port \cite{Locatelli2026}. For the fusion case, however, two operational variables are adjusted to reflect a more plausible first-of-a-kind deployment. The representative trip length is reduced from \SI{8174.8}{nmi} to \SI{2500}{nmi}, corresponding to a dedicated transatlantic liner service operating among a small set of certified home ports. The annual time utilization is reduced from 98.4\% to 70\% to represent longer commissioning periods, more frequent blanket and tritium-system inspections, and longer drydock cycles than would be acceptable for a mature propulsion technology. In this shipping context, utilization should be read more like plant availability, i.e. the fraction of days per year in operation, rather than as capacity factor or throughput. Let $N_{\mathrm{TEU}}$ be nominal ship capacity, $u_{\mathrm{space}}$ the space-utilization factor, $u_{\mathrm{time}}$ the time-utilization factor, $\ell_{\mathrm{trip}}$ the average trip length, $v$ the ship speed, and $p_{\mathrm{port}}$ the port productivity. The sea time per one-way trip is
\begin{equation}
 t_{\mathrm{sea}} = \frac{\ell_{\mathrm{trip}}}{24v},
\end{equation}
and the port time per one-way trip is
\begin{equation}
 t_{\mathrm{port}} = \frac{2N_{\mathrm{TEU}}u_{\mathrm{space}}}{24p_{\mathrm{port}}},
\end{equation}
because the ship is taken to unload and reload the utilized TEU inventory at each stop. The annual trip count is then
\begin{equation}
N_{\mathrm{trip}} = \frac{365u_{\mathrm{time}}}{t_{\mathrm{sea}} + t_{\mathrm{port}}},
\end{equation}
so the annual transport work becomes
\begin{equation}
 k = N_{\mathrm{trip}}N_{\mathrm{TEU}}u_{\mathrm{space}}\ell_{\mathrm{trip}}.
\end{equation}
This annual transport work is the denominator that favors capital-intensive propulsion at higher speed.

\subsection{From propulsion duty to electric duty}

The shipping framework is then reinterpreted as a fusion-electric drivetrain. The paper's VLSFO fuel-consumption regression is used as a proxy for required propulsive duty,
\begin{equation}
\dot m_{\mathrm{fuel}} = 0.000074\,N_{\mathrm{TEU}}^{0.635}v^{2.883}
\end{equation}
in tons per day. That fuel flow is converted to thermal power and then to shaft power with a representative 50\% large-diesel efficiency, used here only to reinterpret the fuel-burn regression as mechanical demand.

The fusion-electric sea-load case is then
\begin{equation}
P_{\mathrm{elec,sea}} = \frac{P_{\mathrm{shaft}}}{\eta_{\mathrm{drive}}} + P_{\mathrm{hotel}},
\end{equation}
with $\eta_{\mathrm{drive}}=0.95$ and $P_{\mathrm{hotel}}=\SI{3}{MWe}$. During port stays and the drydock / idle remainder of the year, shore power is used instead of reactor power. The calendar-year average electric demand is therefore
\begin{equation}
P_{\mathrm{avg}} = \frac{P_{\mathrm{elec,sea}}d_{\mathrm{sea}}}{365}.
\end{equation}
In other words, cold ironing is assumed throughout port operations: the hotel load is supplied by shore power rather than by the reactor while the vessel is berthed. This gives the electric-demand basis quoted in the main text.

\subsection{Fusion-case assumptions}

The fusion case leaves the LCOT framework intact and replaces only the propulsion block. Startup tritium is treated as a one-time CAPEX-like expense,
\begin{equation}
 v_{\mathrm{fusion}} = v_{\mathrm{engine}} + m_{\mathrm{T,start}}c_{\mathrm{T}},
\end{equation}
where $m_{\mathrm{T,start}}=\SI{55.4}{g}$ is taken from the Yinsen fuel-cycle baseline used elsewhere in the paper and $c_{\mathrm{T}}=\SI{30000}{USD/g}$ is the nominal tritium price assumption. Annual tritium make-up is represented as an additional annual fuel-like cost term, but in the present baseline it is set conservatively to zero even though the modeled breeding ratio suggests a surplus. This keeps the break-even result from depending on monetization of surplus tritium.

The reactor fixed-OPEX adder is initially set equal to the fission case used by Locatelli, \SI{26082}{USD/day}. This is only an alignment choice so that the first comparison isolates the reactor-engine CAPEX threshold. The same applies to the 9.5\% learning-rate assumption carried over from the fission case. Annual tritium make-up is conservatively set to zero despite the modeled breeding surplus, which could in practice provide an additional revenue stream.

\subsection{Break-even method and limitations}

The key output is not a single absolute LCOT value for fusion, but the reactor-engine CAPEX at which fusion and a competing propulsion option have equal LCOT. That threshold is obtained by sweeping fusion-engine CAPEX over the chosen grid and solving
\begin{equation}
\mathrm{LCOT}_{\mathrm{fusion}}\left(v_{\mathrm{fusion}}^*\right) = \mathrm{LCOT}_{\mathrm{competitor}}
\end{equation}
by linear interpolation between the two neighboring grid points that bracket the crossing. This break-even use of the recreated model is more robust than quoting a single absolute fusion LCOT because the residual combustion-case offset matters less at the crossing threshold.

Three limits remain important. The present land-based Yinsen capital estimate is not a marine-engine CAPEX, shipboard shielding and integration are outside the current comparison, and the strongest fusion-economic headroom still appears in the high-speed, high-carbon-price cases.